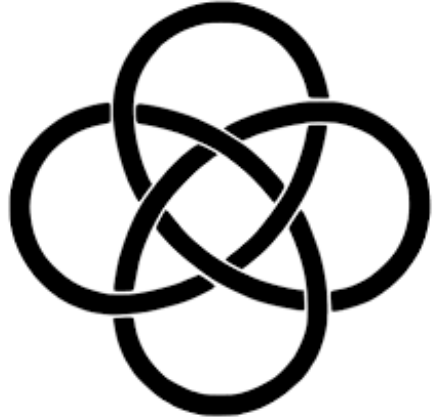

**Inter-University Centre for Astronomy and Astrophysics Post-Bag 4, Ganeshkhind, Pune 411007 MH, India**

# Heating and dynamics of the solar atmosphere

**Candidate: Vishal Upendran**
**Supervisor: Prof. Durgesh Tripathi**

A thesis to be presented for the degree of
Doctor of Philosophy
to
Jawaharlal Nehru University, New Delhi, India

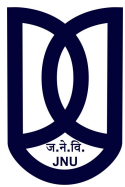

# Declaration from Candidate

I hereby declare that the material present in this thesis is based on work done at the Inter-University Centre for Astronomy and Astrophysics, Pune. This thesis is original and has not been submitted previously for any academic cause. The work of other researchers utilized by us has been properly cited.

**Vishal Upendran**

Inter-University Centre for Astronomy and Astrophysics, Pune, MH - 411007

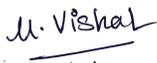

Signature: _______________________________

Date: _____26/12/2022_____________________

# Declaration from Supervisor

This is to certify that this thesis titled 'Heating and dynamics of the solar atmo-sphere' is based on work done by Mr. Vishal Upendran at the Inter-University Cen-tre for Astronomy and Astrophysics, Pune under my supervision. To the best of my knowledge, this thesis is original and has not been submitted fo any academic cause.

**Supervisor: Prof. Durgesh Tripathi**

Inter-University Centre for Astronomy and Astrophysics, Pune, MH - 411007

Signature: 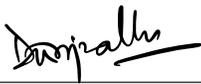 ________________________

Date: _______ 26/12/2022 ________________

# Declaration from the Head of the Institute

This is to certify that this thesis titled 'Heating and dynamics of the solar atmosphere' is based on work done by Mr. Vishal Upendran at the Inter-University Centre for Astronomy and Astrophysics, Pune. To the best of my knowledge, this thesis is original and has not been submitted previously for any academic cause.

**Prof. Somak Raychaudhury**
Inter-University Centre for Astronomy and Astrophysics, Pune, MH - 411007

Signature: _Soma Raychaudh_
Date: _26/12/2022_

SOMAK RAYCHAUDHURY
Director
Inter-University Centre for
Astronomy and Astrophysics
Pune - 411 007.

# Acknowledgments

iṭippārai illāta ēmarā mannaṉ        A king unguarded with reproving counsel
keṭuppā rilāṉuṅ keṭum              needs no foes to come to grief.

-Thirukkural 448 (*On seeking the aid of the great*)

---

Pursuing a Ph.D. is a long, meandering journey – akin to a river flowing from the mountains and finally merging with the ocean. Such a long journey is only possible because of the contribution of numerous other rivers that help maintain a sufficient flow through the journey. And so, I want to thank and acknowledge people who have contributed to this journey and made it a wonderful experience.

First and foremost, I thank my guide and mentor, Prof. Durgesh Tripathi. From academic to non-academic matters, Durgesh has been a constant source of guidance and support. He gave me enough freedom and room for exploration and constantly encouraged me to take up new initiatives like applying for grants and internships. At a personal level, he and his wife, Anupama ma'am, have been a great source of comfort and advice. A general chat with Anupama ma'am over the excellent tea she makes always leaves me with positive vibes and a vastly different perspective on life. Their kids, Jaya and Jai, have been a great source of fun and joy to all of us, especially in the midst of grumbling older people complaining all day!

Coming from a biomedical design engineering background, my journey to astrophysics would not have been possible without three important people. Dr. Mark Cheung has been a great source of knowledge and inspiration in this journey by consistently offering academic/non-academic advice, encouraging me to keep gaining a better understanding of ML/DL/maths, numerical methods alongside astrophysics, and offering critical comments on my work. Prof. Shravan Hanasoge, my first solar physics mentor, constantly encouraged me by putting my work in a global context and motivated me to pursue solar physics by making me *aim high* in life. My master's advisor at IIT-M, Prof. Ganapathy Krishnamurthi (GK), opened up the path to TIFR and solar physics. His infectious enthusiasm toward ML/DL led me to pursue these topics on my own.

I was fortunate at IIT-M to find mentors who encouraged me to explore engineering, pursue physics, and switch to physics. I am very much indebted to Prof. L. Sriramkumar for giving me qualified encouragement while evaluating my strengths/ weaknesses. Similarly, Prof. Suresh Govindarajan was our mentor at the Astronomy club *Horizon*, which played a big role in propelling me into physics, has been another great source of optimism. I thank Prof. Srikanth Vedantam, who created a platform for us to dream big and explore very novel ideas in engineering.

At IUCAA, I first thank my RAC members, Prof. Nishant Singh and Prof. Gulab Dewangan, for providing timely, critical feedback and encouragement. My interactions with Nishant have been also been non-academic due to our love for vegetarian food, freshly brewed filter coffee, and Indian classical music. Similarly, I have had great conversations and good food from Prof. Raghunathan Srianand, whose advice to me in my first year on *the science question driving techniques and not the other way round* has been a guiding principle till now. Numerous science discussions on different topics have given me new perspectives, especially on MHD with Prof. Kandaswamy Subramanian and on ML/DL/computation with Prof. Ajit Kembhavi, Prof. Dipankar Bhattacharya, Prof. Somak Raychaudhury, Kaustubh Waghmere, and Kaushal Sharma.

I am eternally indebted to the late Prof. Thanu Padmanabhan (Paddy) for shaping a major aspect of my thought process. Every sentence from him was an outpouring of nectar – from his course on Quantum mechanics to his books on theoretical astrophysics. Paddy sir left a lasting impact on me through his books on mathematics/science in ancient India and his profound realizations on *The Bhagavad Gita* and *The Upanishads*.

I thank the solar group members for the great academic and non-academic discussions: Abhishek, Megha, Aishawnnya, Sreejith, Sargam, Soumya, Nived, and Biswanath. I also thank Kajal, my master's student, whose motivation pushed me to work *plus ultra*!

No matter how much ever we try to be oblations in this grand sacrifice of science, we are humans in the end. I am truly privileged to have interacted with some of the most amazing people in my peer group. I list them alphabetically, for I find myself unable to weigh this friendship in any measure. My heartfelt gratitude to Anuj Mishra, who *taught me to be less critical of people and give up lethargy to take up travel*; Bhaskar Arya, who *taught me the importance of being cynical in different aspects of life*; Kanchan Soni who *taught me to more empathetic towards others, be mindful of their journey and challenges, and not measure everyone by the same yardstick I use for myself*; and Parisee Shirke who *showed me the importance and impact of a high level of order and organization*. I am very thankful to Prakash Tripathi for showing how simple life inherently is, and for *chilling* me down from high-stress levels. I also thank Yash Bhargava for great conversations and interactions on cultured topics ranging from geek-level coding to anything in science, anime fan theories, and science popularization. I thank all my batchmates and juniors for great interactions and outings and for being the *food tasters and completers* whenever I cook something grand. Kudos to you, folks!

I thank the Nalanda/IUCAA support and reception staff, the canteen staff, and the sysadmin team for smoothly getting things we take for granted done. I particularly thank Mr. Senith Samuel for providing timely help during a grave medical emergency, Mrs. Deepika Susainathan for the smooth handling of international travel logistics, and Mr. Santosh Jagade for always being available on call for fixing issues ranging from the IUCAA network disruption to Zoom meeting shenanigans.

I thank the IUCAA Scipop team, especially Atharva Pathak, for their abundant enthusiasm in performing science outreach and for giving me ample opportunities to perform them. I also thank the enthusiastic team of CosmicVarta, where we have been making cutting-edge astrophysics research accessible to the general public.

Growth through the Ph.D. occurs also through the accretion of varied topics to understand the world, which is possible only in a world with knowledge made free in a qualified manner. I am indebted to our weekly discussions on *The Bhagavad Gita* and Indic philosophy, and thank Saket, Guru, Anusha, Srini, Atul, and Lekha, for making my life much more equanimous and *sattvik*. Similarly, many friends have stuck with me all through this journey, supporting me all the way through – Rakshith, Praneeth, Pawan, Sunil, and Lakshitha.

My family (close and extended) has been a pillar of support all my life. My mother has been a source of inspiration and resilience in adversity. Like the glacier which feeds the river, she has provided me with a passion for science that stems from *her own passion* for science. Her support and encouragement, especially while I moved to science from engineering, were a blessing and privilege for me. As a famous song goes: *"Even if I take numerous births and work hard in all of them, will I be able to repay the debt that I've incurred to you?"*. My little sister Vaishnavi has been the source of fun and motivation for all of us, especially with jokes poorer than mine, wisdom beyond her age, and a more nuanced worldview than many seniors. I have generated wonderful visualization of scientific results from her creative suggestions. I am also truly privileged to have my maternal grandparents always leading the way in front of me, being beacons of ideals in a cruel world.

This thesis would not have been possible if not for open source data policies of different instruments. I acknowledge the use of data from IRIS, AIA, HMI, XSM from Chandrayaan - 2, NASA/GSFC's Space Physics Data Facility's OMNIWeb service, and OMNI data. Finally, I would like to acknowledge my collaborators – Prof. Bhargav Vaidya, Siddha Ganju, Panos Tigas, Bashi Ferdousi, Mithun N.P.S, Téo Bloch, Santosh Vadawale, Asti Bhatt, Ryan McGranaghan, and Yarin Gal for generating great scientific discussions, and helping make the land inundated by this river of thesis fertile!

कृष्णं वंदे जगद्गुरुम्
*Dedicated to teachers who inspire their students and encourage them unconditionally.*

# Abstract

The solar atmosphere shows anomalous variation in temperature, starting from the 5500 K photosphere to the million-degree Kelvin corona. The corona itself expands into the interstellar medium as the free streaming solar wind, which modulates and impacts the near-Earth space weather. The precise source regions of different structures in the solar wind, their formation height, and the heating of the solar atmosphere are inextricably linked and unsolved problems in astrophysics. Observations suggest correlations between Coronal holes (CHs), which are cool, intensity deficit structures in the solar corona, with structures in the solar wind. Observations also suggest the local plasma heating in the corona through power-law distributed impulsive events. In this thesis, we use narrowband photometric, spectroscopic, and disc-integrated emission of the solar atmosphere ranging from Near Ultraviolet to X-rays along with in-situ solar wind measurements to understand (i). the source regions of the solar wind, (ii). the underlying mechanism of solar coronal heating, and (iii). the differentiation in dynamics of CHs with the background Quiet Sun (QS) regions, which do not show any significant signature of the solar wind. We leverage machine learning and numerical modeling tools to develop solar wind forecasting codes using interpretable AI, inversion codes to infer the properties of impulsive events and to understand the differences in the thermodynamics of CHs and QS regions. We finally present a unified scenario of solar wind emergence and heating in the solar atmosphere and discuss the implications of inferences from this thesis.

# List of publications

## Included in thesis

1. **Vishal Upendran**, Mark Cheung, Shravan Hanasoge, Ganapathy Krishnamurthi. 2020. Solar wind prediction using deep learning. Space Weather, 18, e2020SW002478. `https://doi.org/10.1029/2020SW002478` .

2. **Vishal Upendran** and Durgesh Tripathi 2021. On the Impulsive Heating of Quiet Solar Corona. ApJ 916 59. `https://iopscience.iop.org/article/10.3847/1538-4357/abf65a#artAbst`.

3. **Vishal Upendran** and Durgesh Tripathi 2021. Properties of the C II 1334 Å line in Coronal Hole and Quiet Sun as Observed by IRIS. ApJ 922 112. `https://iopscience.iop.org/article/10.3847/1538-4357/ac2575`.

4. **Vishal Upendran** and Durgesh Tripathi 2022. On the formation of solar wind & switchbacks, and quiet Sun heating. ApJ 926 138. `https://iopscience.iop.org/article/10.3847/1538-4357/ac3d88`

5. **Vishal Upendran**, Durgesh Tripathi, Mithun N.P.S, Santosh Vadawale, Anil Bhardwaj, 2022. Nanoflare Heating of the Solar Corona Observed in X-rays, ApJL 940 L38. `https://iopscience.iop.org/article/10.3847/2041-8213/aca078`

6. **Vishal Upendran**, Durgesh Tripathi, Bhargav Vaidya, Takaaki Yokoyama, Mark Cheung, A 2.5D numerical simulation of interchange reconnection at different heights in the solar atmosphere (in-prep).

## Other publications

1. **Vishal Upendran**, Panagiotis Tigas, Bashi Ferdousi, Téo Bloch, M.C.M Cheung, Siddha Ganju et. al. 2022. Global geomagnetic perturbation forecasting using Deep Learning. Space Weather, 20, e2022SW003045. `https://agupubs.onlinelibrary.wiley.com/doi/10.1029/2022SW003045`

2. **Vishal Upendran**, Durgesh Tripathi, Siddha Ganju, Mark Cheung, Solar wind source region estimation using deep learning (in-prep).

3. Abhishek Rajhans, .., **Vishal Upendran**,... Multi-Stranded Simulations Mimicking FOXSI and AIA Observations : A Single Power-Law Distribution for Transients and Steady Background (in-prep)

# Contents

# List of Figures

# List of Tables

# Abbreviations

**AIA** Atmospheric Imaging Assembly

**SDO** Solar Dynamics Observatory

**EUV** Extreme UltraViolet

**HMI** Helioseismic and Magnetic Imager

**IRIS** Interface Region Imaging Spectrograph

**XSM** Solar X-ray Monitor

**CHs** Coronal Holes

**QS** Quiet Sun

**AR** Active Region

**AI** Artificial Intelligence

**ML** Machine Learning

**DL** Deep Learning

# Chapter 1

# Our friendly neighbourhood star: The Sun

Tat saviṭur vareṇyaṃ                           We meditate on the effulgent
bhargo devasya dhīmahi                              glory of the divine Sun.
dhiyo yo naḥ pracodayāt                    May he inspire our understanding
                    -Rigveda 3.62.10

---

The disc of the Sun, rising from the East and setting in the west, has been a source of awe, inspiration, fear, and adoration to humans across space and time. This sheer awesomeness of the Sun is captured in some measure by the verse quoted above, attributed to a sage named Viśvāmitra Gāthinaḥ. The verse comes from one of the oldest texts known to humanity, being recited, interpreted, and meditated upon from the Bronze age till today. Cultures change and evolve in different ways with time, but the inspiration this burning ball of fire provides remains. Indic philosophy attributes various symbolism to the Sun. Sometimes, it is the modulator of our senses of worldly perception, preventing us from looking inwards and contemplating. At other times, it is the neutral witness within us that "observes" the world and that we must realize to lead an equanimous, sustainable life.

Inherently though, the Sun is a physical object with specific attributes. Cultures across geographies sought to study the heavens and kept generating a better understanding of what it entails – including the Sun. In the past couple of hundred or so years, primarily due to the development of telescopes, we have generated an exponential increase in the understanding of what the Sun is.

The Sun is a gravitating ball of plasma with a diameter of $\approx 1.4 \times 10^6$ km, a mass of $\approx 2 \times 10^{30}$ kg, and at a distance of $\approx 150 \times 10^6$ km from us. It is composed primarily of Hydrogen ($\approx 70\%$), Helium ($\approx 28\%$), and other elements (collec-tively called 'metals'). Due to its high temperature, the gas in the Sun is ionized into plasma. Thus, the Sun is also an excellent plasma physics laboratory – particularly due to its proximity to the Earth.

The Sun is powered by the nuclear reactions that take place in the core of the Sun. These reactions primarily convert Hydrogen to Helium, where the mass deficit gets converted to radiation. Nuclear fusion generates very high-energy $\gamma$-ray photons. These photons, as they try to escape the gravity of the Sun, keep undergoing scattering due to the very small mean free path. Thus, these photons may be trapped for millions of years before they start escaping. This region is called the 'Radiative zone' of the Sun, as shown in Fig. 1.1.

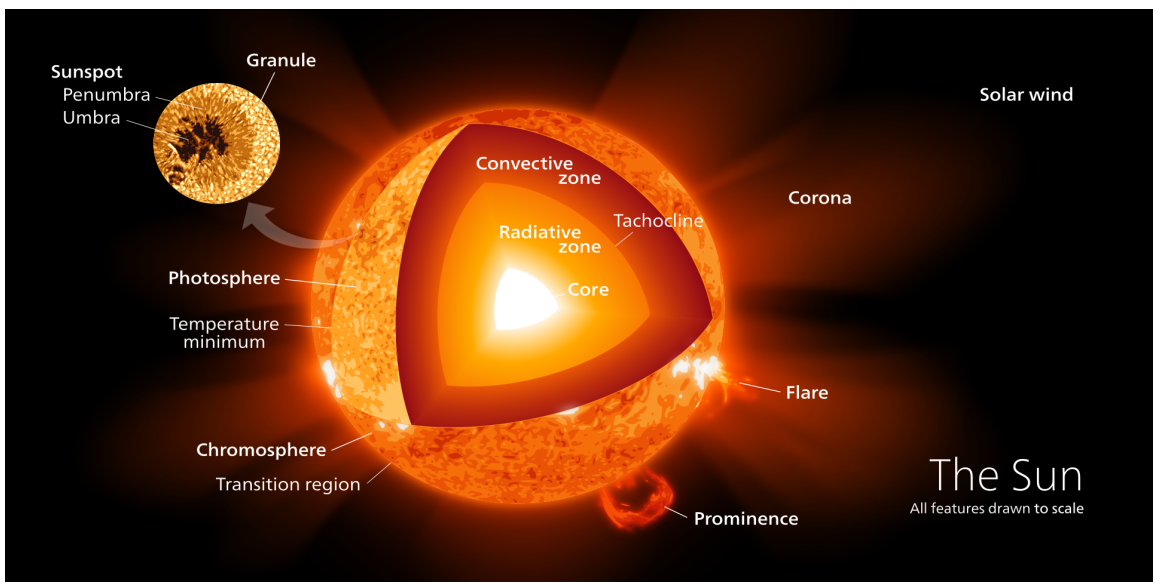

**Figure 1.1**: An artistic rendering of the various layers of the Sun. The solar interior is depicted with the core, radiative, and convective zone with the tachocline. The solar atmosphere starts from the photosphere, consisting of the temperature minimum region, chromosphere, transition region, and the corona, and finally expands into the free streaming solar wind. Various phenomena and features are also depicted in the figure, briefly stated in the text at relevant locations. Art by Kelvinsong - Own work, CC BY-SA 3.0.

As we move outwards from the core, we find that radiation no longer remains the most efficient form of energy transport outwards. Plasma convection starts occurring in the region known as the 'Convection zone' (see Fig. 1.1). In this zone, the gradient of entropy/temperature satisfies the 'Schwarzchild criterion,' rendering the convection of plasma as the dominant energy transport mode. The base of the convection zone experiences a strong shear, for the convective zone marks the start of the Sun's differential rotation. This region, called tachocline, is thought to be the driver behind the activity of the Sun (Charbonneau 2014).

Along with plasma, the magnetic fields are also believed to be transported through to the upper reaches of the Sun. Thus, plasma and magnetic fields start moving outward in this region. Due to its plasma constitution, any small existing magnetic fields are amplified and manifest through the surface. However, the photons still cannot escape – but their mean free paths have increased.

Finally, at some height, the optical-light photons start free streaming: i.e., the photons have a larger mean free path than the length scale of the system. Hence, the photons escape and take about 8 minutes to reach us here on Earth. The height on the Sun from where these optical-light photons escape is called the photosphere or the 'Solar surface.' This is the visible disc of the Sun one sees through a white-light filter on a regular day[1]. An image taken by projecting the Sun onto a sheet of paper is shown in Fig. 1.2 [2] From here springs forth the solar atmosphere – the subject of study in this thesis!

## 1.1  Solar atmosphere

The solar atmosphere is classified into multiple layers – the photosphere, chromosphere, transition region, corona, and the solar wind. This classification is performed based on the characteristic temperature and physical processes dictating the dynamics of the given region. We shall keep referring to the graphic depicted in Fig. 1.1 as a reference for the different layers of the solar atmosphere.

### 1.1.1  Photosphere

We have seen that the photosphere is where optical photons start free streaming. It is $\approx 500$ km thick, at a temperature of $\approx 5500$ K, and exhibits a range of dynamics. Typically, the dynamics are dictated by plasma convection, which regularly brings new plasma and drains the old plasma. This convection forms patterns ($\approx 1$ Mm

---

[1] You must be very careful while seeing the Sun through the naked eye or a white light filter

[2] This image was taken during school-level science outreach conducted as a part of the Young Astronomer's Meeting - 2022 in ARIES, Nainital.

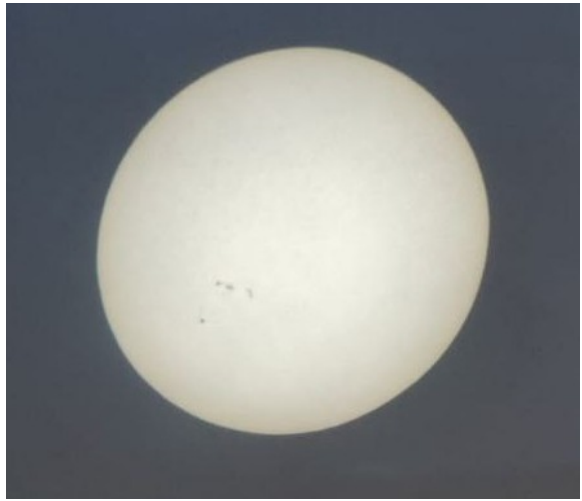

**Figure 1.2**: The photosphere, projected through a telescope onto a sheet of paper. Credit: Srashti Goyal/YAM/ARIES.

in size) called granules, basically 'convection cells' churning out plasma (Title et al. 1989).

The convection brings out hot plasma from the interior in the center of these granules. Like a boiling pot of water, hot upflowing plasma then sinks down along the edges of the granules as it cools down. These edges are called intergranular 'lanes' and appear darker due to lower temperature (Stein & Nordlund 1998).

The magnetic field associated with granulation is not very strong ($\leq 200$ Gauss). However, there do exist regions with very strong magnetic flux densities. These are called sunspots, seen as dark spots on the photosphere. These regions contain a strong magnetic field – of the order of $10^3$ Gauss. The sunspots also exhibit very complex dynamics, resulting in features like faculae (bright spots generally found near sunspots) and light bridges (bright lanes near sunspots), to name a few. A couple of sunspots may be seen in Fig. 1.2.

## 1.1.2 Chromosphere

As we move further upwards, we find the region called chromosphere (*lit.* sphere of color). This wonderful name comes from an equally wonderful observation. During a total solar eclipse where the Moon hides the photosphere, the chromosphere is generally seen as a bright red-colored ring – thus giving it the name. This is shown in Fig. 1.3. The gas temperature here is $\approx 10^4$ K.

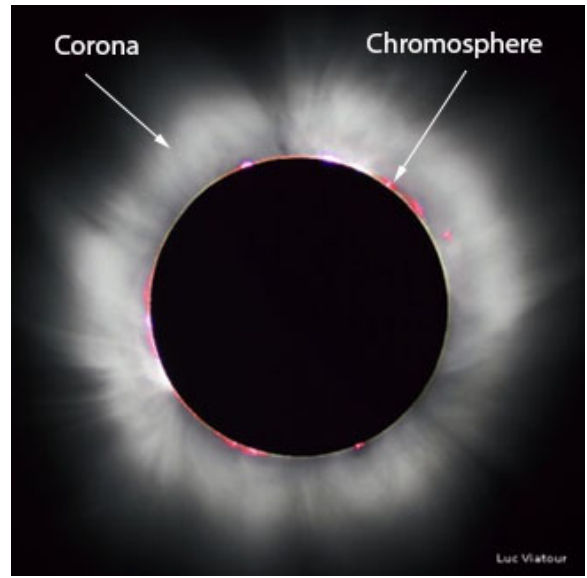

**Figure 1.3**: The solar chromosphere and corona as observed during an eclipse. Image taken by Luc Viatour.

The chromosphere is a region where a lot of spectral lines form – He I 10830 Å, Hα 6562 Å, Ca II H 3968 Å & K 3933 Å, Mg II h 2796 Å & k 2803 Å and C II 1334 Å/ 1335 Å to name a few. The chromospheric emission is primarily in visible, IR, near-UV, and UV. We observe a plethora of structures in this layer like fibrils (thread-like fine structures), filaments (dark streaks of plasma seen over a bright background), prominence (off-limb filaments seen in emission), plage (bright regions accompanying sunspots), spicules (rapidly moving confined plasma), to name a few. The chromosphere is also the region where many ions and neutral species are present and interact with each other. The dynamics here are extremely complicated, with the gas pressure and magnetic field competing for control.

### 1.1.3   Transition region

Further up in the atmosphere is the transition region, where the gas temperature rises to more than $10^5$ K. This is a geometrically very thin layer, spanning just a couple of 100 km in height, within which the temperature difference occurs.

From the transition region onward, the atmosphere becomes optically thin to radiation. This means that the radiation we receive is not coming from just a specific geometric height in the atmosphere. Instead, the radiation is an integral of emission and absorption across the line of sight over which it is measured. The quantity that determines whether the radiation arises from an optically thin/thick atmosphere is called optical depth.

In this region, the magnetic field starts to dominate the dynamics. This is quantified by a term defined as plasma beta:

$$\beta = \frac{\mathsf{P}_{\mathsf{gas}}}{\mathsf{P}_{\mathsf{mag}}}, \tag{1.1}$$

where the denominator corresponds to the magnetic pressure, which is defined (in Gaussian units) as:

$$\mathsf{P}_{\mathsf{mag}} = \frac{\mathsf{B}^2}{8\pi}, \tag{1.2}$$

where B is the magnetic field strength.

The transition region emits primarily in far-UV and Extreme-UV, in lines like the Si IV 1394 Å/ 1403 Å, multiple O IV, S IV lines, and Fe IX 171 Å lines. However, since this region is very dynamic, the optical depth also changes depending on the dynamics, making inference of observations difficult in certain cases.

### 1.1.4 Corona

Even higher up lies the crown of the Sun – the solar corona. The corona can be seen as a diffuse structure extending outwards during a total solar eclipse, seen in Fig. 1.3. The solar corona is home to plasma at $\geq 10^6$ K, $\beta << 1$. This causes the dynamics of the corona to be dominated by the magnetic field dynamics. Furthermore, the very low resistivity in the solar corona causes the magnetic flux in a fluid element to remain constant in time, resulting in the "frozen-flux condition".

The general expectation of temperature trend from a radiating source is a drop in temperature on moving further away from the source. Observation of spectral lines in the corona, however, shows the existence of ions like Fe in high ionization state, for ex: Fe X, Fe XI, Fe XIV to name a few (Swings 1945; Cor 1945). Such a high ionization state is possible only if the source is at a temperature of $10^6$ K or more. Thus, the temperature of the solar atmosphere increases from the photosphere to the chromosphere to the corona, as depicted in Fig. 1.4. Hence, such a steep temperature rise begets the question of a 'heating source' to raise the temperature. This, in a nutshell, is the **coronal heating problem**.

The solar corona is rich in features and appears drastically different from the photosphere. The solar corona may be observed in different wavelengths depending on the different processes that occur at characteristic temperatures and den-sities. The corona is seen in visible light (Fig. 1.3) due to Thompson scattering of photospheric emission by the hot coronal electrons. However, there also exist transitions of specific ions in visible light, for example, Fe X 6375 Å, Fe IV 5303 Å, which give rise to spectral line emission. The EUV and X-ray emissions arise due to radiative de-excitation of specific species – though the X-ray emission may also arise from bremsstrahlung. The corona also has signatures in radio waves that arise due to non-thermal accelerated electrons in coronal loops (see, for e.g. Mondal & Oberoi 2021). Finally, specific transitions of the highly ionized spectral lines also cause emission in infrared. The Fe XIII 10747 Å, for example, is an important coronal diagnostic along with other lines arising from the same ion (Patel et al. 2021). The dynamics of the corona, however, are typically dictated by the magnetic field. Thus, the magnetic field strength and topology strongly affect the morphology of structures that form in the corona. Hence, the characteristics of radiation from the corona are inextricably linked to the underlying structure of the magnetic field.

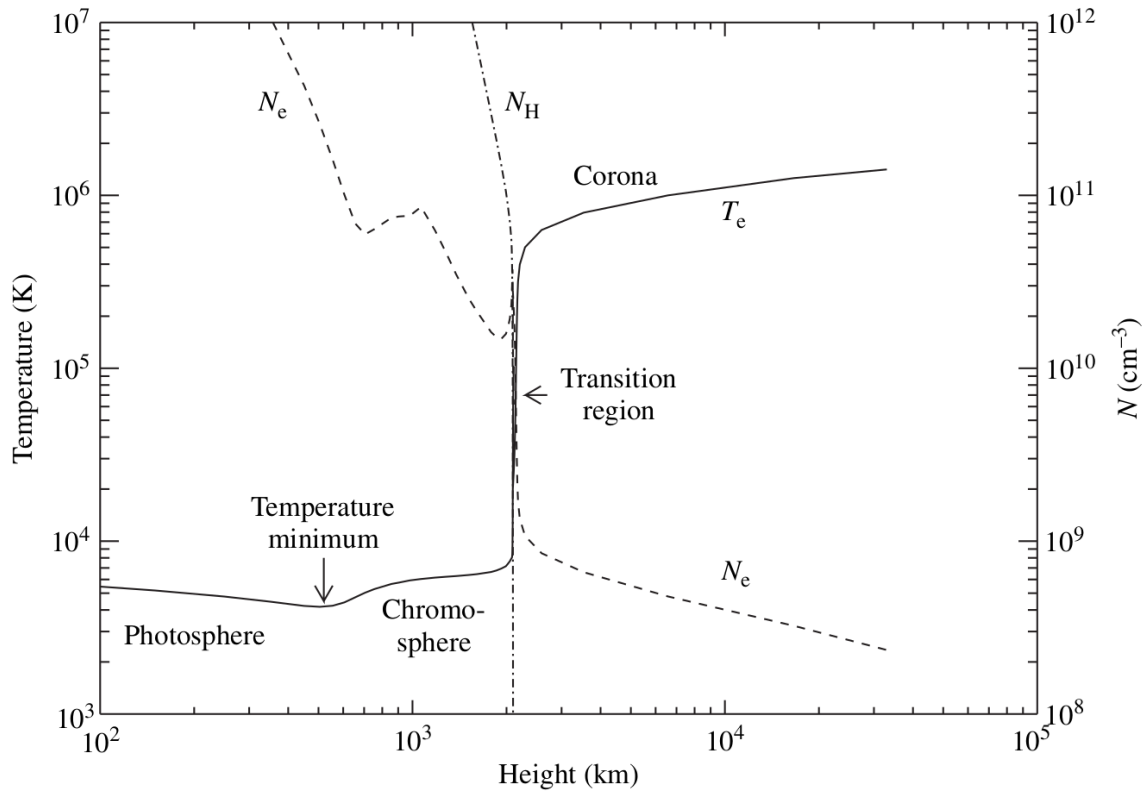

**Figure 1.4**: Variation of temperature (solid curve), electron density (dashed curve), and hydrogen density (dot-dashed curve) with height, as computed using the 1-D model of Vernazza et al. (1981). Figure taken from Phillips et al. (2012).

### 1.1.5 Solar wind

The solar corona keeps expanding outwards until it starts to become free streaming. *It has now become the solar wind, the modulator, and the destroyer of planetary magnetospheres.* The solar wind is just a stream of energetic particles and magnetic field emanating from the Sun. The existence of the solar wind was predicted by Parker (1958) through theoretical considerations based on comet tail observations, for example, by Biermann (1957). It was later proved to be correct by the Soviet probe Luna 1 just a few years later (Zirker 1977; Harvey 2007).

The solar wind is a predominantly radially out-flowing, supersonic motion of plasma. In the solar wind, the gas and magnetic pressure compete for control, with the gas pressure generally winning. However, the solar wind also sweeps with the magnetic field from the Sun and is magnetized plasma.

As it travels outwards, it encounters the magnetic field of different planets and interacts with them. The interaction of the solar wind with Earth's magnetic field results in the modulation of Earth's magnetosphere, ionosphere, and thermosphere. These modulations may be seen in their benign form as the beautiful aurora or destruction of electric grids in their more devastating form. This interaction may, however, be quantified as measurements of perturbations in the geomagnetic field. Overall, the resulting interaction of the solar wind with Earth's magnetosphere results in what is called 'Space weather' (NASA 2017). Thus, from an operational perspective, it becomes imperative to study, characterize, and forecast space weather – and by its extension, its source, the Sun.

## 1.2 The tale of two phenomena

We have seen how varied and dynamic the solar atmosphere is. Various phenomena occur in the corona, of which two, in particular, attract our attention: coronal heating and solar wind emergence.

### 1.2.1 Coronal heating

The coronal heating, as described already in §.1.1.4, asks: "What gives rise to the anomalous temperature of the solar corona?". It has been a long-standing unsolved problem in solar physics for more than 70 years. However, while the precise mech-anism of heat deposition is not well understood, clues may be obtained from observing the large solar flares.

Solar flares are large, dynamic events that occur in the corona – the strongest of which are visible in X-rays but will generally be visible in EUV (Schadee et al. 1983). Solar flares occur when the coronal magnetic field has been stressed enough by motions in the photosphere, resulting in a lot of free energy being held in the configuration (Benz 2008). Such a large amount of free energy causes currents to be generated, leading to the explosive release of this energy as thermal and non-thermal energy as a solar flare. Once the energy is released, the magnetic field tends to go to a more stable configuration, sometimes awaiting more free energy to build up. Typically, flares have a total energy of $\geq 10^{30}$ ergs and last from minutes to hours depending on the wavelength of observation (Benz 2008). Such energy releases are generally scale-free, implying the absence of a specific scale of energy release (Aschwanden 2019). This means a 'flare' may occur over a range of length, time, and energy scales. But do such processes occur at different scales?

It turns out that there is a vast 'zoo' of events that occur in the solar corona, corresponding to different physical scales. Like solar flares, milli- and microflares exist as brightenings in the corona (see, for example Schadee et al. 1983; Chifor et al. 2006; Subramanian et al. 2018; Gupta et al. 2018a). Since solar flares and other such energetic events convert magnetic energy to thermal energy in the corona, one way of generating heating is through ubiquitous impulsive events occurring across the corona. This is called the 'Nanoflare heating paradigm' and was proposed by Prof. E. Parker in 1988 (see Parker 1988a). In this paradigm, the photospheric motions stress the magnetic field, leading to a buildup of free energy in the system. Current sheets may form and dissipate the built-up energy as the field gets tangled. This can occur through both Ohmic heating and magnetic reconnection (Parker 1972). The dominant energy release of the order of $10^{24}$ ergs is needed to maintain the solar corona's temperature.

The nanoflare paradigm, based on reconnection, falls into the 'DC' heating mechanism of the solar corona (Klimchuk 2006a; Parnell & De Moortel 2012; Hansteen et al. 2010). However, the Sun also rings like a bell, and waves are traveling throughout the Sun. Some of these waves may leak into the atmosphere, depositing the wave energy in the corona. Furthermore, various dynamics phenomena occurring in the solar atmosphere may naturally give rise to waves, which may be dissipated into the corona. This wave dissipation typically occurs due to the density gradient with height, leading to localized heating (Alfvén 1947; Osterbrock 1961; Antolin et al. 2008). This paradigm is called the 'AC' heating mechanism (see for example Van Doorsselaere et al. 2020, and references therein). AC heating may also give rise to energy release similar to nanoflares in very short time scales (Antolin et al. 2008). Hence, we shall follow Klimchuk (2006b) and call these small-scale energy releases 'Impulsive events.'

If one were to attempt to constrain the impact of impulsive heating of corona, individual impulsive events would need to be counted and binned with energy to get the total contribution – i.e., counting of individual events, right from flares ($\approx 10^{33}$ ergs) to nanoflares ($\approx 10^{24}$ ergs of energy). Such counting of events results in power-law distribution of these impulsive events, i.e.,

$$\frac{dN}{dW} \propto W^{-\alpha}.$$

If the solar corona were to be sufficiently heated by these impulsive events, there must be more smaller events than larger, resulting in the necessity of $\alpha \geq 2$ (see, e.g. Hudson 1991, for details).

Interestingly, a variety of $\alpha$ was reported in the literature based on the observations made using different instruments sensitive to various energy bands, as shown in Fig. 1.5. Clearly, the Hard X-ray observations show a much steeper slope than the EUV or Soft X-ray (SXR) observations.

There are multiple complications associated with such simple counting statistics. One issue is that plasma emission may be present only in certain energy bands. Hence, an appropriate scaling must be performed to transform the measured luminosity energy to all wavelength integrated energy. Another major issue becomes more prominent as we seek to isolate smaller and smaller events. As events become smaller, our observations may not resolve them enough.

Furthermore, our instrument may even integrate over multiple events as 'one' big event simply because these small events fall in the same pixel or the same time bin! A more technical difficulty is partitioning thermal energy into radiative loss and transport through thermal conduction. Thus, we can determine if impulsive events are viable in heating the corona when such issues may be sufficiently resolved.

### 1.2.2   Formation and acceleration of solar wind

Allied to the above problem of coronal heating is the formation and acceleration of solar wind. We have seen in §. 1.1.5 how Parker (1958) first proposed the existence of the solar wind and observed for the first time by the Soviet spacecraft Luna 1 (Chapter 1 of Harvey 2007). Over the years, many spacecrafts like Vela 3, Voyager, etc., performed observations of plasma parameters in the solar wind.

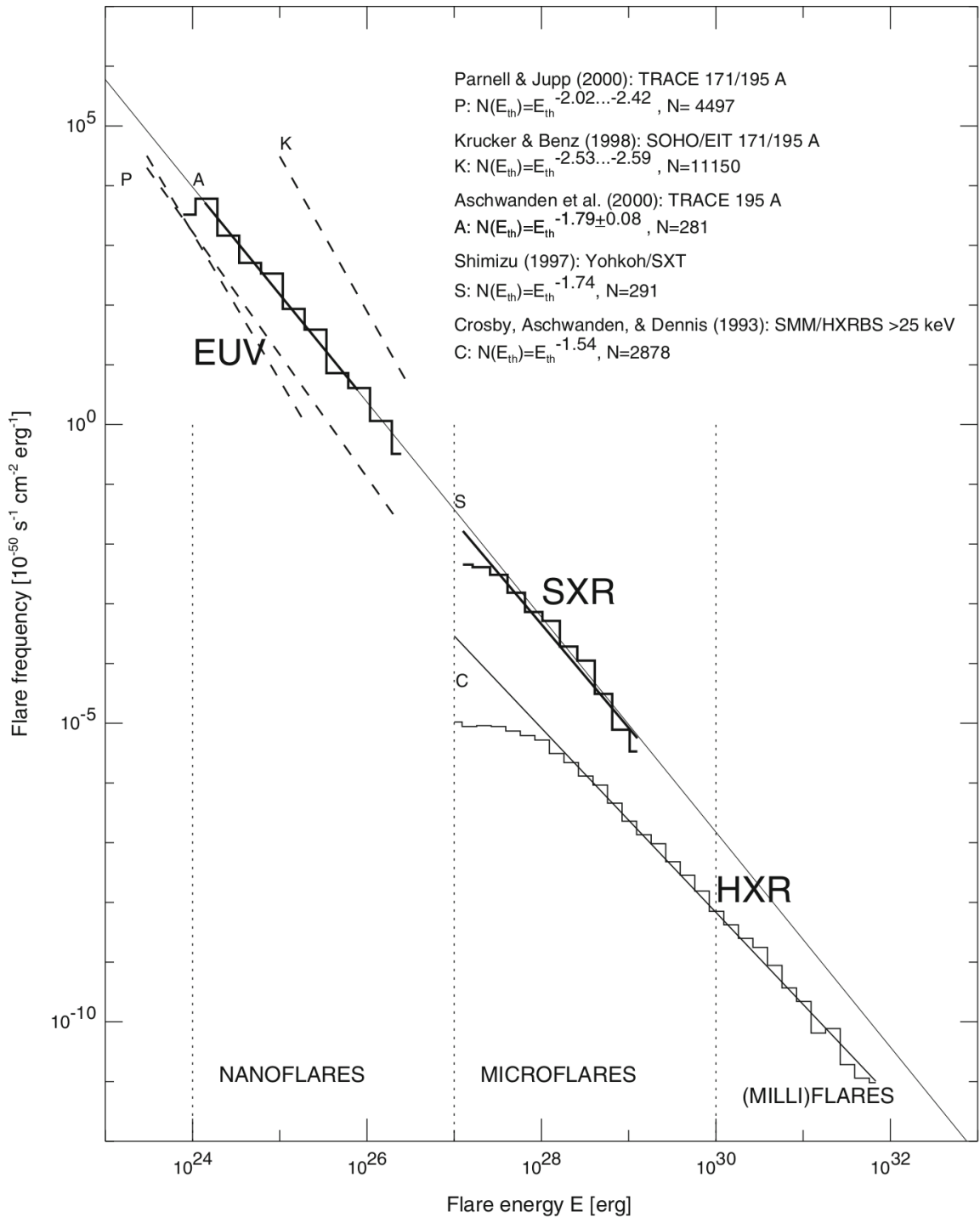

**Figure 1.5**: Collation of measurements of impulsive events from different instruments and temperatures, depicted as a distribution in the energy. Figure taken from Aschwanden (2019).

From then on, attempts have been made to investigate the exact location (on the disc and in height) of the origin and acceleration of solar wind. The earliest correlation of in-situ data with remote sensing observations was found by Krieger et al. (1973), using X-ray images of the corona taken during a sounding rocket mission.

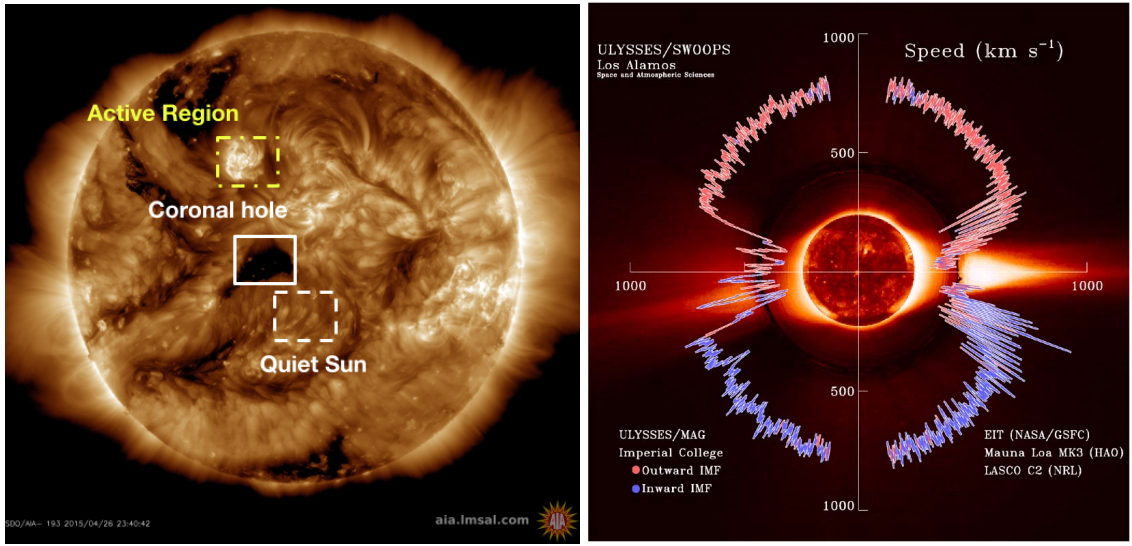

**Figure 1.6**: Left: A representative full disc AIA 193 Å intensity image with the CHs, ARs, and QS depicted with bounding boxes. Right: Solar wind speed (radial axis) measured by Ulysses as a function of heliolatitude over-plotted with the solar coronal images. The color represents the magnetic field measurement at that latitude. Image taken from McComas et al. (1998)

We must get back to the solar corona a bit to understand this observation. In the left panel of Fig. 1.6, we show a typical solar corona image taken in the 193 Å passband of AIA on April 26, 2015. This image shows two interesting, contrasting features on the corona – the dark Coronal Holes (CHs) and the bright Active Regions (ARs). The CHs are regions where the magnetic flux is almost radial and appear dark due to a deficit of emitters. Contrast these with ARs, where the magnetic field lines predominantly form loops. These two regions exist on top of the background – i.e., they appear to be visual increments/decrements of intensity over the background. This background, existing regardless of the presence of CHs or ARs, is called the Quiet Sun (QS).

Krieger et al. (1973) found that the CHs have some form of correspondence with the solar wind measured near Earth. Further developments resulted in a quantification of the influence of polar coronal holes on high-speed streams (Zirker 1977), using data from Skylab. Generally, the magnetic field strength may be assumed to fall off as an inverse square law with distance. This corresponds to the increase in the cross-section of a flux tube with radial distance. However, deviations from this radial expansion lead to super-radial or sub-radial expansion of flux tubes, captured by the flux tube expansion factor ($f_e$) defined as:

$$f_e = \left(\frac{R_\odot}{R_{SS}}\right)^2 \left(\frac{B_r(\odot)}{B_r(SS)}\right),$$

where $R$ is the distance of from the Sun, and $B_r$ is the radial magnetic field strength. Here, $\odot$ corresponds to the solar surface, while $SS$ corresponds to the source surface, the upper boundary from where the magnetic field lines are assumed to become radial. Levine et al. (1977) showed that the solar wind speed is inversely correlated with the flux tube expansion – i.e., the lesser the CH flux tubes expand, the faster the solar wind.

This factor was leveraged by Wang & Sheeley Jr (1990), who further showed that the CH area – a proxy for $f_e^{-1}$, showed a good correlation with the solar wind streams. Thus, the magnetic field topology of the CH showed a strong influence on the observed solar wind. In the 1990s, Ulysses (McComas et al. 1998) performed, for the first time, observations of the Sun from its equatorial to polar regions. It found signatures of two kinds of wind: a fast wind which bellows from the polar regions and the slow wind which coasts from the equatorial regions of the Sun (see the right panel of Fig. 1.6). These two wind modalities do not just have different speeds but also have different compositions, pointing to their origin at different source regions. This was explored by Brooks et al. (2015), who sought to constrain the source regions of slow wind. They thus trace back the slow wind in time while considering only regions showing: (i). open flux, (ii). blueshifts (outflows), and (iii). composition similar to slow wind. They found the edges of ARs to satisfy all three criteria and thus to be potential sources of the slow wind. However, recently Bale et al. (2019) showed that the equatorial CHs are also well correlated with the slow wind measurements. Thus, there is a lack of clarity on the sources of the slow solar wind. However, there is clear evidence of the CHs being the source regions of at least the fast solar wind.

Different regions on the Sun thus have different characteristic solar wind signatures. This suggests that Parker's theory of a radial expansion of the corona into the solar wind is not quite complete. Parker's theory essentially showed that stratified, hot corona in the presence of gravity acts as a de-Laval Nozzle to accelerate plasma to supersonic velocities. However, observations of coronal spectral lines in CHs by SOHO (The Solar and Heliospheric Observatory) indicated that the velocities were much larger than those predicted by Parker's theory (see Domingo et al. 1995). The theory was not enough to account for the differences in the velocities of O VI and H I Ly$\alpha$ (Kohl et al. 1998), for example. Furthermore, Corti et al. (1997)

observed the O VI line in CHs and found the outflow velocity increasing from $\approx 50$ km/s at $1.5 R_\odot$ to $\approx 140$ km/s at $2 R_\odot$. This leads us to two problems: From what height does the solar wind start to form in the solar atmosphere? And what physical processes(es) give rise to this solar wind acceleration?

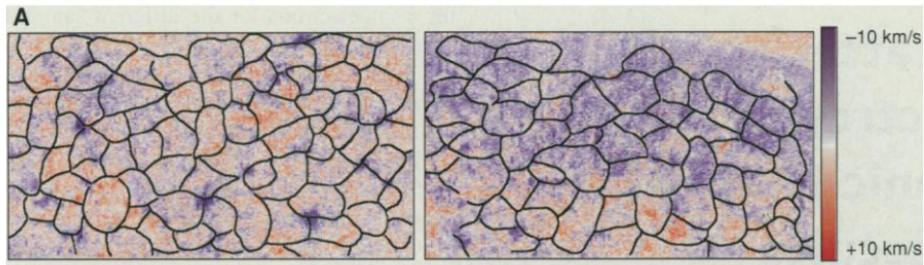

**Figure 1.7**: Chromspheric network, segmented from Si II overlaid (black) on Doppler map from Ne VIII (color). The left figure is from a QS region, and the right is from a polar CH. Figure adapted from Hassler et al. (1999).

SOHO was an extremely fruitful mission in providing major clues to answering these questions. Hassler et al. (1999) studied the Doppler shift of Ne VIII (spectral line in upper transition region) with the chromospheric magnetic network as seen in Si II in a CH region. Since pertinent trends or variations need to be benchmarked against a background, they perform their analysis for the CH and with a nearby QS region. Hassler et al. (1999) found CHs to show excess outflows compared to QS, clearly hinting at the origin of solar wind from the CHs. They also found excess outflow at network boundaries and a clear correlation between network boundaries and outflow (see Fig. 1.7). However, they could not find any clear correlation between Si II intensities and Ne VIII intensities. Hassler et al. (1999) thus present an interesting conjecture: the energy put into the structure is either being utilized for local heating or accelerating the material along open field lines, and hints towards the unification of QS heating and solar wind acceleration. The input energy may be used for QS heating or solar wind acceleration, possibly occurring at similar heights.

Building on these results, Tu et al. (2005) studied the correlations between potential field extrapolated magnetic field inside a CH and Si II intensity, C IV intensity, and Ne VIII Doppler shift as a function of height. They show that these lines form approximately at 4 Mm (Si), 4.5 Mm (C), and 20.6 Mm (Ne), respectively. Tu et al. (2005) find that while the chromospheric lines (Si, C) have a tight correlation of their intensity with the extrapolated magnetic field strength at the formation height, the Ne VIII line showed a similar correlation only for the Doppler shift with the local inclination of the magnetic field. Thus, they found that the solar wind typically comes from open flux 'funnels,' which show a prominent signature much higher in the atmosphere. However, many low-lying loops are present lower in the atmosphere, which give rise to a QS-like intensity structure.

These results give us three critical pieces of information: (i). The solar wind mainly emerges from CHs, (ii). Clear signatures of the solar wind are seen in the upper transition region lines like Ne VIII and (iii). A common underlying mechanism for accelerating the solar wind and heating the corona is possibly at play. The CHs are visibly darker than QS only in the corona or the upper transition region. Lower in the atmosphere, this difference vanishes! Since the solar wind signatures are seen in a spectral line like Ne VIII, we may also ask if similar signatures are seen much lower in height – and, consequently, in much cooler lines.

The average properties of CHs and QS are very similar in the lower atmosphere. However, Tripathi et al. (2021a) perform a comparative study using the Si IV lower transition region line. CH and QS show no visible difference in Si IV line intensity or velocity in an average sense. However, they find intensity and velocity differences for regions with identical magnetic flux densities. Furthermore, they also find the CHs to have subdued intensity for identical magnetic flux density with respect to QS. They then analyzed the blue- and red shifts separately and found the CHs to show excess blueshifts and subdued redshifts w.r.t QS. The intensities and velocities increase with increasing magnetic flux density. Finally, they found the nonthermal width of Si IV to also increase with magnetic flux density, but no significant differences were found between the CHs and QS.

These results showed potential signatures of the solar wind at the lower transition region itself. Similarly, the excess emission-excess blueshift dichotomy manifests when the underlying magnetic flux density is accounted for. This is summarized in Fig. 1.8 and suggests a unified mechanism of QS heating and solar wind formation. Since the differences between QS and CH are markedly seen in higher magnetic flux regions (corresponding to network regions) and not in the lower magnetic flux regions (inter-network regions), the difference arises due to statistics of smaller and bigger loops (Wiegelmann & Solanki 2004). The inter-network region contains short, low-lying loops and is common to CH and QS. However, the longer, higher loops are predominantly in QS rather than the CH, suggesting the difference arises due to a sheer difference in magnetic topology. This is also in line with Hassler et al. (1999).

Tripathi et al. (2021a) explain the observed differences in redshifts due to condensation of plasma from impulsive heating resulting from closed loop-closed loop reconnection and the blueshifts due to open loop-closed loop reconnection result-ing in a cool jet. Thus, they found a signature (albeit not clear) of confined plasma heating in QS and heated plasma escaping in CH, suggesting a unified model of QS heating and solar wind emergence in CHs.

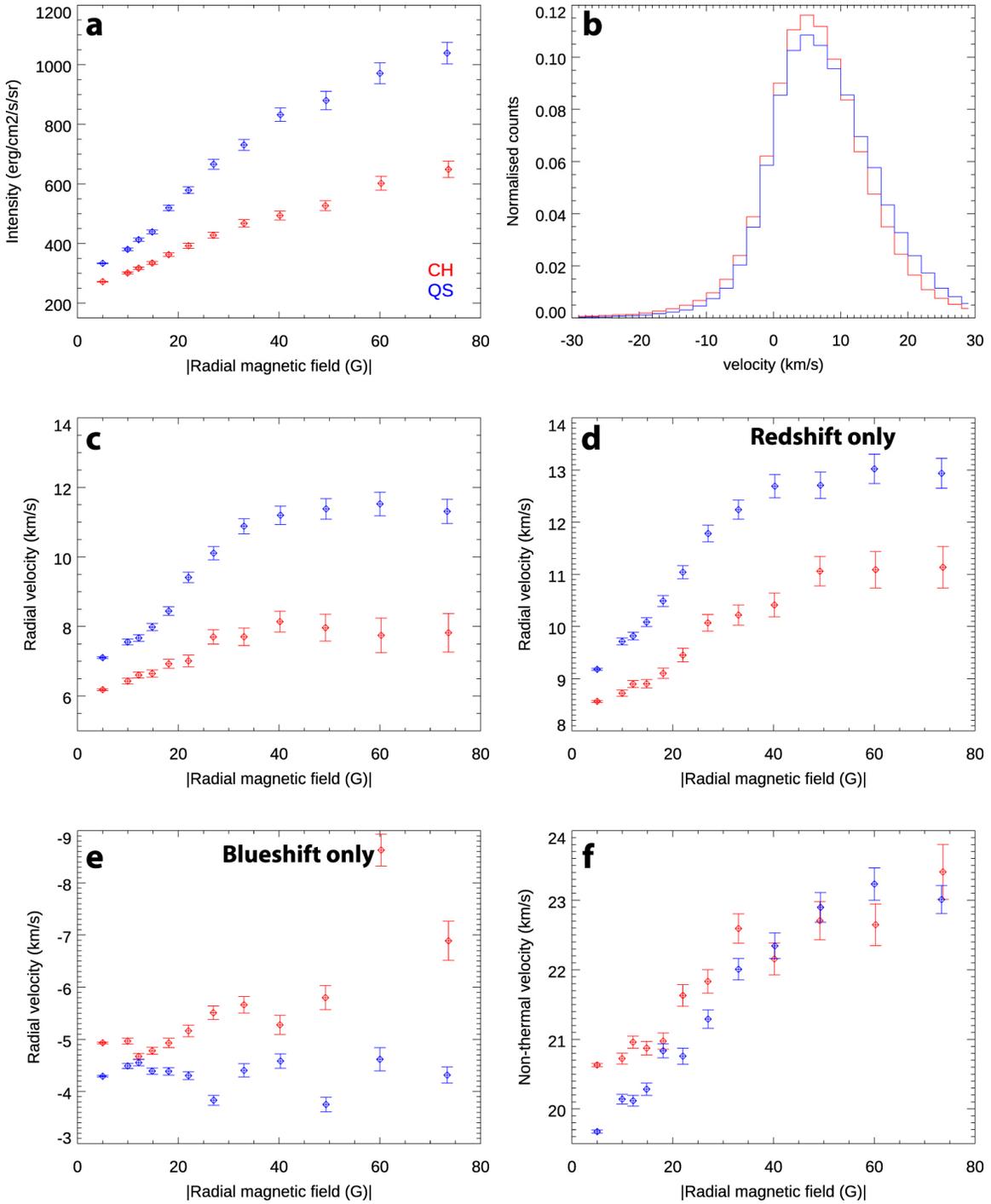

**Figure 1.8**: Per radial absolute B-field bin distribution of QS (blue) and CH (red) data from Si IV: a. Intensity; c. Average velocity; d. Redshift; e. Blueshift; f: Non-thermal width. Figure b shows the distribution of velocity across all the pixels. Image adapted from Tripathi et al. (2021a).

## 1.3 Outline of this thesis

With this backdrop, we seek to answer the following key science questions:

Q1. What are the source regions of the solar wind? Is it possible to forecast solar wind properties given changing conditions in the solar atmosphere?

Q2. What is/are the underlying mechanism(s) of heating up of the solar corona?

Q3. How do the solar wind source regions' dynamics and underlying generation mechanisms compare with potentially non-sources of the solar wind?

Q4. What physical picture gives rise to these physical mechanisms and dynamics?

With these questions in mind, the structure of the remainder of this thesis is described below.

### 1.3.1 Chapter 2: Data and methods

In Ch. 2, we shall discuss three important aspects of our thesis, i.e., observational data, big data methods, and numerical methods. We first discuss the Atmospheric Imaging Assembly (AIA; Boerner et al. 2012a) and Helioseismic and Magnetic Imager (HMI; Scherrer et al. 2012) onboard Solar Dynamics Observatory (SDO; Pesnell et al. 2012) and describe the data products used from these instruments. We then describe the Interface Region Imaging Spectrograph (IRIS; De Pontieu et al. 2014) and the Solar X-ray Monitor (XSM; Mithun et al. 2021a) onboard Chandrayaan–2 (Goswami & Annadurai 2011). We obtain data in different UV and EUV wavelength bands from AIA, while we obtain the magnetic field data of the photosphere from HMI. From IRIS, we get spectral rasters in the NUV and FUV wavelengths along with slit-jaw imaging data, while we obtain disc-integrated X-ray data from XSM. We then provide a broad overview of the area of "big data", with a basic introduction to Machine learning (ML) and Deep Learning (DL). Finally, we briefly introduce the MHD formalism and present the set of equations and approximations needed in this thesis.

### 1.3.2   Chapter 3: The source regions of solar wind

In Ch. 3, we develop a DL model (called *WindNet*) to predict the solar wind speed given full-disc EUV images in 193 and 211 Å of the solar corona from AIA. We evaluate our model against auto-regressive and baseline models using Support Vector Machine (Cortes & Vapnik 1995), Gradient boosted decision trees (Kulkarni 2017), Naive mean model, multi-lag auto-regression, and a 27-day persistence scheme (Owens et al. 2013). We find that *WindNet* outperforms the benchmark models, obtaining the best-fit correlation of 0.55±0.03 with the observed data. We then investigate the importance attributed by *WindNet* to various parts of the input images for fast and slow wind forecasts. We find that *WindNet* gives higher importance to the coronal holes for fast wind prediction and to the active regions for slow wind prediction. Furthermore, *WindNet* deems the CHs important for the fast wind forecast ≈3 to 4 days before forecast. At the same time, the AR importance is attributed closer to the day of forecasts for the slow wind. Thus, it suggests that our model was able to learn some of the salient associations between coronal and solar wind structures without built-in physics knowledge, demonstrating that a DL-based approach with interpretable AI techniques may help us discover hitherto unknown relationships in heliophysics data sets.

### 1.3.3   Chapter 4: The impulsively heated quiescent solar corona

In Ch. 4, the overarching theme is to understand the properties of impulsive events and their possible roles in heating and maintaining the solar corona. We combine the empirical impulsive heating forward model of Pauluhn & Solanki (2007) with a machine-learning inversion model that allows uncertainty quantification. Using this scheme, we infer the statistical properties of impulsive events which give rise to QS light curves. We perform this study using 171 Å, 193 Å, and 211 Å EUV passbands of the AIA and the X-ray emission in 1−1.3 keV, 1.3−2.3 keV, and 1−2.3 keV from XSM. We find that there are ≈2−3 impulsive events per minute in EUV, which increases to ≈25 events per minute in X-rays. These events last for 10−20 minutes in EUV, while the shortest events in X-rays last for 6 minutes. The power law slope $\alpha$ peaks above 2 in the EUV passbands, with the $\alpha$ reducing from the cooler 171 Å passband to the hotter 211 Å passband. This trend continues in X-rays, where the $\alpha$ becomes $\leq 2.0$.

Our exploration of correlations among the various timescales and peak energy of these events suggests that conduction losses dominate over radiative losses. We find the event frequency is inversely related to the peak energy of these events, a relation seen consistently in both EUV and X-rays. This result points to the presence of an energy reservoir, which may be depleted through small frequent events or large intermittent events. Owing to the flux calibrated measurements in X-rays, we find that the typical amplitudes of these events lie in an energy range of $10^{21}$ – $10^{24}$ ergs, with a typical radiative loss of about $\approx 10^3$ erg cm$^{-2}$ s$^{-1}$ in the energy range of 1–2.3 keV. Thus, we find that the properties of these impulsive events depend on the wavelength/energy of observations, with a regular gradation in their properties with energy. These results provide constraints and present new insights into the impulsive events maintaining the quiet corona.

### 1.3.4   Chapter 5: Unifying solar wind origin and coronal heating

In Ch. 5, we ask: How and where does the solar wind start, and what relation does it have with the heating of the solar atmosphere? To this end, we perform a detailed comparative study of CHs and QS to understand the underlying physical processes using the chromospheric Mg II h & k and C II 1334 Å lines and transition region using Si IV 1394 Å line for regions with identical photospheric absolute magnetic flux density ($|$B$|$). We find CHs to have subdued intensity in all lines, with the difference increasing with line formation temperature/height and $|$B$|$. The chromospheric lines show excess upflows and downflows in CH, while Si IV shows excess upflows (downflows) in CHs (QS), where the flows increase with $|$B$|$. The CHs also show excess total widths in the C II line over QS for regions with identical $|$B$|$. However, the C II spectral profiles are found to be more skewed and flatter than a Gaussian, with no difference between CHs and QS. We further demonstrate that the upflows (downflows) in Si IV are correlated with both upflows and downflows (only downflows) in the chromospheric lines. CHs (QS) show larger Si IV upflows (downflows) for similar flows in the chromosphere, suggesting a common origin to these flows. These observations may be explained due to impulsive heating via interchange (closed-loop) reconnection in CHs (QS), resulting in bidirectional flows at different heights due to differences in magnetic field topologies. Finally, the kinked field lines from interchange reconnection may be carried away as magnetic field rotations and observed as switchbacks in the solar wind. We describe a unified model for solar wind emergence, coronal heating, and near-Sun switchback formation based on these results.

### 1.3.5 Chapter 6: 2.5 D self-consistent flux emergence

We have obtained evidence of a unified emergence of solar wind and heating of the solar atmosphere through observations, empirical models, and deep learning. However, we need to understand the underlying processes that dictate the thermodynamics of solar wind emergence and heating in the corona. To this end, we perform numerical simulations of flux emergence in 2.5D in a stratified atmosphere in this chapter. We embed different configurations of background fields in the setup to capture the essential differences in the topology of CHs and QS and endow the system with a flux sheet in the convection zone. The perturbation of the sheet results in the interaction of the rising loop with the different background field topologies, giving rise to different dynamics in these systems. The process's thermodynamics is governed by the reconnected flux, thermal conduction, and optically thin radiative processes. We discuss the experiments with different terms, the thermal structure of the resultant jet, and loops.

### 1.3.6 Chapter 7: Parting thoughts and paths for the future

Finally, in Ch. 7, we present a summary of the results obtained in this thesis and put them into a global perspective of how much more we know of the Sun and Heliosphere. We describe the caveats, the unanswered questions, and the new questions that opened up through this thesis while discussing the many future paths awaiting exploration.

# Chapter 2

# Data and methods

ékaṃ sád             To what is One,

víprā bahudhâ vadanty          the wise give many a title

<div align="center">-Rigveda 1.164.46</div>

---

Living beings interact with their surroundings and go about their daily lives. This interaction involves a perception of various aspects of the near environment of living beings at small scales and of the universe itself at large scales. Critical to perception is the presence of sense organs – the environment produces signals in various forms, of which some may be captured by through our individual senses.

If all we perceive is just a 'lossy projection' of the true environment, what do we even mean by the 'true' environment? Is it possible to unambiguously define something as the absolute truth? Dīrghatamas, the sage of yore to whom the quote above is attributed, echoes a similar take sans the pessimism: the wise realize that one may perceive the 'truth' only through its manifestations captured by our senses. For humans conditioned by our senses, asking for the existence of absolute truth is futile – for the manifestations of this truth and the ability to perceive them change with time. So while we only *tend* towards this absolute truth, we may never even reach it *precisely*.

Perception of the environment is built up through measurements, as *data*. Astrophysics has historically been a data-driven science. Across ages, humans have looked up, mapped the heavens, and sought to understand why things happen the way they happen. By putting together as many observations as possible, we hope to generate a better understanding of the universe at large while uncovering the true picture. Humans have a limited sense of perception and cannot hope to understand the universe just through our sense organs. However, we may translate these non-perceivable signals into something perceivable by our senses through assumptions on how these signals originate. Light, along with different particles like ions, electrons, and neutrinos, are the primary messengers from space we obtain in astrophysics [1]. Within the properties of these messengers are encoded various secrets of the cosmos, which we seek to decipher and understand.

The Sun and the heliospheric systems are very near the Earth and hence provide a massive influx of information in the form of remote sensing (through light) and in-situ (through particle) measurements at various points in space and time. The curiosity and a desire to understand these systems manifest, as a set of hypotheses, on different aspects of these systems, which may then be tested using the data. However, the influx of data in solar physics is very large and falls in the regime of big data. Hence, we employ tools and techniques to digest all this data to develop analysis pipelines, extract features, and understand causal connections within data.

With such tools to reduce, analyze and make the data comprehensible, we may present a hypothesis to understand these observations. Hence, along with the data, we also need modeling tools to generate "virtual systems," which provide us with potential observables to be compared with data.

In this chapter, we shall first go through the data used in this work (§2.1). Then, we shall describe the big data tools (§2.2) which have been used extensively. Finally, we shall have a brief primer on the numerical modeling techniques (§2.3) to present a physical picture of the underlying processes.

## 2.1  Data

As we have seen earlier, solar physics data may be primarily classified into remote and in-situ measurements. The particulars of remote measurements depend on the wavelength of light in consideration. Since the Earth's atmosphere typically only allows optical, microwave, and radio frequencies of light, we may study the Sun in these wavelengths from the ground. As of today, numerous ground-based telescopes are observing the Sun in various wavelengths, like the Udaipur Solar Observatory (USO; optical observations), Swedish Solar Telescope (SST; optical, near-IR), Atacama Large Millimeter Array (ALMA; microwave), and the Daniel K Inouye Solar Telescope (DKIST; optical, near-IR) to name a few.

---

[1] We also have gravitational waves now!

Observations short-ward of visible cannot be conducted using ground-based instrumentation. To conduct such observations, we seek observations from space. From the first small satellite observations of the Sun using the Orbiting Solar Observatories (OSO) in the 1960s, the solar community has a rich heritage of space-based missions. Some of these are the Solar Maximum Mission (SMM), Solar and Heliospheric Observatory (SOHO), Hinode, Solar Dynamics Observatory (SDO), Interface Region Imaging Spectrograph (IRIS), to the recently launched Chandrayaan−2, Parker Solar Probe (PSP) and Solar Orbiter (SolO) to name a few.

The wavelength of light being probed and its generation mechanism is tightly coupled to the particular layer of the solar atmosphere probed. The dynamics of the atmosphere hence leave its imprint on the different properties of the light measured. Therefore, different instruments operate in different observation modes to study different aspects of the light. The remote sensing data in solar physics primarily comes in images in a spectral passband, spectroscopic data, polarimetric data, and full-disc integrated spectra.

On the other hand, the in-situ measurements correspond to proton, electron, and other ion properties, while also including measurements of the interplanetary magnetic field. Contrary to remote observations, the in-situ observations are predominantly space-based since the Earth's magnetic field acts as a shield against these particles. Note, however, that perturbations in the geomagnetic field are still measured on the ground, though we shall not be looking at such measurements as a part of this thesis.

The in-situ measurements also have a rich heritage, starting from the Soviet probe Luna − 1 in the late 1950s. A few of the instruments operating today include the Advanced Composition Explorer (ACE), Wind, and the instruments like SWEAP, IS⊙IS, and FIELDS onboard the PSP.

This thesis uses data from the instruments onboard SDO, IRIS, ACE, Wind, and Chandrayaan − 2. In succeeding sections, we describe the salient features of some of the instruments used in this thesis.

## 2.1.1   Solar Dynamics Observatory

Solar Dynamics Observatory (SDO; Boerner et al. 2012a) is a NASA mission designed to understand the causes of solar variability in various spatiotemporal and wavelength scales and its impacts on Earth and the near-Earth environment. It was launched on the 11$^{th}$ of February 2010 from Cape Canaveral and has been provid-ing near-simultaneous and continuous observations since its launch. It is part of NASA's Living With a Star program and houses multiple instruments onboard to address its numerous scientific goals. Of these numerous instruments, we have used data from two instruments extensively – the Atmospheric Imaging Assembly (AIA) and the Helioseismic and Magnetic Imager (HMI). These two instruments mounted on the SDO are depicted in a schematic shown in Fig. 2.1.

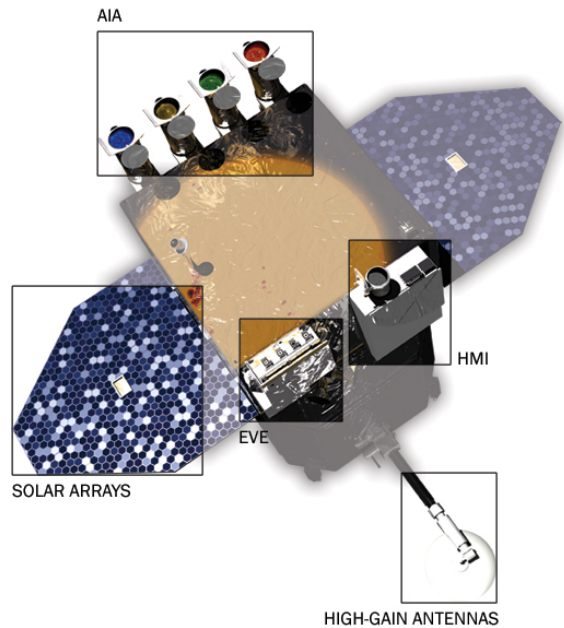

**Figure 2.1**: A schematic of SDO with the different instruments onboard. The AIA is a set of four telescopes, while HMI is a single telescope, as labeled in the diagram. The SDO also has the Extreme Ultraviolet Variability Experiment (EVE), which provides disc-integrated spectral data of the Sun in EUV. Solar arrays and relay communications power the satellite through the High gain antennae. Image sourced from the SDO mission website.

AIA is an imaging instrument in the form of four telescopes observing the various layers of the solar atmosphere in the Extreme UltraViolet (EUV) regime. AIA generates high-resolution 4096x4096 full disc images in multiple passbands with a plate scale of 0.6 arcsecs per pixel centered around specific spectral lines. It provides narrow-band images in seven EUV bands which are centered on transitions from specific ions: 171 Å (Fe IX), 193 Å (Fe XI, XII, XXIV), 211 Å (Fe XI, XIV), 94 Å (Fe X, Fe XVIII), 131 Å (Fe VIII, XXI), 304 Å(He II) and 335 Å (Mg VIII, Fe XVI) (Boerner et al. 2012a; O'Dwyer et al. 2010a). These images are generated at a cadence of

≈12 seconds, with an exposure time of ≈2 seconds in all filters in this work. AIA also provides continuum observations 1700 Å and C IV (1600 Å) at similar spatial resolution but at a lower time cadence. In this thesis, we extensively use the 171 Å, 193 Å, and 211 Å passband data since our science target regions are CHs and QS (O'Dwyer et al. 2010a).

HMI is an instrument that measures the Sun in the form of filtergrams. It images across the Fe I 6173 Å line at six wavelength positions and obtains all four components of Stoke's vector. Using these observations, a variety of data products are derived by the HMI team – including line-of-sight (LOS) magnetic field measurement, vector magnetic field measurement, Doppler velocities, etc. Of these data products, we have used the LOS magnetic field data product, which is available at high spatial resolution (0.5 arcsecs per pixel) and at a cadence of 45 seconds.

Depending on the task, we may limit the resolution and time cadence of the data used. For studies involving long-term variability (in the time scale of years), it is not computationally feasible to use 4Kx4K data at a 12-second cadence. Thus, we also employ a reduced dataset containing 512x512 AIA images generated by Galvez et al. (2019). These images correspond to a plate scale of $4.8$ arcsec/pixel and a time cadence of 6 min.

### 2.1.2   Interface Region Imaging Spectrograph

The Interface Region Imaging Spectrograph (IRIS De Pontieu et al. 2014) is a NASA small satellite explorer mission that observes the dynamics of the lower solar atmosphere. IRIS contains a spectrograph and a slit-jaw imager (SJI) and observes the chromosphere, the transition region, and the lower corona. It mainly observes the Sun in two passbands around 1400 Å and 2800 Å. IRIS provides data at high spatial resolution (0.33 arcsec in FUV and 0.4 arcsecs in NUV), high time cadence (up to 1 second), and high spectral resolution (dispersion of ≈12 or 25 mÅ per pixel). The deployed IRIS image and a schematic of the telescope are depicted in Fig. 2.2.

IRIS has three wavelength bands of observation: two in the Far UltraViolet (FUV) in the 1331.7−1358.4 Å range and 1389.0−1407.0 Å range, and one in the Near UltraViolet (NUV) in the 2782.7−2851.1 Å range. These bands are centered around very strong spectral lines which sample the solar atmosphere.

IRIS provides spectra in two basic modes: (i). raster, and (ii). sit and stare. In the raster mode, the slit is moved across a field of view, and spectra are read from each pixel along the slit. If the displacement between consecutive slit positions is of the order of the slit width, the mode is called dense rastering mode. Otherwise, it is called sparse rastering mode. IRIS may, however, choose to place the slit at some region and observe the Sun. In this mode, it may either choose to let the Sun rotate to sweep across a region or correct for solar rotation while continuing to perform the observations. This mode is called sit and stare mode.

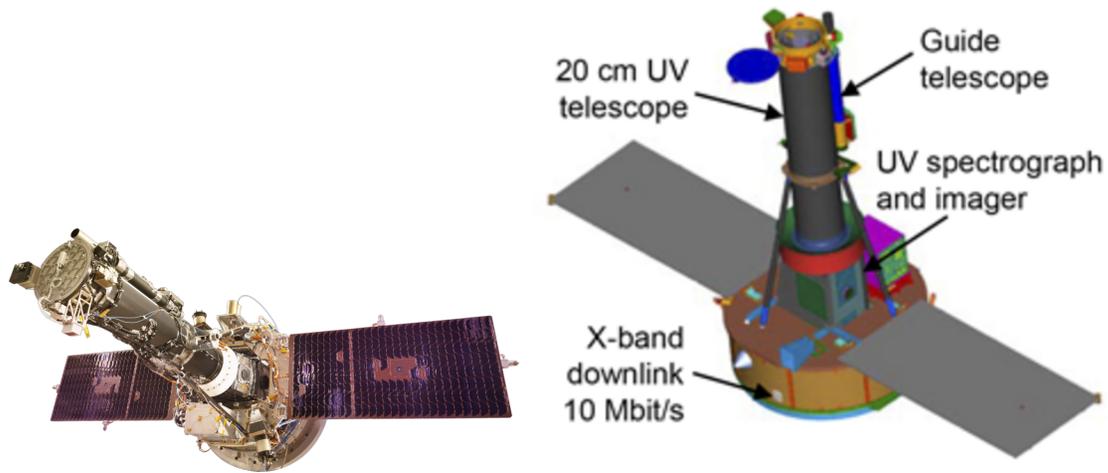

**Figure 2.2**: Left: The deployed configuration of IRIS. Right: A schematic of the satellite depicting the guide telescope, spectrograph, and slit-jaw imager. Image sourced from the IRIS mission website and NASA website.

We have used the spectral dense raster and SJI data corresponding to the Mg II h & k lines from the NUV band, the Si IV 1394 Å and the C II 1334 Å lines from FUV. The spectra provide information in the form of intensity, Doppler shift, and line width for this thesis.

### 2.1.3 Chandrayaan - 2

Chandrayaan - 2 is an Indian mission to the moon launched mid-2019 (Goswami & Annadurai 2011). The mission was primarily focused on showcasing end-to-end lunar mission capability and studying chemistry, thermo-physical characteristics, and the properties of the tenuous lunar atmosphere. The mission consisted of two parts – a lander rover and an orbiter. The lander (named "Vikram") and the rover (named "Pragyan") were intended to land at the southern lunar pole, but the lander deviated from the planned trajectory, resulting in a crash-landing on the moon. The schematic of the complete setup and our instrument of interest is depicted in Fig. 2.3.

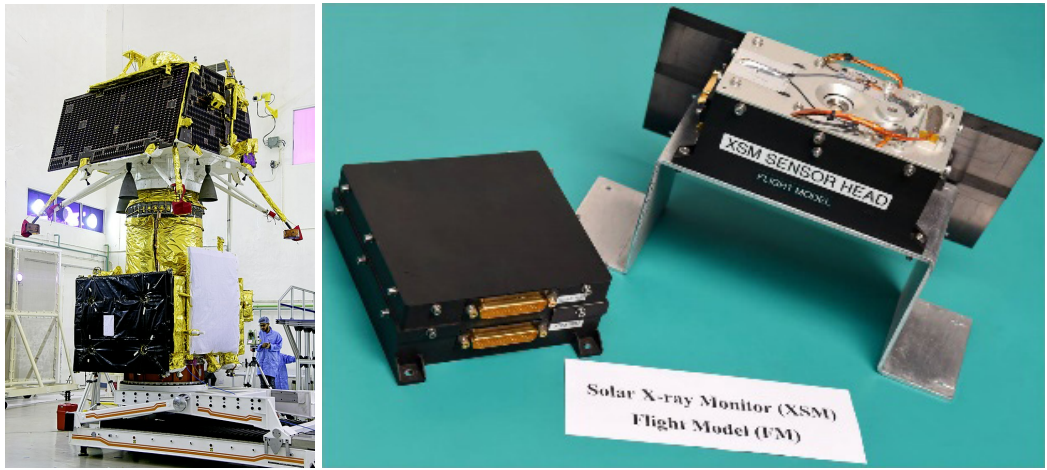

**Figure 2.3**: Left: The deployed configuration of Chandrayaan-2, with the lander mounted on top of the orbiter. Right: The flight model of the X-ray Solar Monitor (XSM) onboard the orbiter of Chandrayaan-2. Image sourced from the ISRO website.

The orbiter, however, has been in a polar orbit around the moon and has been collecting excellent data for a long time. Mounted on the orbiter is the instrument Solar X-ray Monitor (XSM). XSM observes the Sun as a star in the energy range of $1-15$ keV, with an energy resolution of $\sim$175 eV at 5.9 keV and a time cadence of 1 second. The X-ray emission observed by XSM is typical of the very hot corona, and we select only the specific range of 1-2.3 keV for analysis in this thesis.

### 2.1.4 In-situ observations

All of the above instruments provide remote sensing observations of the Sun. As mentioned early on, we have also used in-situ measurements of solar wind in this thesis. The in-situ measurements primarily consist of solar wind and interplanetary magnetic field measurements – wind velocity and magnetic field components. Various instruments have measured these solar wind measurements through time. They have been collated into a common repository by NASA called the "OMNIWEB" data facility at Goddard's Space Physics Data Facility[2].

---

[2] https://omniweb.gsfc.nasa.gov/hw.html

## 2.2 Big data tools

We have used data in timescales ranging from seconds to years, with spatial scales ranging from a couple hundred kilometers to hundreds of megameters. The data is extremely multimodal, and finding salient associations between the data or between data and a presented hypothesis is a non-trivial task. Thus, analyzing this data, finding associations, and extracting features from the data requires using tools that can perform these actions fast and efficiently leverage the available computation.

Machine learning (ML) is a set of techniques that seeks to find associations in a dataset. These associations are found by iteratively optimizing a parameter abstracting a measure of performance or similarities in the data. The keyword here is 'iteratively': the learning power of these techniques comes from a well-founded optimization scheme that recursively updates the selected best solution. Broadly speaking, the "learning paradigms" can be put into three categories: **supervised** learning, **unsupervised** learning, and **reinforcement** learning.

**Supervised** learning involves finding associations between two sets of data, typically between an "input" and an "output". The inputs and outputs can be a bunch of arrays, strings, time series, images, or even time series of images, and the objective would be to translate from one form of data to the other. To put it more precisely, consider the full input dataset to correspond to a 'distribution.' In this distribution, each data point is one sample. Similarly, the set of outputs would also correspond to a distribution. In supervised learning, we are given a set of tuples of data from the two distributions. The machine attempts to learn the correct transformation from the input distribution to the output distribution. Supervised learning comes in two main flavors – **classification** and **regression**.

**Classification** problem can be stated as follows: If there are N unique target categories corresponding to the input dataset, which category will a given datapoint correspond to? For example, given a small input cutout of a part of the Sun, would it correspond to a flare, AR, QS, or CH? Thus, classification seeks to distribute inputs to *finite, discrete, categories*. **Regression** problems can be understood as a generalization of classification problem: what if the number of unique targets $N \to \infty$? Thus, the target output in a regression problem is a continuous-valued variable.

**Unsupervised** learning, on the other hand, involves performing operations on one dataset alone. This would mean we have only one probability distribution, which we may think of as an 'input' distribution. Unsupervised learning comes in different flavors. One application of unsupervised learning is to reduce the dimension-ality of the data. Consider a tabulated dataset with M rows and N columns. This table can be visualized as the bunch of M points in an N dimensional space, where each axis corresponds to one column of the table. Suppose these N columns are not independent in a particular sense. In that case, dimensionality reduction asks if there is a good, low-dimensional representation of the points, which ensures that the resultant 'new' columns are independent in the same sense. Principal Component Analysis (PCA) is one such example of an unsupervised learning method, where the columns are checked for linear independence, and a linear combination of the columns may be presented as a low-dimensional representation. Another application of unsupervised learning comes in the form of a clustering problem. Here, the assumption is that various data points are now closely associated with local clusters. Thus, we ask if it is possible to 'put a box' around different close associations and study the data. There is a zoo of algorithms that perform these operations for both of these applications. These different ML problems are summarized in Fig. 2.4, with simple examples. However, the interested reader may refer to a standard text like Bishop (2006) or Prof. Gilbert Strang's lectures on linear algebra and learning from data[3] for more details on ML.

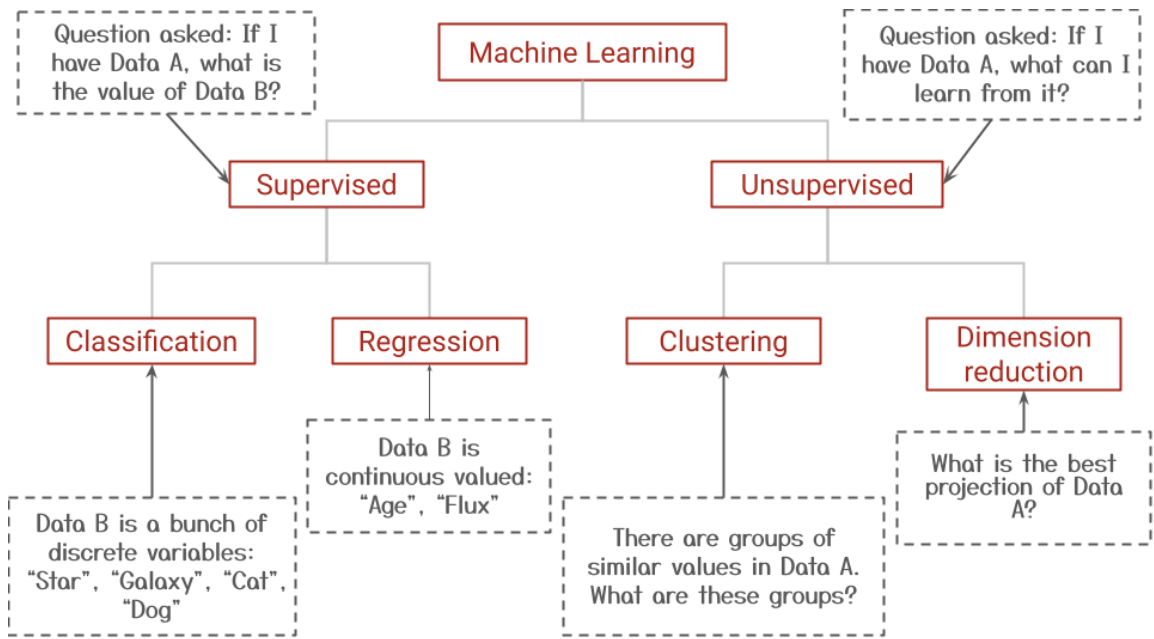

**Figure 2.4**: A summary of the broad types of ML problems to which any general problem can be decomposed.

Now, a subset of ML is called Deep learning (DL), an umbrella term for a broad class of techniques that use neural networks with multiple hidden layers to perform various tasks. A neural network, at its core, is mathematically a set of tensor contractions sandwiched with non-linear function operations. Such sandwiches are layered in series and give rise to the complex 'memory' of a neural network, which readies it to learn from the data. The contractions are performed between the input to the network and the free parameters of the particular layer, called weights and biases. The weights are variables that undergo contraction with the inputs, while biases are variables added as an offset to the product. Henceforth, when we say 'weights,' we shall imply the presence of bias terms unless otherwise explicitly stated.

---

[3]May be found here: `https://bit.ly/3TQwT2n`

Neural networks learn from the data through an iterative update of their weights by minimizing an error or a metric term. The idea behind this iterative update is thus: if the various error values for different weights are considered, they will form an "error surface" in the space of weights. The optimal combination of weights would then be the point that lies at the global minimum of this error surface. Assuming such a minimum exists, we first calculate the local tangent of the error surface for any random starting point on the error surface. This gives us the direction of the largest increase, and hence we seek to move in the *opposite* direction. Eventually, the network weights will converge to the global optimum after taking many steps in this direction. This algorithm is called gradient descent. For a weight $w_i$ at iteration $i$ and the corresponding error term $e_i$ at the same iteration, the weight at the next iteration ($w_{i+1}$) is given by:

$$w_{i+1} \leftarrow w_i - \alpha \frac{\partial e_i}{\partial w_i},$$

where $\alpha$ is called the learning rate of the network, and control the magnitude of weight update. While this is the core idea behind the "training procedure" of *any* neural network, numerous algorithms have been implemented for training that have better convergence properties.

Neural networks grew in prominence in the late 1980s and early 1990s and were extensively employed by the computer vision community (see for example LeCun et al. 1995, 1998, for some classics). However, their usage plummeted with increasing computing needs and limited availability of data. However, there has been a resurgence in neural networks since around the 2010s. This is due to the increased availability of data, and perhaps more importantly, of inexpensive computation in the form of Graphics Processing Units (GPUs) (see, for example Chellapilla et al. 2006; Ciresan et al. 2011a; Krizhevsky et al. 2017). That the hardware used to play *Roadrash* and *Counterstrike* would be crucial for deep learning is a testament to the ingenuity of the human mind!

In engineering and science questions dealing with ambiguous features, DL has been found to outperform ML algorithms that use hand-engineered features (Goodfellow et al. 2016). In supervised learning tasks, DL algorithms often need no prior information regarding the exact input-output mapping but instead try to discover underlying relations through the training procedure outlined above. Prior information can, however, be built into the model in a number of ways, e.g. (1) by providing hand-engineered input features (which are generally physics-based), (2) constructing a neural net with some layers that have been pre-trained and whose weights are kept fixed during training (known as transfer learning, although the kind of pre-training performed limits the amount of prior information shared), and (3) providing external, physics-based constraints on the training procedure of the neural network at hand.

While DL models are able to form associations between data, an increasing complication of these models makes it difficult to inquire into the 'why' of model decision-making. This area of research is known as research into 'Interpretable DL,' which essentially seeks to understand the nature of conditioning of the model parameters and understand why the model is providing a particular output for a given input.

In this thesis, we use ML and DL in three major ways. First, we have used unsupervised ML methods to perform solar image segmentation, which serves to be a preparatory step in performing downstream analysis. Second, we have developed novel supervised DL forecasting codes trained on a large volume of remote sensing and in-situ data. This model is also queried to understand the forecasts using interpretable DL methods, to understand salient associations between the input and output. Finally, we have used DL as an inversion scheme against statistical numerical models. This helps us perform very fast parameter inference across a large dataset very quickly while also providing us with uncertainties in the parameter estimates.

## 2.3   Numerical modelling techniques

With the huge volume of solar observations analyzed, we must also understand them theoretically from the first principles. Such theoretical understanding is possible through the comparison of observations with predictions based on different theories. Typically, this involves solving for the evolution of various particles like electrons, ions, and neutrals with an appropriate set of dynamical equations sup-plemented by consistent initial and boundary conditions. However, we consider the systems where the length scales are larger than the mean free path of these particles & the ion gyro radius, and time scales longer than the corresponding collisional time-scale & ion gyro period. For such a system, the microphysics of interactions may be averaged over, and the system is treated as a fluid continuum. Note that the ion gyro radius is directly proportional to mass and hence is always a larger length scale than the electron gyro radius (Marsch 2006).

Thus, the scales in consideration in this work are far larger than the kinetic scales stated above. Apart from these constraints, let us also assume that the typical velocity scale of the system is smaller than the speed of light in vacuum ($c \approx 10^8$ m s$^{-1}$), i.e., in the non-relativistic regime. Finally, if the ratio of the typical length to the timescale of variation of the electromagnetic field is much smaller than $c$, we operate in the regime of "Magneto-Hydro-Dynamics" (MHD). The typical solar coronal velocities are $\approx 10^4$ m s$^{-1}$, while the solar wind velocities are $\approx 10^5 - 10^6$ m s$^{-1}$, which are far smaller than $c$.

In this framework of MHD, the plasma can yet again be studied as a single fluid or multi-fluid. The plasma is assumed to be fully ionized in the single fluid case. Since the plasma is collisional, the electrons and ions can be considered a single fluid. The plasma contains ions, electrons, and neutral particles in the multi-fluid case. While the ions and electrons undergo Coulomb interaction, the neutrals interact through mechanical collisions. The upper transition region and the solar corona are fully ionized, so a single fluid scheme works well. However, the chromosphere is known not to be fully ionized. Thus, it needs to be treated as a multi-fluid system. However, since the number of equations and variables increases enormously, the MHD approximation is expanded to add new terms that abstract some of the effects of partial ionization of the system (see, for example Martínez-Sykora et al. 2012, for details). In this thesis, we focus only on a single fluid approximation to study the response and dynamics of the upper transition region and the solar corona.

Apart from plasma and the electromagnetic field, there is one more messenger which adds complications – light. The radiation field starts to become free streaming from the photosphere. The processes giving rise to radiation, especially in the form of spectral lines, from the photosphere to the lower transition region are strongly influenced by collisional and scattering processes. Hence, it is important to consider the radiative transfer equation to simulate the lower atmosphere correctly.

The excitation and de-excitation of species may occur through a variety of pro-cesses. Low in the atmosphere, collisions of ions with electrons and other ions are prominent, while spontaneous de-excitation also occurs. Depending on the spectral line of the ion in consideration, the dominant processes may be different. In the corona, however, the low-density results in excitation primarily through collision and de-excitation through radiative transitions. This approximation is able to explain most properties of the observed radiation in the corona and is called the coronal approximation (see, for example Dere et al. 1997). Note that the emission contribution also comes from the recombination of ions with free electrons and bremsstrahlung (Landini & Monsignori Fossi 1970; Gronenschild & Mewe 1978). However, do note that optical thickness depends on the number density of the species along the line of sight and is subject to change depending on the dynamics of the processes occurring in the solar atmosphere.

In this thesis, we concern ourselves with simulating the dynamics and thermodynamics of the upper transition region and corona. Thus, we work in the coronal approximation (Dere et al. 1997). As a consequence, we do not explicitly solve the radiative transfer equation.

We first non-dimensionalise our variables using the unit density ($\rho_0$ in g cm$^{-3}$), unit velocity ($v_0$ cm s$^{-1}$) and unit length ($L_0$ cm). We normalize the pressure through $\rho_0 v_0^2$ and magnetic field through $\sqrt{4\pi\rho_0 v_0^2}$ (Mignone et al. 2007).

The full set of dynamical equations we solve are:

$$\frac{\partial}{\partial t}\begin{pmatrix} \rho \\ \rho\mathbf{v} \\ E \\ \mathbf{B} \end{pmatrix} + \nabla\cdot\begin{pmatrix} \rho\mathbf{v} \\ \rho\mathbf{v}\otimes\mathbf{v} - \mathbf{B}\otimes\mathbf{B} + \overset{\leftrightarrow}{I}\,p_t \\ (E+p_t)\mathbf{v} - \mathbf{B}(\mathbf{v}\cdot\mathbf{B}) \\ \mathbf{v}\otimes\mathbf{B} - \mathbf{B}\otimes\mathbf{v} \end{pmatrix}^{\mathsf{T}} = \begin{pmatrix} 0 \\ \rho\mathbf{g} \\ \rho\mathbf{v}.\mathbf{g} - \nabla\cdot(\eta\mathbf{J}\times\mathbf{B}) \\ \nabla.\mathbf{F}_c - \nabla\times(\eta\mathbf{J}) - n^2\Lambda(T) + S \end{pmatrix},$$
(2.1)

where $\rho$ is plasma density, $\mathbf{v}$ is the velocity, $\mathbf{B}$ the magnetic field, $\eta$ is the resistivity, $\mathbf{g}$ is the gravity, $\rho$ is the density, $\cdot$ indicating contraction and $\otimes$ showing outer product. All bolded quantities denote vectors, while $\overset{\leftrightarrow}{I}$ is the unit tensor. The number density ($n$) and plasma density are related as $\rho = n\mu m_u$, where $m_u$ is the atomic mass unit, and $\mu$ is the mean molecular weight. The total pressure $p_t$ is defined as: $p_t = p + B^2/2$, where $p$ is the gas pressure, and the total energy $E$ is defined as:

$$E = \rho e + \frac{\rho v^2}{2} + \frac{B^2}{2}.$$

The entropy $e$ is defined through the equation of state as $\rho e = f(\rho, p)$. The current $\mathbf{J}$ is defined as $\mathbf{J} = \nabla\times\mathbf{B}$, while $\mathbf{F}_c$ is the thermal conduction flux, $\Lambda(T)$ is the optically thin radiative loss, and $S$ corresponds to various other heating and cooling terms.

Note that we have not included the terms involving fluid viscosity since this viscous loss is expected to be less than the resistive, conductive, and radiative loss in the solar corona (see, for example, the discussion in §3.4 in Marsch 2006).

In the MHD regime, the thermal conduction is highly anisotropic, with the conduction predominantly along the field lines than across. The conduction flux $\mathbf{F}_c$ is then formulated as:

$$\mathbf{F}_c = \kappa_{||}\hat{\mathbf{b}}(\hat{\mathbf{b}}.\nabla T) + \kappa_{\perp}[\nabla T - \hat{\mathbf{b}}(\hat{\mathbf{b}}.\nabla T)], \qquad (2.2)$$

where $\hat{\mathbf{b}} = \mathbf{B}/B$, the unit normal in the direction of the magnetic field.

Finding closed-form solutions to these equations for arbitrary initial and boundary conditions is not tractable yet. Hence, we solve these equations numerically for our problems. In this thesis, we use the PLUTO code (Mignone et al. 2007) developed at Dipartimento di Fisica, Torino University in a joint collaboration with INAF, Osservatorio Astronomico di Torino and the SCAI Department of CINECA [4]. The code is written in C and uses flux-conserving Godunov-based finite volume methods to solve the MHD equations.

The PLUTO code has been written in a very modular format. It also has support for parallel processing using Message Passing Interface (MPI) – thus making it computationally efficient and faster to perform any MHD simulation.

We have developed a set of simulations comparing the effect of the inclusion of different physical processes into the scenario of flux emergence and interaction with a background magnetic field of different topologies. These different topologies are a proxy for different physical regions of interest on the Sun.

*NOTE:* ML and DL techniques would have been very difficult to use if not for Open Source Software using Python. These codes are developed using standard packages like Astropy (Price-Whelan et al. 2018), Cython (Behnel et al. 2011),Jupyter (Kluyver et al. 2016), Matplotlib (Hunter 2007),Multiprocessing (McKerns et al. 2012), Numpy (Harris et al. 2020),OpenCV (Bradski 2000), Pytorch (Paszke et al. 2019), Scipy (Virtanen et al. 2020), Scikit-image (van der Walt et al. 2014), Scikit-learn (Pedregosa et al. 2011), Seaborn (Waskom & the seaborn development team 2020), Sunpy (Mumford et al. 2018), and Tensorflow (Abadi et al. 2016). A major part of this thesis was prototyped using Jupyter notebooks and then ported to python scripts, enabling a fail-fast and log-result working mechanism.

---

[4]`http://plutocode.ph.unito.it/`

# Chapter 3

# Solar wind and its sources

*Emanating from the base of the Sun's corona, the solar wind fills the inter-planetary medium with a magnetized stream of charged particles whose interaction with the Earth's magnetosphere has space-weather consequences such as geomagnetic storms. Accurately predicting the solar wind through measurements of the spatio-temporally evolving conditions in the solar atmosphere is important, and remains an unsolved problem in heliophysics and space-weather research. In this work, we use deep learning to predict the solar wind speed given solar coronal EUV images from AIA. We then demonstrate the potential of deep learning in uncovering the source regions of the solar wind by using Grad-CAM. This thesis chapter originally appeared in the literature as **Solar wind prediction using deep learning** (DOI: 10.1029/2020SW002478 ).*

The solar wind, as we have seen earlier in §. 1.2.2, seems to arise from specific source regions in the corona. Furthermore, these sources in the corona have associations with specific structures in the solar wind. While this science question is inherently interesting, resolving this question to any measure also has a massive human and economic impact due to the effects of space weather. Thus, solar wind sources are intimately tied to understanding and forecasting space weather.

Space weather is defined by the U.S. National Space Weather Plan as the conditions on the sun, in the solar wind, and within Earth's magnetosphere, ionosphere, and thermosphere that can influence the performance and reliability of space-borne and ground-based technological systems and can endanger human life or health (NASA 2017). The influence of the solar wind on space weather arises due to its inter-action with the Earth's magnetosphere, resulting in geomagnetic storms and aurorae Schwenn (2006). This interaction may further induce currents, which may devastate electrical grid distributions, oil pipelines, railway systems, and telecommunication systems, to name a few (see Schrijver et al. 2014; Barlow et al. 1849; Boteler 2001; Pulkkinen et al. 2001; Eastwood et al. 2018, for a non-exhaustive set of references). The first step to forecasting space weather variables starts with forecasting the driver of space weather – the solar wind.

Owens et al. (2008) review solar wind prediction using empirical, physics-based, and hybrid approaches. Typically, a physics-based model uses synoptic magnetograms as the bottom boundary condition. An individual synoptic magnetogram is assembled by sampling the photospheric magnetic flux distribution near the central meridian over the course of a solar rotation. Such magnetograms can be used to extrapolate the surface field into the corona using potential-field source-surface (PFSS) Altschuler & Newkirk (1969) models or magnetohydrodynamics (MHD) models (see Riley et al. 2006, for a comparison between the two). The global coronal magnetic field (or certain derived properties thereof) may then be used as input for physics-based solar wind propagation models (e.g. Linker et al. 1999), or in the case of a hybrid approach, used for estimation of the solar wind at L1 using empirical relations.

WSA-ENLIL and MAS-ENLIL (Owens et al. 2008; Schwenn 2006) are among the most widely used solar wind models. The models provide solar wind properties such as velocity, plasma density, magnetic field, and temperature. Jian et al. (2015) perform a comparison of the various solar wind models through the Pearson correlation between the model forecasts and solar wind speed observations. They present a correlation of 0.57 on hourly prediction using the GONG-MAS Thermo-ENLIL model, and 0.50 on the same dataset using the GONG-WSA-ENLIL model.

In §1.2.2, we presented evidence for the relation between CHs and solar wind properties (see, for example Krieger et al. 1973; Wang & Sheeley Jr 1990). The influence of the CH on solar wind stems primarily from the magnetic field topology of the region. Since the dynamics of the magnetic field dominate the solar corona, the topology has a direct correspondence with the observed intensity structures in the corona. This was exploited by Rotter et al. (2012); Rotter et al. (2015); Temmer, Manuela et al. (2018), who checked for correlations between fractional CH area extracted from EUV imagery data and solar wind speed. These authors obtained correlations from $\approx 0.60$ to $\approx 0.78$ for hourly solar wind speed. More recently, Yang et al. (2018) devised a Neural network-based prediction scheme, taking PFSS model output among other parameters as input, and obtained a correlation of 0.74 on hourly solar wind speed data.

In ML and statistical-learning parlance, the aforementioned traditional empirical models use so-called hand-engineered features as input for their models (e.g., CH area or CH expansion factor). These hand-engineered features are often inspired by some insights from physics-based models or simply from correlations reported in the literature. In the context of validated hypotheses, as we discuss early on in Ch.2, these correspond to specific hypotheses derived from validated first principles. As we have seen in §.2.2, DL is an umbrella term for a broad class of techniques that use neural networks with multiple hidden layers for performing supervised or unsupervised learning tasks. Due to the increased availability of data and, perhaps more importantly, inexpensive computation, DL has been widely applied in many domains. In areas of engineering and science dealing with ambiguous features, DL has been found to outperform ML algorithms that use hand-engineered features Goodfellow et al. (2016). This is because the hypotheses we present on explaining the data are lacking, and the machines learn additional information, updating their hypothesis due to their optimization procedure. We have also seen that DL algorithms try to discover underlying relations in the data by iteratively updating the model parameters and, thus, the hypothesis explaining the data.

The hypotheses or models we keep mentioning are the neural networks appropriate for a given task. Two of the most prominent architectures used in deep learning are Convolutional Neural Networks (ConvNets) and Recurrent Neural Networks (RNNs). ConvNets work by detecting local patterns at multiple scales in the input and mapping them to the appropriate class (classification) or continuous output (regression). They have been successfully applied to different classification and regression problems Ciresan et al. (2011b); Deng & Yu (2014); LeCun et al. (2015) for image data. RNNs, on the other hand, are a class of deep neural nets designed for understanding the structure of data with a sequential ordering. These have been used extensively for text prediction, natural language processing, and regression Hochreiter & Schmidhuber (1997); Sutskever et al. (2014).

In this work, we seek to develop a DL model to forecast solar wind properties. Specifically, we use EUV images in 193 Å and 211 Å from AIA as input and forecast the solar wind speed from the NASA OMNIWEB dataset. We use a ConvNet Szegedy et al. (2015) pre-trained on the ImageNet database Deng et al. (2009) and couple it with a trainable Long-Short Term Memory cell (LSTM) implementation of an RNN Hochreiter & Schmidhuber (1997) to perform this translation. The network is not given any prior information about the physical mapping between the EUV image data and solar wind speed. We thus perform a direct regression from a time series of AIA images to the solar wind speed. While one of the major goals of this work is to develop space weather forecasting models, our core science question is to understand the contribution of the coronal sources to the solar wind ([Q1] in §. 1.3). Thus, we then query the model which has learned the association between EUV data and solar wind speed. In other words, we 'reverse-engineer' the sources given the learned association to understand which regions in the AIA images are important for a fast and slow wind forecast. This provides us with an understanding of the possible source regions of the solar wind.

The remaining chapter is organized as follows: In §. 3.1, we describe the data preprocessing, partitioning into training and testing sets, and then define some control parameters and evaluation metrics. Then, in §. 3.2, we briefly introduce the various algorithms used as benchmarks. We detail our proposed model `WindNet` and the visualization technique used for generating the activation map. The segmentation algorithms used for the generation of binary masks for the computation of mean activation values are also described. In §. 3.3, we summarize our model predictions vis-a-vis our benchmarks, present the trends of mean activation, and draw conclusions in §. 3.4.

## 3.1 Data and Metrics

### 3.1.1 EUV dataset

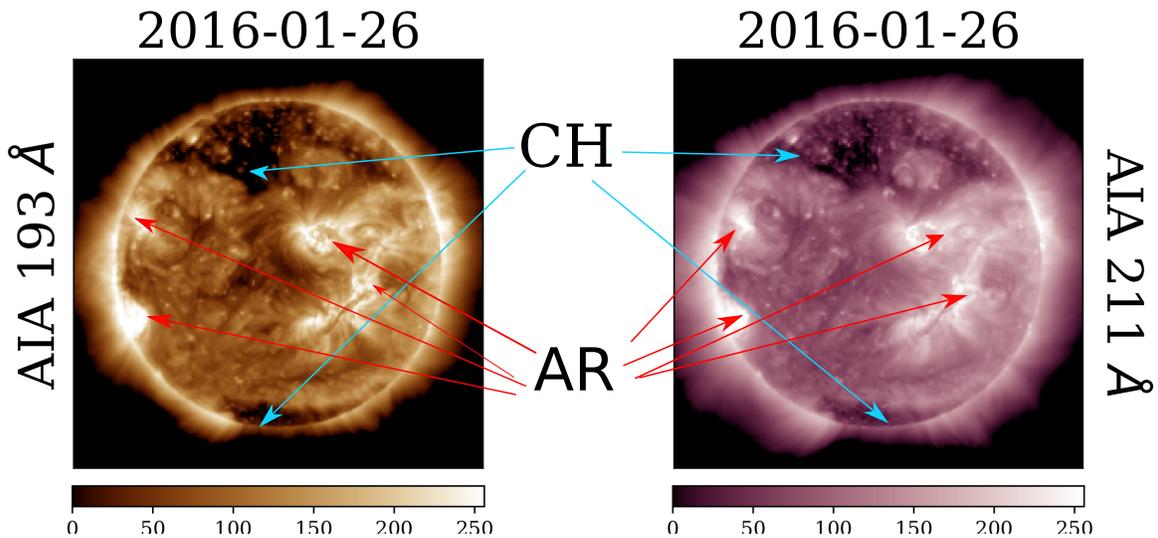

**Figure 3.1**: Representative AIA data in the 193 Å and 211 Å passband with CHs and ARs marked. This is the final data used in our analysis.

We use the data from AIA onboard SDO to represent the conditions in the solar corona. Specifically, we use the SDOML dataset made publicly available by Galvez et al. (2019)[1]. In this dataset, the AIA images have been resampled onto a grid of 512x512 pixels with 4.8 arcsec pixel spacing and are available at a 6 min cadence. SDOML images are stored as binary arrays in the Python numpy format Walt et al. (2011). Our training and testing data include AIA images each day at $00:00$ UTC. The selected image forms a proxy for the whole day of observation. However, if the image at $00:00$ does not exist (as is the case with many days), the closest image to $00:00$ from that day is taken as a proxy for that day.

Even during non-flaring times, solar EUV images can have a dynamic range that greatly exceeds the 8-bits per passband dynamic range typical of most computer-vision datasets. For this reason, the input AIA images are first preprocessed by performing log-scaling to bring out fainter features. The images are then passed through a threshold and saturation, which limits the dynamic range of pixel values. This was done to limit the prediction to contribution from Solar disc alone. Furthermore, we saw that the model performance was better with thresholded and saturated images – thus, the dynamic range was limited. A general sweep of threshold and saturation was performed for a particular combination of History and delay (3.1.4) for the 193 Å data. The correlation of predicted solar wind speed with observed solar wind speed $0.48\pm0.03$ for the best set, with higher thresholds (log($250$), log($10000$)) giving us $0.46\pm0.02$ and lower thresholds (log($100$), log($1000$)) giving us $0.35\pm0.02$. A coarse search was performed to find the threshold values. The thresholds for 193 Å data were scaled to 211 Å through a ratio of maximum intensities on a given day selected randomly. Eq (3.1) and (3.2) specify the threshold and saturation operations for log scaled 193 Å and 211 Å passband images, respectively. AIA 193 Å and 211 Å data, with CHs and ARs, marked after preprocessing, are shown in Fig. 3.1.

$$x(193) = \begin{cases} \log(125.0) & \text{if } x \leq \log(125.0) \\ \log(5000.0) & \text{if } x \geq \log(5000.0) \\ x & \text{else} \end{cases} \qquad (3.1)$$

$$x(211) = \begin{cases} \log(25.0) & \text{if } x \leq \log(25.0) \\ \log(2500.0) & \text{if } x \geq \log(2500.0) \\ x & \text{else} \end{cases} \qquad (3.2)$$

The pixel values are then rescaled between $0.0$ and $255.0$.This is done since our feature extractor expects inputs within this range of values.

---

[1] `https://purl.stanford.edu/jc488jb7715` and links therein

### 3.1.2   Solar wind dataset

The target output of the prediction models is daily-averaged solar wind speed measured at L1. We use daily averages since the variation of wind speed over a day is not large, and the variation across the mean value sets the uncertainty in wind speed value. The variation (or the standard deviation $\sigma$) is calculated as the variance in hourly measurements over the day, at the OMNIWEB archive[2].

A representative variation in solar wind speed data over 10 days is plotted in Fig. 3.2. The distribution of solar wind speed and the corresponding $\sigma$ for the entire dataset is shown in Fig. 3.3.

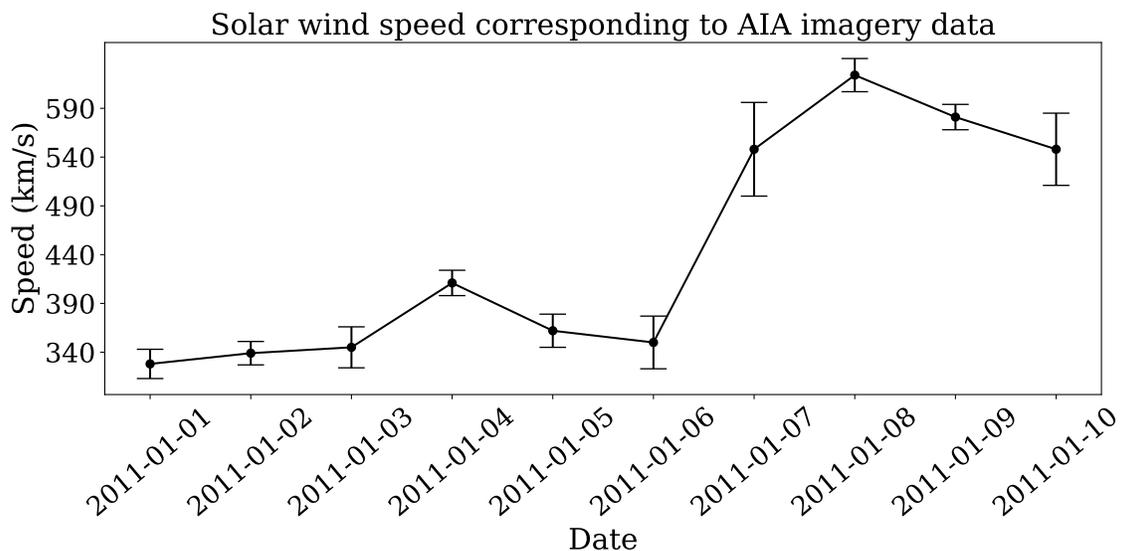

**Figure 3.2**: 10 days of solar wind speed from NASA OMNIWEB dataset.

There are gaps in the AIA EUV data (30 days of missing data in 211 Å and 31 days of missing data in 193 Å) for **00:00 UTC**, owing to various reasons ranging from calibration maneuvers to recoveries from instrument anomalies. Thus, the solar wind speed during these gaps has been removed to form sets of {image, wind speed}.

---

[2]available online at `https://omniweb.gsfc.nasa.gov/form/dx1.html`

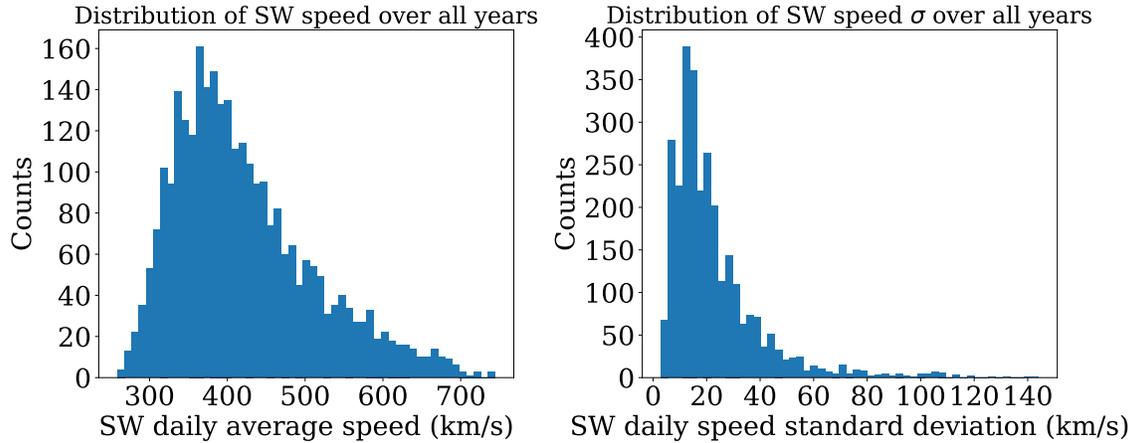

**Figure 3.3**: Left: Distribution of solar wind speed. Right: Distribution of the associated $\sigma$. The distributions are computed over the entire dataset.

### 3.1.3 Dataset partitioning and Cross Validation

Data are available from 1 January 2011 to 9 December 2018. Given the presence of a background solar cycle and events in the Sun, which might systematically bias our model to perform only for a particular phase of the cycle, the whole dataset was partitioned into batches comprising 20 contiguous days of data. If during batch formation, there exists a single discontinuity in the batch, the data from the day prior to discontinuity to 20 days prior is sampled, and placed in the same place as the previous batch, thereby removing any data leak. If there exist multiple discontinuities (there are only 2 instances of such an event in either of the datasets), that particular window between the discontinuities is discarded. This results in 157 batches for 211 Å and 158 batches for 193 Å data (courtesy of the one missing data, which resulted in a new batch). These batches were randomly sorted into 5 folds with equal probability, and these 5 folds were used to perform cross-validation. The dataset partitioning scheme is shown in Fig. 3.4.

In cross-validation, if there are [1, N] folds of data, a cross-validation set is constructed by holding the fold *i* as the test set and the remaining in the training set. Such a construction is done for all folds of the batches. Our models are evaluated against this cross-validation dataset, thereby providing us with a mean value of the metric and a standard deviation. Henceforth, any standard deviation associated with the predictions is to be taken as evaluated on the cross-validation dataset.

The image data are centered using the mean pixel value of the training dataset per cross-validation fold. The images are resized to $224 \times 224$ pixels using Lin-ear Interpolation with OpenCV default values (Bradski 2000), and each image is replicated into $3$ RGB channels. This was performed as our pre-trained network demands the input images to be of dimensions $224 \times 224 \times 3$ since terrestrial images generally have Red, Green, and Blue as the color basis. These images are then finally used for training our network. The solar wind speed data are scaled between 0 and 1 using the training data statistics (max and min values) of each cross-validation fold.

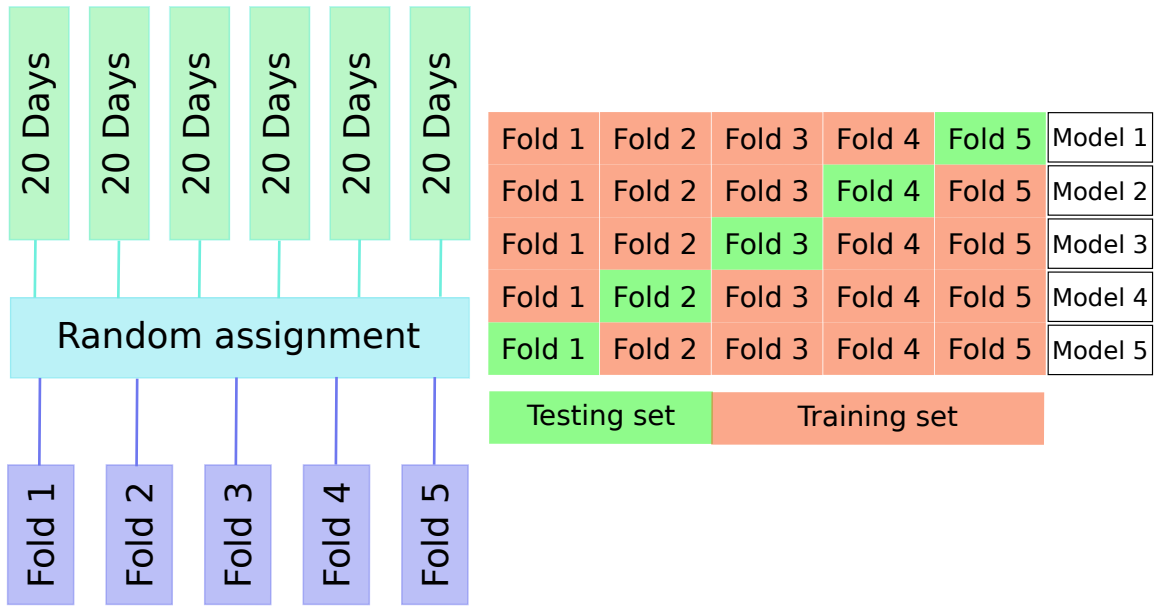

**Figure 3.4**: Training and test split for cross-validation. First, the data are split into batches of 20 days each, and each batch is randomly assigned to one of the cross-validation folds. Then, for one particular model (i.e., for one particular combination of history and delay), one of the folds is marked as the test set and the remaining as training sets. A circular permutation is performed till each fold is used as a test set. In our case, at the end of a training exercise, we will have 5 different variants of the particular model, from which we derive the mean and standard deviation of fitting metrics.

### 3.1.4 Control Hyperparameters

Hyperparameters are free parameters that give a handle in controlling the whole algorithm. We define two control hyperparameters: **history (H)** - number of days of input data required for one prediction, and **delay (D)** - the time from the latest input datapoint to the day of solar wind prediction. For example, if the day of prediction is $T$, and data from $T$-3 to $T$-6 are used as input, our *history* is defined as 4 and the *delay* as 3. We have trained models with different combinations of delay ($D$ =1 to 4) and history ($H$ =1 to 4), resulting in 16 variants of the `WindNet` model. These two control hyperparameters are illustrated in Fig. 3.5.

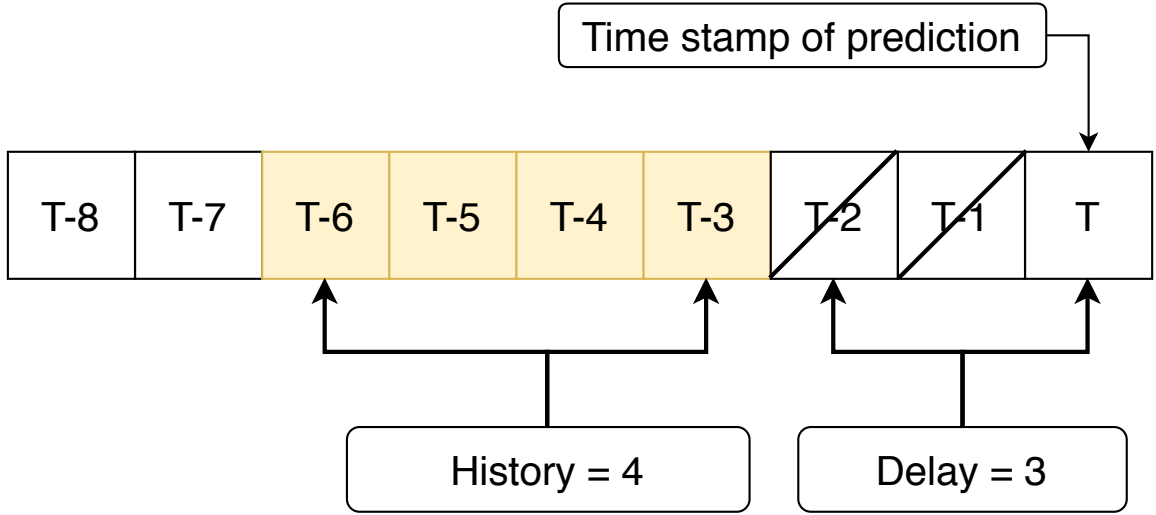

**Figure 3.5**: Each variant of the `WindNet` model is trained to predict the solar wind speed on the day $T$, using input data (SDO/AIA) from days in the range [$T$-$H$-$D$+1,$T$-$D$], where $H$ and $D$ denote the history and delay control hyperparameters, respectively.

### 3.1.5 Metrics for Comparison

Quantitatively, a set of metrics need to be defined to unambiguously quantify if the fit is good or bad. We define three metrics to estimate the goodness of fit ($\hat{y}$ = Prediction, $y$ = Observation):

1. Mean square error ($\chi^2$ value):

$$\chi^2 = \frac{1}{N} \sum_i^N (\hat{y}_i - y_i)^2, \tag{3.3}$$

where N is the no. of data points. However, we present the Root Mean Square Error (RMSE), defined as $\sqrt{\chi^2}$, in the units of $km/s$.

2. Reduced mean square error ($\chi^2_{red}$):

$$\chi^2_{red} = \frac{1}{N} \sum_i^N \frac{(\hat{y}_i - y_i)^2}{\sigma_i^2}, \tag{3.4}$$

where $\sigma_i$ denotes the standard deviation associated with each observation $y_i$. The standard deviation is computed over OMNI measurements for each day and reported in the dataset.

3. Pearson correlation coefficient ($r$): The standard definition of correlation is

$$r = \frac{\sum_i^N (y_i - \bar{y})(\hat{y}_i - \hat{\bar{y}})}{\sigma_y \sigma_{\hat{y}}}, \tag{3.5}$$

where $\bar{y}$ and $\hat{\bar{y}}$ represent the mean values and $\sigma_y$, $\sigma_{\hat{y}}$ represent the standard deviation of the dataset in consideration. To perform an average of correlation across all folds, we transform the data to Fischer's z-space, perform the averaging, and then transform back – to prevent bias while performing average Corey et al. (1998). The standard deviations are calculated in Fischer's z-space and propagated back to correlation.

The three metrics defined have their own advantages and drawbacks.

1. The predicted data are scaled between *0* and *1*. Hence, even a large deviation, when squared, seems very small if both the prediction and observation are $<$ 1. Thus, while $\chi^2$ is a good minimizing function for training, it fails to perform well as a metric for a good fit.

2. To counter the above case, the $\chi^2_{red}$ metric is used. This takes into account the inherent error in each measurement and scales the fit accordingly. A bad fit for a high error datapoint is acceptable, as the observation itself has high uncertainty, while a bad prediction on a low error datapoint is bad since it serves as a much better point of comparison.

3. If the output were a naive mean value of the batch, the $\chi^2$ and $\chi^2_{red}$ would still be reasonable – however, there would be no variance in the fit. Hence, the Pearson correlation $r$ is used to understand the trend captured by the fitted curve. The exact fit values may not match, but if the trend is captured, the model is fairly good according to this metric.

To summarize, Pearson $r$ captures the trend but ignores any scaling error. $\chi^2$ captures scaling errors but doesn't perform well on scaled data $<$ 1. And $\chi^2_{red}$ captures the errors by weighing them vis-a-vis the variance of observed data. These three metrics are used for comparing the models - i.e., the $r$ value of our proposed model should be higher, and the $\chi^2$ and $\chi^2_{red}$ lesser than the benchmark models. Please note that errors (or spread) reported (for both the metrics and activation evaluation) later on correspond to Standard Error (or uncertainty in the estimated mean), defined as

$$S(x) := \frac{\sigma(x)}{\sqrt{N(x)}},$$

where $\sigma(x)$ is the standard deviation derived from the sample, and $N(x)$ is the number of samples in the set.

Model performance, while accounting for timing errors, is an important marker for capturing the response to dynamic events in the solar wind. Thus, we also compare the performance of our models through their ability to capture High-Speed Enhancements (HSE), as used in several texts (Owens et al. 2005; Reiss et al. 2016; Bu et al. 2019). We use the method as outlined in Jian et al. (2015) for finding out HSE. This is performed as:

1. Mark all time points which are more than 50 km/s faster than 1 day earlier.

2. Group each contiguous block of marked points as a distinct high-speed enhancement (HSE) and find the start and end time of each HSE.

3. For each HSE, find the minimum speed starting 2 days ahead of the HSE till the start of the HSE, and mark it as the minimum speed (Vmin) of the HSE; find the maximum speed starting from the beginning of the HSE through 1 day after the HSE and mark it as the maximum speed (Vmax) of the HSE.

4. For each HSE, find the last time reaching Vmin and the first time reaching Vmax and mark them as the start and end time of a Stream Interaction Region (SIR).

5. For the regrouped SIRs, find the Vmin and Vmax for each SIR and mark the last time of the highest speed gradient as the stream interface (SI), the boundary between slow and fast wind. Eliminate SIRs with redundant SI time.

6. Reject any SIRs with Vmin faster than 500 km/s, or Vmax slower than 400 km, or speed increase less than 100 km/s.

Each HSE present in the observation, and captured by the model is called a True Positive (TP), and those not captured by the model are called False Negative (FN). Spurious HSE predictions by the model are called False Positives (FP). With these, we define the metric of comparison Threat Score (TS) as:

$$TS = \frac{TP}{TP + FN + FP}. \tag{3.6}$$

The threat score is a proxy for the accuracy of the forecast of any model. A model which predicts all the HSE perfectly (while not predicting any spurious HSE) has a TS of 1 – thus, the lower the TS, the worse the model. For every cross-validation set per model, we identify the HSEs and calculate the TS – thereby giving us a mean TS and its uncertainty per model. Note that if the HSE (i.e the peak of the enhancement) occurs very near the boundary, it would be missed by the algorithm due to our data partitioning scheme. Such HSE are discarded by benchmarking the H=1, D=1 Persistence model to give a TS = 1.0.

This study does not account for the effect of ICMEs (Near-Earth Interplanetary Coronal Mass Ejections). There are 170 ICMEs reported within the time range considered in this study, affecting solar wind measurements in 336 days. In both model training and evaluation, we did not remove days for which there were ICMEs. The prediction of solar eruptions leading to CMEs and ICMEs is outside the scope of this study. Nevertheless, their occurrence impacts the solar wind measurements at L1. So for the evaluation of the solar wind models in this paper, we decided to include even the days when ICMEs were present.

## 3.2 Modelling and methods

### 3.2.1 Benchmark Models

We next describe various models taken as benchmarks for our proposed `WindNet` model. These benchmark models all operate as autoregressive models on the solar wind data only and do not use AIA images as input. The models (except 27-day persistence) are all corrected for the data gaps, thereby making the comparison reasonable.

- Naive mean value model.

- N day and 27-day Persistence model.

- Autoregression with XGBoost (Chen & Guestrin 2016).

- Autoregression with Support Vector Machines (SVMs).

**Autoregression using a 'Mean value'**

One of the most basic benchmarks for any model is the comparison of the fit with a mean value model. This benchmark takes in the solar wind data and outputs the mean value of the whole batch. This model serves as the lowest benchmark that the proposed model should surpass since untrained models output mean values.

**Persistence Model**

The second benchmark model is persistence. The solar wind speed is fed in as input, and the same output is obtained. Such a model would show how long the data persists through time.

The N-day persistence is calculated from $H + D - 1$ days prior to prediction, to the day of prediction. As such, there is no individual dependence of the persistence model on $H$ or $D$ – rather, the dependence is on the combined value, thereby having degeneracy. This model is primarily used for determining how far into the future our models consistently give a good prediction, given an observation today, or observations starting today.

We also benchmark our results against 27-day persistence for 1 Carrington rotation, as it has been shown to be a good benchmark model in Owens et al. (2013). The 27-day persistence model operated on the complete solar wind dataset (devoid of any gaps).

**Autoregression using XGBoost**

The solar wind speed is autoregressed for different H and D using the XGboost algorithm Chen & Guestrin (2016). That is, the prediction $\hat{y}_{T+1}$ is given as $\hat{y}_{T+1} = f(\mathbf{x})$, where model input is $\mathbf{x} = (y_{T-H-D+1}, y_{T-H-D}, ..., y_{T-D})$, and the function $f()$ comprises the gradient-boosted decision trees. The various parameters set for the algorithm are shown in Table. 3.1. The best model from the swept set of parameters is selected based on the lowest $\chi^2$ value.

**Autoregression using support vector machines (SVMs)**

SVM is also used as a benchmark for a good fit since it has more non-linearity than decision trees due to the presence of kernels. Three kernels are used for bench-marking - Radial Basis function, Linear, and Polynomial kernel of degree 5. We use the Scikit-learn Pedregosa et al. (2011) implementation of SVM in this work. The parameters were selected by grid search using the $\chi^2$ value as the comparison metric. The best fitting parameters are shown in Table. 3.2.

Table 3.1:: XGBoost parameter selection using grid search.

| Parameter | Value |
| --- | --- |
| eta | [0.001,0.01,0.1,0.8,0.9,1.0] |
| seed | 0 |
| objective | reg:linear |
| max_depth | 200 |
| lambda | [50,10,5,1,0.5,0.05] |

Table 3.2:: Support-vector regression-parameter selection.

| Kernel | Parameter | Value |
| --- | --- | --- |
| RBF | C | 1e+4 |
| RBF | gamma | 0.001 |
| Linear | C | 1e+4 |
| Polynomial | C | 1e+4 |
| Polynomial | degree | 5 |

### 3.2.2 Proposed solar wind model

In this work, we have a time series of images that must be translated to wind speed measurements. To this end, we first reduce the dimensionality of the AIA images into a set of generic features. Then, we feed this representation of the image as a time series to a regressor, which regresses against the solar wind speed. We propose the DL model **WindNet**, constructed using a ConvNet and an RNN. We use a pretrained GoogLeNet (Szegedy et al. 2015) model as a ConvNet feature extractor and then feed the obtained embeddings into a variant of an RNN, called Long-Short Term Memory (LSTM; Hochreiter & Schmidhuber 1997) model[3].

GoogLeNet is a ConvNet (Szegedy et al. 2015) developed for the ImageNet (Deng et al. 2009) competition. This competition provides a huge database of labeled images with the objective of classifying them into different categories. As mentioned earlier, ConvNets work by detecting patterns at multiple scales in the input. The size of the convolution kernel gives this scale over which patterns are detected. ConvNets generally have convolutions performed sequentially – thus, at a given layer, the sensitivity is only to a particular scale. However, GoogLeNet, for the first time, introduces us to the concept of the *Inception module* (Szegedy et al. 2015). Essentially, this module has, at each layer, convolutions using different kernel sizes in parallel. Thus, it provides sensitivity at multiple scales at the same time. This has been shown to outperform other models on the ImageNet14 dataset Szegedy et al. (2015).

---

[3]GoogLeNet weights were obtained from: http://www.deeplearningmodel.net/)

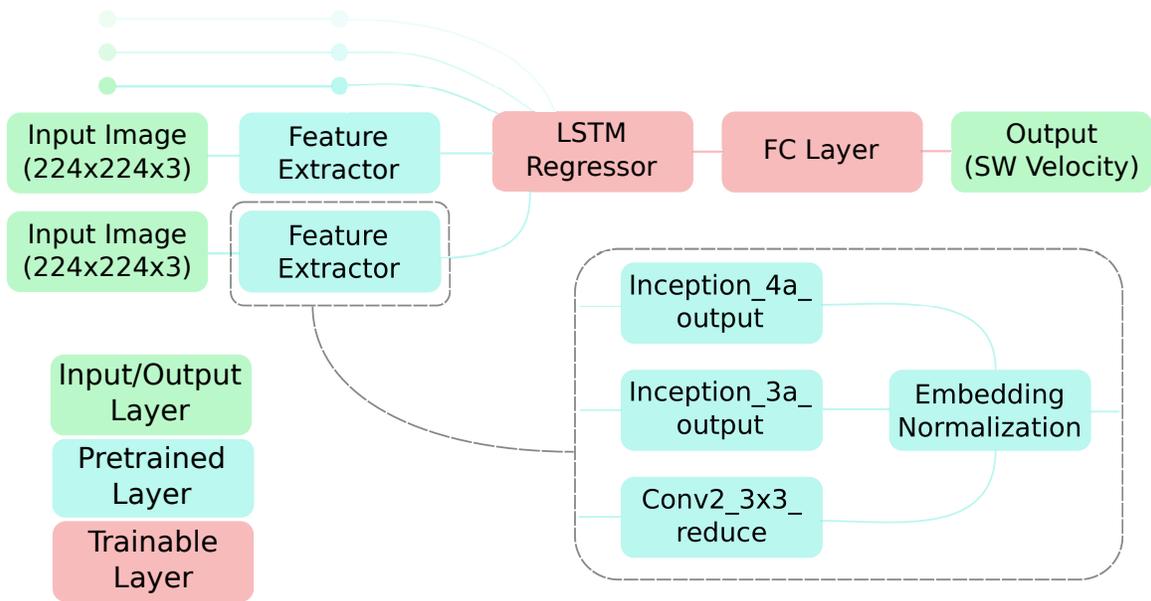

**Figure 3.6**: `WindNet` architecture using GoogLeNet and an LSTM.

GoogLeNet has been trained on everyday objects. However, given the large volume of training data in ImageNet14, the initial layers of the network capture generic global features in the images Goodfellow et al. (2016). Thus, the first couple of layers essentially generates a generic, non-linear global transformation of the input images – like edge detection, curve detection, high- and low-pass filtering, etc. As one goes deeper, the network captures features specific to the dataset – which is not relevant to our dataset. Thus, we use this pretrained network to generate a low-dimensional representation of the AIA data in the form of embeddings. This technique is known in the literature as Transfer learning Yosinski et al. (2014). We adopt a 'multi-resolution approach' to generate the embeddings – i.e., responses from layers at different depths are taken, normalized, and concatenated. The embeddings are then fed to an LSTM for regression against the solar wind speed. GoogLeNet has its weights fixed, while the LSTM (and a fully connected layer at the end) are trained. We use a single LSTM cell in our work. The model is developed using the Tensorflow package for Python Abadi et al. (2015), and is summarized in Fig. 3.6.

The training details for the algorithm are summarized in Table. 3.3.

Table 3.3:: WindNet parameter selection

| Parameter | Value |
| --- | --- |
| Cost function | $\chi^2(\hat{y},y)+\chi^2_{red}(\hat{y},y)$ |
| Optimizer | Adam |
| Learning rate | 5e-4 |
| Dropout for LSTM | 0.5 |
| L2 Norm coefficient | 1e-6 |
| No. of hidden units in one LSTM cell | 400 |
| No. of iterations | 300 |
| Feature length from GoogleNet | 832 |

### 3.2.3 Activation Visualization

There exist techniques in the DL literature to visualize neurons in hidden layers which are preferentially activated for a given input - this activation can be extrapolated back to the given input to understand which regions of the input data have a large impact on the prediction. These methods rely primarily on the gradient of output w.r.t each input pixel, thereby providing an approximation of regions most responsible for an increase or decrease in the output. The methods, while not being perfect visualizers, are a window into the workings of the network. In this work, we use Grad-CAM (Selvaraju et al. 2017) maps as a visualization technique.

Grad-CAM, or Gradient Class Activation Maps, are maps generated by pointwise multiplication of the average gradient per channel of output vis-a-vis a given convolution layer with the corresponding ConvNet layer activation. The obtained map is then passed through a Rectified Linear Unit (ReLU, namely $f(x) = \max([0,x])$) activation function to obtain the activation map. The maps are averaged across channels and then scaled up to the dimensions of the input image for comparison. This method produces activation maps of the model on the input data. These activation maps are subsequently used to generate a metric for the determination of the influence of the CHs and ARs.

### 3.2.4 Generating binary masks

A simple metric for understanding the influence of a particular set of features for a regression problem would be to look at the mean value of the activation on that particular set of features across all data points and look for the variation of this mean value over days leading to prediction. Therefore it is of great importance to segment out the CHs and the ARs to generate binary maps.

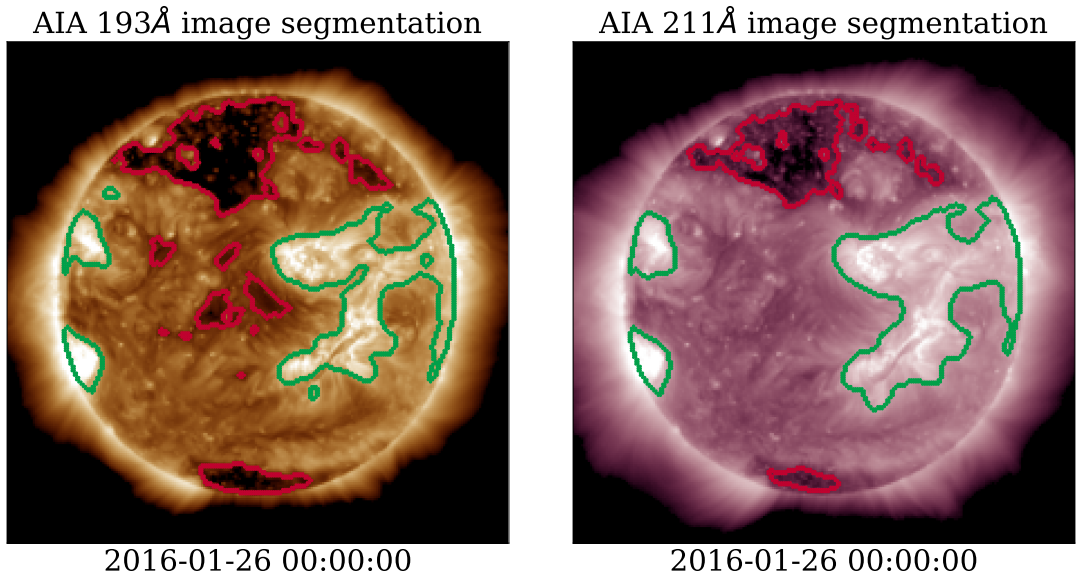

AIA 193Å image segmentation    AIA 211Å image segmentation

2016-01-26 00:00:00              2016-01-26 00:00:00

**Figure 3.7**: A representative visualization of the segmentation map of 193 Å passband (left) and 211 Å passband (right) using classical computer vision algorithms. The overplotted green contours enclose AR, and the red contours CH. The segmentation maps are created separately for the AR and CH.

To obtain the CH segmentation map, we use Otsu thresholding (Otsu 1979). This thresholding assumes the presence of two distinct classes of pixel intensities, essentially Gaussian, and tries to find an intensity value that would maximize the inter-class variance (or alternatively, minimize the intra-class variance). We use stacked thresholding – i.e, a preliminary threshold to segment out the approximate region of the coronal holes first, and then another threshold to segment out the coronal holes from this subset of the image.

The AR segmentation is far more non-trivial. Otsu thresholding picks out spurious areas as 'active regions'. Hence, we apply a 5-class Gaussian Mixture Model (Pedregosa et al. 2011) on the pixel intensities to segment out the ARs. The Gaussian with the highest mean is found to segment out the ARs well. A representative set of segmentation maps overplotted on the EUV data is shown in Fig. 3.7.

With these binary maps, we simply perform a pointwise multiplication of our activation values on a given image with its CH and AR map, respectively, while also scaling by the total area of segmentation. The scaling by area of CH and AR is done to remove dependence on the absolute size of these regions and obtain a normalized quantity. We then take the mean value over the image and across all datasets to obtain a single scalar to quantify the activation at ARs and CHs, across the days of history for both fast and slow solar wind. The activation plots are constructed for the training set (for better statistics) since the generalizability of the model is captured in its performance on the test set (or the cross-validation set).

## 3.3  Results

### 3.3.1  Model benchmarking

Table 3.4:: Correlation comparison of our model predictions with the Benchmark models. 27-day persistence gives a correlation of $0.456 \pm 0.02$. Models which do not have a correlation value are given '−'. The p-values are all less than $10^{-7}$ for WindNet variants and less than $10^{-2}$ for the benchmark models.

| (H,D) | WindNet 193 | WindNet 211 | XGBoost | Persistence | SVM Linear | SVM RBF | SVM Polynomial | Naive mean |
|---|---|---|---|---|---|---|---|---|
| (1,1) | $0.28 \pm 0.03$ | $0.34 \pm 0.02$ | $0.73 \pm 0.01$ | $0.76 \pm 0.01$ | $0.76 \pm 0.01$ | $0.76 \pm 0.01$ | $0.56 \pm 0.01$ | − |
| (1,2) | $0.37 \pm 0.03$ | $0.42 \pm 0.03$ | $0.36 \pm 0.02$ | $0.43 \pm 0.02$ | $0.43 \pm 0.02$ | $0.43 \pm 0.02$ | $0.34 \pm 0.01$ | − |
| (1,3) | $0.47 \pm 0.01$ | $0.48 \pm 0.02$ | $0.12 \pm 0.02$ | $0.19 \pm 0.02$ | $0.19 \pm 0.03$ | $0.19 \pm 0.02$ | $0.18 \pm 0.03$ | − |
| (1,4) | $0.46 \pm 0.03$ | $0.52 \pm 0.04$ | $0.02 \pm 0.03$ | $0.07 \pm 0.03$ | $0.08 \pm 0.03$ | $0.07 \pm 0.03$ | $0.07 \pm 0.03$ | − |
| (2,1) | $0.37 \pm 0.05$ | $0.42 \pm 0.02$ | $0.76 \pm 0.01$ | $0.43 \pm 0.02$ | $0.79 \pm 0.01$ | $0.79 \pm 0.01$ | $0.56 \pm 0.01$ | − |
| (2,2) | $0.47 \pm 0.02$ | $0.39 \pm 0.06$ | $0.40 \pm 0.02$ | $0.19 \pm 0.02$ | $0.47 \pm 0.02$ | $0.47 \pm 0.02$ | $0.34 \pm 0.02$ | − |
| (2,3) | $0.46 \pm 0.03$ | $0.53 \pm 0.03$ | $0.16 \pm 0.02$ | $0.07 \pm 0.03$ | $0.22 \pm 0.02$ | $0.21 \pm 0.02$ | $0.18 \pm 0.03$ | − |
| (2,4) | $0.51 \pm 0.03$ | $0.48 \pm 0.03$ | $0.01 \pm 0.03$ | $0.03 \pm 0.03$ | $0.08 \pm 0.03$ | $0.08 \pm 0.03$ | $0.05 \pm 0.03$ | − |
| (3,1) | $0.41 \pm 0.04$ | $0.51 \pm 0.03$ | $0.76 \pm 0.01$ | $0.19 \pm 0.02$ | $0.79 \pm 0.01$ | $0.79 \pm 0.01$ | $0.57 \pm 0.01$ | − |
| (3,2) | $0.46 \pm 0.03$ | $0.47 \pm 0.02$ | $0.39 \pm 0.02$ | $0.07 \pm 0.03$ | $0.47 \pm 0.02$ | $0.47 \pm 0.02$ | $0.34 \pm 0.01$ | − |
| (3,3) | $0.47 \pm 0.03$ | $0.53 \pm 0.03$ | $0.15 \pm 0.02$ | $0.03 \pm 0.03$ | $0.22 \pm 0.02$ | $0.22 \pm 0.02$ | $0.16 \pm 0.02$ | − |
| (3,4) | $0.46 \pm 0.03$ | $0.54 \pm 0.03$ | $0.03 \pm 0.04$ | $0.01 \pm 0.03$ | $0.08 \pm 0.03$ | $0.09 \pm 0.03$ | $0.05 \pm 0.03$ | − |
| (4,1) | $0.47 \pm 0.04$ | $0.54 \pm 0.03$ | $0.75 \pm 0.01$ | $0.07 \pm 0.03$ | $0.79 \pm 0.01$ | $0.79 \pm 0.01$ | $0.57 \pm 0.01$ | − |
| (4,2) | $0.48 \pm 0.03$ | $0.52 \pm 0.02$ | $0.38 \pm 0.01$ | $0.03 \pm 0.03$ | $0.47 \pm 0.02$ | $0.47 \pm 0.02$ | $0.34 \pm 0.01$ | − |
| (4,3) | $0.45 \pm 0.04$ | $0.55 \pm 0.03$ | $0.16 \pm 0.02$ | $0.01 \pm 0.03$ | $0.22 \pm 0.02$ | $0.22 \pm 0.01$ | $0.15 \pm 0.03$ | − |
| (4,4) | $0.48 \pm 0.04$ | $0.50 \pm 0.03$ | $0.04 \pm 0.04$ | $-0.02 \pm 0.03$ | $0.09 \pm 0.03$ | $0.09 \pm 0.03$ | $0.06 \pm 0.04$ | − |

From Table. 3.4 through Table. 3.7, we have summarized the performance of `WindNet`, as well as the benchmark autoregressive models for the metrics defined − Correlation ($r$), RMSE, $\chi^2_{red}$ and TS respectively. We see that `WindNet` outperforms the benchmarks over combinations where the delay is generally more than 1 - i.e, where the autoregressive models do not have the immediately preceding solar wind speed available. In fact, for larger delays and histories, `WindNet` shows consistent performance, while other models fail to perform a reasonable prediction. The best performance of `WindNet` is for a history-delay combination of $(4, 3)$, wherein the correlation is ≈ 0.55, and the spread is 0.03 – this is for 211 Å. Similarly, the best fit using 193 Å data occurs for a combination of $(2, 4)$, with a correlation of 0.51 and a spread of 0.03.

Table 3.5:: RMSE comparison of our model predictions with the Benchmark models. 27-day persistence gives an RMSE of 93.14±4.43.

| (H,D) | WindNet 193 | WindNet 211 | XGBoost | Persistence | SVM Linear | SVM RBF | SVM Polynomial | Naive mean |
|-------|-------------|-------------|---------|-------------|------------|---------|----------------|------------|
| (1,1) | 97.01 ± 4.02 | 96.64 ± 4.27 | 60.31 ± 0.62 | 62.02 ± 1.05 | 57.68 ± 0.72 | 57.68 ± 0.73 | 74.17 ± 1.32 | 88.05 ± 2.08 |
| (1,2) | 92.13 ± 2.88 | 89.45 ± 2.68 | 83.47 ± 1.00 | 95.35 ± 1.60 | 80.60 ± 1.28 | 80.55 ± 1.28 | 83.74 ± 1.75 | 87.77 ± 2.20 |
| (1,3) | 83.70 ± 1.77 | 87.34 ± 3.85 | 90.33 ± 1.30 | 113.81 ± 2.39 | 87.77 ± 1.74 | 87.78 ± 1.75 | 87.97 ± 1.91 | 88.04 ± 2.42 |
| (1,4) | 84.33 ± 2.31 | 85.94 ± 4.67 | 92.14 ± 1.73 | 122.20 ± 3.13 | 89.06 ± 1.98 | 89.07 ± 1.97 | 88.91 ± 1.94 | 88.29 ± 2.46 |
| (2,1) | 96.31 ± 4.87 | 91.12 ± 2.30 | 57.87 ± 0.65 | 95.35 ± 1.60 | 54.27 ± 0.93 | 54.19 ± 0.92 | 74.64 ± 1.80 | 87.77 ± 2.20 |
| (2,2) | 90.80 ± 2.85 | 102.85 ± 9.00 | 83.48 ± 0.83 | 113.81 ± 2.39 | 78.94 ± 1.58 | 78.98 ± 1.57 | 84.36 ± 1.90 | 88.04 ± 2.42 |
| (2,3) | 86.21 ± 2.12 | 83.38 ± 2.78 | 91.86 ± 1.19 | 122.20 ± 3.13 | 87.09 ± 1.73 | 87.17 ± 1.70 | 87.91 ± 1.85 | 88.29 ± 2.46 |
| (2,4) | 86.24 ± 2.63 | 86.53 ± 2.27 | 93.11 ± 1.14 | 125.16 ± 3.12 | 88.68 ± 1.95 | 88.72 ± 1.97 | 88.77 ± 1.88 | 88.86 ± 2.46 |
| (3,1) | 93.35 ± 4.33 | 82.60 ± 1.75 | 57.80 ± 0.80 | 113.81 ± 2.39 | 54.40 ± 1.01 | 54.34 ± 1.00 | 73.84 ± 1.70 | 88.04 ± 2.42 |
| (3,2) | 88.17 ± 1.81 | 85.46 ± 2.63 | 84.10 ± 0.86 | 122.20 ± 3.13 | 78.59 ± 1.67 | 78.55 ± 1.63 | 83.91 ± 1.85 | 88.29 ± 2.46 |
| (3,3) | 87.04 ± 1.25 | 83.97 ± 3.04 | 91.58 ± 1.05 | 125.16 ± 3.12 | 86.68 ± 1.76 | 86.75 ± 1.73 | 87.91 ± 1.79 | 88.86 ± 2.46 |
| (3,4) | 87.21 ± 2.17 | 81.21 ± 1.86 | 92.72 ± 1.28 | 126.79 ± 2.92 | 88.72 ± 1.75 | 88.62 ± 1.80 | 88.96 ± 1.67 | 89.14 ± 2.41 |
| (4,1) | 84.19 ± 2.83 | 80.27 ± 2.07 | 59.14 ± 0.82 | 122.20 ± 3.13 | 54.52 ± 1.03 | 54.48 ± 1.04 | 74.28 ± 2.07 | 88.29 ± 2.46 |
| (4,2) | 86.42 ± 1.98 | 83.06 ± 2.51 | 83.78 ± 0.74 | 125.16 ± 3.12 | 78.47 ± 1.81 | 78.45 ± 1.78 | 84.16 ± 1.88 | 88.86 ± 2.46 |
| (4,3) | 88.32 ± 1.93 | 80.28 ± 3.05 | 91.00 ± 1.25 | 126.79 ± 2.92 | 86.81 ± 1.59 | 86.82 ± 1.63 | 88.25 ± 1.68 | 89.14 ± 2.41 |
| (4,4) | 82.93 ± 1.72 | 85.34 ± 3.10 | 92.34 ± 1.34 | 128.23 ± 2.96 | 88.87 ± 1.46 | 88.78 ± 1.58 | 89.41 ± 1.40 | 89.43 ± 2.29 |

Table 3.6:: $\chi^2_{red}$ comparison of our model predictions with the Benchmark models. 27-day persistence gives a $\chi^2_{red}$ of 51.69±9.14.

| (H,D) | WindNet 193 | WindNet 211 | XGBoost | Persistence | SVM Linear | SVM RBF | SVM Polynomial | Naive mean |
|-------|-------------|-------------|---------|-------------|------------|---------|----------------|------------|
| (1,1) | 33.38 ± 4.10 | 29.63 ± 1.36 | 23.18 ± 0.82 | 24.50 ± 0.73 | 21.20 ± 0.77 | 21.20 ± 0.77 | 35.17 ± 1.91 | 41.98 ± 3.71 |
| (1,2) | 28.43 ± 1.98 | 30.80 ± 2.05 | 44.10 ± 1.75 | 57.63 ± 1.76 | 41.13 ± 1.77 | 41.09 ± 1.79 | 44.54 ± 2.65 | 39.16 ± 3.92 |
| (1,3) | 25.82 ± 1.39 | 26.27 ± 1.97 | 51.67 ± 2.08 | 81.61 ± 2.74 | 48.79 ± 2.17 | 48.80 ± 2.20 | 49.09 ± 2.57 | 40.34 ± 3.25 |
| (1,4) | 26.83 ± 2.31 | 26.01 ± 2.47 | 54.21 ± 2.55 | 95.19 ± 4.25 | 50.63 ± 2.40 | 50.64 ± 2.39 | 50.48 ± 2.48 | 44.99 ± 3.65 |
| (2,1) | 48.07 ± 11.96 | 43.75 ± 7.58 | 21.17 ± 0.67 | 57.63 ± 1.76 | 18.63 ± 0.77 | 18.59 ± 0.77 | 35.44 ± 2.38 | 39.16 ± 3.92 |
| (2,2) | 47.32 ± 8.43 | 90.87 ± 38.04 | 44.05 ± 1.05 | 81.61 ± 2.74 | 39.41 ± 1.49 | 39.45 ± 1.50 | 45.15 ± 2.44 | 40.34 ± 3.25 |
| (2,3) | 30.09 ± 3.42 | 31.60 ± 3.84 | 53.88 ± 2.23 | 95.19 ± 4.25 | 48.38 ± 2.00 | 48.46 ± 1.99 | 49.37 ± 2.45 | 44.99 ± 3.65 |
| (2,4) | 31.41 ± 3.97 | 37.90 ± 4.07 | 54.86 ± 1.92 | 99.64 ± 5.18 | 49.71 ± 1.90 | 49.76 ± 1.91 | 49.83 ± 1.96 | 49.46 ± 3.69 |
| (3,1) | 47.56 ± 6.01 | 29.16 ± 2.87 | 21.13 ± 0.65 | 81.61 ± 2.74 | 18.71 ± 0.63 | 18.67 ± 0.62 | 34.60 ± 1.90 | 40.34 ± 3.25 |
| (3,2) | 33.51 ± 3.20 | 36.66 ± 3.87 | 45.06 ± 1.14 | 95.19 ± 4.25 | 39.34 ± 1.41 | 39.31 ± 1.41 | 44.97 ± 2.29 | 44.99 ± 3.65 |
| (3,3) | 43.29 ± 4.10 | 37.54 ± 7.33 | 53.09 ± 1.86 | 99.64 ± 5.18 | 47.47 ± 1.55 | 47.56 ± 1.57 | 48.89 ± 1.92 | 49.46 ± 3.69 |
| (3,4) | 42.35 ± 5.76 | 31.52 ± 5.16 | 53.40 ± 2.06 | 101.18 ± 5.53 | 48.83 ± 1.79 | 48.73 ± 1.85 | 49.12 ± 1.90 | 50.96 ± 5.72 |
| (4,1) | 31.16 ± 1.76 | 31.18 ± 3.36 | 22.25 ± 0.40 | 95.19 ± 4.25 | 18.92 ± 0.54 | 18.89 ± 0.54 | 35.33 ± 2.23 | 44.99 ± 3.65 |
| (4,2) | 27.99 ± 3.35 | 29.59 ± 3.94 | 44.34 ± 0.80 | 99.64 ± 5.18 | 38.87 ± 1.24 | 38.86 ± 1.25 | 44.81 ± 1.93 | 49.46 ± 3.69 |
| (4,3) | 36.98 ± 3.82 | 26.83 ± 2.20 | 51.47 ± 2.06 | 101.18 ± 5.53 | 46.73 ± 1.53 | 46.76 ± 1.62 | 48.36 ± 2.02 | 50.96 ± 5.72 |
| (4,4) | 31.90 ± 3.38 | 32.58 ± 4.83 | 52.66 ± 2.21 | 103.16 ± 5.66 | 48.70 ± 1.81 | 48.62 ± 1.94 | 49.35 ± 2.06 | 52.29 ± 4.11 |

The Naive mean model has no variance, so there would be no correlation associated with it – however, it is presented for the sake of completeness. Autoregressive SVM using an RBF kernel seems to perform better given the solar wind speed closer to the day of prediction but falters as more delay is induced. The linear SVM performs as well as the non-linear RBF kernel, but the polynomial kernel fails to get a good fit. The 27-day persistence is a set of just 5 models – thus, this performance is stated in the caption of the respective Tables.

Table 3.7:: HSE Threat Score comparison. The 27-day persistence model gives a TS of $0.506 \pm 0.029$. Cases with TS $0.0$ imply a value less than $1e-3$.

| (H,D) | WindNet 193 | WindNet 211 | XGBoost | Persistence | SVM Linear | SVM RBF | SVM Polynomial | Naive mean |
|-------|-------------|-------------|---------|-------------|------------|---------|----------------|------------|
| (1,1) | 0.081±0.023 | 0.112±0.040 | 0.776±0.037 | 1.000±0.000 | 0.748±0.038 | 0.748±0.038 | 0.263±0.026 | 0.0 |
| (1,2) | 0.042±0.022 | 0.150±0.042 | 0.329±0.012 | 0.858±0.037 | 0.288±0.027 | 0.271±0.028 | 0.162±0.035 | 0.0 |
| (1,3) | 0.167±0.008 | 0.140±0.041 | 0.061±0.015 | 0.351±0.021 | 0.0 | 0.0 | 0.029±0.013 | 0.0 |
| (1,4) | 0.212±0.042 | 0.206±0.047 | 0.036±0.018 | 0.199±0.024 | 0.0 | 0.0 | 0.0 | 0.0 |
| (2,1) | 0.203±0.064 | 0.227±0.029 | 0.711±0.036 | 0.858±0.037 | 0.850±0.022 | 0.845±0.021 | 0.292±0.027 | 0.0 |
| (2,2) | 0.293±0.022 | 0.297±0.040 | 0.423±0.096 | 0.351±0.021 | 0.461±0.017 | 0.449±0.017 | 0.148±0.033 | 0.0 |
| (2,3) | 0.225±0.036 | 0.198±0.037 | 0.030±0.027 | 0.199±0.024 | 0.0 | 0.0 | 0.020±0.011 | 0.0 |
| (2,4) | 0.239±0.045 | 0.282±0.044 | 0.043±0.038 | 0.215±0.022 | 0.0 | 0.0 | 0.0 | 0.0 |
| (3,1) | 0.310±0.051 | 0.259±0.031 | 0.753±0.024 | 0.351±0.021 | 0.850±0.026 | 0.844±0.026 | 0.323±0.026 | 0.0 |
| (3,2) | 0.237±0.048 | 0.292±0.029 | 0.472±0.107 | 0.199±0.024 | 0.426±0.018 | 0.408±0.024 | 0.113±0.029 | 0.0 |
| (3,3) | 0.328±0.038 | 0.287±0.017 | 0.116±0.047 | 0.215±0.022 | 0.0 | 0.0 | 0.0 | 0.0 |
| (3,4) | 0.357±0.031 | 0.294±0.026 | 0.024±0.022 | 0.236±0.026 | 0.0 | 0.0 | 0.0 | 0.0 |
| (4,1) | 0.286±0.037 | 0.309±0.027 | 0.737±0.034 | 0.199±0.024 | 0.849±0.032 | 0.845±0.034 | 0.292±0.031 | 0.0 |
| (4,2) | 0.298±0.040 | 0.200±0.039 | 0.428±0.083 | 0.215±0.022 | 0.431±0.024 | 0.428±0.025 | 0.157±0.037 | 0.0 |
| (4,3) | 0.289±0.070 | 0.200±0.056 | 0.115±0.044 | 0.236±0.026 | 0.0 | 0.015±0.009 | 0.011±0.010 | 0.0 |
| (4,4) | 0.251±0.049 | 0.314±0.080 | 0.035±0.022 | 0.307±0.033 | 0.0 | 0.0 | 0.007±0.006 | 0.0 |

## 3.3.2 `WindNet` **prediction**

In this section, we investigate the variation in prediction for our `WindNet` models. The model with the highest correlation, as mentioned previously, is for a history of 4 and a delay of 3 for 211 Å. As can be seen in the Table. 3.4, there seems to be a subtle trend of an increase in correlation with history for a given delay for short delays. The performance of models with a delay smaller than history seems mostly consistent within the error bars. For the 193 Å model in Table. 3.4, it can be seen that an increase in delay for a given history results in almost a consistent prediction correlation for high history models (again, within the errorbars – though the mean values do not seem to follow an ordered trend), except in the case of 1-day history, where the correlation increases. This trend of increase in delay for a given history is largely followed in the 211 Å data, though the $4$-day history seems to be the most consistent in this case within the errors and the best performing. In general, the expectation would be an increase in correlation with increasing history and some form of variation due to an increase in delay. The variation in performance with history for small delays is fairly consistent between both 193 Å and 211 Å with only the actual correlation values being different – however, larger delay models do not have the same variation in performance for 193 Å and 211 Å. 211 Å, in fact, seems to be a better passband for solar wind prediction since the corresponding models have higher correlation means and smaller standard deviations. Short-delay and short-history models (for example, 1-day history and delay) do not perform as well as models with larger history and delay (for example, 4-day history and 3-day delay) since the solar wind is yet to arrive at L1. 193 Å data shows a peak in correlation at 2-day history and 4-day delay. The 211 Å data shows a similar peak at 4-day history and 3-day delay.

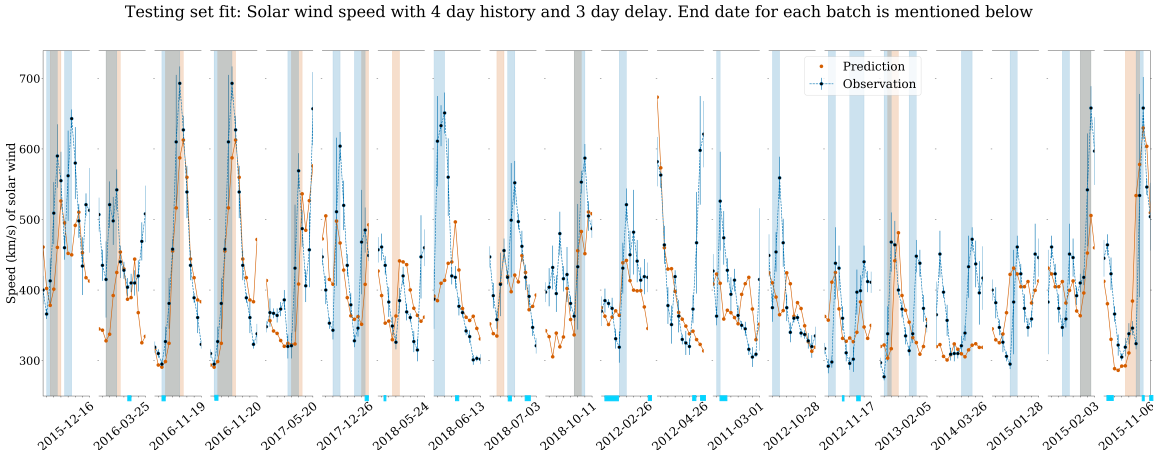

**Figure 3.8**: Wind-speed prediction plot from one of the cross-validation models, using 4 days of image data with 3 days of delay. On the x-axis, the ending date of each batch is shown. Since batches are randomly assigned to each cross-validation fold, the dates are not kept in order. The model has a correlation value of 0.61, RMSE of 76.4 km/s, and $\chi^2_{red}$ of 19.35. The error bars are measurement errors of the wind-speed observations. The HSE are highlighted by their start and end times – blue for the observed wind speed and red for the predicted wind speed. The blue bars below the plot indicate ICMEs.

A summary of RMSE is shown in Table. 3.5, and a similar summary of $\chi^2_{red}$ is shown in Table. 3.6. The TS is tabulated in Table. 3.7. The TS table shows that our proposed model has a maximum of 0.357±0.03. The low TS may be explained better by a careful observation of Fig. 3.8. This is a plot of one of the cross-validation models using 211 Å, having the highest correlation, with 4-day history and a 3-day delay of data. With 10 TP, 2 FP, and 13 FN, the model is seen to have a TS of 0.4, a correlation with the observation of 0.61, RMSE of 76.4 km/s, and $\chi^2_{red}$ of 19.35. Here, we see that there are many more HSE present in the observed wind speed, which seems to be missing from the prediction. However, upon careful observation, it may be seen that many of the observed HSE do correspond to an enhancement in the wind speed of the predictor – either at the exact time step or with a lag/lead of 3 to 4 days. However, the predicted values do not show a drastic enhancement more than the prescribed thresholds. Thus, these events are not marked as HSE.

The `WindNet` performance on the error metrics, though, largely complements the correlation performance and shows `WindNet` has better performance than the benchmark models for delays larger than 1 day in most cases.

### 3.3.3 Activation visualization

We next analyze our models' activation for various data days using Grad-CAM activation maps (§. 3.2.4). Fig. 3.9 shows a sample Grad-CAM map from a fast and slow wind prediction using 211 Å data and a similar map are shown from 193 Å prediction in Fig. 3.10, for comparison. We see that the CHs are activated for the prediction of the fast wind, and the ARs are predominantly activated for the slow wind prediction. Note that some CH activation is also seen in the 193 Å slow wind map. The CH peak activation for fast wind occurs 3 to 4 days prior to prediction, which seems to corroborate with the correlations independently obtained Vršnak et al. (2007).

The slow wind activation peaked at the AR close to the day of prediction (and also at the earliest day prior to prediction for 211 Å data), with activation at other regions of the Sun further away from prediction. We hypothesize this might be due to bias of the LSTM to the most recent input to the network – but this is still a hypothesis.

To understand the statistics of activation given to CHs and ARs, we look at the mean activation value (as described previously) and plot it for 'fast-wind' and 'slow-wind' predictions from the model. While each cross-validation model set will have its own activation plot, we present the plot for both 193Å and 211Å models using 4-day history of data with 3 days of delay. We also plot the activation for the models using 4-day history of data with 2 days of delay since it shows consistent (and good) performance using both 193 Å and 211 Å data. These trends are shown in Fig. 3.11 and Fig. 3.12 respectively.

It can be seen from these plots that the fast solar wind induces greater activation at the CHs closer to the day of prediction, and the activation (at CH) decreases as we go farther into the past – however, for the 211 Å model, the activation shows a slight increase. The fast wind also seems to activate the AR at much further times – for both 193 Å and 211 Å. Note, however, that the peak CH activation is larger than the peak AR activation for 193 Å – for 211 Å data, they are consistent within the errors in Fig. 3.11. For the same parameters in Fig. 3.12, CH peaks at 3 days prior to prediction for both the passbands and then goes down. Interestingly, however, the AR also seems to be activated to a similar level but much further away from prediction time.

For the slow wind, activation for ARs remains high for much longer than the CHs – however, the peak occurs closer to the day of prediction rather than further away from prediction. This trend is seen in both Fig. 3.11 and Fig. 3.12.

## 3.4 Discussion

Identification of solar wind source regions and the problem of solar wind prediction can be approached in two ways. The first is through purely theoretical modeling of the mechanism, while data is accumulated to constrain the physical parameters. The second method is to let the data speak for itself using purely empirical modeling and then attempt to extract the physics.

We propose the `WindNet` to empirically model solar wind speed using AIA imagery data. We are able to predict the solar wind speed better with the 211 Å data and obtain a correlation of $0.55$ with the observed wind speed in the cross-validation. The best-performing models using 193 Å and 211 Å outperform most of the larger delay benchmark models and the 27-day persistence model. The $\chi^2_{red}$, which accounts for uncertainty in the measurement itself, indicates that our best models outperform the 27-day persistence and are only slightly worse off than an autoregressive model with a single day delay – more so for lead time predictions of 3-4 days.

We then study the possibility of uncovering the associations between coronal sources and the solar wind speed using `WindNet`. To this end, we use Grad-CAM as an 'explainable AI' tool to understand the activation at different spatial locations in the EUV images, given the solar wind modality.

The Activation plots suggest that the `WindNet` pays attention to certain solar features consistent with heuristic expectations from solar wind theory. We see that CHs are deemed important 3-4 days prior to the prediction of a fast wind, while the ARs are deemed important for a slow wind prediction with misplaced timing. The CH-fast wind association is seen with strong significance in the 193 Å passband. The slow wind association is seen in both the passbands, albeit with misplaced timing. These are indicative of CHs and ARs potentially being sources of these two different kinds of the solar wind, as also known from literature (Krieger et al. 1973; Brooks et al. 2015). However, we must note that all we are observing and interpreting is one aspect of the real process occurring underneath. Thus, care must be taken while trying to understand these observations and results in the context of our hypothesis and models. Especially the significance of the interpretation of activation values depends on a couple of other factors:

- **Fitting error of our model**: We still have a maximum correlation of 0.51(0.55) for the 193 Å(211 Å) data. A higher correlation points to a more confident estimation of the source region of the solar wind.

- **Visualization**: The Grad-CAM used in this work gives a very coarse localization of activation and thus may not point to the precise origin of the particular kinds of wind.

- **Segmentation**: Defining a region as CH or AR accurately is difficult with intensity values alone – ideally, one would require extrapolated magnetic field lines to check for these structures.  Thus, an accurate definition of CHs based on intensity is required.  In this work, we attempted to automate the CH and AR definitions using histogram analysis.  Thus there is bound to be some form of uncertainty.  Hence, better segmentation methods may accurately capture the entire activation within a CH or an AR, and give a much better estimate of activation per unit area.  Furthermore, the slow wind is known to arise from the boundaries of ARs (as the outflow regions) – thus, there is a need to segment out the inner core and the boundary regions of the ARs.

From a purely forecasting perspective, at first glance, `WindNet` may appear to not outperform existing models in terms of the metrics used.  However, comparing our model to existing models (like the regressive models of Rotter et al. (2015), or Wang & Sheeley Jr (1990)) would be an apples-to-oranges comparison since:

- We perform predictions over multiple Carrington rotations on the 8-year dataset.

- Our prediction target is the daily averaged solar wind speed, which must be compared to daily averaged predictions by other models.

- We perform 5-fold cross-validation on this dataset.  However, due to a lack of confidence intervals in the previous results, we are unable to check if our results are statistically different from the existing models.

Thus, any benchmarking of our model must be done with models undergoing the same data preparation procedure, the same span of data, and at the same cadence.  We thus do not compare our results with the existing aforementioned models.

To overcome this limitation, we propose empirical benchmark models, not unlike the existing empirical solar wind prediction models.  In this regard, `WindNet` shows reasonable performance vis-a-vis the benchmark models; however, numerous improvements are possible.

- Data preparation: As H+D increases, more samples are discarded (as explained in §. 3.1.4).  This may be made more efficient by performing the Cross

Validation (CV) first, then splitting it into folds later with the downside of high memory consumption.

- ICME mitigation: Our random assignment of 5-fold cross-validation is to ensure the ICMEs are distributed uniformly across all the folds, thereby influencing all the CVs equally. Due to an inadequate number of ICME samples, we do not characterize them.

- Network architecture: Better architectures may be designed to improve the prediction vis-a-vis the observations, or more novel ML methods may be employed for a direct prediction.

- Visualization: Visualization of ML models is a hot area of research in the ML community – thus, more accurate visualization techniques may be expected to emerge in coming years.

- TS evaluation: As seen in §. 3.3, the HSE capturing algorithm misses many potential enhancements due to the speed increases not satisfying the absolute speed change criteria. Hence, the TS evaluation should be taken with caution.

This work serves a twofold purpose. One, it is the first step toward training and testing various ML models for predicting other solar wind target parameters, such as proton density, temperature, and magnetic field (specifically, $B_z$). Two, we have demonstrated the potential exhibited by DL to probe and uncover salient associations between different processes using techniques of explainable AI. Such techniques serve as independent verification and validation of conventional 'source mapping' techniques, which depend on global extrapolation, radial backtracking, and abundance matching. All of these techniques can be potentially merged together in the near future to generate stronger constraints on the possible sources of solar wind.

To this end, the code and data used in this work are open-sourced on GitHub: `https://github.com/Vishal-Upendran/WindNet`. Our publicly released source code promotes reproducible research by allowing others to reproduce the results presented here. This includes data partitioning, cross-validation, model training, and evaluation. This code base can be built upon by other researchers to further improve the performance of solar wind prediction models. Furthermore, with the ever-increasing research on Interpretable AI, this codebase may be used by researchers to come up with various methods of visualizations to quantify the source regions of solar wind.

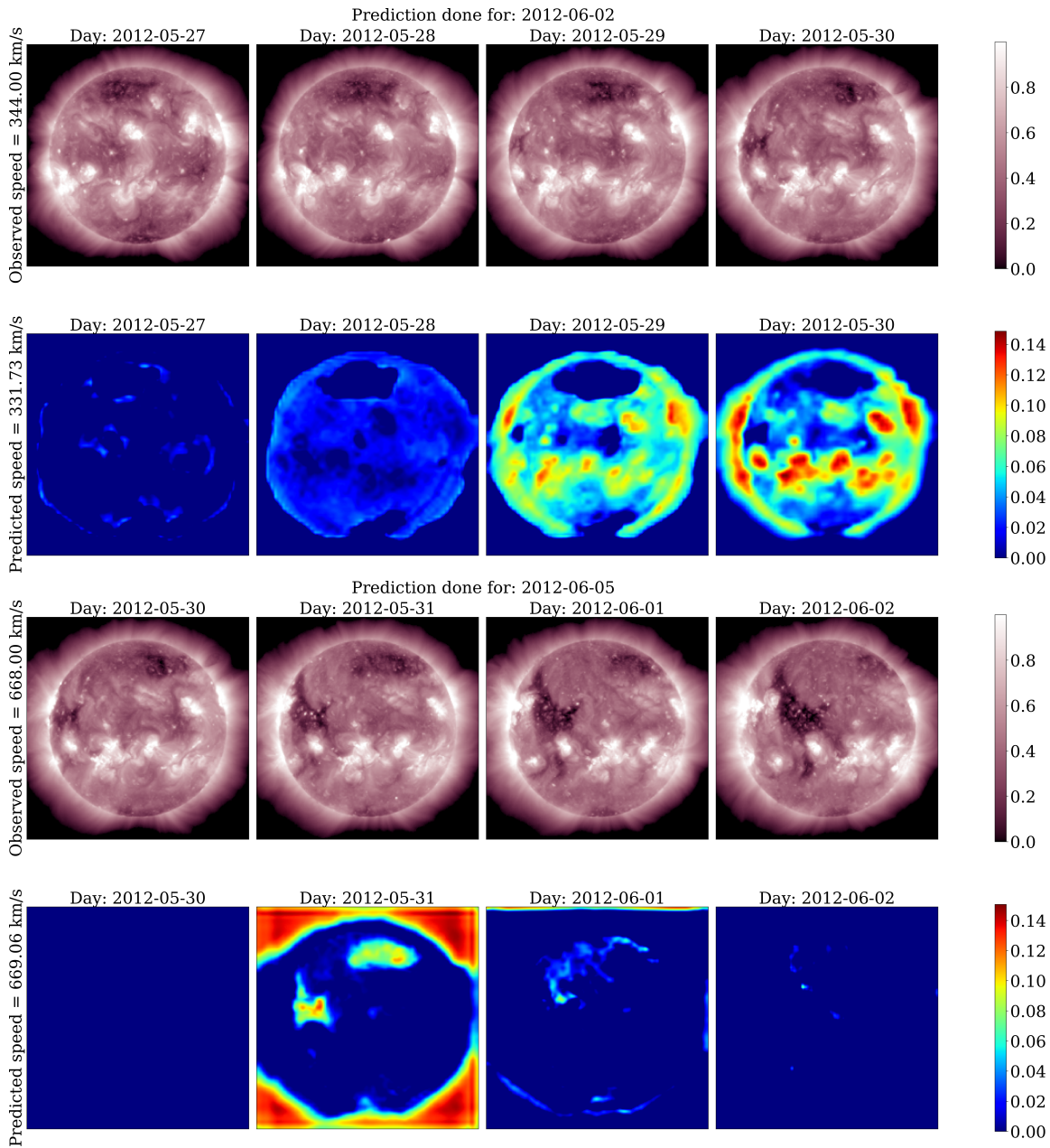

**Figure 3.9**: GC activation maps for a slow (top) and fast wind (above) prediction using 211 Å data, with the colourmap corresponding to each row given on the right. The activation maps and the images have been rescaled between 0 and 1 row-wise for ease of comparison. For the fast wind prediction, note how the maximum activation occurs at the CH, 3 to 4 days prior to prediction, which seems to match with the correlations obtained in the literature Vršnak et al. (2007). The slow wind, on the other hand, activates the AR closer to the prediction, with no activation at the CH.

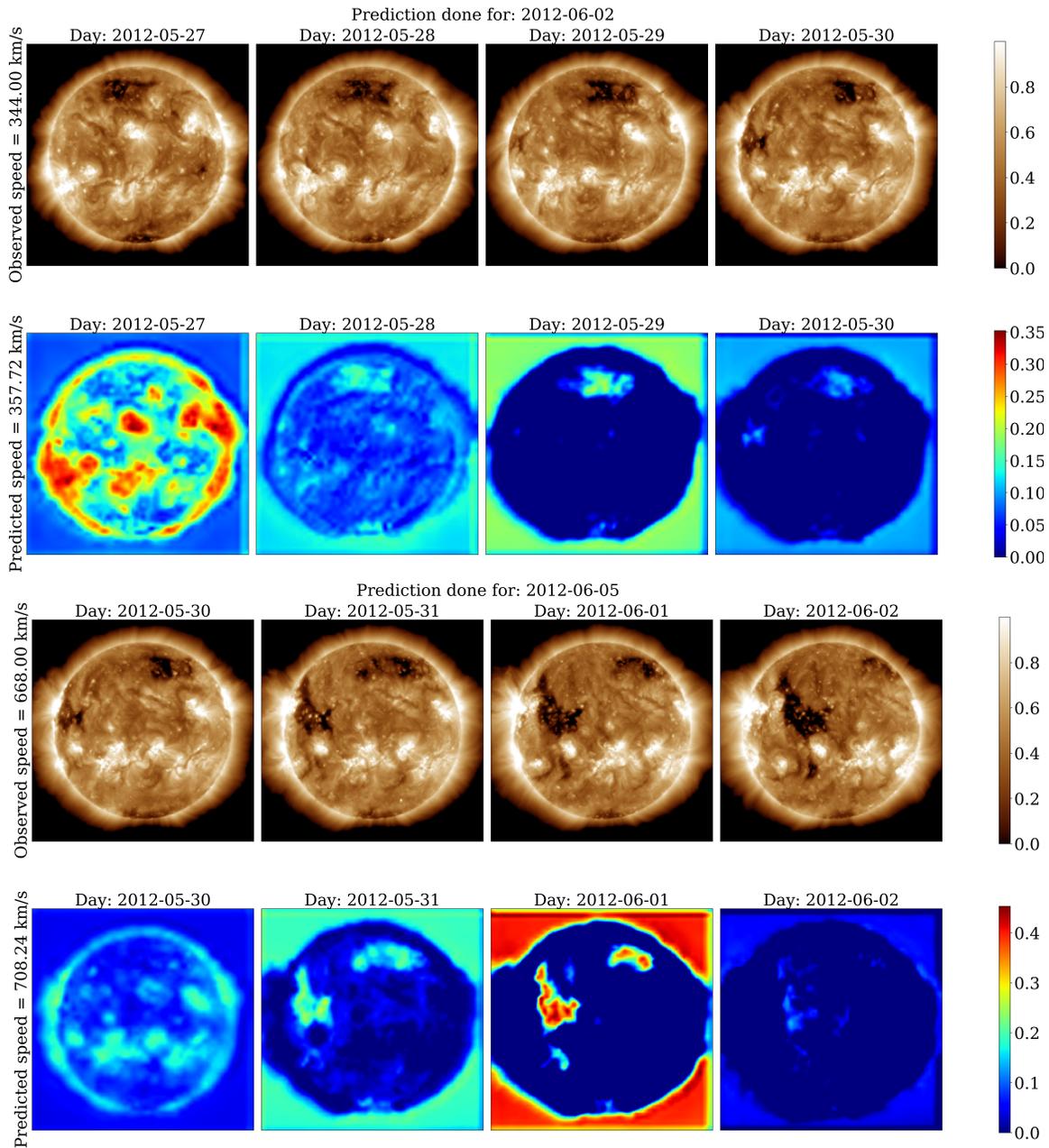

**Figure 3.10**: GC activation maps for a slow (top) and fast wind (above) prediction using 193 Å data, with the colourmap corresponding to each row given on the right. The activation maps and the images have been rescaled between 0 and 1 row-wise for ease of comparison. For the fast wind prediction, note how the maximum activation occurs at the CH, 3 to 4 days prior to prediction, which seems to match with the correlations obtained in the literature Vršnak et al. (2007). The slow wind, on the other hand, activates the AR both closer and further away from prediction and activated at the small CH on the closest day to prediction. However, other regions of the quiet Sun show a higher activation further away from the day of prediction. The slow wind activation is quite mixed and unclear when compared with the fast wind activation.

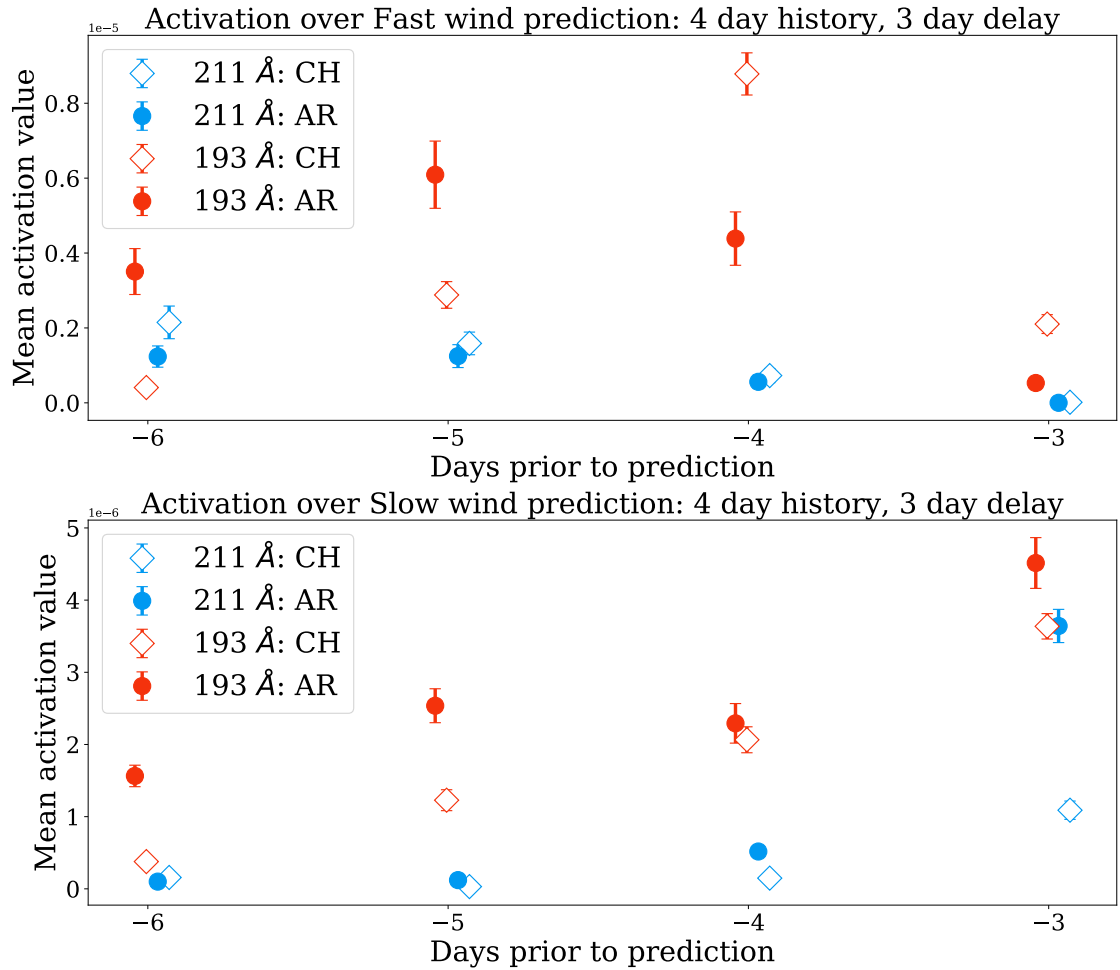

**Figure 3.11**: Variation of mean activation for the 4-day history and 3-day delay model, for fast and slow solar wind prediction, respectively. The activation is shown for models using 193Å and 211Å data respectively. The activation is shown over CH and AR alone. The error bars indicate the standard error on the mean value, estimated from the standard deviation of the sample of activations. Please note that the error bars here represent $3S$, i.e., thrice the standard error to make sure they are visible. Those activations with seemingly no error bars have very small errors.

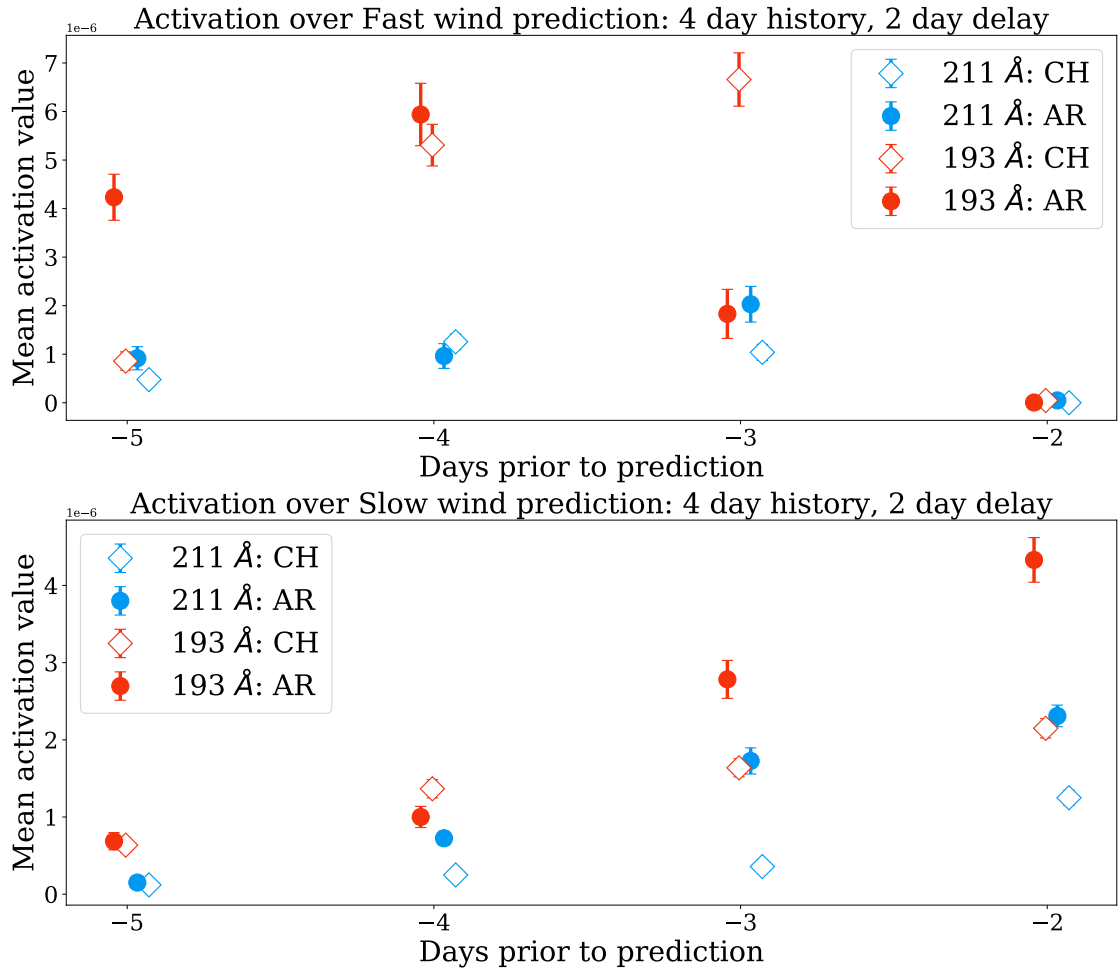

**Figure 3.12**: Variation of mean activation for the 4-day history and 2-day delay model, for a fast and slow solar wind prediction, respectively. The activation is shown for models using 193Å and 211Å data respectively. The activation is shown over CH and AR alone. The error bars indicate the standard error on the mean value, estimated from the standard deviation of the sample of activations. Please note that the error bars here represent $3S$, i.e., thrice the standard error to make sure they are visible. Those activations with seemingly no error bars have very small errors.

# Chapter 4

# The impulsively heated quiescent solar corona

*A complete understanding of the million-degree Kelvin solar corona demands the study of QS regions. In this work, we first study QS regions in the 171 Å, 193 Å and 211 Å EUV passbands of the AIA, and in the 1–1.3 keV, 1.3–2.3 keV, and 1–2.3 keV X-ray channels of XSM. We perform these studies by combining the empirical impulsive heating forward model of Pauluhn & Solanki (2007) with a machine-learning inversion model that allows uncertainty quantification. Through such an exercise, we provide constraints on the impulsive event frequency ($p_f$), timescales ($\tau$), and the distribution slope ($\alpha$) across all the energy bands. Furthermore, we also provide constraints on the absolute event amplitudes (in erg) and the radiative loss flux in X-ray owing to flux-calibrated measurements. This thesis chapter is adapted from a set of two papers that originally appeared in the literature as **On the Impulsive Heating of Quiet Solar Corona** (DOI: 10.3847/1538-4357/abf65a ) and **Nanoflare Heating of the Solar Corona Observed in X-rays** (DOI: 10.3847/2041-8213/aca078).*

We have seen the solar wind and its sources in §. 1.2.2 and Ch. 3. However, the solar wind itself owes its existence to the anomalously hot solar corona, which we had a glimpse of in §. 1.2.1, bringing us to the coronal heating problem.

We have seen in Ch. 1 that while the solar corona is at a million degrees Kelvin, it typically depicts three major morphological features – Coronal Holes (CHs), Active Regions (ARs), and Quiet Sun (QS). All three regions are approximately at million degrees Kelvin. The ARs are predominantly found during the maximum of the solar cycle, while the CHs dominate during the minimum. The QS, however, is present irrespective of the activity phase. Conventional studies on the heating of the corona predominantly focus on ARs (see Reale 2014, for a review) – with studies in QS and CHs being sparse (see, e.g. Tripathi et al. 2021b). Thus, to address the physics of coronal heating in general, a comprehensive study of the heating in QS regions that exist independent of the solar cycle is of utmost importance(Aschwanden 2014).

One of the most popular mechanisms to heat the corona is via impulsive events, for example, through Nanoflares (see, e.g. Parker 1988a). Impulsive events are transients generated through the dissipation of magnetic stresses or waves (see, e.g., Antolin et al. 2008). Solar atmosphere presents us with a plethora of impulsive events at various energies and time scales, viz. flares (Benz 2008; Tripathi et al. 2004), microflares (Hannah et al. 2008; Schadee et al. 1983; Subramanian et al. 2018; Chifor et al. 2006, 2007), active region transient brightenings (Testa et al. 2013; Gupta et al. 2018a; Tripathi 2021; Vilangot Nhalil et al. 2020; Rajhans et al. 2021), transition region blinkers (Harrison 1997), UV bombs (Peter et al. 2006; Gupta & Tripathi 2015), Ellerman bomb (Ellerman 1917; Pariat et al. 2004; Isobe et al. 2007a) and other activities such as jets (Mulay et al. 2016a; Raouafi et al. 2016). It is also observed that the properties of loops in ARs are well described by the impulsive heating scenario (Ghosh et al. 2017; Tripathi et al. 2009a; Warren et al. 2008; Gupta & Tripathi 2015; Winebarger et al. 2013b, and references therein). Such a scenario has also been studied independently in ARs using a variety of techniques like Time lag analysis (Viall & Klimchuk 2012, 2013, 2015, 2016, 2017), Differential Emission Measure (DEM) and Doppler shifts analysis (see, e.g., Tripathi et al. 2010, 2011, 2012; Winebarger et al. 2011; Warren et al. 2012; Winebarger et al. 2013a; Subramanian et al. 2014; Del Zanna et al. 2015), hydrodynamic modeling (see, e.g. Bradshaw et al. 2012; Reep et al. 2013; Cargill et al. 2015; Barnes et al. 2016), Magneto hydrodynamic modeling (see, e.g. Rappazzo 2015; Rappazzo et al. 2017; Knizhnik et al. 2018, 2019; Knizhnik & Reep 2020) and empirical models (see, e.g. Jess et al. 2019, 2014). Thus, it is natural to assume that the coronal heating in QS may also be governed by impulsive heating. However, since the QS regions have a very diffuse structure, it is not possible to count individual events and understand the energetics of these events in the QS.

We have seen in §. 1.2.1 that to maintain the corona at a million degrees, the frequency distribution of impulsive events must follow a power law distribution in energy – i.e., $\frac{dN}{dW} \propto W^{-\alpha}$, with $\alpha > 2$ (see Hudson 1991). Observations do show that impulsive events in the corona follow a power-law distribution. However, there is a range of $\alpha$ values reported in the literature (see, for example, Fig. 6.14 in Aschwanden 2019, or Fig. 1.5 earlier in this thesis).

One of the most significant caveats in these results occurs due to the assumption that each detected event is a single entity. But, we know that small-scale impulsive events may happen at a sub-resolution scale (Pauluhn & Solanki 2007). Hence, individual brightenings may not be a single entity but may consist of many tiny events. This may lead to the undercounting of events, particularly at lower energies. These individual events, however, leave a collective imprint on the entire light curve in a statistical manner. These imprints have been shown to be statistically reflected in the intensity distribution of light curves from Active Regions (see, for example, the analysis by Vekstein 2009; Terzo et al. 2011; Jess et al. 2014, 2019) and coronal loops seen in X-rays (Sakamoto et al. 2009). While individual events may not be measured, their cumulative effect on the statistical properties of intensity light curves can be leveraged to understand these events. Thus, the existence of such small-scale events may only be inferred statistically.

Typically, a 'statistically-realistic' simulation would reproduce some salient properties of the observations well. A statistical and impulsively heated mechanism may leave signatures in the distribution of intensity, the characteristic temporal features, or in the thermal structure of plasma (see, for e.g. Sturrock et al. 1990; Hudson 1991; Sylwester et al. 2019; Rajhans et al. 2021). Hudson (1991), for example, show that the relative interplay of frequency of occurrence of events and the time scale of the events reflect in the temporal power spectrum of the emergent light curves.

However, some observations may be used to develop simple, empirical models. Typically, the observations of QS radiance in UV and EUV, both in space and time, show log-normal distribution (Pauluhn et al. 2001; Andretta & Del Zanna 2014). Thus, the QS radiance might be generated due to some form of a Markovian process (Pauluhn & Solanki 2007; Gorobets et al. 2016).

Using the Markovian property of the QS light curve, Pauluhn & Solanki (2007) proposed an empirical model for heating the QS corona. Hereafter, we call it the `Pauluhn and Solanki Model (PSM)`. In brief, `PSM` approximates the response of the plasma to a unit heating event as an exponential rise and fall of intensity. The amplitude of the heating event is sampled from a power-law distribution, with the frequency of occurrence of the events being kept as a free parameter. The resultant light curve is then a combination of a multitude of these events. Hence this addresses the sub-pixel resolution scale of these structures. Moreover, the resulting intensities are also log-normal, mimicking the observations. Finally, the observed power-law distribution of energetic events is also incorporated into this model, enabling us to understand the viability of impulsive events maintaining the observed intensity in a given light curve. Thus, `PSM` may provide an excellent proxy for the generation of the QS coronal intensities.

The overarching goal of any such statistical model is to constrain its free parameters given observations. Once again, we circle back to the fact that we have a hypothesis, which we then try to constrain/update/discard using available observations. Pauluhn & Solanki (2007) generate the parameters for different observations by comparing the similarity of intensity distribution and the Global Morlét wavelet power (Torrence & Compo 1998) of simulated light curves with those of observed light curves. The comparison is qualitative – a sufficiently good match in distribution and the frequency with excess power in the power spectrum was taken to represent a good match between the observation and simulation. Although the comparison had a sound basis, it was done by eye and needed a more quantitative foundation.

The problem of obtaining parameters from the observed light curve thus becomes an inversion problem (or "parameter estimation"). In ML language, we have a 'supervised learning' problem at hand. In general, the inversion approaches depend primarily on generating important "features" from the light curves, which then have a one-to-one mapping with the parameter set of PSM. This is performed qualitatively by Pauluhn & Solanki (2007). Since it is not trivial to objectively pick out features for inversion, one trains an inversion model to pick out abstract features and perform the inversion. In this case, an optimization principle guides the mapping from light curves to the parameter sets developed by the inversion model.

Tajfirouze & Safari (2012); Bazarghan et al. (2008) employed this method by using a Probabilistic Neural Network (PNN). Under this scheme, every simulated light curve is classified, where each class has a unique combination of free parameters. The PNN is trained on the full set of simulated light curves. Finally, the observed light curves are fed into the PNN, which assigns each to one of the learned classes. For their study, Tajfirouze & Safari (2012) used $\approx 10,000$ light curves (at max). These light curves corresponded to CHs, QS, and ARs on the Sun obtained once again from AIA and data from the Extreme UltraViolet Imager (EUVI) on board STEREO (Kaiser et al. 2008). On average, they obtained $\alpha > 2$ for all the regions.

This method is a great start to a tricky stochastic inversion problem. But, such an approach must be well-validated by an exhaustive testing set. Moreover, the classification of light curves imposes a discretization constraint on the parameter set, which depends on the grid resolution of the simulation. Thus, any assigned class to a given light curve may change on improving the grid resolution. In other words, we do not know the confidence level of the PNN for each inversion.

In this chapter, we develop an inversion scheme using machine learning called the `iPSM`. This model is a CNN that takes in the light curves under consideration and performs a regression on the target parameter set. We have seen CNNs earlier, in §. 2.2 and §. 3. Recapping, a CNN may be thought of as a set of 'kernels,' which perform convolution on the input and return a convolved output. The kernel size can be interpreted as the scale over which information is summarized. Several such kernels are operated on a given input, and these outputs are passed through a non-linear function called the 'Activation function.' Successive kernels are thus sensitive to local scales of the corresponding input, but that input itself may be an extremely non-linear transformation of the original light curve. We use a CNN in this work to preserve information on the time scales of features in the light curve along with the distribution of intensity values

Furthermore, we employ the Inception module used in Ch. 3 for developing `iPSM`. Since the Inception module makes it possible to perform multi-scale analysis at each level, we hope to capture the multiple scales of variation in the data, thereby circumventing grid resolution issues by considering the target as a regression problem. We can effectively interpolate between simulated grid points while performing the inversion. We also obtain associated inversion uncertainties with the CNN by perturbing the network.

Now, note that Tajfirouze & Safari (2012) used poor resolution of AIA 171 Å data ($90$s cadence and $2''$spatial resolution) and further binned the data spatially by $3 \times 3$ and $5 \times 5$ window. Such a binning will average out the small-scale events and, thus, will not allow the use of the full potential of AIA observations. Hence, we perform analysis with two sets of data. First, we use the full spatial, and temporal resolution observations that were taken using AIA and perform the parameter estimation using `iPSM`, thereby circumventing the potential caveats in the work of Tajfirouze & Safari (2012); Bazarghan et al. (2008) as detailed above. However, the AIA data also come with their own caveats – the biggest caveat comes due to a lack of absolute flux calibration. Thus, we also apply `iPSM` with an updated parameter search on full-disk integrated, flux-calibrated data from the Solar X-ray Monitor (XSM) onboard Chandrayaan-2 mission (Vadawale et al. 2014; Shanmugam et al. 2020; Mithun et al. 2021a) of the Indian Space Research Organization (ISRO).

This chapter is structured as follows: we first describe the `PSM` in §. 4.1, which is common to all analysis, following which we describe the DL inversion scheme `iPSM` in §. 4.2. Then, we describe the AIA dataset and associated noise in §4.3.1 & §4.3.2, while we present the corresponding results in §4.3.5. We then change gears to study the X-ray data by describing the data in §4.4. Specific to these X-ray observations, we have a different way of incorporating the measurement uncertainties while also updating the `iPSM` to include a parameter search. These we explain in §4.4. We then showcase the results for X-ray observations in §4.4.6. Finally, we discuss the consequences of our results in §4.5.

## 4.1 The PSM

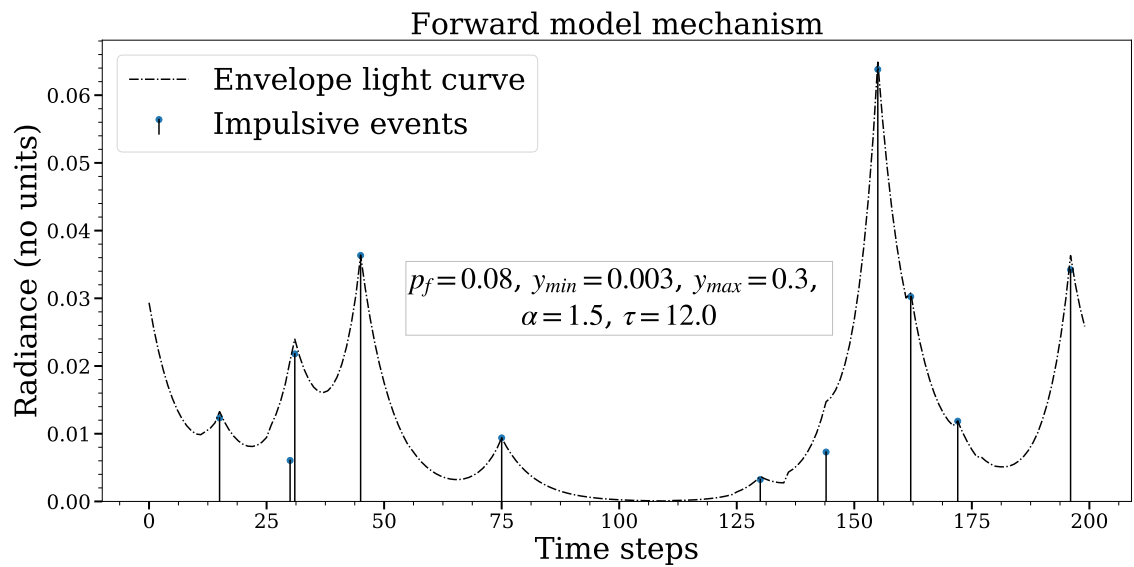

**Figure 4.1**: An example showing light curve generated using the Nanoflare generation model, similar to Fig. 2 of Pauluhn & Solanki (2007).

The forward model employed here, as mentioned earlier, is the PSM. This is an empirical model based on two key observations, i.e., log-normal distribution of spatial and temporal distribution of QS intensities (Pauluhn et al. 2001), and power law distribution of energies from flares to micro-flares (see, e.g. Aschwanden 2019). The algorithm may be summarised by asking the following questions:

1. What is the probability of a flare occurring at a given time step?

2. If a flare is meant to occur at the given time step, what would its peak energy be?

3. How long will the flare last once it has occurred, i.e., the duration of the flare?

In this model, there are 5 free parameters: the event or flaring frequency ($p_f$), i.e., the probability of a flare to occur at a given time; the duration of the individual flare event ($\tau$); the power law slope ($\alpha$) and the minimum ($y_{min}$) and the maximum ($y_{max}$) amplitude that is allowed for an individual flare event, which provides the bounds of the power law. An example simulated light curve, depicting the formation of the light curve from individual events, is shown in Fig. 4.1. It may be seen from here that given only the envelope light curve, the individual events may not be inferred. The simulations are performed over a grid of parameter space while fixing $y_{min}$ and $y_{max}$, in preparation for the inversion of the observed light curves.

## 4.2 The `iPSM`

We perform the inversion using a 1-D CNN. In this approach, generally, there are convolution layers followed by an activation layer. The activation function is a nonlinear function that forms the core of the complex learning ability of any NN. We use the function `Elu` as defined in Eqn 4.1 as activation for all layers except the last since `Elu` enables the network to train faster and generalize better (Clevert et al. 2015). For the final layer, we have no activation since this is a regression problem mapping to a continuous variable.

$$Elu := \left\{ \begin{array}{ll} e^x - 1 & \text{, if } x \leq 0 \\ x & \text{, else} \end{array} \right. \tag{4.1}$$

A graphical representation of the model architecture is shown in Fig. 4.2. As de-picted in the figure, there are input/output layers, convolution layers (where we implicitly assume the activation function to be present), and fully connected (FC) layers. In the tensor shape, 'None' is generally used to denote a variable size, representing the number of light curves to be inverted during a single forward pass. The convolution layers (marked in red) are given a 4-dimensional shape, representing [height, width, input channel, output channel], and the stride size given by an integer. Note that "channel" here is not to be confused with AIA passbands and that since we have used a 1-D signal, the height is set to 1. After a suitable number of convolutions, the array is unrolled to 1D and captured by fully-connected layers (marked in yellow).

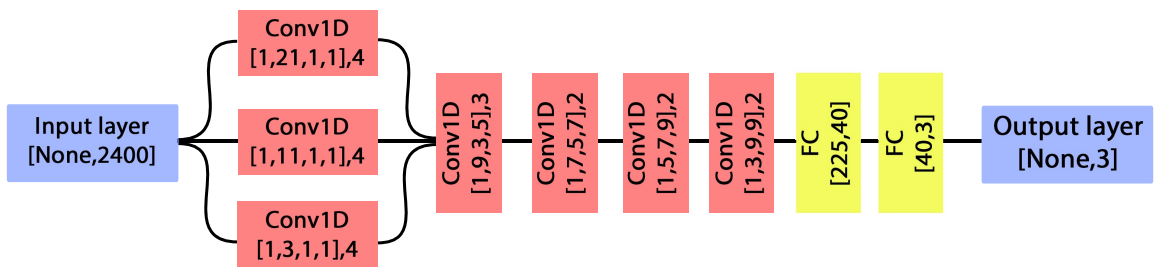

**Figure 4.2**: CNN architecture used in `iPSM`. The blue boxes indicate input/output layers, and the other colors indicate trainable layers. The tensor shapes are given in square brackets as [ ], and the number outside, for the Convolution layers, is the stride size. FC denotes the fully connected layers.

Any CNN would need the number of features in the input and output to be fixed – i.e., the length of the light curve and the number of parameters to be inferred must be fixed. We fix these to be of length 2400 time steps and 3 parameters – i.e., we infer only $p_f$, $\tau$, and $\alpha$ with this scheme.

### 4.2.1   Data preparation and training

For a given grid of parameters – which depends on the dataset used –, we divide the simulated light curves into a training set (80%) and a testing set (20%). However, before feeding any data to the CNN, it must be prepared appropriately to ensure all the parameters are of the same scale. The training set parameters are rescaled between 0 and 1, and the testing set parameters are rescaled using the training statistics, as is the standard procedure in machine learning. All light curves are also rescaled between 0 and 1.

We train the CNN by feeding in the training set light curves and obtain three free parameters ($p_f$, $\tau$, and $\alpha$). This obtained parameter set is then compared with the original target parameter set using an error metric, which is used to update the kernel values of the CNN. The error metric is the sum of the two terms defined as:

$$\mathcal{L}_1(\hat{\mathbf{x}}, \mathbf{x}) := \Sigma_i \left| x_i - \hat{x}_i \right|, \mathcal{L}_2(\hat{\mathbf{x}}, \mathbf{x}) := \Sigma_i \left( x_i - \hat{x}_i \right)^2, \tag{4.2}$$

where $\mathbf{x}$ is the target parameter set, $\hat{\mathbf{x}}$ is the predicted parameter set, and the summation is performed over all observations and all parameters.

In Ch. 3, we have seen how any ML/DL exercise involves both model and some free parameters called 'hyperparameters.' Such a case holds for this work too. The hyperparameters for generating the `iPSM` are fixed. For training the `iPSM`, we have used the "Adam Optimizer" (Kingma & Ba 2014), which is a stochastic optimization algorithm. The "size" of the update at each step is controlled by the hyperparameter called the learning rate.

Overfitting is a serious issue in NN training whereby the model starts fitting the noise in the model and stops generalizing. This may result in erroneous results and interpretations of inversion. To prevent overfitting, we use dropout (Hinton et al. 2012), which switches off neurons randomly with a fixed probability for every forward pass. The training hyperparameters are summarized in Table. 4.1.

## 4.2.2 Uncertainty estimation

A CNN generates a single prediction for a given forward pass. However, in general, two kinds of uncertainties – Epistemic and Aleatoric- are associated with any such predictions (Kendall & Gal 2017). Epistemic uncertainty relates to model uncertainty due to the unexplored weight space of the neural network. Aleatoric uncertainty relates to the inherent uncertainty in the target parameter values. Our study has no Aleatoric uncertainty because the parameters are predefined grid points. Hence, we have only Epistemic uncertainty.

Deficiencies in model training, resulting in unexplored weight space, can occur if not enough data is provided during model training. Hence, this effect can be simply minimized by increasing the size of the training set. However, while informing us about the deficiencies in fitting, the epistemic uncertainty measure may also inform us about any outliers in the dataset. Throughout the analysis in this work, the PSM is assumed to be the ground truth, i.e., it fully describes the observed light curves to infer parameters. This is never the case with any model. Hence, departures of the observations from simulations, where the PSM does not fully explain the given observation, would behave as outliers. Hence higher uncertainties associated with the parameters inferred from the observations tell us either there are deficiencies in model fitting in certain regimes or the light curve is not explained well by the PSM. However, we note that it is practically impossible to disentangle these effects (see, e.g., Kendall & Gal 2017).

The epistemic uncertainty may be estimated by application of dropout (Hinton et al. 2012; Díaz Baso et al. 2019). In addition to being used to prevent overfitting,

Table 4.1:: Training Hyperparameters

| Hyperparameter | Value |
|---|---|
| Cost function | $\mathcal{L}_1$(prediction,target)+$\mathcal{L}_2$(prediction,target) |
| Optimizer | Adam Optimizer with default values |
| Learning rate | $1e-3$ |
| Dropout rate | $0.2$ |
| No. of iterations | $3000$ |

Dropout can also be used to create perturbations and obtain the variability in the predictions (Gal 2016). Since the neurons switched off in every forward pass are random, we perform a Monte Carlo forward pass to obtain multiple realizations of the CNN and present the mean and standard deviation from the passes. Thus, the error bars reported for the estimated parameters of an individual light curve are the standard deviation obtained from Dropout.

## 4.3   Analysis on AIA EUV data

### 4.3.1   Observations and Data

As we have seen in §. 2.1.1, AIA observes the Sun's atmosphere in UV and EUV bands using eight different passbands sensitive to plasma at different temperatures (O'Dwyer et al. 2010b; Boerner et al. 2012b). For our analysis, we consider the data taken using 171 Å, 193 Å, and 211 Å passbands. These images are taken with a pixel size of $\sim 0.6''$ and an approximate time cadence of 12 s. We have chosen these particular passbands since the count rates in these passbands for QS is large when compared to others (see Table 2 in O'Dwyer et al. 2010c).

Monitoring the EUV images on `Solar Monitor` [1], we identified QS patches during 2011 and 2019, where no activity was observed. Details of the two data sets (DS1 & DS2) are given in Table. 4.2. We have obtained eight continuous hours of data for each set, corresponding to 2400 time steps. All the images are aligned to the first image and are exposure time normalized. The full FOV (single snapshot) for DS1 and DS2, as observed in 171 Å, is displayed in the left panel of Figs. 4.3 & 4.4 with the corresponding spatial distribution of intensities in the right panels.

To study the distribution of the intensity for each pixel, we create light curves of intensity in all three passbands. We plot sample light curves from both the datasets for one passband and their corresponding distribution in Figs. 4.5 & 4.6 for a single pixel. Note that we also show time series of magnetic flux density of the corresponding pixel taken from the Helioseismic and Magnetic Imager (HMI; Scherrer et al. 2012) on board SDO corresponding to the AIA Field of View (FOV). The LOS magnetograms are obtained by HMI at approximately 45 s cadence with a pixel size of $0.5''$. We map the HMI data to the same plate scale as that of AIA. The error in HMI LOS measurements is estimated to be $\pm$ 10G (Yeo et al. 2014; Couvidat et al.

---

[1]https://www.solarmonitor.org/

Table 4.2:: Dataset details for QS heating in EUV

| Identifier | DS 1 | DS 2 |
|---|---|---|
| Start time | 2011-08-14 T00:00:00 | 2019-05-02 T00:00:00 |
| End time | 2011-08-14 T08:00:00 | 2019-05-02 T08:00:00 |
| Reference time | 2011-08-14 T00:00:00 | 2019-05-02 T00:00:00 |
| Xcen,Ycen | 192″, 749″ | 19.0″, 211.5″ |
| FOVx, FOVy | 230″, 116″ | 346.0″, 269.0″ |
| Instrument | AIA | AIA |
| Passband | 171,193,211 | 171,193,211 |
| Exposure normalize | True | True |
| Cadence | 12 sec | 12 sec |

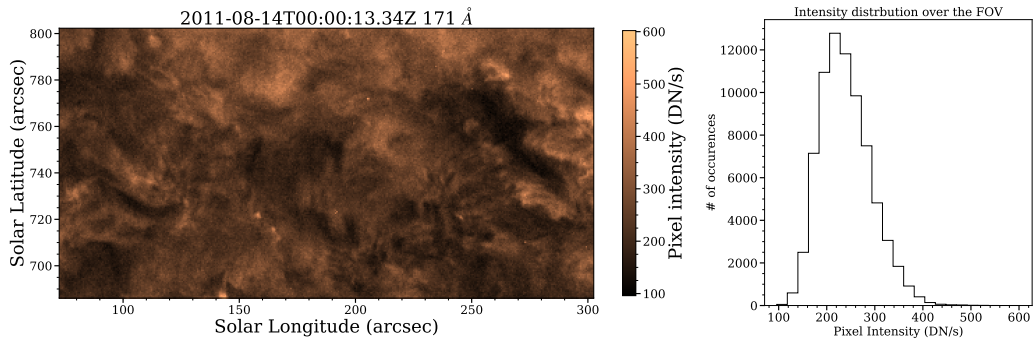

**Figure 4.3**: 171 Å image of quiet Sun corresponding to DS1, and the corresponding histogram of intensity.

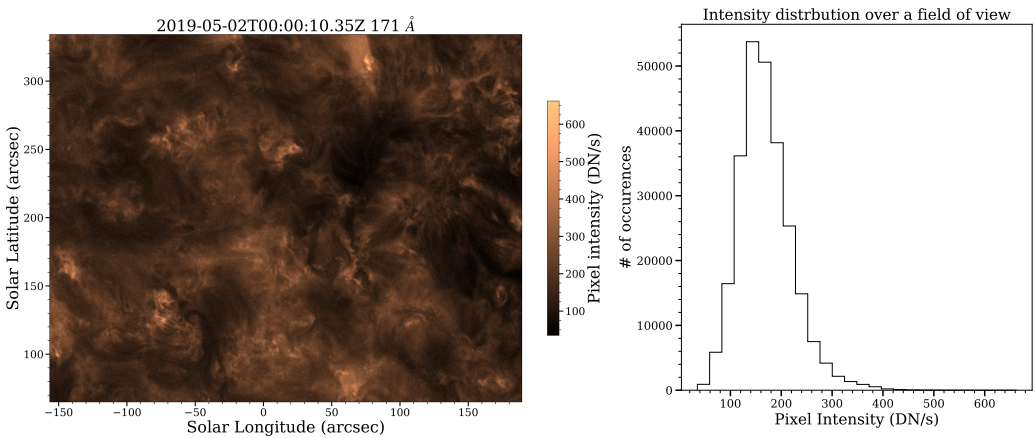

**Figure 4.4**: Same as Fig. 4.3 but for DS2.

2016), and this is depicted in the figure as the black horizontal line. The time series for intensity and magnetic flux density are shown in panels b and c. The distribu-tions are shown in panels a and d, respectively, which demonstrate the log-normal distribution, as was previously observed by Pauluhn & Solanki (2007) in SUMER observations and Tajfirouze & Safari (2012) in AIA observations.

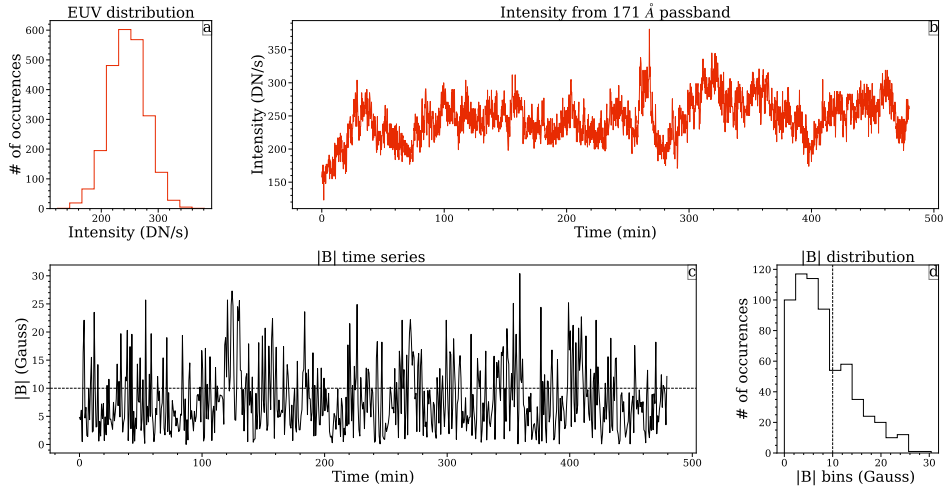

**Figure 4.5**: Intensity and Magnetic field intensity time series of 1 pixel from the FOV of DS 1, and their corresponding distributions, as labeled. The 10 G noise level has been indicated in (c) and (d).

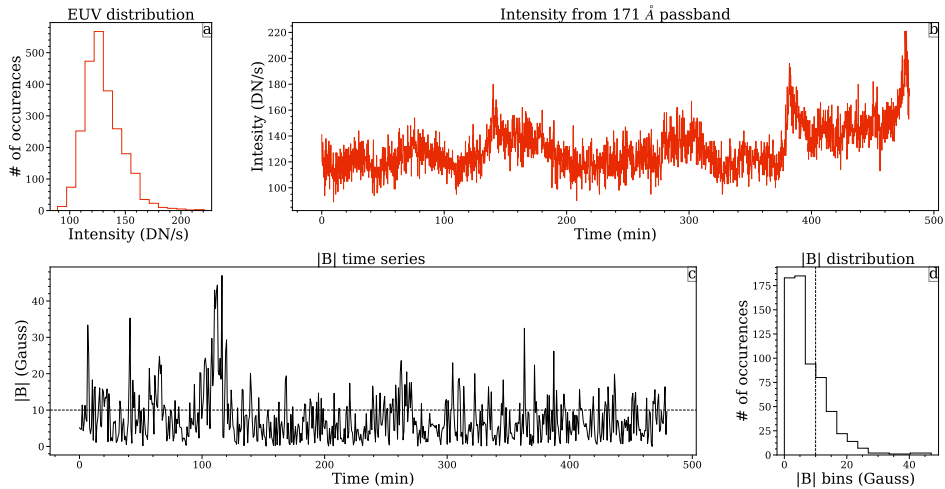

**Figure 4.6**: Same as Fig. 4.5 but for DS2.

### 4.3.2 Noise characterization

The observed EUV light curves, shown in Figs. 4.5 and 4.6, have inherent noise, which is essentially dominated by photon shot noise. Thus, we would need to mitigate the presence of this noise. Generally, the noise would be incorporated into the simulation set so that the neural network learns to differentiate between noise augmentation and signal. However, we have $\approx 900,000$ light curves across all the passbands, and incorporating the specific noise characteristics of each light curve into the simulation would increase the dataset manyfold. Due to a lack of computation power, we do not perform such a task. Instead, we come up with a scheme to reduce this noise while preventing over-smoothing (and thus averaging over real events). We call this procedure `Finding kneemo`.

`Finding kneemo` is based on existing knee analyses performed in Machine learning. Broadly, the goal of the algorithm is to monitor a performance metric against the free parameter of smoothing, which, in our case, is the size of the smoothing window. The window size for which we observe a drastic improvement in performance metric is taken as the box-car window. The change point is generally known as "the knee".

The knee determination is extremely qualitative, though some methods exist which quantify this well (see, for example Salvador & Chan 2004). In our analysis, we consider a random light curve for a pixel in our data set, along with its error time series, obtained from `aia_bp_estimate_error.pro`. We smooth the light curve using a box-car of box size varying between 1 and 100 and obtain its modified SNR (Signal-to-Noise Ratio). We then plot the obtained SNR against the box size in Fig. 4.7, along with the asymptotes of the SNR, and find their point of intersection. This point is then selected as the box-car window size. From Fig. 4.7, we find that the asymptotes intersect at a box-car size of 5-time points. Therefore, we use this value to enhance SNR. Note that we have run this analysis on several lightcurves within our dataset and have found a consistent result. Thus, we have performed box-car averaging with a box size of 5-time points for all the light curves in our dataset. An example plot with the original and the smoothed light curve is shown in Fig. 4.8. Our observed data is now ready. We next need to ready the simulation grid and get the `iPSM` trained and ready.

### 4.3.3 `PSM` grid for EUV data

We perform the simulations over a large parameter space, as detailed in Table 4.3. We generate the light curves for a length of $5L + 1600$, where $L$ is the length of the light curve (2400 in our case), and reject 800 samples from either side to remove boundary effects. The remaining light curve is folded 5 times to get a final light curve of length $L^2$. This is done to minimize the effects of the initial seed for random number generators. The observed light curves are normalized by their median values, following Bazarghan et al. (2008). Hence, the radiances reported from hereon have no associated units. Finally, we train our model with the simulations and perform inference on all three passbands with the same inversion model.

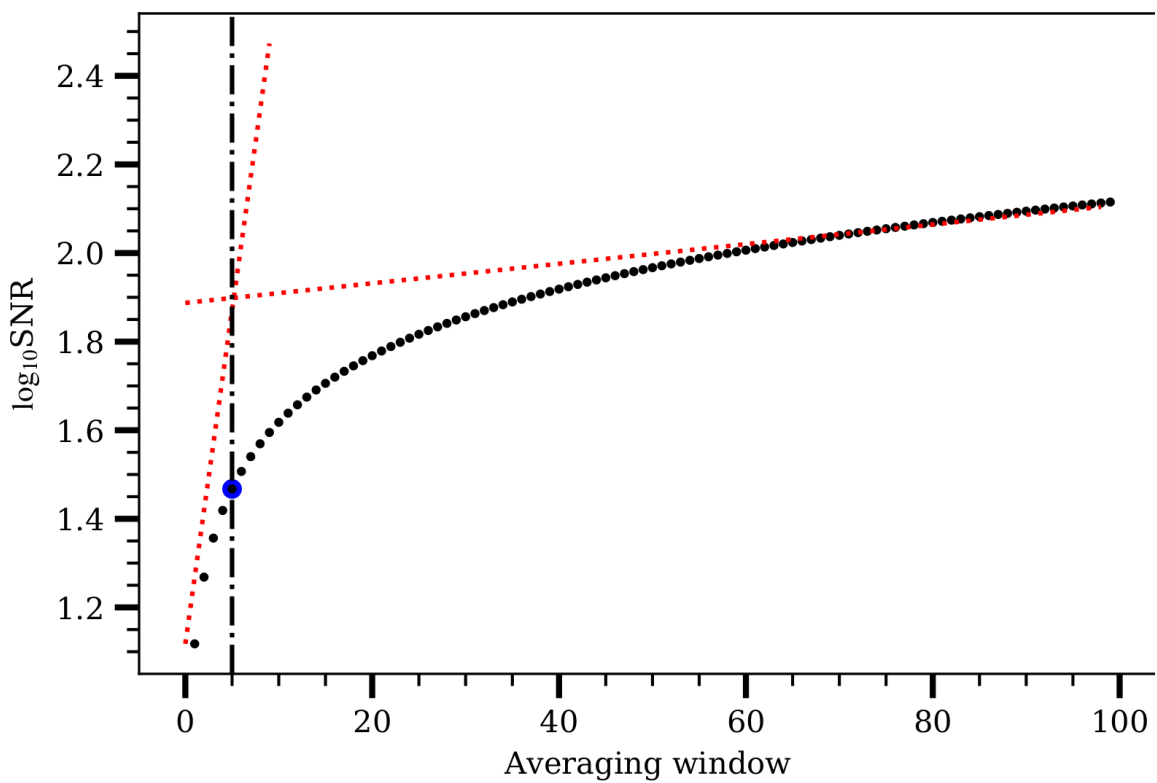

**Figure 4.7**: SNR variation with box-car window. The black dots represent the variation, while the red dotted lines represent the asymptotes. The vertical black line shows the approximate point of intersection (marked as blue)

Table 4.3:: Simulation grid parameters. Note that all parameters are in code units.

| Parameter | Range | Stepsize |
|---|---|---|
| $p_f$ | $[0.05, 0.95]$ | steps of 0.05 |
| $\alpha$ | $[1.1, 3.0]$ | steps of 0.1 |
| $\tau$ | $[1, 100)$ | steps of 2.0 |
| $y_{max}$ | 0.3 | |
| $y_{min}$ | 0.03 | |

---

[2]Folding essentially divides up the light curve in 5 equal chunks of length L and gets the average curve from these chunks

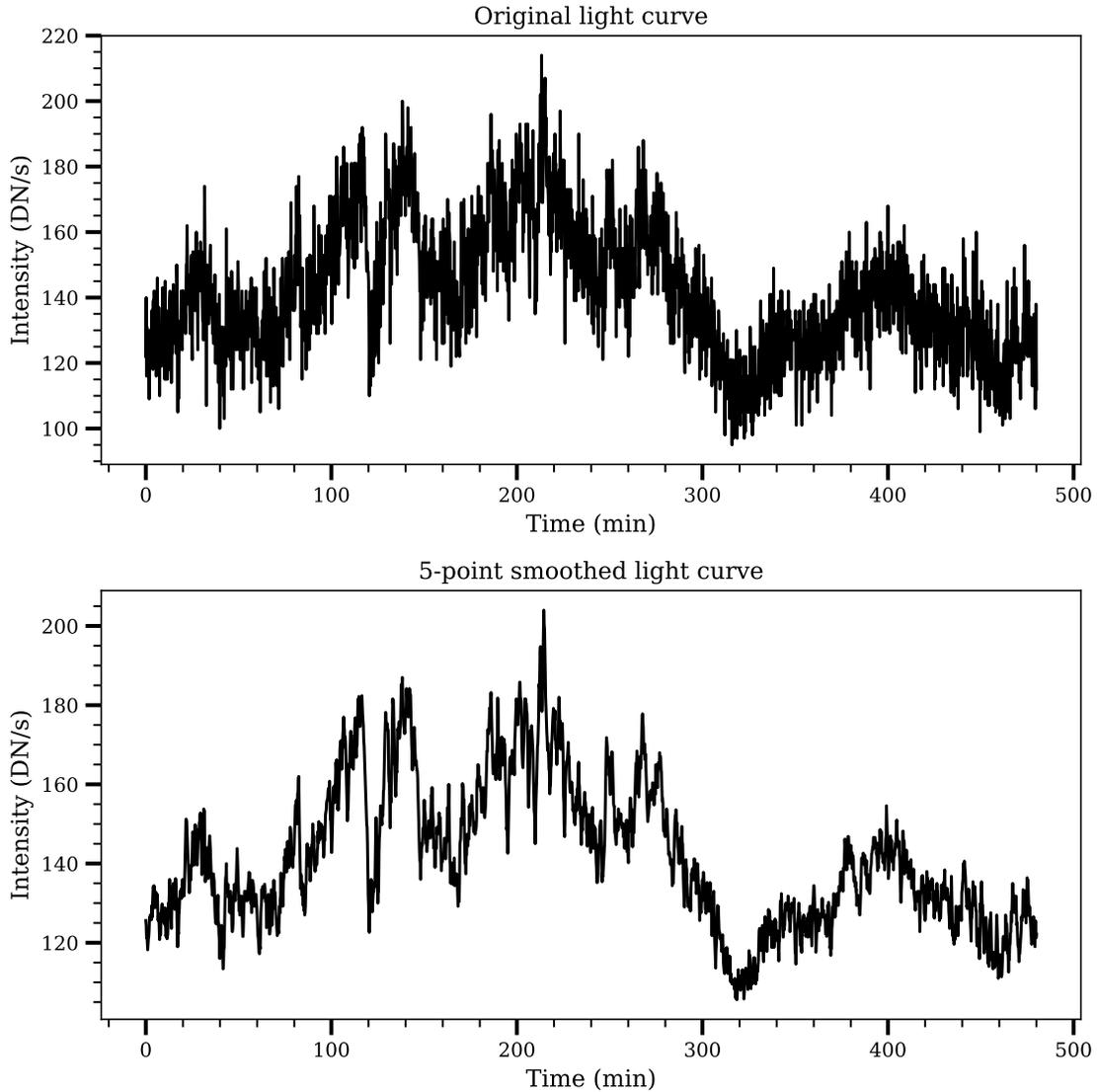

**Figure 4.8**: Comparison of original and smoothed light curves by `Finding kneemo`.

### 4.3.4   EUV `iPSM` inversion performance

To assess the performance of `iPSM`, we display scatter plots between the target and predicted values of $p_f$ (left panel), $\tau$ (middle panel), and $\alpha$ (right panel) in Fig.4.9. As can be readily noted, the predicted values lie very close to the target values, thereby validating the performance of our network on a test set. This may be quantified using the coefficient of determination ($R^2$) defined as:

$$R^2 := 1 - \frac{\Sigma_i \left(x_i - \hat{x}_i\right)^2}{\Sigma_i \left(x_i - <x>_i\right)^2},\tag{4.3}$$

where $<x>$ represents the mean of target set, $\mathbf{x} = \{x_i\}$ represents the target values, and $\hat{\mathbf{x}} = \{\hat{x}_i\}$ represents the predicted values. In this case, $i$ corresponds to the number of points in the test set, i.e., $R^2$ is computed separately for each target parameter. The $R^2$ values are 0.990, 0.999 and 0.97 for $p_f$, $\tau$ and $\alpha$, respectively, showing excellent performance of our network. Our network is now ready to be fed with the observed intensity light curve.

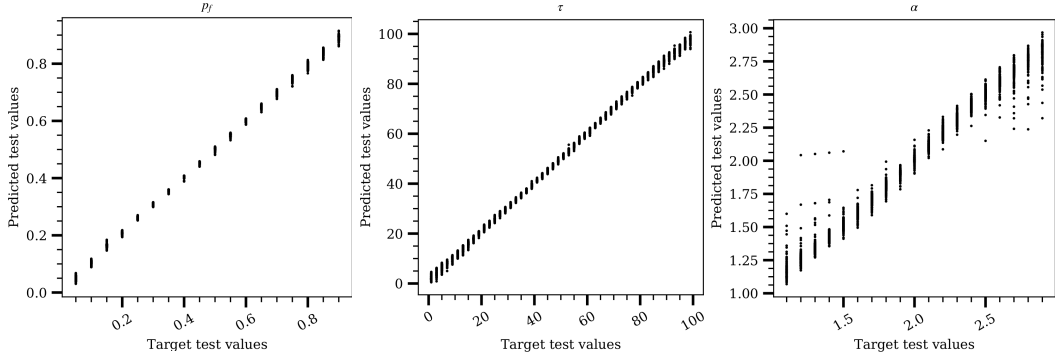

**Figure 4.9**: Correlation plot of target and predicted parameter values from our CNN, for $p_f$ (left), $\tau$ (center), and $\alpha$ (right). This quantifies the generalizability of the CNN from the training set.

### 4.3.5   EUV: Results

Now we discuss the application of the network on the EUV light curves. We first discuss the results obtained for a single light curve in §4.3.5. Then, we discuss the results obtained for all light curves obtained for all three AIA passbands. We finally explore the various correlations between our parameters and perform an analysis of the involved energetics.

**Application of the CNN on a Single Light curve**

For representation purposes, we choose the intensity light curve for a random pixel from DS1 and DS2 and obtain the corresponding simulated light curve. It is important to note that since `PSM` is a statistical model which generates a representation of the observations "statistically", one should not perform a point-by-point comparison of the simulations with the observations. A simple change in the random number generator's seed can change the exact times when events occur. Furthermore, since the amplitude of events is sampled from a distribution, the random seed value can also change the event amplitude at particular times. However, these seeds cannot change the overall statistical properties of the light curve. Thus, a comparison of the observation and simulation must be made using statistical properties of light curves (e.g., intensity distribution, frequencies showing enhanced power) rather than a pointwise comparison of light curves. Once a good representative simulated light curve is obtained, the corresponding parameter set is taken to characterize the observed light curve.

In Figs. 4.10 and 4.11, we show the comparison between observed (orange) and simulated (blue) light curves (panels a), Kernel Density Estimation (KDE) of intensity distribution (panels b), the Global Morlét power spectra (panels c) and the cumulative distribution function (CDF; panels d). Note that the observed and simulated light curves are normalized by their median values. The parameter sets for the simulated light curves are the mean values of the obtained parameter distribution by performing 1000 Monte Carlo simulations and are denoted at the bottom of the figures. The KDE can be understood to be essentially a continuous extension of histogram (see, e.g. Chen 2017). Note that $p_f$ and $\tau$ are defined as per time step in Fig. 4.9, and may be converted into real units as

$$p_f(\text{per min}) = \frac{p_f(\text{inferred})}{\text{Cadence(min)}},$$

and

$$\tau(\text{min}) = \tau(\text{inferred}) \times \text{Cadence(min)}.$$

The $p_f$ denoted henceforth is given as the number of events per minute, while $\tau$ is the timescale in minutes. The $\alpha$ remains a dimensionless parameter. From Fig. 4.10a and 4.11a, we can see an excellent statistical correspondence between the observed and simulated light curve. This is corroborated by the match in their corresponding KDEs (panel b) and CDFs (panel d). Furthermore, the Morlét wavelet power spectrum shows peaks at corresponding frequencies for both the observed and simulated light curves. These results confirm that the Inversion model was able to learn both the time series and distribution properties corresponding to the 3 free parameters. We further emphasize that the value of $\alpha$ inferred in these two cases is $\geq 2$, which in turn suggests that events with smaller energy are dominantly responsible for generating the radiance of these particular examples.

From Figs. 4.10 and 4.11, we find a clear relationship between the goodness of representation of simulated light curve (using CDF and Morlét Power) and the spread of parameters obtained by Monte Carlo simulations. Consider the percentage uncertainty (i.e., uncertainty/mean value) − we find this quantity is approximately 3% in $p_f$ for DS1, 6% in $p_f$ for DS2; ∼ 5% in $\tau$ for DS1 and ∼ 4.5% DS2; and ∼ 12% for DS1 and ∼ 15% for DS2. The Inversion model is hence more certain of the parameters of DS1 than DS2, which is also reflected in the relative mismatch of Morlét power between the observation and simulation for DS2 over DS1 at the first two peaks (note the difference in y-axis limits in panels c). Thus, such an uncertainty measure, along with the Monte Carlo forward pass, can help us explain which parameters are strongly influencing the quality of a given inversion, assuming `PSM` as the ground truth.

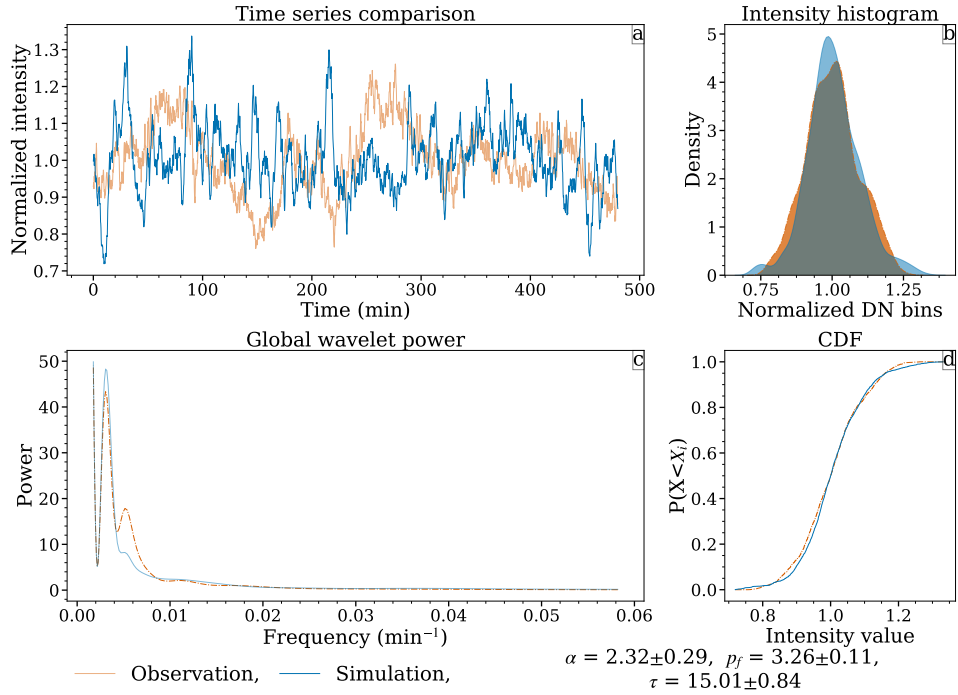

**Figure 4.10**: The comparison of a representative light curve obtained for 171 Å passband from DS1 with a simulated light curve. Observations are shown in orange translucent, and simulations are shown in blue. Panel a: Normalised observed and simulated light curves; Panel b: KDE of observed and simulated light curves; Panel c: Global Morlét power for observation and simulations; panel d: Cumulative Distribution Function (CDF) comparison of simulation and observation.

**Multi-light curve - multi-passband analysis**

Since our network gives reliable results for the light curve obtained for a random pixel in both the data set, we now take all of our light curves (331967 light curves per passband) for the three AIA passbands and pass them through the network to obtain the relevant parameter set for each light curve. Due to operational constraints, we perform only 100 Monte Carlo forward passes in this case. We emphasize that the obtained parameter set for 100 and 1000 Monte Carlo forward passes are statistically the same for a limited, handpicked set of representative light curves. For both data sets, we first divide each light curve by its median value, rescale between 0 and 1, perform the Monte Carlo forward pass through the CNN, and obtain the mean parameter set. Finally, we concatenate the parameter set across the whole field of view for both data sets separately for each AIA passband to improve our statistics. This concatenation can be done since all light curves are from QS regions and are evolving independently. Fig.4.12 displays the distribution of flaring frequency $p_f$ (panel a), duration $\tau$ (panel b), and power-law slope $\alpha$ (panel c) for this concatenated dataset. The solid blue curves are for 171 Å, dashed-dotted red for 193 Å, and dashed green are for 211 Å observations.

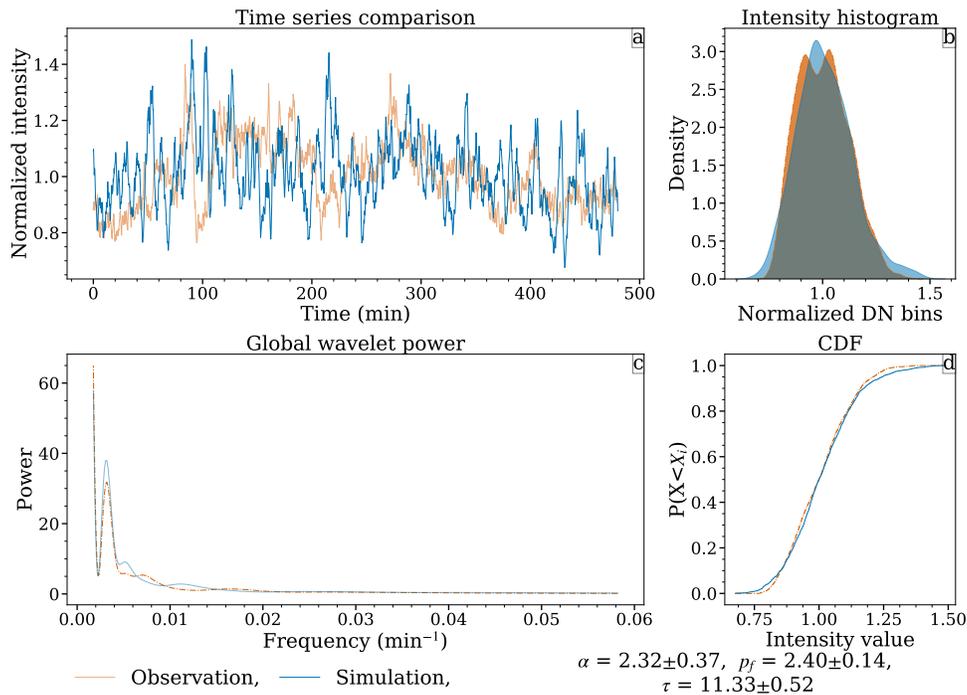

**Figure 4.11**: Same as Fig. 4.10, but for DS2.

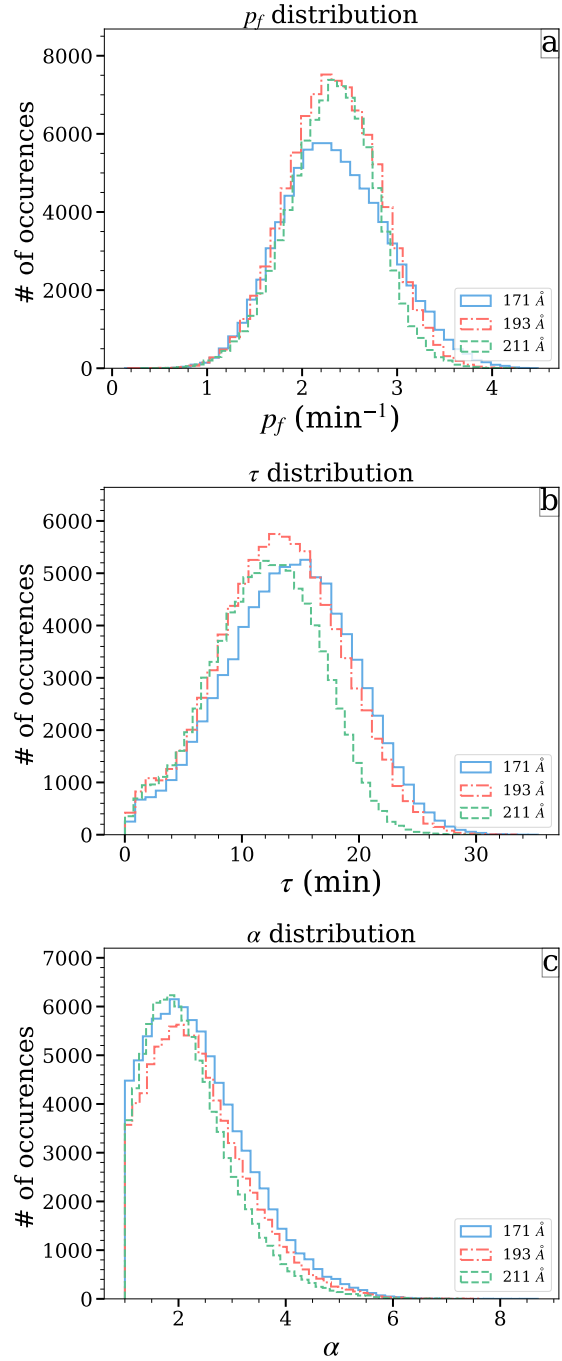

**Figure 4.12**: Distribution of inferred parameter set for $p_f$ (panel a), $\tau$ (panel b) and $\alpha$ (panel c) over both the datasets. The colours are distributed as blue (171 Å), red (193 Å) and purple (211 Å).

The plots reveal that the distribution of all three parameters for all passbands is remarkably similar. The $p_f$ distribution peaks at $\sim 2.2$ events per minute for 171 Å and at $\sim 2.4$ events per minute for 193 and 211 Å, with a range of values between 1 and 4 events per minute. The distribution of $\tau$ peaks near 12 minutes for 211 Å, 14 minutes for 193 Å and 16 minutes for 171 Å, implying a slight temperature dependence. However, we emphasize that since the AIA passbands are multi-thermal, this inference should be taken with caution. The distribution of $\alpha$, plotted in panel c, ranges between $1.0$ to $8$, with a peak at $\sim 2.3$ for all three passbands. When we consider the whole set of light curves, we find that $\sim 62\%$ of light curves in $171$ Å have $\alpha \geq 2$, while the fraction is $\sim 61\%$ and $\sim 54\%$ respectively for $193$ Å and 211 Å passbands respectively. Thus, we find a reduction in the dominance of lesser energy events as we progressively probe the QS in hotter passbands. We also note that the fall-off for all three parameters goes from the hotter passband (i.e., 211 Å) falling off first, followed by progressively cooler channels (193 and 171 Å respectively).

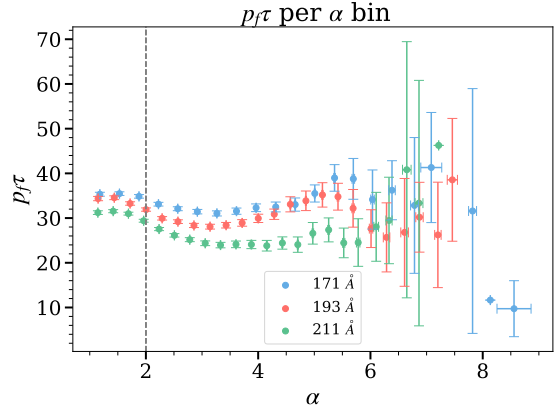

**Figure 4.13**: Variation of $p_f\tau$ with $\alpha$. The vertical dashed line marks $\alpha = 2$. The error bars are $3\sigma$ standard errors.

To further understand any peculiarities exhibited by the QS in the two regimes of $\alpha$, we plot the variation of $p_f\tau$ with $\alpha$ in Fig. 4.13. The factor $p_f\tau$ may be interpreted as the ratio of excitation (i.e., intensity generation by $p_f$) to damping(i.e., intensity dissipation by $\tau$) for a given pixel, following Pauluhn & Solanki (2007). Here, we investigate the dominance of one over the other. To boost the SNR, we have averaged $p_f\tau$ within a constant bin of $\alpha$. Note that the bin size for averaging is obtained from Doane (1976). The error bars shown are $3\sigma$ standard error[3].

The plot reveals that there is almost no change in the excitation to damping

---

[3]Standard error is defined as $\sigma/\sqrt{N}$, where $\sigma$ is the standard deviation in the sample present in the bin, and $N$ is the number of points in the sample

ratio for $\alpha < 1.8$ and is independent of $\alpha$ in this regime. However, for $\alpha \geq 1.8$, the dynamics changes, and we find a reduction in the ratio with increasing $\alpha$. The larger error bars at the end are due to poor statistics. But even considering the variation till $\alpha = 4$, we find a considerable reduction in the ratio with $\alpha$, presenting an increasing dominance of damping overexcitation. Thus, there is either a reduction in excitation, or an increase in damping, or both, which comes into play once the smaller events start dominating radiance generation.

**Energetics**

With the parameter set obtained, we can now investigate the involved energetics. A simple way to understand the energetics is by comparing the behavior of the slope $\alpha$ vis-a-vis the other free parameters. A large slope implies a predominance of smaller energies, while a small slope implies a predominance of larger energies.

To quantify relations in terms of the peak intensity of a nanoflare, Pauluhn & Solanki (2007) defined the average peak nanoflare radiance ($E_{avg}$) as:

$$E_{avg} := \left( \frac{1-\alpha}{2-\alpha} \right) \cdot \left( \frac{y_{max}^{2-\alpha} - y_{min}^{2-\alpha}}{y_{max}^{1-\alpha} - y_{min}^{1-\alpha}} \right) \tag{4.4}$$

$E_{avg}$ is a measure of the average peak nanoflare radiance value for a given time series. Using Eq. 4.4, we estimate the values of $E_{avg}$ for each pixel and study its relationship with flaring frequency ($p_f$) as well as flare duration ($\tau$) through scatter plots between these parameters.

In Fig. 4.14, we plot $p_f$ (left panel) and $\tau$ (right panel) as a function of $E_{avg}$ for all three wavelengths. We have averaged the free parameters within a constant bin of $E_{avg}$, and our error bars are $3\sigma$ standard errors. The plots reveal a slight tendency of decreasing $p_f$ as a function of $E_{avg}$, albeit there is a sharp decline in the beginning. However, $\tau$ monotonically increases with increasing $E_{avg}$ till about $E_{avg} = 0.085$ and shows saturation thereafter. Moreover, the plot further suggests a systematic lowering of flaring duration, being largest for the coolest passband (171 Å).

These are the dynamics revealed by $\approx 900,000$ light curves from EUV across three passbands. However, quantities like $E_{avg}$ are left dimensionless, thereby preventing a determination of the exact energetics involved in the impulsive heating paradigm. To this end, we now apply `iPSM` on the X-ray data from Chandrayaan−2.

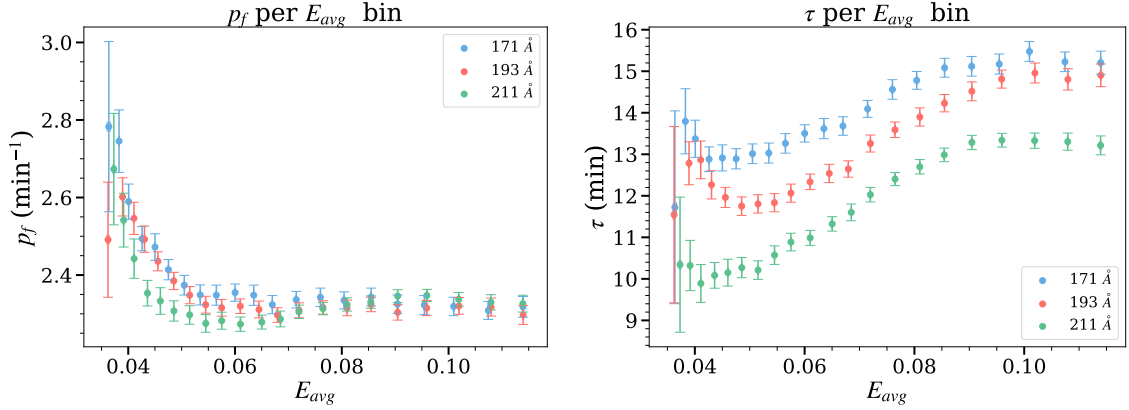

**Figure 4.14**: Inferred dependence of free parameters (binned) on $E_{avg}$, over different passbands. Blue colour represents 171 Å, red for 193 Å and green for 211 Å. Left: $p_f$ is plotted per $E_{avg}$ bin; Right: $\tau$ is plotted per $E_{avg}$ bin. The error bars indicate $3\sigma$.

## 4.4 Analysis on X-ray data

We shall now look at the analysis and results from the X-ray data. We shall first have a look at the Data in §. 4.4.1, an overview of the modeling in §. 4.4.2 along with generating bounds of energies in §. 4.4.3. We shall then look at the statistical uncertainty model in §. 4.4.4, with the updated parameter search in §. 4.4.5. We shall look at the results in §. 4.4.6.

### 4.4.1 Observations and Data

We now use the observations recorded by the XSM on-board Chandrayaan-2 mission. XSM observes the Sun as a star and provides the measurement of X-ray spectra in the energy range of 1−15 keV. It has been demonstrated that XSM has the sensitivity to carry out spectral measurements even when the solar activity is well below A-class (Mithun et al. 2020). Thus, it is possible to use XSM observations to obtain X-ray flux from the Sun during quiet phases.

We have selected XSM observations for two time periods (Oct 17−21, 2019 and Feb 14−21, 2020) when there were no active regions on the solar disk as confirmed from Solar Monitor [4]. In this work, we are interested in studying the contri-bution of unresolved impulsive events to the quiet coronal light curves and not the well-resolved events like microflares. Thus, by visual inspection of the X-ray light curves, we removed the microflare-like events studied by Vadawale et al. 2021b so that the selected observations form a true representation of quiescent solar corona, similar to Terzo et al. (2011). This step inevitably gave rise to data gaps. However, since we are interested in a statistical study of the QS light curves, we have concatenated the light curves by ignoring the gaps and obtained a continuous time series.

---

[4]https://solarmonitor.org

For the selected duration, we generated effective area-corrected and time-resolved X-ray spectra from the raw data using XSM Data Analysis Software (XSMDAS; Mithun et al. 2021b). Given the very low solar X-ray flux during these observations, the time bin size for spectra was chosen to be 2 minutes so that uncertainties on the flux due to counting statistics are typically less than 5%. The X-ray flux light curve, $F(t)$, in the energy range $E_1$ to $E_2$ is then computed from the time-resolved spectra $S(E, t)$ as:

$$F(t) = \sum_{E=E_1}^{E_2} \frac{S(E, t)\, E}{A(E)} \tag{4.5}$$

where $A(E)$ is the on-axis effective area of the XSM (Vadawale et al. 2021b). For both the observations, we generated light curves using eq. 4.5 for the energy ranges of 1.0−1.3 keV, 1.3−2.3 keV, and 1.0−2.3 keV. The light curves so obtained are shown in Fig. 4.15. Spectra above 2.3 keV are not considered as no appreciable flux is observed above that energy by XSM during QS observations.

## 4.4.2   Overview of modelling

We need to perform inference of $p_f$, $\tau$, $\alpha$, $y_{min}$ and $y_{max}$ given the X-ray light curves from XSM as shown in Fig. 4.15. Since the XSM light curves have absolute flux calibration, we would like to perform a coarse parameter sweep to provide constraints also on $y_{min}$ and $y_{max}$. While the `iPSM` model forms the core inference block of our work, the optimization for all 5 parameters is inherently difficult to perform due to degeneracy in the parameter space. Hence, `iPSM` performs inference of only three of the free parameters while keeping $y_{min}$ and $y_{max}$ fixed. Hence, we breakdown the inversion scheme into two steps (see Fig. 4.16), i.e., determining $p_f$, $\tau$ and $\alpha$ in the first step and $y_{max}$, $y_{min}$ in the second step by performing a fine search over the exact range of amplitude of the events. As shown in Fig. 4.16, each step further consists of several parts. However, the first step requires we already have a reasonable estimate of $y_{max}$ and $y_{min}$. Hence, we first put reasonable bounds on the range of values $y_{max}$ and $y_{min}$ can take by using prior observations and fixing an initial guess. Using this $y_{max}$ and $y_{min}$, we generate a bank of the PSM simulations sweeping across a range of $p_f$, $\tau$ and $\alpha$. Note that this bank of simulations is **different** from the one used in our EUV analysis. We then use this simulation bank to train the iPSM (the "Training iPSM" block in Fig. 4.16) and learn the mapping from the simulated light curves to their corresponding parameters. Finally, we perform a forward pass of the XSM light curves through the trained model and infer the corresponding values of $p_f$, $\tau$, and $\alpha$ ("Inference time", pink colored section in Fig. 4.16).

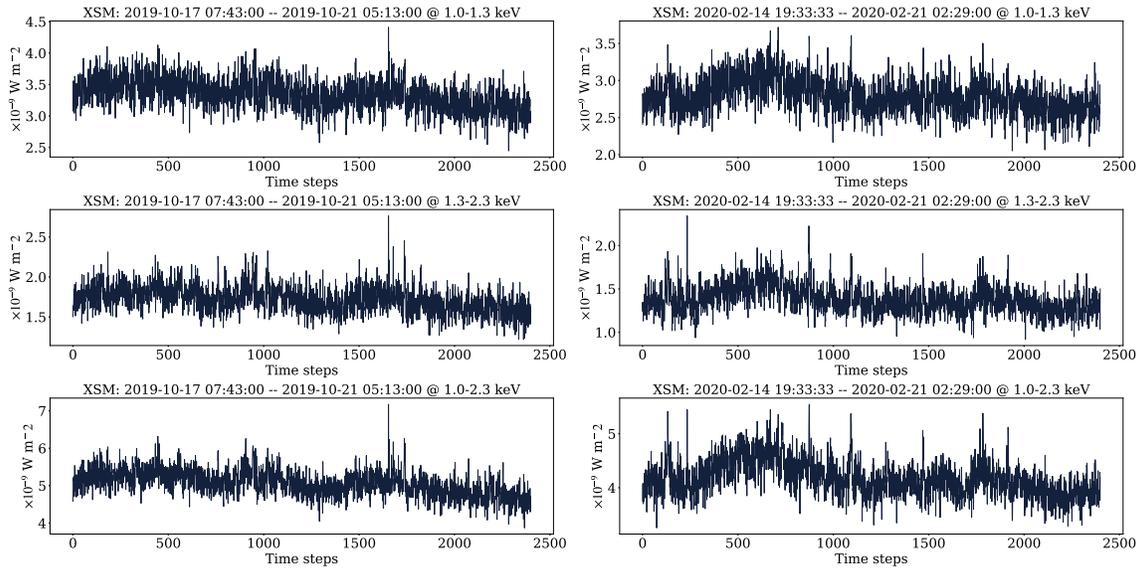

**Figure 4.15**: The XSM observed light curves considered in this work. The left(right) column is for 2019(2020) observations. The top row is the light curve for 1.0−1.3 keV, the middle row is for 1.3−2.3 keV, and the bottom row is for 1.0−2.3 keV.

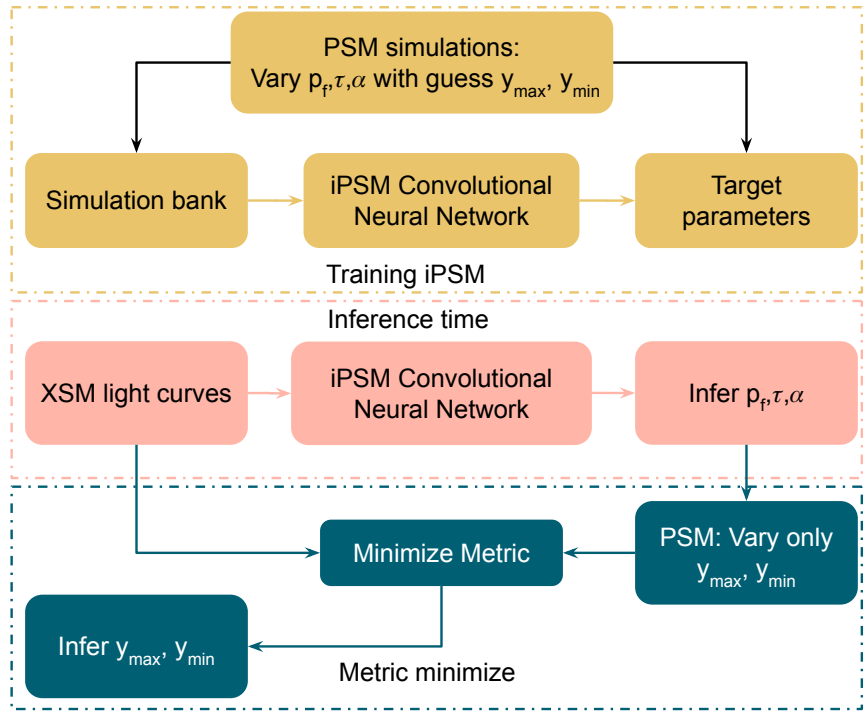

**Figure 4.16**: Flow chart detailing the various steps in our algorithm. First, the `iPSM` is trained on simulations. Next, the trained model is used to infer $p_f$, $\tau$, and $\alpha$. Finally, the $y_{max}$ and $y_{min}$ are inferred by minimizing an error metric.

Using the inferred values of $p_f$, $\tau$ and $\alpha$ from step−1, in step−2, we generate another bank of simulations, this time sweeping on $y_{max}$ and $y_{min}$. Note that the range of $y_{max}$ and $y_{min}$ is within the bounds as described in step−1. Finally, by minimizing an appropriate metric, we perform a parameter sweep considering the XSM light curves to infer $y_{max}$ and $y_{min}$ ("Metric minimize", sea green section in Fig. 4.16). Thus, through a two-step process, we infer all the 5 free parameters of the PSM.

### 4.4.3 Fixing $y_{max}$ and $y_{min}$

As described above, for generating the simulation bank for iPSM in step−1, we need to fix $y_{max}$ and $y_{min}$. Furthermore, we need to define bounds of $y_{max}$ and $y_{min}$ over which the step−2 search is performed. To do so, we first fix the upper bound of $y_{max}$ and lower bound of $y_{min}$ approximately, and then fix the $y_{max}$ and $y_{min}$ values within this range for step−1. We first define the integrated energy per impulsive event as:

$$E = 4\pi R_{1AU}^2 \cdot \tau \cdot F_{median} \cdot F_{code}, \tag{4.6}$$

where E is luminosity in a given energy band, $F_{code}$ is the amplitude of an event in code units, $\tau$ is the associated timescale (in seconds), $R_{1AU}$ the distance from Sun to Earth in meters, and $F_{median}$ the median intensity of the XSM light curve in $Wm^{-2}$. Note that since we divide the observed light curves by their median values during training and inference time, the event amplitudes in code units would need to be multiplied by the same scaling to get the correct dimensional values. This conversion factor in Eq. 4.6 helps us translate from the energy of an event in code units to real units. Since we want to generate bounds on $y_{max}$ and $y_{min}$, we fix the bounds for $F_{code}$, given other terms in Eq. 4.6.

Eq. 4.6 has terms on the right-hand side (except $F_{code}$) common for both the upper bound of $y_{max}$ and lower bound of $y_{min}$. Let us consider the median intensity in the 1−2.3 keV energy band, which is $\approx 5 \times 10^{-9}$ Wm$^{-2}$ (see Fig. 4.15), 1 AU to be $\approx 1.5 \times 10^{14}$ m, and a maximum time scale of $\approx 720$ seconds. We obtain this timescale from the `iPSM` inversions of light curves in the 211 Å passband of QS as seen in §.4.3.5. The AIA 211 Å passband corresponds to a temperature of $\log T \approx 6.2$, while the X-ray measurements typically lie in the range of $\log T \approx 6.2 − 6.8$ (Vadawale et al. 2021a). Thus, we use the 211 Å results as a proxy for the X-ray measurements. Hence, an event with unit amplitude event (i.e $F_{\text{code}} = 1$) would correspond to an energy of $\approx 10^{25}$ ergs.

First, we generate an upper bound for $y_{max}$. We note that all the microflares studied by Vadawale et al. (2021a) have been removed from our dataset. Hence, an individual event in our simulation cannot be larger than the smallest flare observed by Vadawale et al. (2021a). Since we are operating in a particular energy band, we redo the energy distribution computation in Vadawale et al. (2021a) for the energy band of 1−2.3 keV. The lowest energy thus inferred by Vadawale et al. (2021a) for this energy band corresponds to $10^{24}$ ergs. This corresponds to 'E' in Eq. 4.6. Thus, $F_{\text{code}}$ should be $< 0.1$ for the maximum amplitude condition to be satisfied. Thus, we obtain an upper bound on $y_{max}$ − i.e, $y_{max} < 10^{-1}$.

Having fixed the upper bound for $y_{max}$, we turn our attention to fixing the lower bound on the $y_{min}$. We consider the energetics of fluctuations observed in soft-X ray light curves derived by Katsukawa & Tsuneta 2001; Katsukawa 2003. These authors found the energies of impulsive events of $\approx 10^{20-22}$ ergs consistent with the distribution of fluctuations of soft-X ray light curves in active regions. Labonte & Reardon (2007), however, showed that the fluctuations in the light curves as obtained by Katsukawa & Tsuneta (2001) are consistent with noise. Thus, we take the lower limit of the possible energies and set a lower bound on $y_{min}$ as $10^{20}$ ergs, where the lower limit corresponds to "noise" events. This value would correspond to $y_{min} > 10^{-5}$ in code units, following Eq. 4.6. Thus, the event amplitudes may lie only between $10^{-5}$ and $10^{-1}$. Hence, these physical observations set the general bounds of the range of the expected energies of events.

We have obtained the lower bound on $y_{min}$ and the upper bound on $y_{max}$. To fix the values of $y_{max}$ and $y_{min}$ in step−1, we prototype on a very limited combination of $y_{max}$ and $y_{min}$, generating one `iPSM` model for each combination. Through visual inspection, we find $y_{max}$ and $y_{min}$ of $5 \times 10^{-3}$ and $10^{-4}$ (code units) to give us simulations which show a good match in the intensity distribution & wavelet power spectrum with the XSM observation. Thus, we fix $y_{max}$ and $y_{min}$ to be $5 \times 10^{-3}$ and $10^{-4}$ for generating the bank of simulations for step−1 of our inversion.

### 4.4.4 Statistical uncertainty model and simulation bank generation

Using all the parameters discussed above, we generate simulated light curves that can be compared with the observed light curves from XSM. However, the QS is known to have weak emission in X-rays (see, for example, Brosius et al. 1997; Katsukawa & Tsuneta 2001; O'Dwyer et al. 2010b), and is expected to have a non-negligible contribution of counting statistics. Therefore, the associated simulated light curves must be incorporated with these statistical uncertainties for an objective comparison.

For the EUV data, we developed the algorithm `Finding kneemo` to smooth the light curves. We performed this smoothing simply because we did not have enough compute to generate a humongous dataset of simulations incorporating the observed statistical uncertainties. Now, however, we only have 6 X-ray observations – thus, we can incorporate these statistical uncertainties in the simulations themselves and do not need to resort to the smoothing by `Finding kneemo`.

For this purpose, we estimate the statistical uncertainties on the light curves by propagating the Poisson error on the observed count for each light curve. Hence, for each observed light curve to be inverted, we know the signal and the associated uncertainty at each time step.

To get a non-dimensional estimate of the uncertainty as a single number for the full light curve, we first calculate the uncertainty-to-signal ratio $r_t$ at each time step for a given XSM light curve. This provides a measure of the "uncertainty fluctuation" as a fraction of the signal. The mode of $r_t$, denoted as r, is the estimate of uncertainty as a fraction of the observed signal for the full light curve. We obtain r for each observed light curve.

To incorporate this uncertainty into each simulated light curve, we replace the intensity at each time step with a sample from a Gaussian distribution with a mean of the simulated intensity (from the `PSM`), and a standard deviation of r times the intensity at that time step. This is justified as while the original photon counts follow a Poisson distribution, the flux values after integration over two minutes are expected to follow a Gaussian distribution. Since there are 6 light curves, we have 6 associated sets of simulations for each light curve.

Finally, we simulate the light curves by taking care of all the steps explained above. We generate simulated light curves for a duration of 2400×120 =288000 seconds at a time cadence of 1 second, i.e., 1 code time step = 1 second. To minimize the effect of starting seed, we generate the simulations with extra 1600 seconds and throw away the first and last 800 seconds. The simulations are then re-binned at a 120-second cadence, giving rise to 2400 time points to match the observations.

We incorporate the statistical uncertainties to each simulated light curve and normalize each by its median value as a pre-processing step to the `iPSM`. After parameter inference, we construct the best-matching simulations by multiplying the median normalized simulations with the median value of the corresponding observation. This gives us simulation light curves in the units of $W\ m^{-2}$. Since we are scaling the intensities in the simulations, we also scale the corresponding $y_{max}$ and $y_{min}$ values, which determine the amplitude of these events in the same way.

For step$-1$, we generate the bank of simulations by varying $p_f$ between 3 to 57 events per minute, which translates to $p_f$ between 0.05 to 0.95 events per second in steps of 0.01 events per second. The time scale $\tau$ is varied between 1 second and 500 seconds in steps of 10 seconds, while $\alpha$ is varied between 1.5 and 3.0 in steps of 0.1. This parameter space is similar to Upendran & Tripathi (2021a). We have, however, reduced the maximum value of $\tau$ (in seconds) since we expect X-ray observations to show much shorter time scales than EUV observations, as seen by Upendran & Tripathi (2021a).

We then perform the step$-1$ inference using the `iPSM` model, as detailed in §. 4.2. However, since the simulations in this work need to be uncertainty-incorporated, we retrain the model from scratch for the new set of simulated light curves. We generate 6 inversion models in total corresponding to each light curve. All of our models show $R^2 > 0.98$ for $p_f$ and $\tau$, while the $R^2$ for $\alpha$ are more than $0.91$. This step is depicted graphically as the yellow and pink flow diagrams in Fig. 4.16. Thus, we infer $p_f$, $\tau$ and $\alpha$ for each XSM light curve from step$-1$.

### 4.4.5   Beyond `iPSM`: Metric minimization

In step$-1$, we have inferred three parameters $p_f$, $\tau$ and $\alpha$ for fixed values of $y_{min}$ and $y_{max}$. In step$-2$, we fix these three parameters and generate a new set of light curves by sweeping $y_{max}$ and $y_{min}$. We sweep $y_{max}$ between $9 \times 10^{-4}$ and $5 \times 10^{-2}$ with 45 steps in $\log_{10}$, and $y_{min}$ between $1 \times 10^{-5}$ and $5 \times 10^{-4}$ for 36 steps in $\log_{10}$. We incorporate the photon counting uncertainties on these light curves as described in §4.4.4. These light curves serve as a bank from which we may perform an inexpensive, simple search to generate better constraints on $y_{min}$ and $y_{max}$. However, to do so, we need to define a metric that we may then minimize. Since our qualitative

"best fit" is determined by a good match between the simulation and observation in terms of intensity distribution and power spectrum, we define a simple metric in Eq. 4.7 as:

$$m = \max\left((\mathsf{CDF_O} - \mathsf{CDF_S})^2\right) + \max(((\mathsf{P_O} - \mathsf{P_S})/\mathsf{P_O})^2). \quad (4.7)$$

Here the subscripts O and S correspond to observation and simulation, respectively. The first term finds the maximum the absolute difference between the cumulative distribution function of the two light curves. The second term finds the maximum relative wavelet power mismatch between the two light curves.

With this metric, we then perform a grid search and find the combination which gives us the lowest possible metric value. The corresponding $y_{max}$ and $y_{min}$ are then taken up as the 'inferred' final values.

## 4.4.6   X-ray: Results

We now apply our two-step updated parameter estimation scheme to the X-ray data. We next showcase the results for each light curve and then present the results on the energetics of these events.

**Light curve inversions**

On applying our two-step procedure described in §4.4.2, we obtain the "best fit" parameters of the PSM simulations. In Fig. 4.17, we present the metric surface from step−2 as a function of the swept range of $y_{max}$ and $y_{min}$, where the metric value is lower for the darker color. Note that we have displayed the metric in log scale. The blue circle represents the originally pre-fixed $y_{max}$ and $y_{min}$ for step−1, while the green star is the $y_{max}$ and $y_{min}$ solution inferred from step−2 parameter search.

Fig. 4.17 reveals a number of salient features about our inferred solution(s). First, there is a whole diagonal of "good" solutions, showcasing the degeneracy between $y_{max}$ and $y_{min}$. Second, the pre-fixed $y_{max}$ and $y_{min}$ lie very close to the diagonal ridge of good solutions, thereby also justifying our choice of the initial guess for $y_{max}$ and $y_{min}$. Third, the final good solutions are sometimes quite close to the pre-fixed values, while sometimes they change by order of magnitude. The final amplitudes, however, would depend on the median flux value. Therefore, the constraint is strongly performed for the ratio of $y_{max}$ and $y_{min}$. A strong global minimum is not seen for constraining $y_{max}$ and $y_{min}$. However, the solutions we shall see next give rise to a good representation of the observed light curves. We now present the inversion results for all the light curves obtained by integrating the signal between 1−2.3 keV energy band in Fig. 4.18− 4.20. For the sake of discussion, we only focus on Fig. 4.18, while the results for the other light curves are similar.

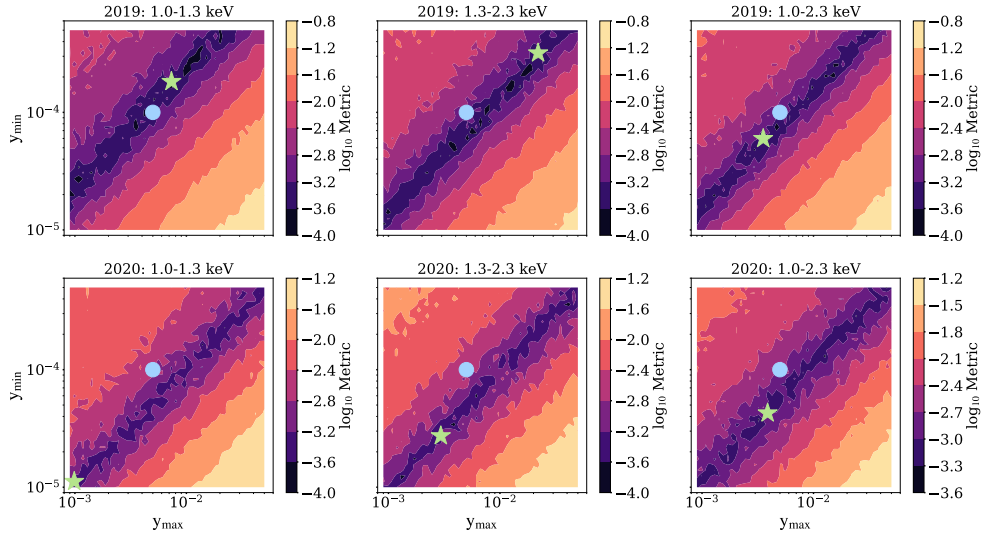

**Figure 4.17**: The variation of metric with $y_{max}$ (x-axis) and $y_{min}$(y-axis). The two parameters are presented here in code units (note the log scale), while the metric is presented in a scale of $\log_{10}$. The blue circle shows the originally selected $y_{max}$ and $y_{min}$ (as used by the `iPSM`), while the green star corresponds to be $(y_{max}, y_{min})$ with the lowest metric value.

In Fig. 4.18, we show the light curves (panel a), intensity distributions (panel b), wavelet power spectrum (panel c), and cumulative distribution function (CDF; panel d). The orange represents the observation and the black represents the `PSM` forward model of the best-fit parameters. Note that the simulated light curve is uncertainty-incorporated. The uncertainty bands in the power spectrum corresponding to 1-$\sigma$ standard deviation in time. The top four panels are for data recorded in 2019 and the bottom four panels are for that in 2020. Note again that a statistically accurate simulation must capture the intensity distribution well. Similarly, such a simulation must also capture the essential frequencies in the time series which have excess power. These are represented by the histogram (and CDF) and the wavelet power spectrum. The presence of peaks at similar frequencies in the power spectrum gives us the scales of importance, though we emphasize that the exact amount of power need not exactly match. The plots reveal a good correspondence be-tween the observed and simulated light curve, both in matching the distribution and wavelet power at different scales. Thus, the two-step inversion scheme with `iPSM` successfully captures the necessary information from the presented observations. We summarize the inversion parameters for all the six light curves in Table. 4.4.

Table 4.4:: Summary of the inferred parameters for the six light curves.

| Parameter | 1.0 − 1.3 keV | | 1.0 − 2.3 keV | | 1.3 − 2.3 keV | |
|---|---|---|---|---|---|---|
| | 2019 | 2020 | 2019 | 2020 | 2019 | 2020 |
| $p_f$ (events min$^{-1}$) | $27.89 \pm 1.67$ | $33.18 \pm 1.87$ | $28.00 \pm 2.16$ | $34.17 \pm 1.97$ | $25.42 \pm 1.57$ | $24.95 \pm 5.15$ |
| $\tau$ (min) | $10.56 \pm 0.88$ | $9.12 \pm 0.73$ | $11.80 \pm 1.05$ | $8.26 \pm 0.67$ | $9.29 \pm 0.79$ | $6.56 \pm 0.56$ |
| $\alpha$ | $2.00 \pm 0.12$ | $1.74 \pm 0.15$ | $1.87 \pm 0.15$ | $1.58 \pm 0.12$ | $1.94 \pm 0.13$ | $1.56 \pm 0.13$ |
| $y_{max}$ (W m$^{-2}$) | $8.14 \times 10^{-11}$ | $7.30 \times 10^{-11}$ | $1.26 \times 10^{-10}$ | $9.07 \times 10^{-11}$ | $7.71 \times 10^{-11}$ | $6.86 \times 10^{-11}$ |
| $y_{min}$ (W m$^{-2}$) | $2.03 \times 10^{-12}$ | $8.28 \times 10^{-13}$ | $2.12 \times 10^{-12}$ | $1.00 \times 10^{-12}$ | $1.12 \times 10^{-12}$ | $6.36 \times 10^{-13}$ |

We note that the flaring frequency $p_f$ ranges from $24 - 35$ events per minute, and the time scale $\tau$ ranges from $\approx 6 - 12$ minutes with a maximum uncertainty of the order of a minute. For all the light curves, the inversion gives us power law slopes of $\leq 2.0$. Finally, $y_{max}$ generally ranges from $7 \times 10^{-11} - 1.26 \times 10^{-10}$ W m$^{-2}$, while $y_{min}$ ranges from $6 \times 10^{-13} - 2 \times 10^{-12}$ W m$^{-2}$. Overplotting the typical energies inferred from our analysis with the results from Vadawale et al. (2021a) in Fig. 4.21, we clearly see that the events making up these light curves are much smaller than the microflares observed by Vadawale et al. (2021a).

**Energetics**

We now have a train of events giving rise to each of the observed light curves. Our goal is to study the energetics of these events. For this purpose, we first convert the obtained intensities into fluxes and energies following Eq. 4.6. Since we would be integrating only in a particular energy band, they would correspond to a "lower bound" of energy. The energy estimates are better representatives of the energy content of these events if larger energy bands are considered. Hence we consider the energies in the widest $1-2.3$ keV passband. We find that our energies typically range between $10^{21} - 2 \times 10^{23}$ ergs for this passband, with $\alpha$ shallower than $2.0$. These events will thus correspond to the nanoflare or even picoflare energy range.

To understand the average radiative loss flux, we consider the average amplitude of flare in a given time series ($E_{avg}$). Note that this is in code units, which can be converted into real units of energies following Eq. 4.6. Inherently, we assume that the corresponding energy obtained is emitted isotropically by the Sun. To estimate the amount of energy emitted across the whole time series, we also need the frequency of occurrence of these events ($p_f$). Hence, for a given flaring frequency of $p_f$ (events per second), the amount of energy radiated per unit time would be $p_f \cdot E$. Thus, the radiative flux loss from the unit solar area (since we are performing full-disk integration) would be

$$\text{RL} := \frac{p_f 4\pi \text{R}_{1\text{AU}}^2 \cdot \tau}{A_{\odot, disk}} \text{E}_{\text{avg}} \tag{4.8}$$

The full set of radiative flux losses is presented in Table. 4.5. We find the radiative flux losses to be $\approx 5 \times 10^3$ erg cm$^{-2}$ s$^{-1}$ in the 1-2.3 keV energy band, while they are $\approx 3.5 \times 10^3$ and $\approx 2 \times 10^3$ erg cm$^{-2}$ s$^{-1}$ in the 1−1.3 and 1.3−2.3 keV energy bands, with errorbars on each term. Thus, the losses are typically of the order of $10^3$ erg cm$^{-2}$ s$^{-1}$.

Table 4.5:: Radiative losses in erg cm$^{-2}$ s$^{-1}$ for the 3 energy passbands and two years.

| Energy band (keV) | 2019 | 2020 |
|---|---|---|
| 1.0-1.3 | $4.18 \pm 0.65 \times 10^3$ | $3.02 \pm 0.74 \times 10^3$ |
| 1.0-2.3 | $6.29 \pm 1.37 \times 10^3$ | $4.39 \pm 0.92 \times 10^3$ |
| 1.3-2.3 | $2.26 \pm 0.43 \times 10^3$ | $1.82 \pm 0.57 \times 10^3$ |

## 4.5  Summary, Discussion and Conclusion

QS coronal region provides a wealth of data to narrow down and understand the underlying heating processes. Assuming the underlying heating mechanism to be impulsive, coronal light curves may be approximated using empirical statistical models to infer physics-inspired parameters. To this end, we have developed a DL inversion model called the `iPSM` using CNN and validated it using a separate test set. We have applied this model to perform the inversion on coronal datasets and infer the free parameters for (i). The three AIA pass-bands viz. 171 Å, 193 Å and 211 Å, and (ii). The three energy bands of XSM, viz. 1−1.3 keV, 1.3−2.3 keV, and 1−2.3 keV.

We find that the light curves inverted using the CNN and observed light curves are statistically in excellent statistical agreement, considering the CDF and PDFs (see Fig. 4.10, 4.11, and Figs. 4.19− 4.20). Note that we are not concerned with a point-to-point match at each timestep between the simulation and the observation.

A change in the seed of the random number generator is enough to shift the location of individual events and their amplitudes in simulation. However, this does not cause any changes in the statistical properties of the light curve, which is what we are primarily interested in. The Global wavelet power shows the simulations and observations to have peaks at similar frequencies, validating that the simulation and observation both have enhanced power in similar frequencies. The quality of approximation may be understood by the Epistemic uncertainty of our CNN, obtained by the application of Dropout to perturb the model.

### 4.5.1 EUV results

We find the distribution of all parameters to be similar for light curves from all three EUV passbands. The flaring frequency lies within the range of 1 to 4 events per minute, with a peak at $\sim 2.3$ events per minute for all three passbands. Similarly, the flaring duration has a range of values between 5 and 30 minutes. However, unlike the flaring frequency, the peak of the distribution of flaring duration shows temperature dependence, being highest ($\sim$16 min) for the coolest 171 Å channel, $\sim$14 min for 193 AA and lowest ($\sim$12 min) for the hottest 211 AA channel.

The power-law index $\alpha$ has a range of values between 1 and 8. The distribution of $\alpha$ peaks at 2.3 for all three passbands of AIA viz. 171, 193, and 211 Å. This finding strongly suggests that nanoflare heating is indeed a viable source of energy in the quiet corona (see, e.g., Hudson 1991). We also find the fraction of light curves giving $\alpha \geq 2$ progressively reduces from cooler to hotter passbands. We further note that there are a significant number of pixels where $\alpha < 2$. This is suggestive that low-energy events may not be dominant everywhere. However, note that the viability of these low-energy events also relies on the flaring frequency $p_f$, requiring a full analysis of the energetics.

Our finding further suggests that there is a change of dynamics for pixels with $\alpha < 1.8$ and $\alpha \geq 1.8$ (from Fig. 4.13). In the former case, $p_f\tau$ is nearly constant, while in the latter case, it reduces with increasing $\alpha$. This may also be observed in Fig. 4.14. Note that $\alpha = 2$ corresponds nearly to $E_{avg} = 0.08$ and the right tail of $E_{avg}$ represents $\alpha < 2$. Here, a definite increase of $p_f$ with decreasing $E_{avg}$ is seen, which is interpreted as excitation increasing with decreasing $E_{avg}$ (and thus increasing $\alpha$). However, we also find that $\tau$ reduces with reducing $E_{avg}$. Thus, the increase in damping counters that in excitation in $\alpha \geq 2$ regime, causing a reduction in the ratio in Fig. 4.13. Thus, the increase in excitation is essentially nullified by the increase in damping.

We note that our inferred $p_f$ distribution peaks at $\sim 2 - 3$ events per minute, while Tajfirouze & Safari (2012) obtained a mean $p_f$ of $\sim$ 0.33 events per min. This may be explained by the better temporal cadence and spatial resolution of our data, whereby our simulation captures much smaller transient events. However, note that our inferred $p_f$ corresponds to an average waiting time of $\sim 30$ sec, which is much smaller than the waiting times of $\sim 230$ s suggested from simple geometric arguments (Klimchuk 2015). magnetoac There is a discrepancy in the $\tau_d$ (decay time) and $\alpha$ derived here with those obtained by Tajfirouze & Safari (2012). The $\tau_d$ being about a factor of six smaller in our case ($\sim 60$ min in Tajfirouze & Safari (2012) to $\sim$10 min in our case) and $\alpha$ peaking near $2.3$ in our case, compared to a mean $\alpha$ of 2.6 in Tajfirouze & Safari (2012). This discrepancy may be attributed to the fact that the data used in this study are at much higher spatial and temporal resolution than Tajfirouze & Safari (2012). Therefore, we must have captured smaller events with much shorter timescales (as also alluded to by Tajfirouze & Safari (2012)). Similarly, the discrepancy in power-law index $\alpha$ may be attributed to the high spatial resolution data used in the present work. Moreover, in our simulation, we are sampling flares within a larger energy range with larger $y_{max}/y_{min}$ than those by of Tajfirouze & Safari (2012)). However, note that our obtained cooling time scales ($\sim 600$ sec) are indeed of the order of cooling time scales in the corona obtained by Klimchuk (2015)($\sim 1000$sec).

Next, we find that the $p_f$ decreases with $E_{avg}$ (see the left panel in Fig. 4.14). The decrease of $p_f$ with $E_{avg}$ is interpreted as a decrease in peak energy released per flare with increasing frequency. Thus, we can either have intermittent, high-energy events or sustained low-energy events. This variation of $p_f$ with $E_{avg}$ is similar to the observation of the relation obtained between peak flare flux and waiting times by Hudson (2020) (see also Sarkar et al. 2019). Furthermore, the inverse relation between $p_f$ and $E_{avg}$, (or a direct relation between the waiting time and succeeding nanoflare energy) was a necessary ingredient needed to reproduce observed EM distribution with temperature in ARs by Cargill (2014). Since the rise time is given as a fraction of the decay time in this setup, we do not distinguish between pre-flaring and post-flaring times. Thus, this may point to the presence of a reservoir of energy that may be exhausted by frequent, small-energy events or intermittent, large-energy events. However, we emphasize that the change in $p_f$ is tiny (2-3 events per min, as can be seen from Fig. 4.14) when compared to the total time scale of these events (10-15 min across all passbands). Thus, we must take this interpretation with caution.

The time scale $\tau$ is seen to increase with $E_{avg}$ (see the right panel in Fig. 4.14). This shows an increase in flare time scale corresponds to an increase in average flare energy. This weak correlation is similar to the weak correlation observed between peak flux and flare time scale by Veronig et al. (2002, see Fig. 3). However, we may seek to explain this relation qualitatively as below:

For an iso-thermal, optically thin plasma, the observed intensity is directly proportional to electron number density (O'Dwyer et al. 2010c), i.e $DN \propto n_e^2$. From Cargill (1994), we find that:

$$\tau_c \propto n_e : \text{Conductive cooling dominated plasma} \tag{4.9}$$

and

$$\tau_r \propto n_e^{-1}; \text{Radiative cooling dominated plasma} \tag{4.10}$$

where $\tau_c$ and $\tau_r$ are conductive and radiative cooling times, respectively.

Combining the equations of timescale and intensity, we obtain:

$$\tau_r \propto \frac{1}{\sqrt{DN}}, \tau_c \propto \sqrt{DN}, \tag{4.11}$$

Thus, for a conduction cooling-dominated plasma, the timescale $\tau$ should increase with the emitted intensity, while for a radiative cooling-dominated plasma, we expect the opposite. Our results show a direct relation between $\tau$ and the peak flare intensity $E_{avg}$ and qualitatively suggest that in such events, conduction losses are dominant over radiative losses, assuming a constant flaring frequency. This is similar to the results obtained by Rajhans et al. (2021); Gupta et al. (2018b); Subramanian et al. (2018) for tiny transient brightenings.

Finally, the flaring time scales are seen to be largest in the coolest passband and decrease from the cooler to hotter passbands. This indicates the decreasing dominance of conduction loss over the radiative loss (but the conduction loss still dominates), as would be the case for cooling loops (see e.g., Klimchuk 2006b; Viall & Klimchuk 2012; Tripathi et al. 2009a; Gupta et al. 2015). We emphasize that this is true under the assumption of constant flaring frequency since only $\sim$ 2-3 events are happening per minute, whereas our total (rise+fall) time in consideration $\sim$ 15 min.

## 4.5.2 X-ray results

In the X-ray regime, the flaring frequency is $\approx 24 - 35$ events per minute. This flaring frequency is $10\times$ larger than those we found in the EUV observations ($p_f \approx 2.5$ events per minute; see §. 4.3.5). These two results may be reconciled by noting that $E_{avg}$ defined in Eq.4.4, is $\approx 10^{-3}$ in the X-ray regime, while it is $\approx 10^{-2}$ in EUV. This shows that an approximately $10\times$ reduction in the flare amplitude has resulted in an approximately $10\times$ increase in flaring frequency ($p_f$). Thus, the X-ray and EUV results are consistent in that the $p_f$ is found to reduce with increasing event amplitude. These results strongly indicate the presence of an energy reservoir that may be depleted by large events occurring infrequently or small events occurring more frequently.

We find that the X-ray event timescale ranges from $\approx 6-11$ minutes. In the EUV regime, we saw that the event time scales reduce with increasing temperature, i.e., from $\approx 16$ minutes in 171 Å (log T $\approx 5.85$) to $\approx 12$ minutes in 211 Å (log T $\approx 6.2$) (see §.4.5.1 above). Since the observations reported here are at a higher temperature, the obtained results are consistent with those from EUV, though note that these values reported corresponding to the mean values of a distribution.

We may also compare the properties of these unresolved X-ray events with those of resolved microflares.Sylwester et al. (2019), for example, studied microflares in the 1.2−15 keV energy range using data from SphinX (Sylwester et al. 2008), with similar events studied by Vadawale et al. (2021b). They find the median temperatures of log T $\approx 6.3$, while the time scales range from $\leq 1$ minute to $\approx 10$ minutes. Thus, the timescales we obtain are typical of the order of, or even slightly longer than those obtained by Sylwester et al. (2019) − though we emphasize that timescales are consistent within the uncertainties.

Finally, we obtain $\alpha$ between $2.0$ and $1.56$ in the X-ray regime, which are far flatter than those obtained in EUV ($\alpha \geq 2$, see §. 4.3.5). However, note that the median $\alpha$ in EUV varies from 2.26 in 171 Å to 2.07 in the 211 Å passband. Consistent with this trend, we also find the $\alpha$ from X-rays to be smaller. Moreover, from Table. 4.4, we see that the $\alpha$ value reduces with increasing energy (from 1−1.3 keV to 1.3−2.3 keV). However, we note that the increase is only in the mean value, but within the error bars, they are consistent. On the whole, there appears to be a particular flattening of $\alpha$ with the increasing temperature of plasma when the EUV and X-ray observations are taken together. We emphasize, however, that the smaller $\alpha$ for higher temperatures is intriguing, i.e., for a given range of amplitudes, a larger $\alpha$ would have infrequent outlier intensities. However, a smaller $\alpha$, as inferred here, implies that these outlier events start to become the norm, implying that the typical amplitude of events is nearly constant. Moreover, Vadawale et al. (2021a) find the power law slope to be larger than 2 for the microflare observations using the same instrument and for the same time periods. This raises the question of a possible change in the underlying mechanism of heating from higher energies to lower energies, which reflects differences in the power law exponent.

We note that the obtained energy of the impulsive events in X-rays are typically in the nanoflare/picoflare regime and vary in the range of $10^{21} - 2 \times 10^{23}$ ergs. These values are typical of the scale of thermal energy as measured by Sylwester et al. (2010) for typical solar quiet times, though we note that we report only a conversion of luminosity to energy, and not the thermal energy itself. Considering a very small range of energies the actual value of $\alpha$ may not even have a strong meaning. It may simply suggest that the events of $\approx 10^{23}$ ergs are dominant over events with energy $\approx 10^{21}$ ergs. Therefore, it is imperative to not consider just the parameter $\alpha$, but also consider the radiative flux in the events to get a better estimate.

Due to the flux-calibrated data of XSM, we can estimate the radiative energy loss from the quiet corona. We find the flux to be $\approx 10^3$ erg cm$^{-2}$ s$^{-1}$ for the full energy range of $1-2.3$ keV. This flux is two orders of magnitude lower than the radiative loss estimates in the quiet corona by Withbroe & Noyes (1977). Prima-facie, it suggests that such sub-pixel impulsive events may not have enough energy to maintain the quiet corona. However, we must note that the energy estimate presented here only provides a lower bound since the energy is radiated away in many wavelengths. A better way would be to estimate the "thermal energy" content of the impulsive events, which is not possible in our case due to lack of spatial content, i.e., a length measure along the line of sight.

There do exist some caveats in these set studies. It is important to emphasize that the PSM is very well suited for explaining the observed light curves from the quiet Sun region obtained at much higher temporal and spatial resolution than was initially studied by Pauluhn & Solanki (2007). We note that although we have improved the inference by quantifying the uncertainties, the PSM may further be developed by incorporating the plasma filling factor and effective area in the forward model, as has also been suggested in the original paper (Pauluhn & Solanki 2007). Note that in the PSM, the radiance distribution is considered to be the same as the energy distribution. However, note that there may be a number of further smaller events that may or may not produce detectable signatures in the intensity images. It is also possible that many events produce signatures in one passband and not in another. Therefore the distribution reported using the radiance is just a lower limit of the total number of events. Hence, the incorporation of filling factors in the PSM is an important next step in improving the model.

These exercises tell us that impulsive event trains can *statistically* explain QS light curves – either from a single pixel or from full disc integration. Typically, these events occur in the corona across a range of temperatures, and their properties also vary depending on the characteristic temperature of the plasma emitting them. There is no reason not to expect an impulsively driven chromosphere, similar to an impulsively driven corona. Thus, it would be interesting to apply our inversion model to chromospheric observations (see also Jess et al. 2014), possibly using data from IRIS or using chromospheric observations in Near-UltraViolet (NUV) from the Solar Ultraviolet Imaging Telescope (SUIT; Tripathi et al. 2017; Ghosh et al. 2016), on-board the Aditya-L1 mission (Seetha & Megala 2017; Tripathi et al. 2022) of Indian Space Research Organization (ISRO).

XSM: 2019-10-17 07:43:00 -- 2019-10-21 05:13:00 @ 1.0-2.3 keV

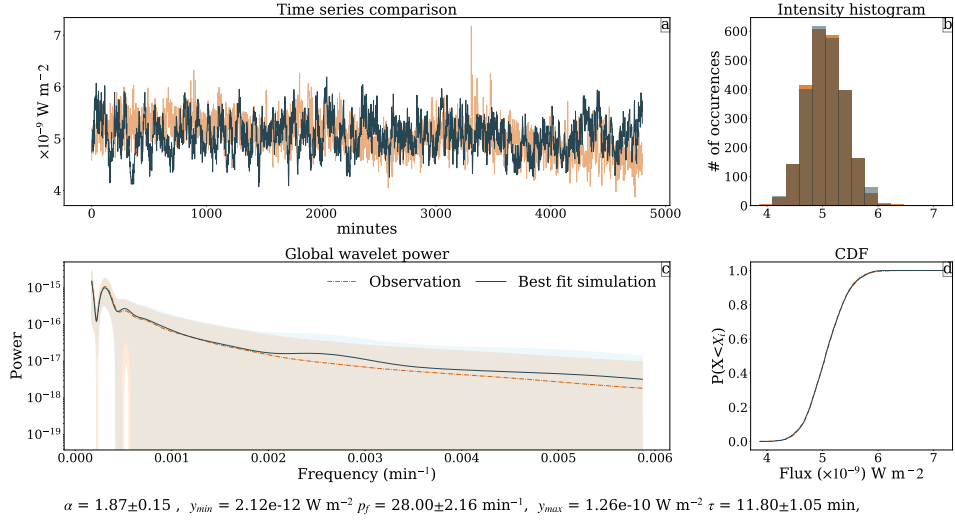

$\alpha = 1.87 \pm 0.15$, $y_{min} = 2.12\text{e-}12$ W m$^{-2}$ $p_f = 28.00 \pm 2.16$ min$^{-1}$, $y_{max} = 1.26\text{e-}10$ W m$^{-2}$ $\tau = 11.80 \pm 1.05$ min,

### I. 1-2.3 keV from October 2019

XSM: 2020-02-14 19:33:33 -- 2020-02-21 02:29:00 @ 1.0-2.3 keV

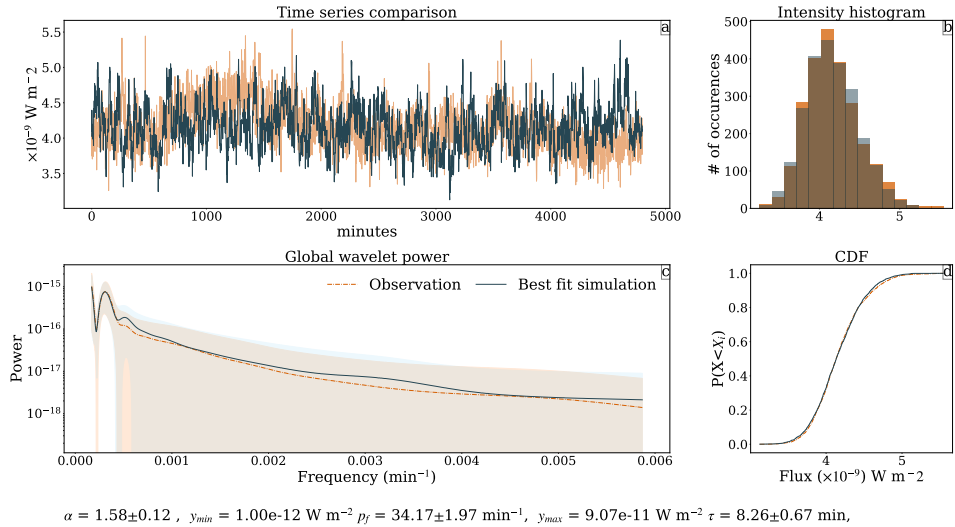

$\alpha = 1.58 \pm 0.12$, $y_{min} = 1.00\text{e-}12$ W m$^{-2}$ $p_f = 34.17 \pm 1.97$ min$^{-1}$, $y_{max} = 9.07\text{e-}11$ W m$^{-2}$ $\tau = 8.26 \pm 0.67$ min,

### II. 1-2.3 keV from February 2020

**Figure 4.18**: Comparison of the observed light curve from XSM (orange), and the PSM forward model of best-fit parameters inferred from our inversion code (black) in 1-2.3 keV energy band from 2019 (subfigure: I) and 2020 (subfigure: II). Each subfigure has four panels depicting: Panel (a): Observed and simulated light curves; Panel (b): Distribution of observed and simulated light curve intensities; Panel (c): Global Morlét power for observation and simulations, with the uncertainties presented in orange and blue bands; Panel (d): Comparison of simulation and observation intensity CDF. The inset reports the inferred parameter set for the respective data.

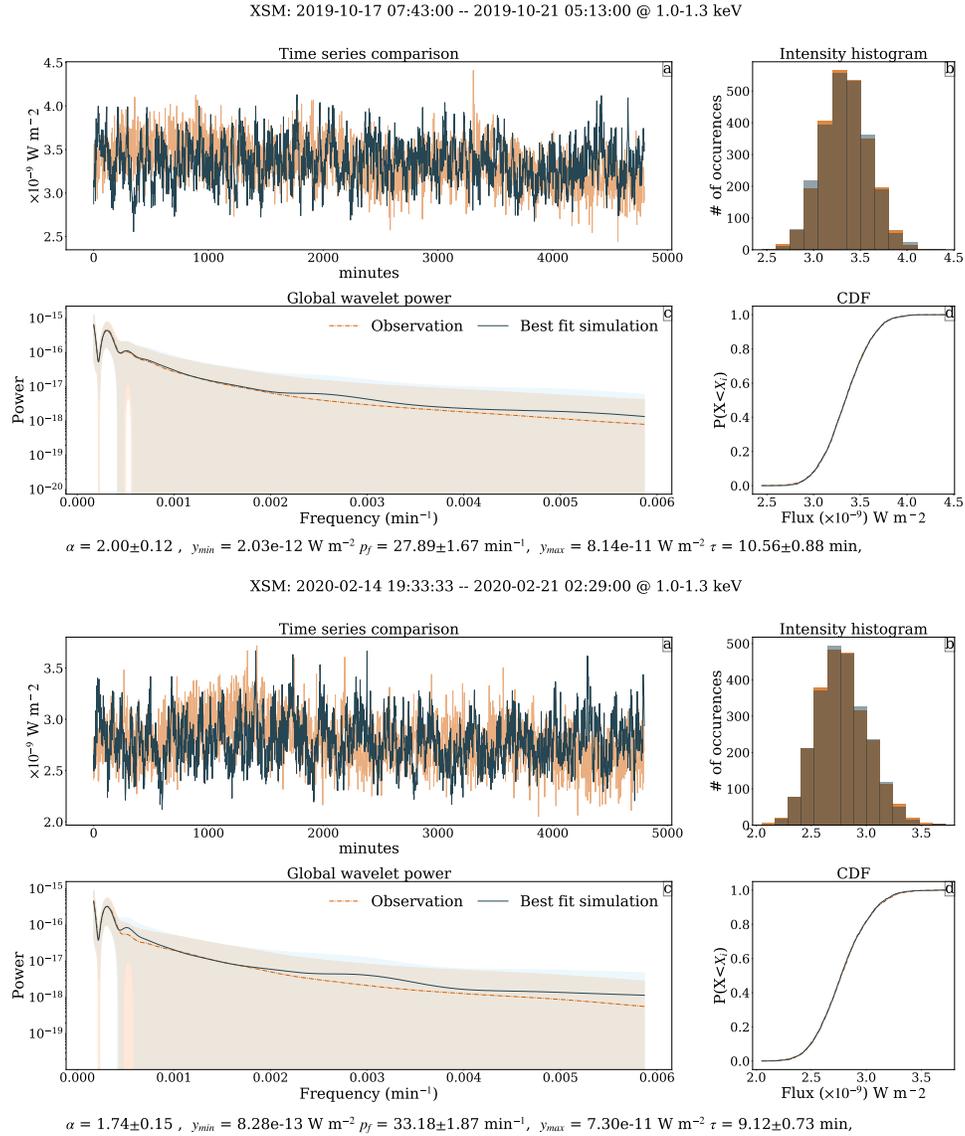

**Figure 4.19**: Same as Fig. 4.18, but for 1.3−2.3 keV band.

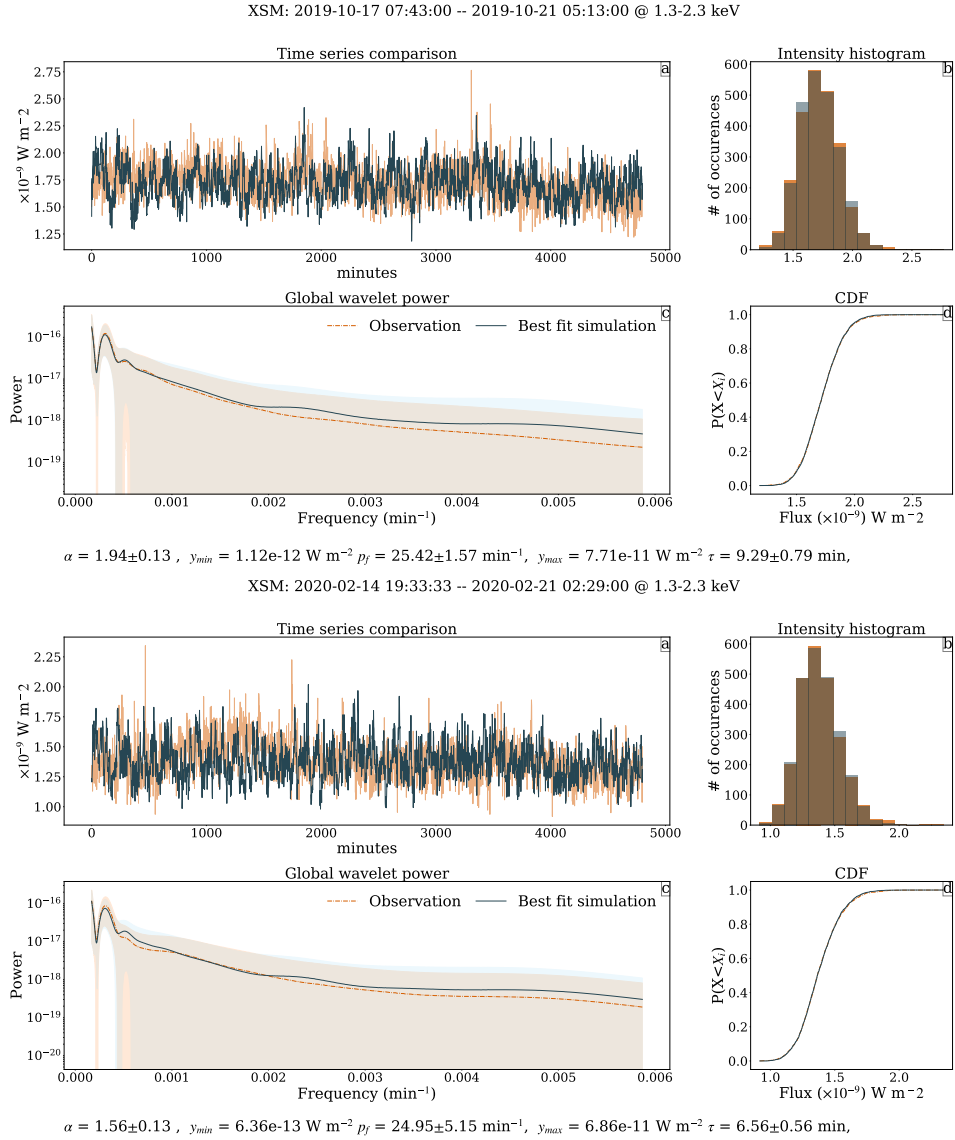

**Figure 4.20**: Same as Fig. 4.18, but for 1.3−2.3 keV band for the segment from year 2020.

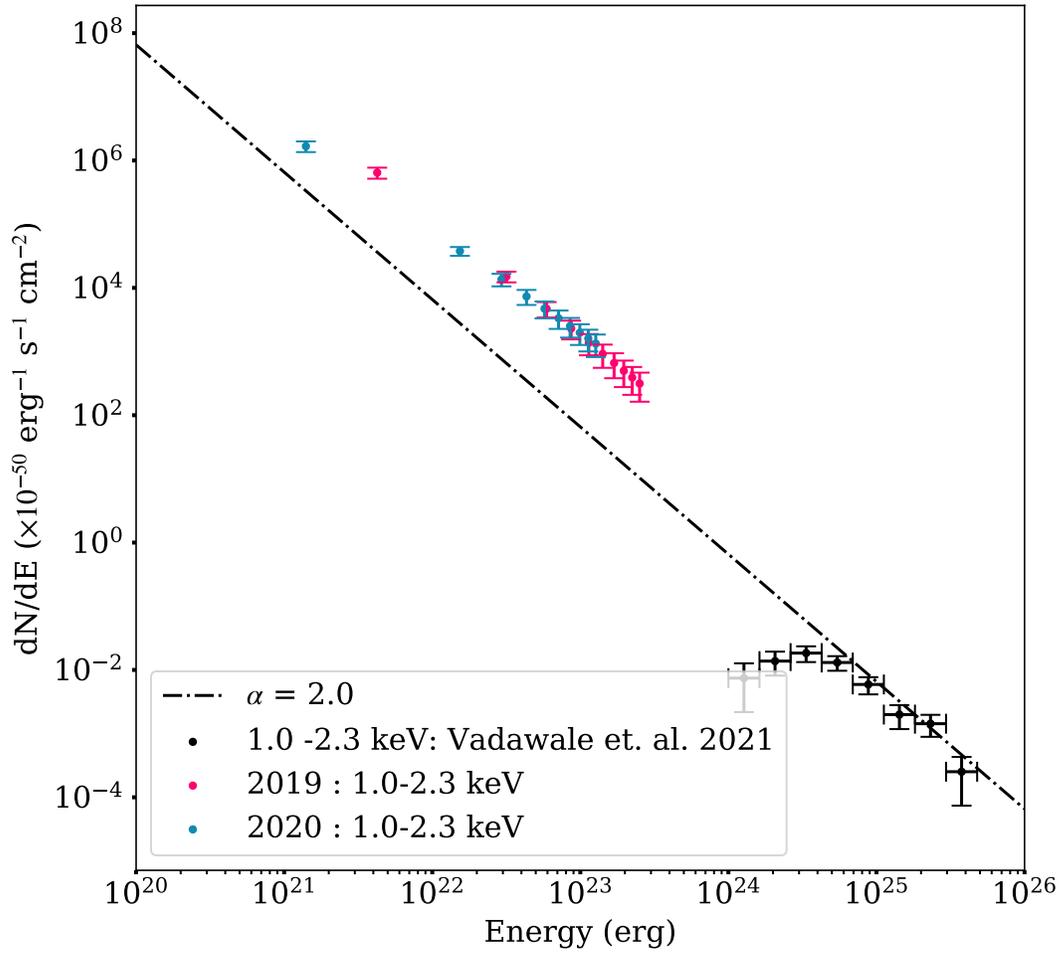

**Figure 4.21**: Frequency distribution of impulsive events inferred from the observations in the 2019 year (pink) and 2020 (cyan). The scatter is the frequency distribution of event energies from the model. The error bars on inferred parameters are obtained by propagating the Monte Carlo uncertainties. The black scatter shows the inferred frequency distribution from Vadawale et al. (2021a), while the black dot-dashed line corresponds to $\alpha = 2.0$.

# Chapter 5

# Unifying solar wind origin and coronal heating

*The solar coronal heating in quiet Sun (QS) and coronal holes (CH), including solar wind formation, are intimately tied by magnetic field dynamics. Thus, a detailed comparative study of these regions is needed to understand the underlying physical processes. In this work, we investigate the similarities and differences between CHs and QS in the chromosphere using the Mg II h & k, C II lines, and transition region using Si IV line for regions with identical absolute magnetic flux density ($|B|$). This thesis chapter is an adapted from a set of two papers that appeared in the literature as **Properties of the C II 1334 Å Line in Coronal Hole and Quiet Sun as Observed by IRIS** (DOI: 10.3847/1538-4357/ac2575), and **On the Formation of Solar Wind and Switchbacks, and Quiet Sun Heating** (DOI:10.3847/1538-4357/ac3d88).*

Over the course of the last few chapters, we have seen how the solar corona is anomalously hot when compared to the photosphere. Furthermore, we have also seen how different morphological structures exist in the corona, which must be studied to understand the heating of this corona. These different morphological structures – i.e., CH, QS, and AR – are also tied intimately with the outflow coming from the Sun, i.e., the solar wind.

The CHs appear dark in the corona, while in the lower atmosphere, they do not show any such differentiation. Thus, on average, the CHs have lower temperatures than either QS or AR. However, the CHs are also strongly associated with solar wind streams.  Is there a connection?

It is interesting to note that while CHs are clearly distinguishable from QS in the EUV and X-ray images, these two regions appear extremely similar at lower heights *viz* the chromosphere and photosphere (see, e.g., Stucki et al. 2000, 1999; Kayshap et al. 2018; Tripathi et al. 2021a).  The He I 10830 Å is a chromospheric absorption line, which, however, shows excess intensity (and thus lower absorption) in CHs (Harvey & Sheeley 1977; Kahler et al. 1983).  The He I 584 Å is an emission line that shows lower intensity (Jordan et al. 2001) in CHs, while showing excess blueshift and line widths in the CHs when compared to QS (Peter 1999).  However, these differences may be attributed to the sensitivity of these lines to coronal radiation, reflecting conditions in the corona.  Furthermore, at 17 GHz in the microwave, CHs are found to be brighter than QS (Gopalswamy et al. 1999), while this difference is not observed in radio wavelengths at 1.2 mm (Brajša et al. 2018).  Thus, a gross differentiation of a given region in QS or CH is markedly seen predominantly in the coronal observations and not so lower in the atmosphere.

We have seen earlier in §. 1.2.2 that Hassler et al. (1999) found a relation between the network regions and blueshifts of Ne VIII, with more blueshifts in CHs.  Similarly, comprehensive studies of CHs and QS were undertaken by (Stucki et al. 1999, 2000) using spectral lines sensitive to a range of temperatures from $\approx 8 \times 10^3$ K to $\approx 1.4 \times 10^6$ K. While CHs showed a clear deficit in intensity, excess blueshift, and excess line width with respect to QS for spectral lines forming at a temperature higher than $\approx 4 \times 10^5$ K, at chromospheric temperatures, the differences were negligible and within the measurement error.  Similarly, Xia et al. (2004) studied the relationship between Doppler shifts of C II, H I Ly$\beta$, and O VI in CHs, and found a direct relationship between the Doppler shifts of O VI with that of C II and H I Ly$\beta$.  These correlated shifts led Xia et al. (2004) to conclude that they are signatures of solar wind in the chromosphere.  However, note that the associated uncertainties in the velocity scatter obtained by Xia et al. (2004) were large. Moreover, while the average chromospheric velocities in bins of the O VI velocities were studied, the systematic associations between red and blue shifted pixels, separately, for these lines, was not performed by Xia et al. (2004).

The correspondence between network region and outflows in the CHs, using Ne VIII line, demonstrated by Hassler et al. (1999) was further investigated by Tu et al. (2005) by mapping the formation heights of Si II, C IV and Ne VIII in a CH. On further detailed investigation, Tu et al. (2005) showed a clear relation between the Ne VIII blueshifts and the underlying magnetic field configuration, obtained using the potential field extrapolation of the photospheric magnetic field.  Thus, Tu et al. (2005) suggested modulation of the solar wind velocities due to the underly-ing magnetic field configuration.

More recently, Kayshap et al. (2018) investigated the intensity differences between CH and QS in the Mg II k line, observed by the Interface Region Imaging Spectrometer (IRIS, De Pontieu et al. 2014). They find a clear deficit of intensity in CHs over QS for regions with similar absolute photospheric magnetic flux density (|B|) and with larger differences for larger |B|. A similar analysis for the intensity, velocity, and non-thermal widths for Si IV was performed by Tripathi et al. (2021a) (henceforth referred to as Paper I). Similar to the results of Kayshap et al. (2018), intensity deficit in CHs over QS for regions with similar |B| was observed. Moreover, CHs (QS) were more blueshifted (redshifted) for identical |B|. However, no significant difference was observed in the non-thermal width between CH and QS. The excess CH blueshifts were interpreted to be signatures of the nascent solar wind at Si IV formation heights in Paper I. Thus, while a clear signal of solar wind was reported in the hotter Ne VIII line by Tu et al. (2005), the signatures are already present in the upper TR line Si IV, if the underlying photospheric magnetic flux density distribution is taken into account. Furthermore, since the regions with identical |B| were compared, the deficit in intensity in CHs over QS would mean energy to be either used to accelerate the solar wind or heat up the corona. Thus, a unified picture of solar wind formation & coronal heating was presented in Paper I.

Thus, we come to ask one of this thesis's most important science questions: What is/are the underlying mechanism(s) of heating up of the solar corona and subsequent generation of the solar wind? ([Q3] in §. 1.3).

The differentiation between CH and QS starts becoming statistical in nature in the lower solar atmosphere. Thus, we first study the Mg II h & k line dynamics and the C II line dynamics in CH and QS. We then go ahead and explore the correlations between the Mg II, C II, and Si IV lines in CH and QS, with the intention of explaining the observations. The remainder of this chapter is structured as follows: In §5.1, we describe our observations, with feature extraction for the Mg II line in §5.1.1, and for the C II line in §5.1.2. In §5.2, and §5.3, we present results of the Mg II lines and C II line on one dataset, while we present the results across all the datasets in §5.4 and §5.5 for the two lines. We then recapitulate the results for the Si IV line from Paper I across the extended dataset in this work, in §5.6. In §5.7, we present the correlations between the velocities of different lines, while we summarize all of our results in §5.8. Finally, we provide an interpretation in the context of the origin of the solar wind, switchbacks (Bale et al. 2019), and QS coronal heating §5.9.

# 5.1 Data

In this study, we use the observations recorded by IRIS, AIA, and HMI. We consider spectra from all three windows of IRIS, while we use the SJI data centered around 1330 Å and 2796 Å for co-alignment purposes. From AIA, we consider the 193 Å images to distinguish between CHs and QS and the 1600 Å images to co-align the IRIS, AIA, and HMI observations. We obtain the information on the photospheric absolute magnetic flux density (i.e. |B|) from the line-of-sight (LOS) magnetograms obtained with HMI. In the datasets we used, IRIS provides photometric context images in NUV and FUV with a pixel size of $\approx 0.16''$ and at a cadence of $\approx 63$s. The spectra have a pixel size of $\approx 0.16''$ along the slit and sample at $\approx .33''$ across the field of view (FOV). The spectral pixel size in these rasters is $\approx 25.9$ mÅ, while the time cadence between successive slit positions is $\approx 30$ s. The AIA images used are taken with a pixel size of $\approx 0.6''$, with the EUV images a time cadence of $\approx 12$ s, while the Ultraviolet images are taken at $\approx 24$ s cadence. HMI obtains the $B_{LOS}$ magnetograms at $\approx 45$ s cadence with a pixel size of $0.5''$.

For our study, we analyzed five sets of observations recorded by IRIS in spectroscopic mode. The main criteria used to select these observations are that the raster must include CH and QS within the same FOV and that they must be taken within latitude and longitude of $\pm 60°$. The IRIS observation details are given in Table. 5.1. Out of these, three of the observations *viz.* DS1, DS2, and DS5 were also studied in Paper I to characterize the similarities and differences in QS and CHs in TR using Si IV line. We use corresponding coordinated AIA data cubes with cutouts from the full disk data used from HMI.

The Mg II h & k lines form near 2803.53 Å and 2796.35 Å, respectively, while the C II and Si IV lines form near 1334.53 Å and 1393.755 Å, respectively. The Mg II and C II lines form in an optically thick chromosphere under non-local thermodynamic equilibrium conditions (see, for e.g. Leenaarts et al. 2013; Rathore et al. 2015b).

Table 5.1:: Details of the IRIS rasters used in this study. Note that average $\mu$ is mentioned for each Field of View.

| Dataset name | Time range | (Xcen,Ycen) | Raster FOV | $\mu$ |
|:---:|:---:|:---:|:---:|:---:|
| DS1 | 2014-07-24 11:10:28 − 14:40:53 | (128″,-180″) | (141″,174″) | 0.97 |
| DS2 | 2014-07-26 00:10:28 − 03:40:53 | (469″,-167″) | (141″,174″) | 0.85 |
| DS3 | 2014-08-02 23:55:28 − 03:25:53 +1d | (332″,-152″) | (141″,174″) | 0.92 |
| DS4 | 2015-04-26 11:39:31 − 15:09:56 | (-288″,45″) | (141″,174″) | 0.95 |
| DS5 | 2015-10-14 11:07:33 − 14:37:58 | (215″,-165″) | (141″,174″) | 0.97 |

Thus, these lines show extremely complex features and have non-trivial associations with local plasma properties. They have been explored in detail in Rathore et al. (2015b); Leenaarts et al. (2013). For all practical purposes, the Si IV line, forming in QS TR, can be considered to be formed in optically thin conditions (Tripathi et al. 2020; Gontikakis & Vial 2018), and its properties in QS and CH are studied in detail in Paper I.

Fig 5.1.**a** displays a portion of the solar disk obtained from AIA 193 Å full disk image. The over-plotted white box represents the IRIS raster FOV. Panels **b** and **c** display the pseudo-rasters of AIA 193 Å and HMI LOS magnetogram, while the unsigned magnetic flux density is shown in panel **d**. We apply the segmentation algorithm from Upendran et al. (2020) to the AIA 193 Å pseudo-rasters to obtain a demarcation of CHs from QS. This algorithm has also been explained in §. 3.2.4. In Fig. 5.1.**b** & **c**, the green contours demarcate CH from QS. We see that the HMI pseudo-raster does not show any visual difference between CHs and QS, similar to the results obtained by Tripathi et al. (2021a); Kayshap et al. (2018).

### 5.1.1 Feature extraction: Mg II

The Mg II lines offer crucial information on the plasma conditions in the formation region, encoded into the line intensities and Doppler shifts of the line core (k3 & h3) and the peaks (k2v, k2r, h2v & h2r). For a detailed analysis and discussion of these lines, see Leenaarts et al. (2013); Leenaarts et al. (2013); Pereira et al. (2013).

We first extract the positions and intensities of these different spectral line features. For this purpose, we develop a peak finding algorithm based on Leenaarts et al. (2013); Pereira et al. (2013) that locates the zero-crossing of $dI/d\lambda$ within a window of $\pm 40$ km s$^{-1}$ from the reference wavelength (taken to be 2796.350 Å and 2803.529 Å for the k and h lines respectively, see Pereira et al. 2013).

The line core is identified to be the location with minimum intensity at the zero crossing. If the procedure is unable to locate such a minimum, e.g., in case of single-peaked or noisy profiles, we assign a default velocity of 5 km s$^{-1}$ following Leenaarts et al. (2013), since the remaining procedure rests on the identification of line core. Note, however, that the Mg II spectral profiles in this study, i.e., for QS and CHs, are predominantly double-peaked, as also noted by Leenaarts et al. (2013).

The two peaks closest to the line core on either side are the k2 (h2) peaks. Since the line core and peaks form at the local extrema of the line profiles (as a function of wavelength), they may be approximated to be a parabola close to the peak value. Thus, we may fit a parabola near the maximum/minimum and obtain a better estimate of the real extremum. This is called sub-pixel centroiding (similar to Teague & Foreman-Mackey 2018). Thus, the velocities and intensities for the core and peaks are then determined by fitting a parabola to the points near the feature extremum. Profiles that contain missing values of $-200$ are discarded. This procedure provides us with the intensities and Doppler shifts of the peaks & core of Mg II h & k lines. The line peak Doppler shifts are determined by taking the signed average of shifts of the blue and red peak (Leenaarts et al. 2013).

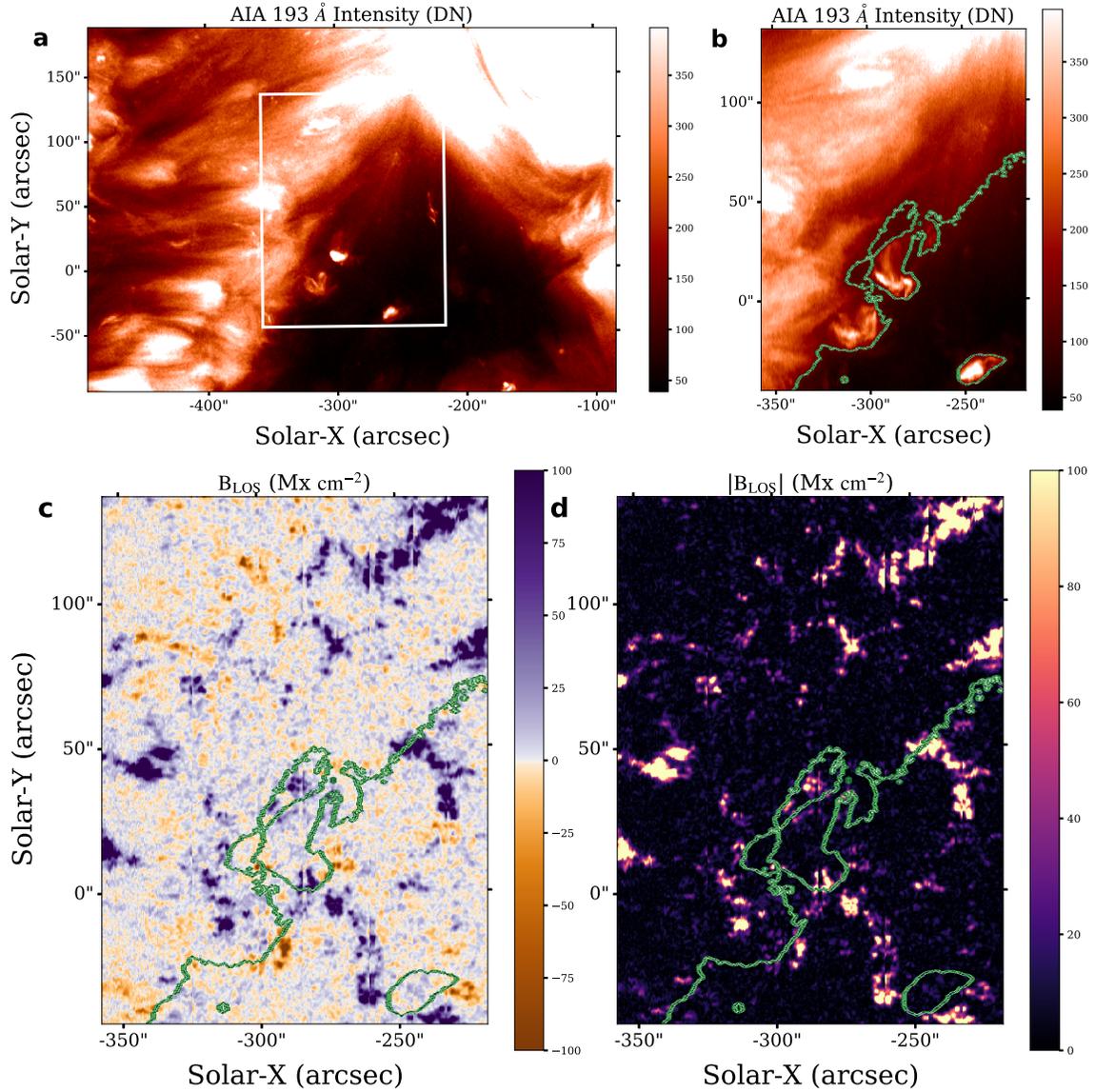

**Figure 5.1**: AIA 193 Å context image (Panel **a**). The over-plotted white box corresponds to the IRIS raster FOV. The pseudo-rasters obtained from 193 Å images and HMI LOS magnetograms corresponding to the IRIS raster for DS4 are shown in panels **b** & **c**, respectively, while |B| is displayed in panel **d**. The green contours in panels **b**, **c**, and **d** demarcate the CH and QS, obtained from the segmentation algorithm.

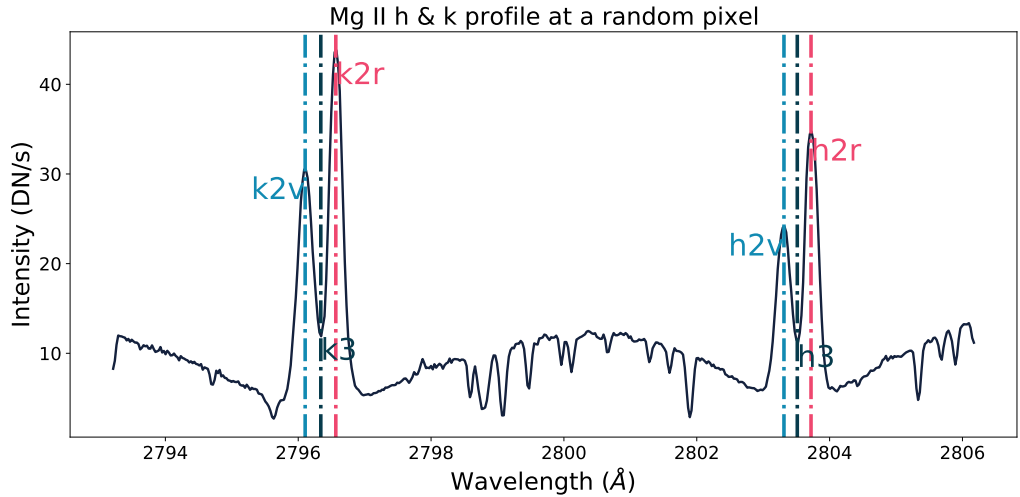

**Figure 5.2**: Mg II spectrum at a random QS location of DS4. The Mg II line features, along with their locations, are labeled as k2v or h2v (blue), k3 or h3 (black), and k2r or h2r (red).

Fig. 5.2 displays a spectrum obtained at a random pixel in DS4 centered at the two Mg II lines. The two lines and their associated features are labeled. The core & peaks have been identified using the algorithm presented above. The black vertical line denotes the line core. The red (blue) vertical line corresponds to the line's red (blue) peak. This convention is followed for both the h and the k lines.

## 5.1.2   Feature extraction: C II

For extracting the different properties from the C II line, we first smooth the spectral profiles following Rathore et al. (2015a). This smoothing marginally increases the number of converged fits, especially in regions with low intensity. The smoothing

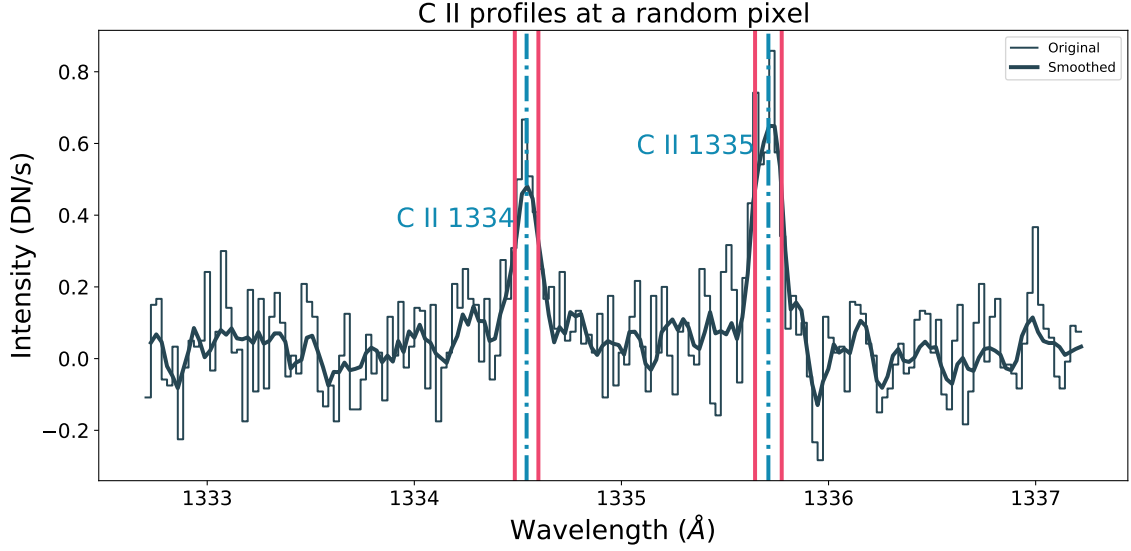

**Figure 5.3**: An example spectrum centered at the two C II lines obtained at a random QS location of DS4. The step plot shows the original spectrum, while the black solid line is the smoothed locally-averaged spectrum (as described through Eq. 5.1.2). The two C II lines are marked, with the dot-dashed blue lines depicting the line center and the red solid lines depicting $\pm\sigma$, both obtained from a single Gaussian fit to each line.

filter taken from Rathore et al. (2015a) is:

$$S_{\text{filt}} = \begin{cases} \frac{\sigma_s^2}{\sigma_s^2} m_s + \left(1 - \frac{\sigma_s^2}{\sigma_s^2}\right) s & \sigma_s^2 \geq \sigma^2 \\ m_s & \sigma_s^2 < \sigma^2 \end{cases}$$

where $s$ is the original signal, $S_{\text{filt}}$ is the filtered signal in a 3×3 window, where $m_s$ and $\sigma_s^2$ are the local means and variances, while $\sigma^2$ is average of local variances. For regions with a strong signal, the $S_{\text{filt}}$ tends to the local mean $m_s$, while the weaker regions are smoothed out. This operation is performed in slices of the 2-D spectrogram [coordinate along the slit, wavelength]. We perform a single Gaussian fit on the obtained spectra with a constant continuum to the C II line profiles following Rathore et al. (2015b). This scheme, while having the disadvantage of being influenced by the whole line profile in providing line core information, was our best bet due to the relatively large noise in using a peak-finding algorithm.

The fit is performed within a spectral window of $\pm 50\,\text{km s}^{-1}$ with respect to the reference wavelength of $1334.532\,\text{Å}$, as taken from Rathore & Carlsson (2015); Kelly & Palumbo (1973). From this fitting, we obtain the line core intensity, Doppler shift (i.e., the centroid), and width. The smoothed spectrum (solid) and the fitted line centroid, $\sigma$, are depicted in Fig. 5.3.  As mentioned earlier, the line core intensity is a proxy for the strength of the source function, as shown in  Rathore et al. (2015b). Similarly, the Doppler shift is a measure of the plasma velocity at the formation height. The line width, however, is a function of the line formation temperature and opacity broadening factor, as shown in Rathore et al. (2015b).  Double-peaked profiles are formed due to a local maximum in the source function, while line profiles become asymmetric due to the presence of velocity gradients in the chromosphere. For further information on the formation of C II lines and their general properties, see Rathore et al. (2015b,a); Avrett et al. (2013).

We also estimate the third and fourth moments, namely the skew and kurtosis, respectively, of the spectral profiles following Jeffrey et al. (2016).  These are computed since the observed spectral profiles are known to have marked departures from a Gaussian profile (Rathore et al. 2015a).  The skew and kurtosis for a perfectly Gaussian profile are expected to be 0 and 3, respectively.  Hence any departures would indicate a significant difference from a Gaussian profile.

The skew (S) and the excess kurtosis (K) are defined as:

$$S = \frac{1}{\sigma^3} \frac{\int_\lambda I(\lambda)\,(\lambda - \lambda_D)^3 d\lambda}{\int_\lambda I(\lambda)\,d\lambda}, \tag{5.1}$$

$$K = \frac{1}{\sigma^4} \frac{\int_\lambda I(\lambda)\,(\lambda - \lambda_D)^4 d\lambda}{\int_\lambda I(\lambda)\,d\lambda} - 3.0, \tag{5.2}$$

where $\lambda_D$ is the centroid estimated from the Gaussian fits and the integral is performed over the range $\pm 50$ km/s of our spectral window in wavelengths, around the reference wavelength.  The $\sigma^2$ is the second moment of the line given by:

$$\sigma^2 = \frac{\int_\lambda I(\lambda)\,(\lambda - \lambda_D)^2 d\lambda}{\int_\lambda I(\lambda)\,d\lambda}. \tag{5.3}$$

Note that the moments are computed for the Gaussian fit to the line.  For the spectral line, the continuum is subtracted, and then the moments are computed, following Jeffrey et al. (2016)

We shall first present the analysis and results obtained from one dataset (DS4) for Mg II and C II lines in §5.2 and §5.3.  In the end, we average the results obtained for all five data sets in §5.4 and §5.5 for the two lines.  For the Si IV line, we present results obtained from the extended dataset based on the analysis performed in Paper I in §5.6.

Following the procedure outlined in Paper I, to improve the signal-to-noise ratio (SNR) and statistics, we consider the derived quantities in the bins of |B| and report

the average values in these bins. We use a constant |B| bin size of 0.1 in log space to account for the fewer pixels at high |B|. Note that the LOS |B| and Doppler shifts are converted to the radial field and flows by dividing with $\mu$ (the heliocentric coordinate, see Thompson 2006) of the respective pixel. Furthermore, the errors reported in all the plots are the standard errors on the mean. The standard error is defined as $\sigma/\sqrt{N}$, where $\sigma$ is the standard deviation for the samples present in the bin, and $N$ is the number of samples. Note that while we are interested in and report the variation of mean value in each bin, we present the distribution of samples in each bin with $1$ and $90$ percentile bounds in the §. 5.10.

## 5.2  Results from the analysis of the Mg ɪɪ: Single dataset

We now investigate the dependence of the following features on |B| through scatter plots: 1) core & peak intensities of the two lines, 2) intensity ratios of the two peaks, 3) line core velocities, and 4) average peak velocities. Note that we consider 10 G as the noise floor of |B| (Yeo et al. 2014; Couvidat et al. 2016).

### 5.2.1  Intensities

First, we consider the intensities obtained from the two Mg ɪɪ lines. We have six intensity measurements: four from the peaks and two from the cores of h & k lines. In Fig. 5.4, we display the intensity maps obtained in these features for DS4. The over-plotted blue contours demarcate the QS and CH. We see no visible difference between CH and QS in any of the features of the Mg ɪɪ line. However, a clear relation is seen with the photospheric magnetic flux density in Fig. 5.1.**c**, inline with the results of Kayshap et al. (2018) for Mg ɪɪ k line.

In Fig. 5.5, we plot the intensities of different Mg ɪɪ h& k features in bins of |B|. In the plots, black (orange) data points represent CH (QS), with the k (h) line features in the top (bottom) row. We see that the intensity increases with |B| for both CH and QS for all the line features. Furthermore, the QS shows excess intensity over CH for |B| $\geq$30 Mx cm$^{-2}$. However, there is a mild difference in the intensities already at 10 G for the k line. We further note that the difference in intensities between QS and CH increases with increasing |B|, with an apparent saturation at higher |B|. These results are in agreement with those reported by Kayshap et al. (2018).

Another key inference from Fig. 5.5 is the larger intensities of the blue peaks

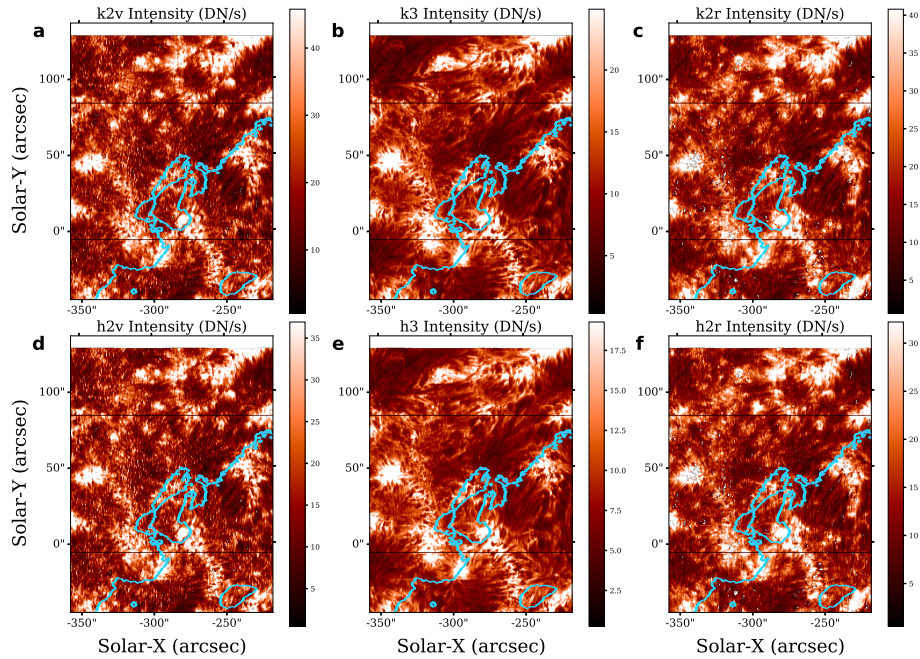

**Figure 5.4**: Intensity maps obtained in Mg Ⅱ k (top row) and h (bottom row) line features from DS4. The blue contours represent the CH-QS boundary, as shown in panel Fig. 5.1.**b**.

(k2v and h2v) over the red peaks (k2r and h2r; see panels **b**, **c**, **e** and **f**). Note that the peak ratio $(I_v\text{-}I_r)/(I_v\text{+}I_r)$ is a proxy for the average chromospheric velocity, as has been suggested by Leenaarts et al. (2013). A positive peak ratio corresponds to down-flowing plasma in the atmosphere, while a negative ratio corresponds to up-flowing plasma. A preferentially larger blueward or redward peak arises due to increased absorption on the side of the smaller peak (see Leenaarts et al. 2013, for details). The enhanced intensities in the blue peaks over red peaks suggest that the chromosphere is more redshifted on an average, resulting in increased redward absorption at the height corresponding to Mg Ⅱ formation. Note that unless stated otherwise, redshift means plasma moving toward the Sun, and blueshift means plasma moving away from the Sun.

In the following, we consider pixels with only positive and negative ratios separately and the variation of the ratio with |B|. This would consider only pixels with downflows (or upflows) as a function of |B|. Fig. 5.6 plots positive (panel **a** & **c**) and negative ratios (panels **b** & **d**) for k2 and h2 line features. From the plots, we find that the peak ratios vary between 0.1 and 0.2, which is in a sufficiently linear regime of the scatter between peak ratio and average $v_z$ (as may be seen in Fig. 8.e and f of Leenaarts et al. 2013). Thus, we may consider the peak ratio as a proxy for the average middle chromospheric velocities in CH and QS. The plots show that the peak ratio becomes increasingly positive or negative, rising with |B| till 50 Mx cm$^{-2}$ and saturating thereafter. Also, note that the positive and negative peak ratios are larger in CHs than in QS for identical |B|. This intriguing finding is indicative of larger downflows as well as upflows in CH over QS for the regions with identical |B|.

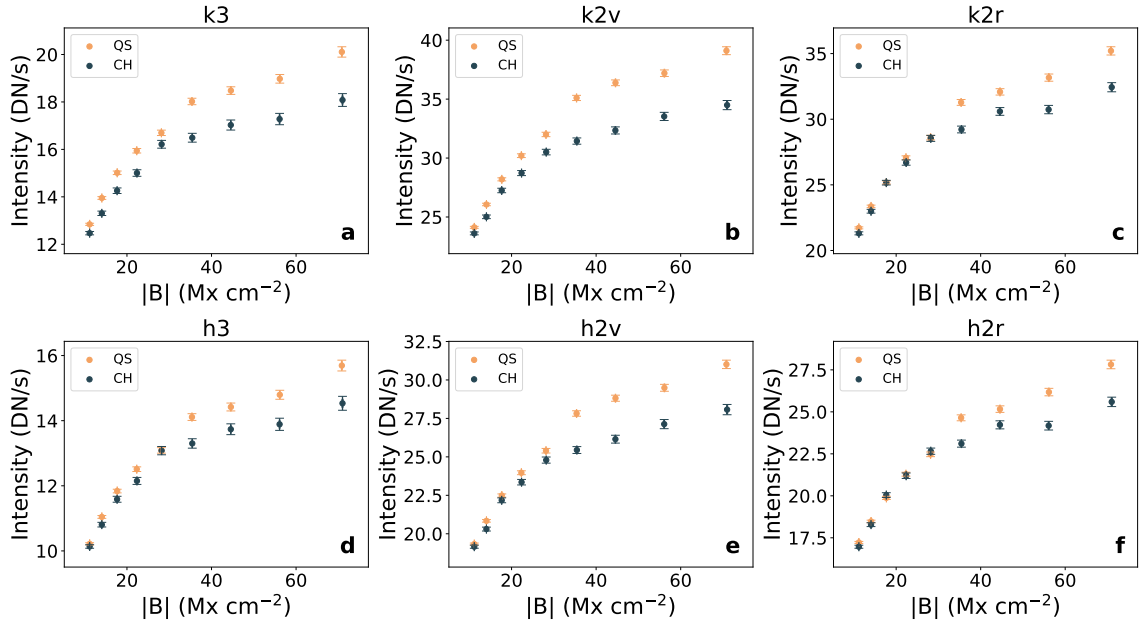

**Figure 5.5**: Variation of intensities in Mg II k (top row) and h (bottom row) line features with |B|. The orange color indicates QS, and the black indicates CH. Note that the standard errors in |B| have also been plotted in this figure and all subsequent figures, but they are too small to be seen.

## 5.2.2  Doppler Shifts

To further explore and understand the chromospheric velocities, we now consider the velocities derived from Doppler shifts, which have a tight correlation with local plasma velocity at the height of formation (Leenaarts et al. 2013). Fig. 5.7 displays the velocity maps obtained for k3 (panel a), k2 (panel b), h3 (panel c), and h2 (panel d). Note that while the core velocities are the straightforward shifts from the reference wavelength, the peak velocities are a signed addition of the peak shifts from the reference wavelength. The red contours demarcate CH from QS. The velocity maps for both k and h lines reveal that, on average, the chromosphere is redshifted in both QS and CH., as observed in Mg II lines. Moreover, there are no conspicuous differences between CH and QS in the Doppler maps obtained in k3/h3 as well as k2/h2.

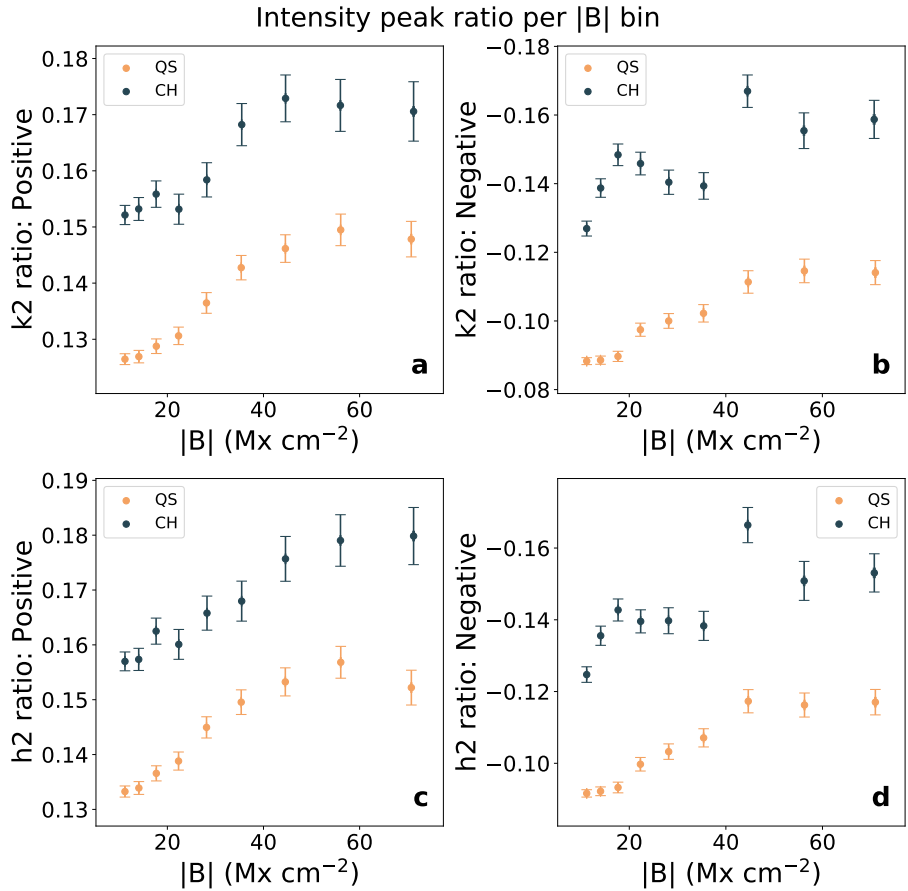

**Figure 5.6**: The peak ratios as a function of |B| for the k (top row) and h (bottom row) lines. The positive ratios (downflowing plasma) are depicted in panels **a** & **c**, while the negative ratios (upflowing plasma) are depicted in panels **b** & **d**. Note the absolute values of the ratio increase along the y-axis for all the plots.

In Fig. 5.8, we plot the variation of velocities obtained in k3 (top row) and h3 (bottom row) with |B|. Following Paper I and Paper II, we analyze this data in two ways. On the one hand, we consider the signed average velocities in every |B| bin and plot the variation with |B| (panels **a** & **d**). On the other hand, for each bin of |B|, we consider the redshifted and blueshifted pixels separately and plot the variation of velocities with |B| (panels **b** & **e** for upflows and panel **c** & **f** for downflows). While the former provides us with the average velocities, the latter gives us a systematic variation of downflows and upflows with increasing |B| in CH and QS. This is akin

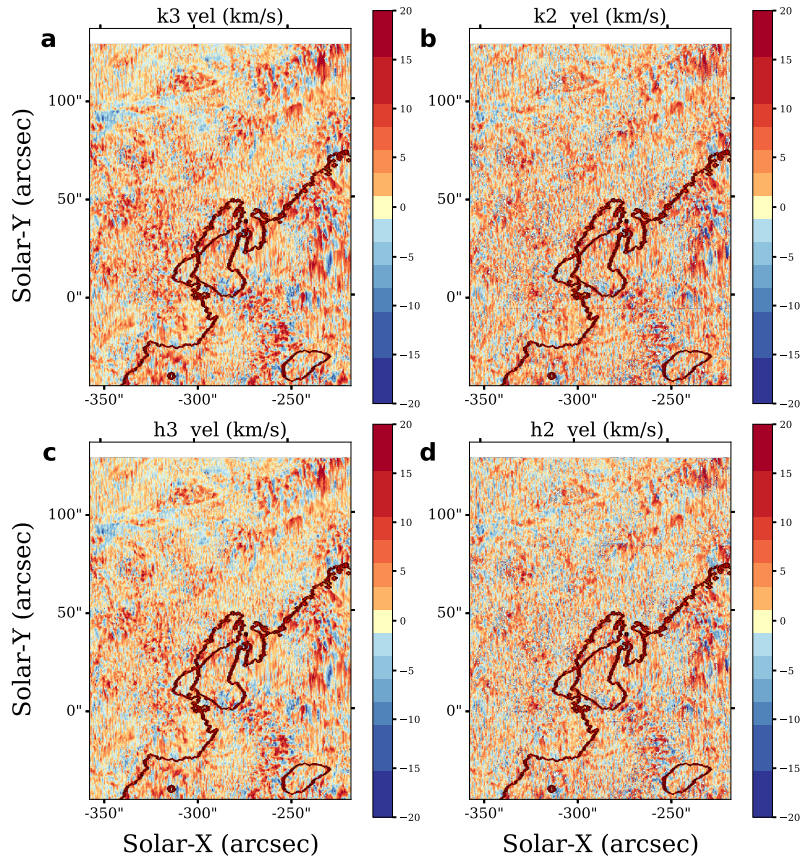

**Figure 5.7**: Velocity maps obtained in Mg II k3 (**a**), k2 (**b**), h3 (**c**) and h2 (**d**) from DS4. The red contour shows the CH-QS boundary.

to the systematic variations seen in Fig. 5.6. Such an exercise can tell us if the dynamics of the magnetic field cause any preferential effect on the redshifts and blueshifts.

Figs. 5.8.**a** & **d** clearly show that, on average, the chromosphere is redshifted in both QS and CH, similar to what is inferred from the maps shown in Fig. 5.7. This result is consistent with the known observations (see e.g., Stucki et al. 2000, 1999; Avrett et al. 2013, and references therein). Moreover, CHs show a larger redshift than QS for $|B| \leq 30$ Mx cm$^{-2}$, beyond which there are no differences in the velocities. At $|B| \geq 80$ Mx cm$^{-2}$, there is some hint for the CHs to show a larger redshift. However, note that the average velocities are quite small in both regions.

When we consider the blue/red-shifted pixels separately, both in CH and QS, we find a definite increase in the upflow (see Fig. 5.8.**b** & **e**) and downflows (see Fig. 5.8.**c** & **f**) with increasing $|B|$. Moreover, the magnitudes of upflows and downflows are larger in CHs than in QS for the regions with identical $|B|$. Such a trend is consistent with the inference made using the ratios of the two peaks shown in Fig. 5.6. Note that the magnitude of the downflows in QS and CH is much larger than that of the upflows, explaining the predominant downflows. Finally, the veloc-ity differences between CHs and QS increase with increasing |B|, with an apparent saturation of velocities for $|B| \geq 60$ Mx cm$^{-2}$.

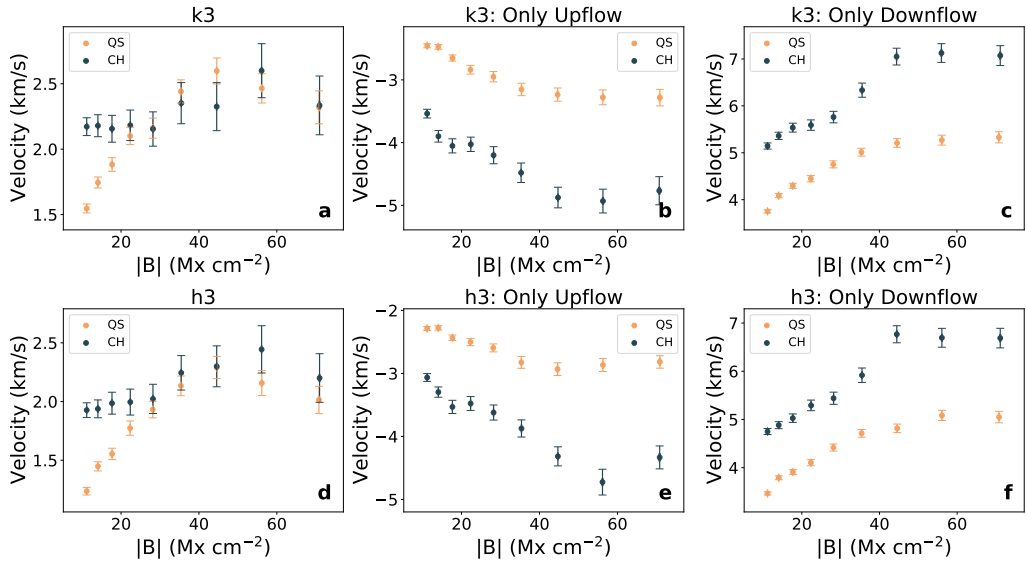

**Figure 5.8**: Mg ɪɪ k3 and h3 velocity variation with |B|. Panels **a** and **d** show the variation of signed average velocities in k3, and h3 binned in |B|. Similarly, panels **b** and **e** show the variation of only blueshifted pixels, while panels **c** and **f** show the variation of only redshifted pixels. The black (orange) scatter corresponds to CHs (QS).

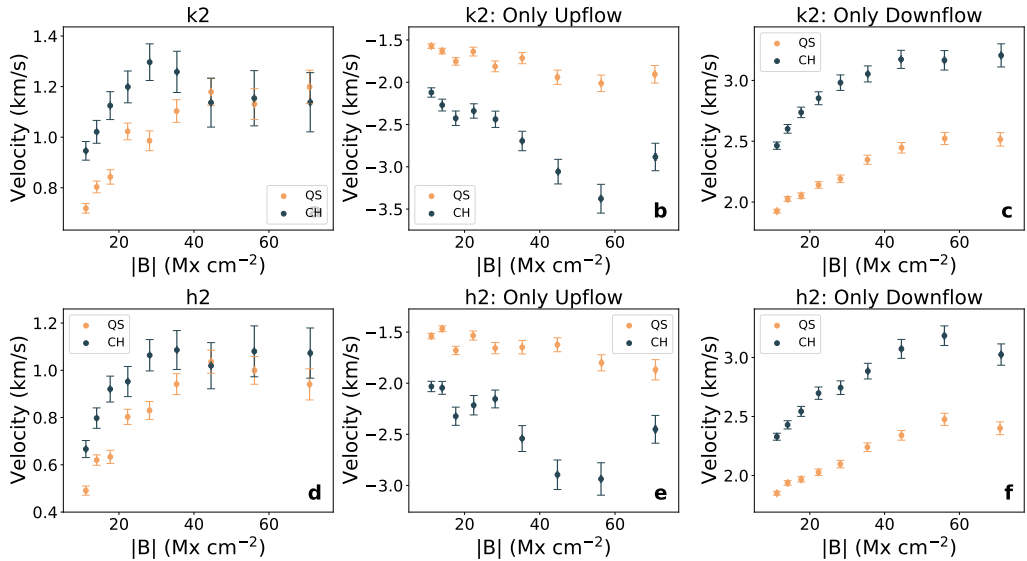

**Figure 5.9**: Same as Fig. 5.8 but for k2 and h2.

To investigate if these variations are also seen at the average formation height of k2 and h2, we perform the same analysis with the average velocity obtained from the k2 and h2 peaks and display the results in Fig. 5.9. The plots (panels **a** & **d**) reveal that the average velocities obtained at k2/h2 peaks are much smaller than the core velocities. The CHs show excess redshifts than QS for regions with $|B| \lesssim 30$ G, beyond which the difference in velocities cease to exist. Moreover, the velocities in both CHs and QS increase with |B| till 30 Mx cm$^{-2}$ and saturate thereafter.

We further note that CHs show excess upflows (**b** & **e**) as well as downflows (**c** & **f** of Fig. 5.9) over QS for regions with identical |B|. Both upflows and downflows in CHs show a monotonic increase with increasing |B| till about 60 Mx cm$^{-2}$ and saturate thereafter. For QS, however, variation in upflows is very tiny, while downflows show an increase with increasing |B| that also saturates beyond $\approx$60 Mx cm$^{-2}$. The velocities obtained from the peaks largely follow the velocities obtained using the core of the line, with the former being smaller than the latter.

## 5.3 Results from the analysis of the C II: Single dataset

We shall now investigate the dependence of intensity, velocity, line width, skew, and kurtosis of the C II line on |B|.

### 5.3.1 Intensity, velocity and line width

In Fig. 5.10, we display the intensity, velocity, and line width across the full FOV of DS4 for the C II line. Panel. **a**, **b**, and **c** display the intensity, velocity, and line width map, respectively. The over-plotted green contours are the same as those plotted in Fig. 5.1.b. Note that the intensity map and all subsequent maps show a white space at the bottom of the raster that corresponds to missing data. There is no visual difference between the CH and QS in Fig. 5.10, like the differences seen in the coronal image of Fig. 5.1. From the intensity maps shown in Fig. 5.10.a, and the photospheric magnetic field maps shown in Fig. 5.1.c, we find a clear correspondence between the |B| and intensities. Furthermore, there is a clear correspondence between the C II and Mg II intensities, though the C II line intensity structure appears to be more diffuse. In Fig. 5.11.a, we plot the variation of intensities as a function of |B|. We find that for both CHs and QS, the intensities of the C II line increase with increasing |B| till about 50 Mx cm$^{-2}$ and show a reduced rise thereafter in CHs. The intensities in the QS are larger than those in CH for the regions with identical |B| for larger flux densities. We further note that with increasing |B|, the difference in intensities increases slightly. This is similar to results from the Mg II lines from § 5.2, the findings of Kayshap et al. (2018) for Mg II lines and Tripathi et al. (2021b) for Si IV line.

The Doppler velocity map in C II 1334 Å is shown in Fig. 5.10.b, with the green contours demarcating CH and QS. The velocity maps shown in panel **a** reveal that both the C II line is predominantly red shifted in CH as well as QS, similar to the

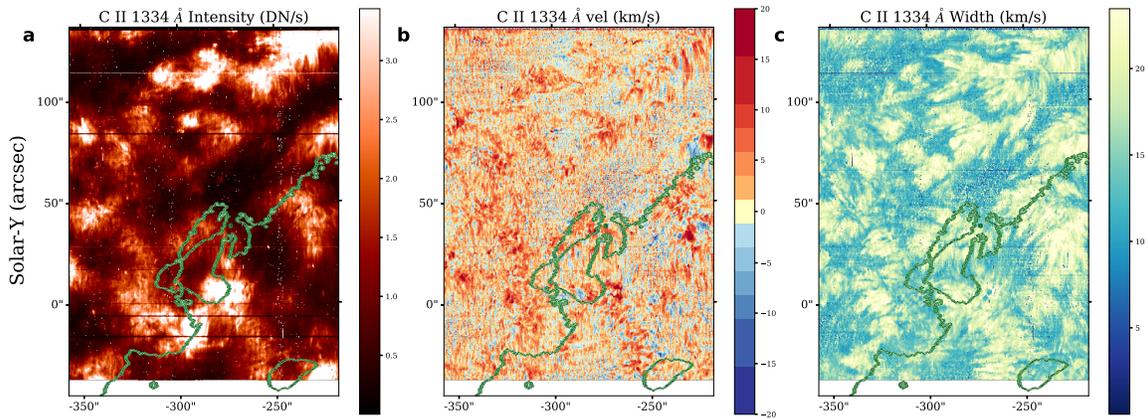

**Figure 5.10**: Intensity, velocity and line width map of C II 1334 Å for DS4 is shown in panel **a**, **b**, and **c** respectively. The green contours show the boundary between the CH and QS.

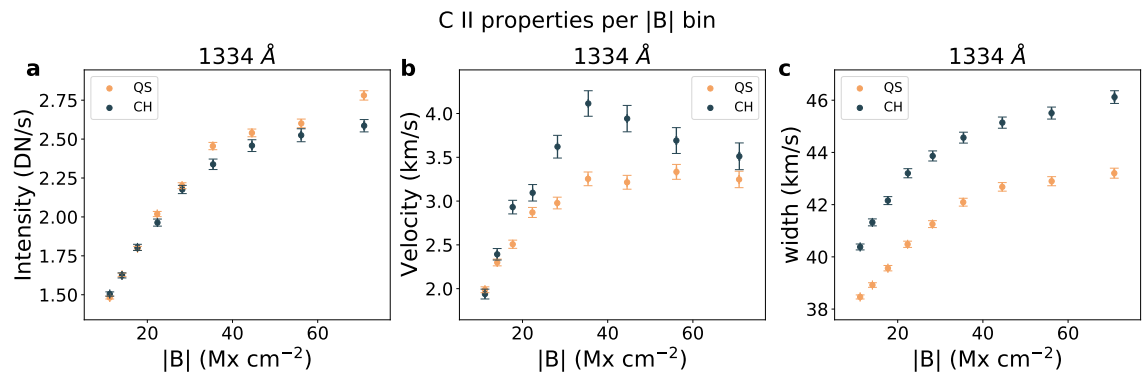

**Figure 5.11**: Intensity, average velocity and line width as in bins of |B| for CH (black) and QS (orange), as computed in the C II 1334 Å line for DS4 is shown in panel **a**, **b**, and **c** respectively.

Mg II line. Similar to the intensities, we find no visual difference in the Doppler shift in the CH and QS. Black (orange) curves in panel **b** denote CH(QS).

Similar to the intensities, we study the Doppler velocity in QS and CH as a function of |B|. Similar to the analysis in §. 5.2, we analyze the dependence of the average shift, redshifts, and blueshifts individually on |B|.

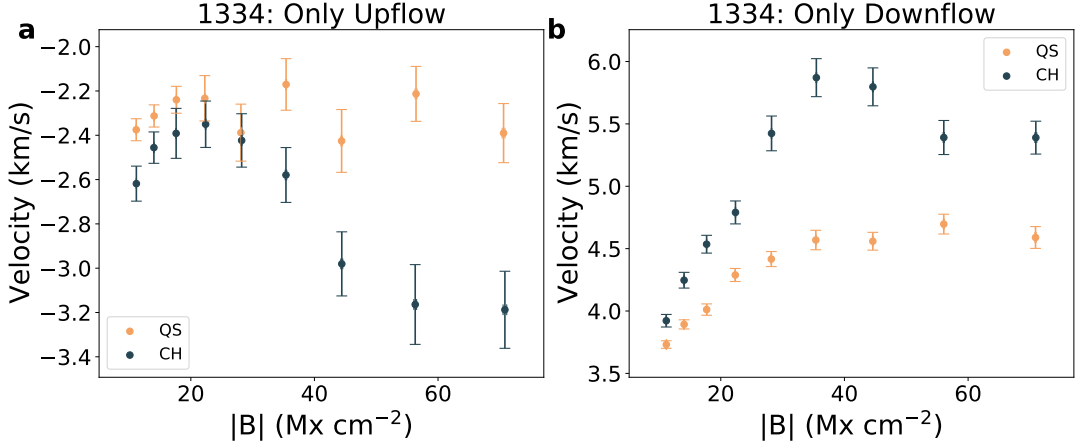

**Figure 5.12**: Doppler shift as a function of |B| for DS4. Panel **a** shows the variation of velocities obtained for only blue-shifted pixels, while panel **b** shows the variation of only red-shifted pixels.

In Fig. 5.11.b, we display the variation of signed average velocities obtained within bins of |B|. This reveals that, on average, both QS and CH are red-shifted in the chromosphere, and this velocity increases with |B|. We then consider the blueshifted and redshifted pixels separately in Fig. 5.12. The blue shifted pixels in Fig. 5.12.**a** show higher velocities in CHs than QS, where the CH blueshift increases with |B|. However, the QS blueshift appears independent of |B| for QS. Finally, when considering only the red-shifted pixels (see Fig. 5.12.**b**), we find that redshifts increase with |B|. Furthermore, the CHs have excess redshifts when compared to QS. We further note that the magnitude of the downflows is much larger than that of the upflows in both CHs and QS, which explains the predominant downflows in the chromosphere.

We next study the total line width obtained from the Gaussian fit. Note that the line width (see e.g. Rathore et al. 2015b) is defined as:

$$W_{\mathsf{FWHM}} = 2\sigma\sqrt{2\ln(2)} \tag{5.4}$$

where $W_{\mathsf{FWHM}}$ is the line width, and $\sigma$ is the standard deviation obtained from the Gaussian fits to the spectral line.

The line width map obtained for DS4 is shown in Fig. 5.10.c, with the green contours demarcating the CH and QS. Similar to the intensities and Doppler shifts, we do not see any conspicuous difference between the CH and QS.

In Fig. 5.11.c, we plot the line widths as a function of |B|. Note that the bin size of the |B| is the same as those used for intensity and Doppler shift. The line width increases rapidly with increasing |B|. Beyond 30−40 Mx cm$^{-2}$, for CH, the width still increases, albeit slowly. However, QS shows saturation beyond 40 Mx cm$^{-2}$ and a slight reduction thereafter.

### 5.3.2 Skew and Kurtosis

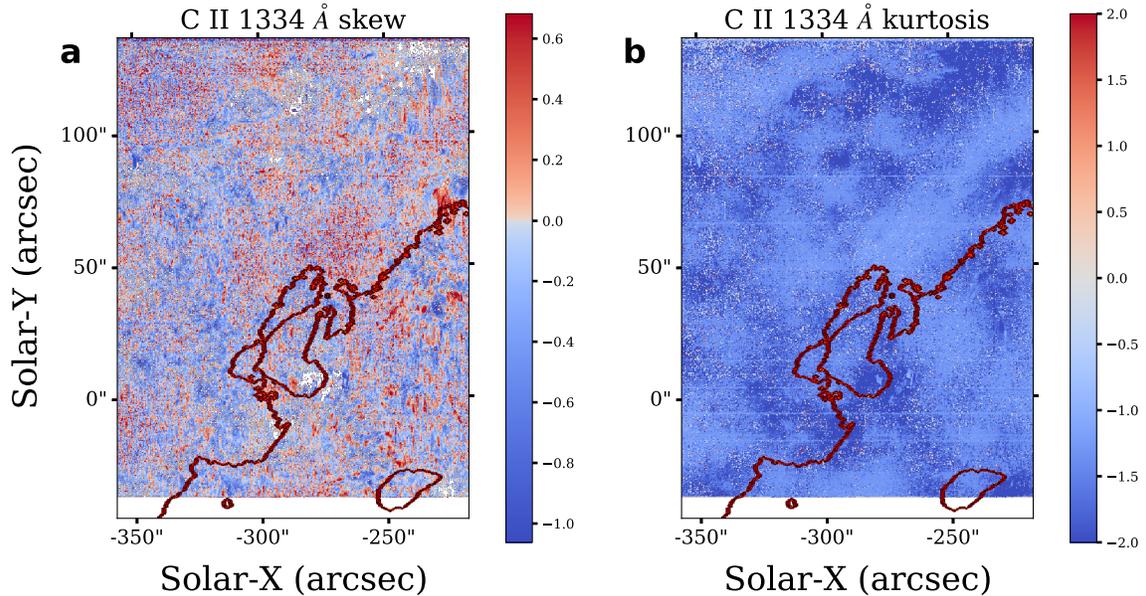

**Figure 5.13**: Skew (panel **a**) and kurtosis (panel **b**) maps for C II 1334 Å line obtained for DS4. The red contours depict the boundary between CH and QS.

Finally, we study the skew and kurtosis for the C II line using Eqn. 5.1 & 5.2. Fig. 5.13.a (b) displays the skew (kurtosis) maps. The skew maps show a good correspondence with that of the magnetic field in Fig. 5.1. This structure, however, is far more prominent as a deficit of kurtosis in Fig. 5.13.**c**.

In Fig. 5.14, we plot the variation of skew (panels a−c) and kurtosis (panels b−d) with |B|. In the plots, the skew and kurtosis obtained for the lines are shown as a scatter with dots, while the filled star bands correspond to the moments computed on the single Gaussian fit, with the bands and errors representing one sigma stan-

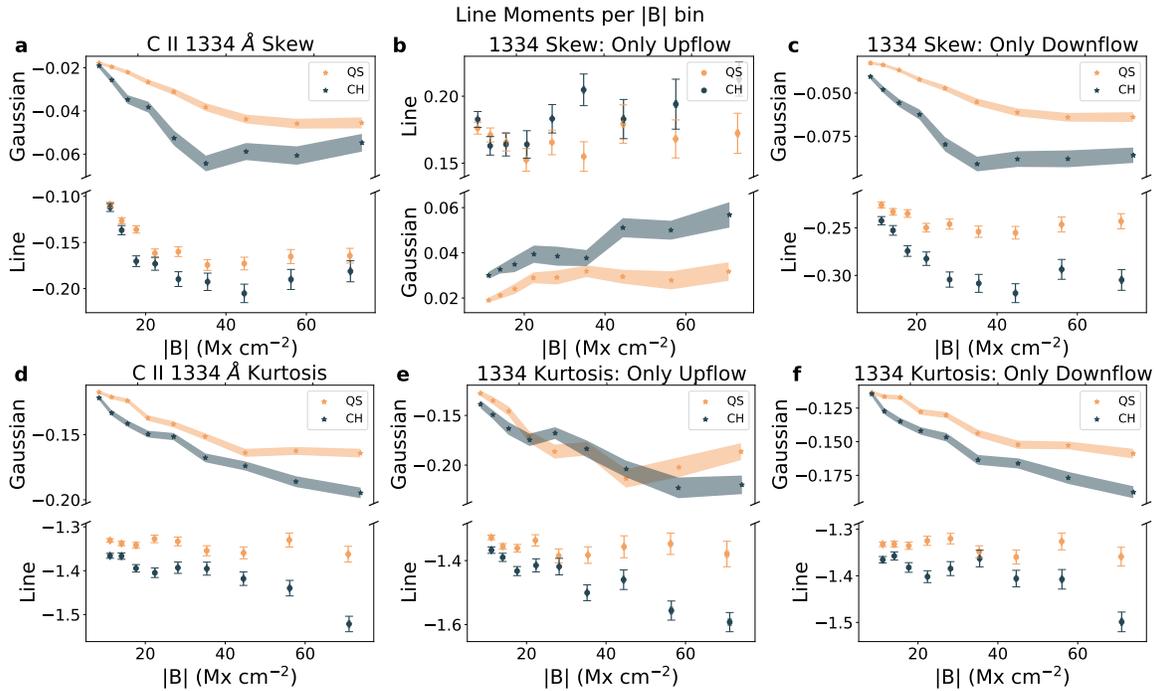

**Figure 5.14**: Skew (panels **a**–**c**) and kurtosis (panel **d**–**f**) variation with |B| for 1334 Å for DS4. The average quantities are shown in panels **a** and **d**, while the average for pixels with blueshifts (redshifts) alone are shown in panels **b** and **e** (**c** and **f**). The scatter plot with dots shows the moments for the spectral profiles, while the filled star plots are computed for the single Gaussian fits over the same wavelength range. Note that the y-axis has been broken to show clearly the variation of moments with the |B|.

dard error. Theoretically, a Gaussian's skew and excess kurtosis are zero. However, this need not be the case for a Gaussian profile sampled at specific wavelength locations. Hence, to get a handle on the significance of the computed profile moments, we also compute the moments for the Gaussian fit as a benchmark. The deviations of obtained moments of the Gaussian profile from its theoretical values quantify the effects of a discrete wavelength grid. To make these clearer, we have split the plots for the moments obtained from Gaussian fits and those directly computed from the spectral profiles themselves. The plots reveal that the Gaussian fit and spectral line have significantly different moments. The spectral profiles are negatively skewed with respect to the Gaussian fits (Fig. 5.14.a), indicating a general tendency to have a longer blue tail (or a steep red-ward rise) in the observed spectrum. Moreover, the line gets more skewed with increasing |B|. However, for the pixels showing blueshifts (redshifts) in Fig. 5.14.b and c, the lines show excess positive (negative) skew. This tells us that the blueshifted spectra have a steeper blueward rise, while the redshifted ones have a steeper redward rise. The kurtosis plot ( (Fig. 5.14.d)) shows that the spectral lines are flatter and have lesser outliers than a Gaussian due to the kurtosis deficit. This flatness is seen for both redshifted and blueshifted pixels (Fig. 5.14.e and f). Significant differences indicate that these are not just due to sampling artifacts but also to physical processes. However, the skewness and kurtosis values are similar in CH and QS. Thus, the spectral profile shapes are generally similar and not significantly different in these regions.

Having demonstrated the analysis and results obtained for a single dataset, we now consider all the five datasets listed in Table. 5.1 to increase the statistical significance of our results. We emphasize that the results for each dataset are similar to the results reported for DS4 in the previous section. For this purpose, we average the obtained parameters from all five sets of observations and study the dependence of intensities and velocities on |B|. Combining all the datasets is possible because the observations are taken at similar values of $\mu$. We further note that we present the results only for the Mg II k line features for brevity, as the results for both k and h lines are extremely similar.

## 5.4   Results from the analysis on the combined Dataset:: Mg II

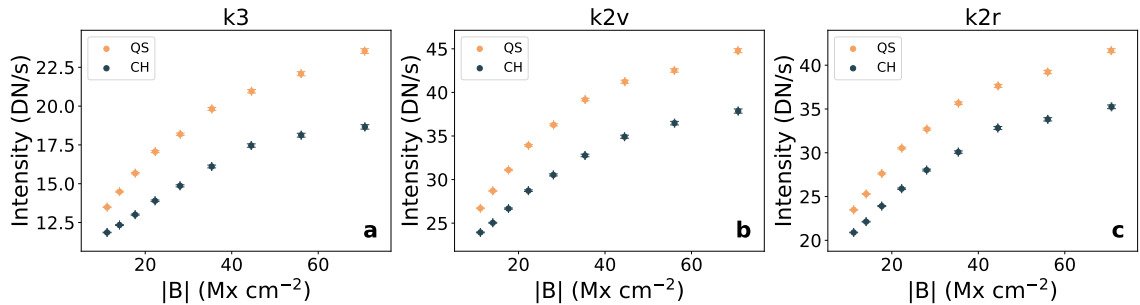

**Figure 5.15**: Same as the top row of Fig. 5.5, but for combined dataset.

In Fig. 5.15, we plot the variation of averaged intensities obtained in k3 (panel **a**), k2v (panel **b**) and k2r (panel **c**) as a function of |B|. For all three features of Mg II k line, we find that the intensity increases with increasing |B|, albeit some sign of saturation at higher |B|. We also find that QS regions show excess intensity over CHs for the regions with identical |B| and that the difference in intensities increases with increasing |B|.

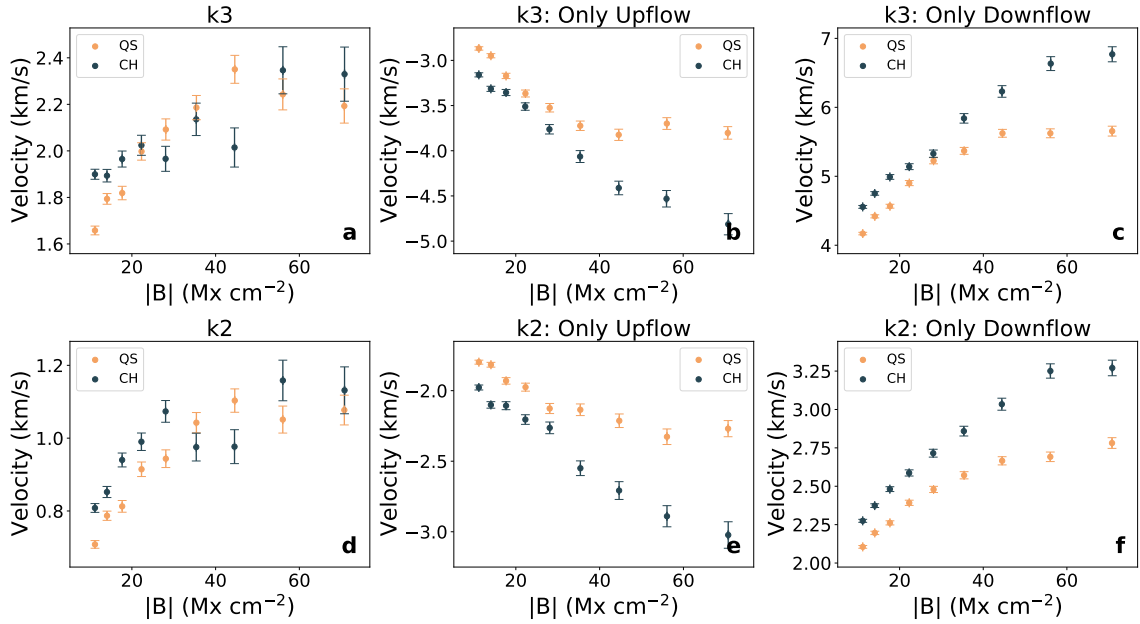

**Figure 5.16**: Same as the top rows of Fig. 5.8 and Fig. 5.9, but for the combined dataset.

We study the behavior of Doppler shifts as a function of |B| in Fig. 5.16. We plot the signed average of the Doppler shifts of k3 (k2) in Fig. 5.16.**a** (Fig. 5.16.**d**). The plots clearly show that both QS and CHs are redshifted on average and that the redshift increases with increasing |B|. Moreover, for $|B| \leq 30$ Mx cm$^{-2}$, the CHs show marginally excess redshifts, which disappear at higher |B|. We plot the velocity variation of pixels showing upflows in panels **b** and **e** and of downflows in panels **c** and **f**. There is a clear signature of monotonic increase of upflows and downflows in CHs with increasing |B|. However, such clear monotonicity is not seen for QS regions. While the flows increase for QS till about 30 Mx cm$^{-2}$, they get saturated thereafter. Moreover, the CHs show larger excess upflows as well as downflows over QS for larger |B|. Finally, the magnitudes of the flows in k3 are larger than that in k2.

## 5.5 Results from the analysis on the combined Dataset:: C II

We display the results for intensity, velocity, and width of the C II line as a function of |B| in Fig. 5.17. The plots reveal that the intensities increase in both QS and CHs

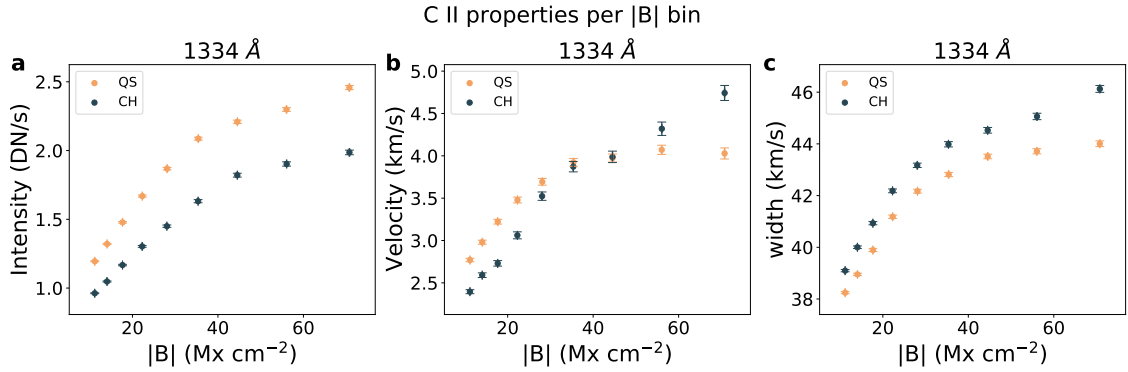

**Figure 5.17**: C II intensity (panel **a**), velocity (panel **b**) and line width (panel **c**) variation with |B| for all data sets taken together.

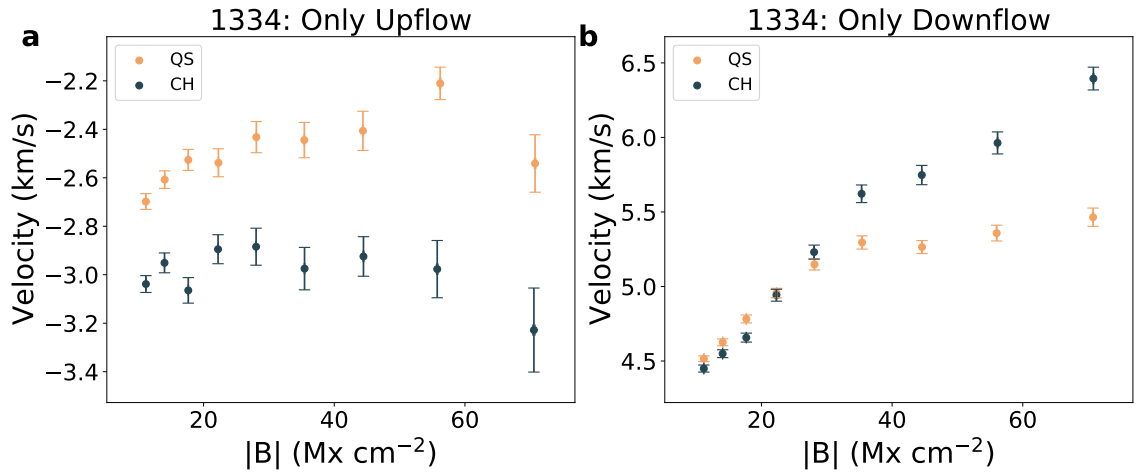

**Figure 5.18**: C II upflow and downflow velocity variation with |B| for all data sets taken together. Panel **a** shows the variation of upflows binned in |B|, while panel **b** shows the variation for downflows.

as a function of |B| (see panel a), similar to the Mg II intensities. Once again, the QS regions have higher intensities than CHs for the regions with identical |B|, with the difference in the intensities increasing with |B|. The Doppler shifts plot (panel b) suggests that both QS and CH are, on average red-shifted and that the magnitude of the Doppler shift increases with increasing |B|. We also note that for the smaller |B| (<30 Mx cm$^{-2}$), QS is slightly more redshifted than CH. Between 30−50 Mx cm$^{-2}$, both show similar redshifts. At higher |B| (>50 Mx cm$^{-2}$), CHs are slightly more redshifted than QS. Panel c shows that the line width increases with |B| and that CHs exhibit larger widths than QS regions.

In Fig. 5.18, we plot the velocity results for upflows and downflows separately as a function of |B|. We find that the CH pixels are blue-shifted relative to the QS pixels with identical |B|. The blueshifts in CH show a marginal relation with the |B|. Such a relation is not seen for QS, which in fact, shows a marginal reduction in blueshift with |B| (see panel **a**). Figs. 5.18.**b** shows that the redshifts in both CH and QS are almost the same till $\approx 30$ Mx cm$^{-2}$, following which the CHs show excess redshifts and QS show saturation.

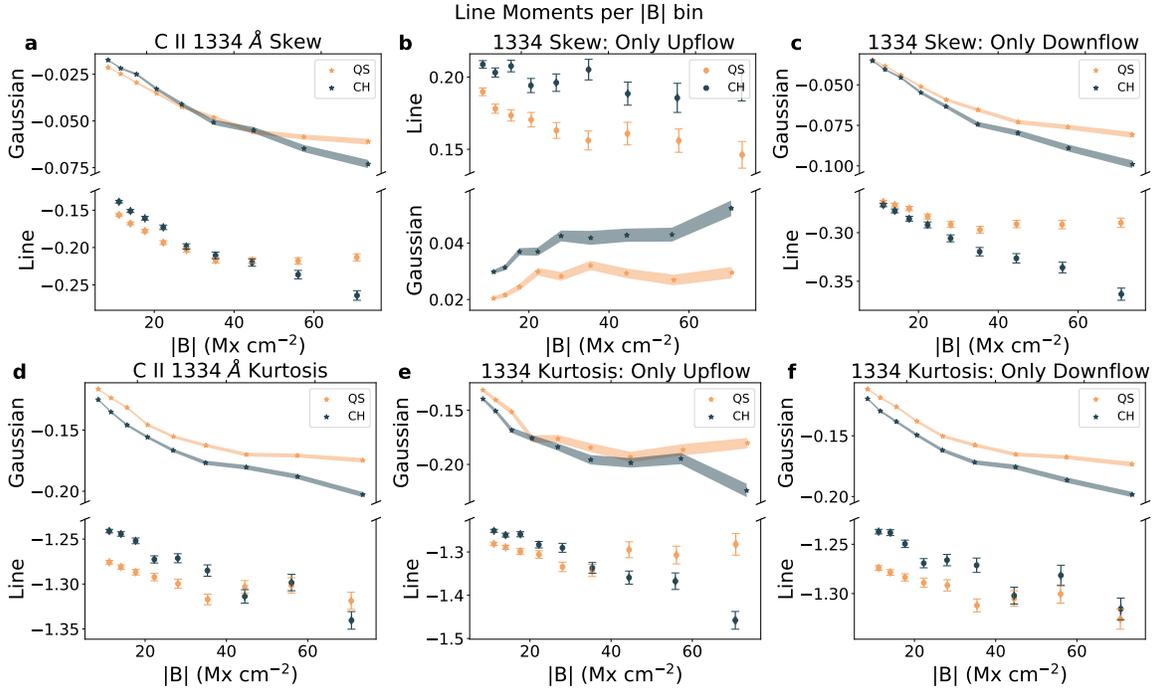

**Figure 5.19**: C II skew and kurtosis as a function |B|, for all data sets taken together. The top row corresponds to line skew, while the bottom row corresponds to kurtosis. Panels **a** and **d** correspond to the variation of moments of all profiles, while **b** and **e** (**c** and **f**) correspond to moments of blueshifted (redshifted) profiles. The bands of black and orange, with stars over-plotted, correspond to the respective moment of a single Gaussian fit. The y-axis has been broken to depict the variation with |B| better.

In Fig. 5.19, we plot the skew (panel **a**) and kurtosis (panel **d**) averaged over the five sets of observations as a function of |B|. Similar to Fig.5.14, the star and banded plots depict the moments for the Gaussian fit, while the dots depict moments for the spectral profile. We find a clear signal of kurtosis deficit and negative skew of the lines vis-a-vis a Gaussian profile. Furthermore, we also study the moments for red- and blue-shifted pixels separately (see panel **b** and **c**). We find that blueshifted (redshifted) pixels are positively (negatively) skewed. Such behavior suggests that the spectra with blue (red) shifts rise more steeply than a Gaussian on the blueward (redward) side and fall off gradually on the opposite side. Finally, the kurtosis shows no dependence on the line shift, as seen from panels **e** and **f**, which have kurtosis as a function of redshift and blueshift of the line. This implies that the spectral profiles themselves are flatter than a Gaussian profile, irrespective of whether they are shifted to the blue or red side.

## 5.6   Si IV:: Results From the Combined Dataset

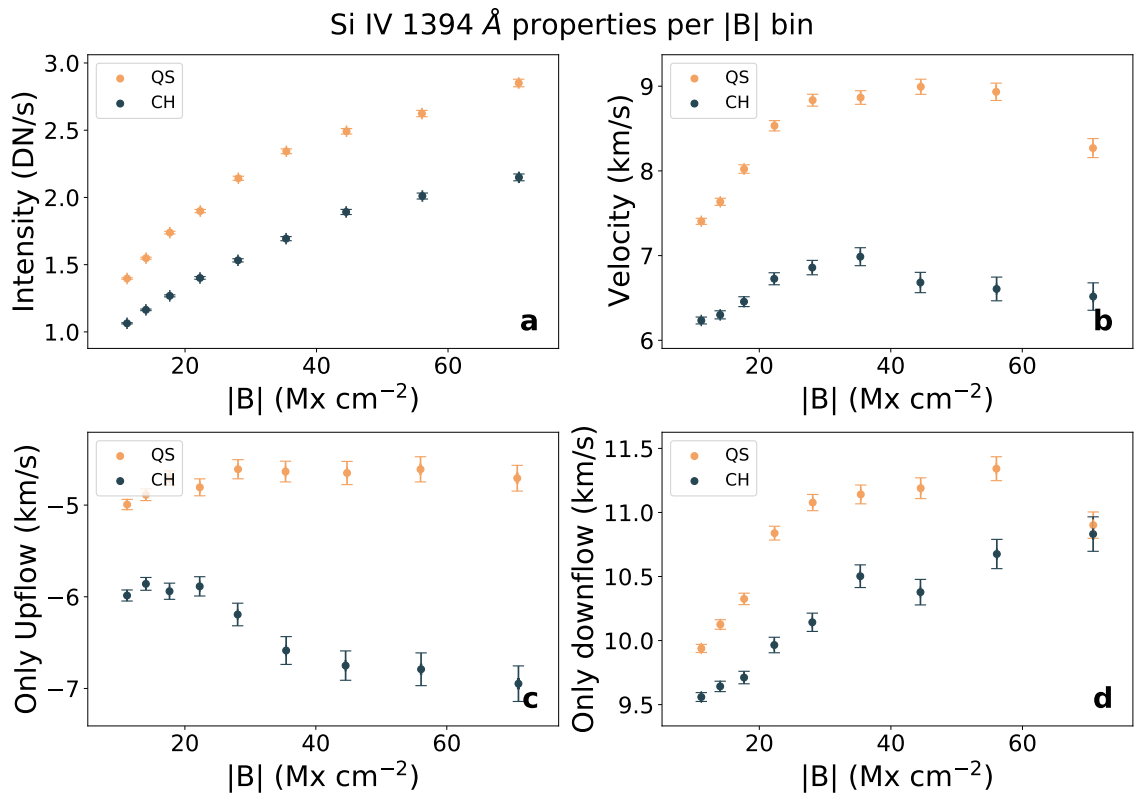

**Figure 5.20**: Variation of Si IV line intensity (panel.**a**), average velocity (panel.**b**), blueshifts (panel.**c**) and redshifts (panel.**d**) as a function of |B| in CHs (black) and QS (orange). These results are computed across all the datasets and are a more statistically significant version of the results by Tripathi et al. (2021a).

We now present the results from the intensity and Doppler shift of the Si IV line for all the datasets. The only difference between the results presented here and those from Paper I is the inclusion of two additional observations in this work, thereby increasing the statistical significance of the results. We obtain the Si IV line parameters by fitting the spectra with a single Gaussian and constant continuum. The relevant results are graphically summarized in Fig. 5.20. These results are in complete agreement with those obtained in Paper I. Moreover, the results obtained for Si IV bear some similarities with those obtained for C II and Mg II lines.

The intensity (Fig. 5.20.**a**) and blueshift (Fig. 5.20.**c**) differences between CHs and QS, and their relations with |B| are consistent across all three lines, though more enhanced for Si IV line. However, the signed average velocities indicate reduced average redshifts in CHs over QS (Fig. 5.20.**b**). These average redshifts increase with |B| and saturate at $\approx 40$ Mx cm$^{-2}$. The CHs show excess blueshifts over QS (Fig. 5.28.**c**), with the variation similar to those exhibited by the C II line (Fig. 5.17.**c**). The redshifted pixels alone also show a direct relation to |B|, but the QS is more redshifted than CHs (Fig. 5.20.**d**). Note also that the upflow and downflow velocities obtained for Si IV are much larger than those inferred from Mg II (Fig. 5.16) or the C II lines (Paper II).

## 5.7 Correlations between Doppler shifts of Mg II, C II and Si IV

The velocities and intensities show a highly non-trivial relation as a function of the formation height of different spectral lines. Note here that ascribing exact formation heights to each of these spectral lines is difficult, and it would be better to use 'formation temperature' of these lines. However, the solar atmospheric stratification may allow us to make a qualitative association between formation height and formation temperature, and we shall consider them to be almost equivalent henceforth. We however also note that these associations are adapted from various numerical simulations which do not quite incorporate the dynamics of Type II spicules, for example. Therefore, to investigate if there is any correlation between Doppler signatures observed in the three different spectral lines viz. Mg II, C II, and Si IV, we consider the approximate formation height of these lines obtained from numerical simulations. It has been suggested that, on average, C II lines form slightly higher in the atmosphere than the Mg II lines. Within the Mg II lines, the k line forms higher than the h line. Moreover, it has also been found that the line cores of both k & h lines form higher than their respective peaks (Leenaarts et al. 2013; Rathore et al. 2015b). The Si IV line forms in optically thin conditions, so ascribing an exact formation height is not possible. However, it forms at a higher temperature in the TR. We may, therefore, ascribe a greater height to Si IV than the Mg II and C II lines. With this prior, we may assume that the formation height (ascending order) is approximately Mg II h2 $\leq$ Mg II k2 $<$ Mg II h3 $\leq$ Mg II k3 $\approx$ C II $<$ Si IV.

The obtained velocities in different line features of Mg II (Fig. 5.16. **b**, **c**, **e** & **f**), C II (Fig.5.18) and Si IV (Fig. 5.20. **c** & **d**) clearly show that the velocity magnitude increases with increasing formation height. Considering mass flux conserving flows (Avrett et al. 2013), and that the density decreases as a function of height in the solar atmosphere, it is plausible to hypothesize that the upflows (downflows) at lower (greater) heights are enhanced (reduced) while traveling towards greater (lower) heights.

To check this hypothesis, we investigate the correlations between Mg II, C II, and Si IV velocities. Since Mg II and C II form at approximately the same height, we expect these two lines to have similar properties vis-à-vis the Si IV line. In Paper I, it is suggested that the increase in Si IV blueshift with increasing with |B| may indicate the signatures of the solar wind emergence. This motivates us to explore if the observed flows in chromosphere detected in Mg II and C II lines are in any way related to those obtained from Si IV. For this analysis, we use the results obtained from the combined dataset. Note that we only consider the Mg II k2, Mg II k3, C II 1334 Å and Si IV lines in this analysis. We emphasize that the results from Mg II h follow the results from the k feature and are not shown for brevity.

We split the velocities observed in Mg II, C II, and Si IV into sets of pixels containing upflows and downflows. Then, we consider scatter plots between flows in intersection of these sets, e.g., the relation between pixels showing upflows in Mg II k3 & upflows in Si IV, upflows in Mg II k3 & downflows in Si IV and so on. These scatter plots are obtained for Si IV velocities in Mg II and C II velocity bins to improve statistics. Note that the bins here are selected in deciles, i.e., every 10% of the data for each Mg II or C II feature is considered to be in one bin.

In Fig. 5.21, we plot the correlations between downflows (top row) and upflows (bottom row) observed in Si IV with those observed in Mg II k2 (panels **a** & **d**), Mg II k3 (panels **b** & **e**) and C II (panels **c** and **f**). Panels **a**, **b**, and **c** demonstrate that the downflows observed in Si IV are strongly correlated with those observed in Mg II k2, k3 and C II. For a given value of downflow in Si IV, the downflows are stronger in k3 and C II than those in k2. Note, though, that the downflows in k3 and C II are very similar. Moreover, Si IV displays excess downflows in QS vis-à-vis CH for similar C II and Mg II downflows. These differences in QS-CH Si IV downflows are also observed to increase with increasing Mg II and C II downflows. Similarly, Panels **d**, **e** and **f** suggest that the upflows in Si IV have slightly better correlation with upflows in Mg II k3 and C II than those in Mg II k2.  Furthermore, the correlation is stronger between C II and Si IV lines, possibly reflecting the influence of an optically thin C II component forming much higher in the atmosphere (Rathore et al. 2015b).  Like downflows, we find that for the similar upflows in Mg II and C II, the CH exhibit larger upflows vis-à-vis QS in Si IV. We further note that there is a slight hint of an increase in the difference of upflows observed in CH and QS in Mg II and C II lines.

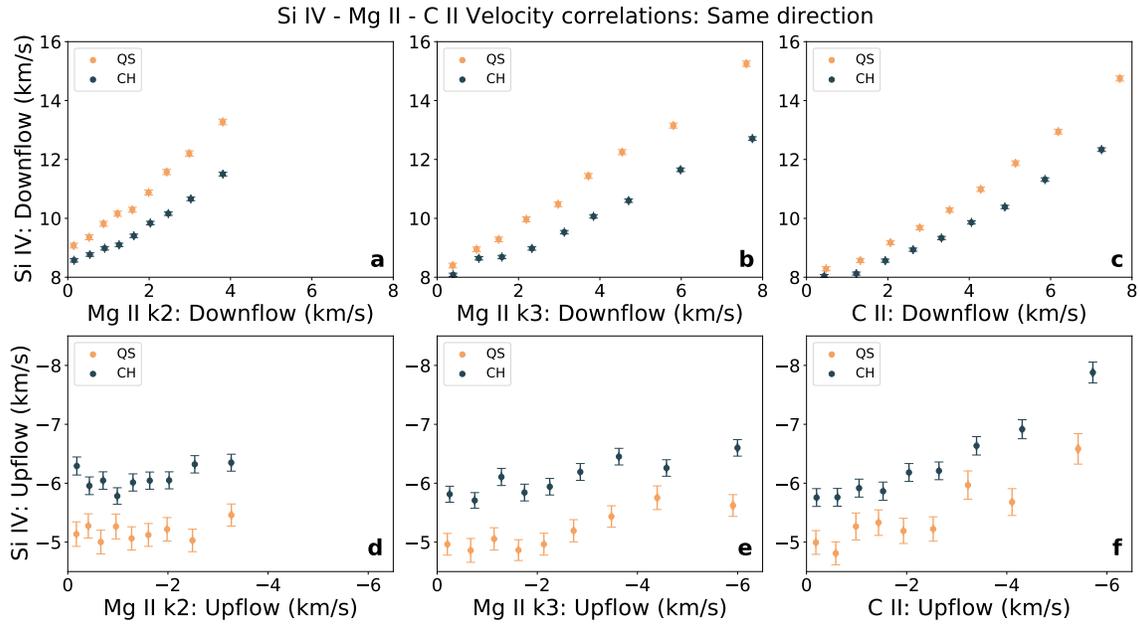

**Figure 5.21**:  Inter correlations between Si IV and Mg II k2, Mg II k3, C II Doppler shifts.  The top row depicts correlations between the downflows in Mg II, C II, and Si IV, while the bottom row depicts correlations between upflows in Mg II, C II, and Si IV. The columns follow approximate formation height from Mg II k2 to C II.

In Fig. 5.22, we study the correlations between the upflows in Si IV with downflows in Mg II and C II and vice-versa as shown in the top and bottom rows, respectively.  We find that the upflows in Si IV have a monotonic relation with the downflows in Mg II and C II (top row).  In addition, we note that for similar downflows observed in Mg II and C II, Si IV shows stronger upflows in CH and than in QS. On the other hand, the Si IV downflows do not show any particular correlation with Mg II and C II upflows.  Furthermore, the Si IV downflows in CH and QS remain consistent for similar upflows in Mg II and C II. We do not find any relation between the downflows in Si IV with upflows in Mg II and C II.

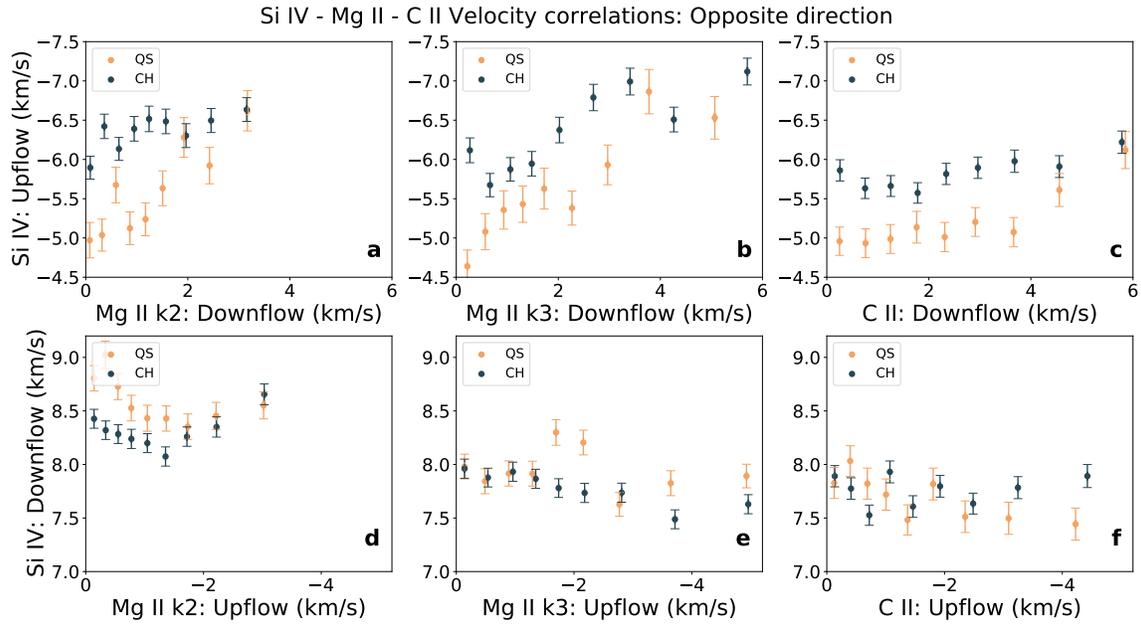

**Figure 5.22**: Inter correlations between Si IV and Mg II k2, Mg II k3, C II Doppler shifts. The top row depicts correlations between the downflows in Mg II & C II with upflows in Si IV, while the bottom row depicts correlations between upflows in Mg II & C II with downflows Si IV. The columns follow approximate formation height from Mg II k2 to C II.

## 5.8 Inferences from intensity, profile shape, and velocity diagnostics

The problems of coronal heating in QS and CHs and the formation and acceleration of solar wind are intimately tied to the structure and dynamics of the magnetic field in the respective regions. Therefore, comparative studies between CHs and QS become important in understanding the plasma dynamics and the underlying processes. The Mg II, C II, and Si IV lines observed by IRIS probe different layers in the chromosphere & TR, have provided us with a unique opportunity to understand the dynamics of these regions as a dynamically coupled system. In this paper, we characterize, in detail, the dynamics of the Mg II line by combining the information related to the plasma dynamics with that of the magnetic field. In addition, we also probe the correlations between the Doppler shift obtained for Mg II lines, the C II line, and those obtained for the Si IV line from Paper I, to investigate if a common origin may be ascribed to the observed dynamics in the different lines. Below we summarize and discuss our results followed by an interpretation in §5.9.

## 5.8.1 Intensity differences

The intensities in the Mg II h, k lines (both in the core and peaks) & C II lines formed in the chromosphere, Si IV line formed in the TR increase with |B| (see Fig. 5.15, Fig. 5.17.a and Fig. 5.20.a). For all three lines, CHs show reduced intensity over QS for the regions with identical |B|. Moreover, the difference in the intensities increases with increasing |B|. The observed differences in the chromospheric intensities in CHs and QS suggest that CHs have lower source function values over QS for regions with identical |B| (Rathore et al. 2015b).

Our results further show that the differences between CH and QS intensities exist already at the chromospheric level for a given magnetic flux region (see also Kayshap et al. 2018, & Paper II). We note that the ratios of QS to CH intensities in the largest |B| ($\approx 80$ Mx cm$^{-2}$) bins are smallest in Mg II k2 and increase through C II, Mg II k3 and Si IV lines as 1.18, 1.22, 1.26, and 1.32, respectively, suggesting an increasing differentiation between CH and QS from the low chromosphere to TR.

The intensity differences in the CHs and QS in the corona are well known (see, e.g., Krieger et al. 1973). However, such differences are not seen in either chromospheric or TR above noise level (Stucki et al. 1999; Xia et al. 2004). These observations led Wiegelmann & Solanki (2004) to attribute the CH-QS intensity differences to loop statistics in these regions. Based on potential field extrapolations, Wiegelmann & Solanki (2004) found that the QS has an excess of longer closed loops over CH, while similar numbers of shorter closed loops are present in both regions. Using the scaling laws, valid for optically thin plasma, they proposed that the reduction in CH intensity over QS naturally results from a deficit of longer loops in CHs.

While the scenario proposed by Wiegelmann & Solanki (2004) is used to explain the intensity difference in Si IV in Paper I, it may not be directly applicable to the chromosphere, as also argued by Kayshap et al. (2018). However, since the loop statistics of Wiegelmann & Solanki (2004) is derived from the extrapolations of photospheric magnetograms, it may be plausible to suggest that the statistics itself (and not the scaling relation for plasma emission) is also valid for the chromosphere. Therefore, we may conclude that at a relatively higher |B|, a deficit of shorter loops in the CHs with respect to the QS is observed already in the low chromosphere. The source function of the chromospheric lines may in part be influenced by this loop statistics at chromospheric heights and may explain the marginal deficit of intensities in CHs over QS. However, given the complexity of structure in the chromosphere, it may also be possible that magnetoacoustic shocks and differences in properties of spicules reflect into the intensity differences in CHs and QS (see, for example Vecchio et al. 2009; Pereira et al. 2012).

### 5.8.2 Spectral profile behaviour of C II line

Our observations show larger line width in CH over QS for regions with similar |B|. From simulations, it has been observed that opacity plays an important role in broadening the lines (Rathore & Carlsson 2015; Rathore et al. 2015b). Opacity broadening may be qualitatively explained by Eq.25 of (Rathore & Carlsson 2015), where in the absence of any flows, the opacity broadening is proportional to the ratio of column mass at the line wing to the column mass as line center. From Eq. 23 and 20 of (Rathore & Carlsson 2015), we find:

$$\frac{1}{m_c(0)} = \frac{\chi_{l0}}{\rho} + \frac{1}{m_c(\infty)}, \tag{5.5}$$

where $m_c(\Delta\nu)$ is the column mass at a shift $\Delta\nu$, $\Delta\nu = 0$ representing the line core, $\Delta\nu = \infty$ representing the continuum, $\chi_{l0}$ the opacity at line core per unit volume, and $\rho$ being the density. Thus, with other terms being constant, $m_c(0)$ depends directly on the density $\rho$, and any reduction in density reduces the line core column mass, thereby increasing the opacity broadening. Assuming the line intensity to be directly related to density, a reduction in density would be seen as a reduction in the core intensity. Thus, density reduction in the line core of CH over QS can neatly explain the observed intensity and line width differences. Note that Eq. 5.5 has been derived under a static atmosphere, while in the real solar atmosphere, there would be components of non-thermal velocities and micro-turbulence that will affect the line width. Furthermore, the enhanced width may also have a component from increased non-thermal width. This may occur due to spicular activity, which may give rise to enhanced widths due to high velocity Alfveń waves, occuring especially in the network regions (see, for example Van Ballegooijen et al. 2011; Tian et al. 2014).

Our observations further show that the spectral profiles have less kurtosis than a Gaussian and are negatively skewed vis-à-vis a Gaussian profile. To understand these profiles further, we look at the skew and kurtosis of redshifted and blueshifted profiles separately and attempt to disentangle their properties in Fig. 5.19. The skew is observed to change sign depending on the line shift. The observed profiles are observed to be positively (negatively) skewed if the profile is blueshifted (redshifted). Since the comparison is performed with respect to a Gaussian fit, it would mean that the blueshifted (redshifted) profiles have a steeper blueward (redward) rise than a Gaussian. Such asymmetric C II profiles have been observed in 1-D simulations by Avrett et al. (2013). Moreover, the authors have observed increasing asymmetry with increasing atmospheric velocities. It has been suggested by Avrett et al. (2013) that the asymmetry arises if the flows are column mass con-

serving – implying that the vertical velocity is inversely proportional to the density. Hence, the part of the line that is emitted higher shows a greater shift than the part of the line emitted lower. This may be a possible explanation for the observed skew of the line. However, note that in general the spectral profile shape depends on the variation of source function with wavelength, which further depends on the coupling of the source function to local conditions and velocity gradients among other effects.

Finally, we see that the kurtosis is independent of whether the profile is blueshifted or redshifted and is significantly different from a Gaussian. It also shows a distinct variation with |B|. Thus, in general, C II profiles are flatter than a Gaussian, and the flatness increases with increasing |B|. The Ca II lines in spicules have been shown to change from having a central reversal to a flat-topped to a peaked profile with increasing formation height by Zirker (1962). Such changes in profiles were explained by a reduction in opacity in these lines. A similar picture may also hold with the C II line, which may show such flat-topped profiles due to opacity variations. Note, however, that similar kurtosis-deficit profiles have been seen as the presence of a "box-shaped" profile by Rathore et al. (2015b). From 3D simulations, Rathore et al. (2015b) assert this to be a consequence of a steep rise in the source function near the continuum, with a more gradual rise near the core formation region. Also note that such a source function variation would also give rise to broader lines, as shown in Rathore et al. (2015b). Thus, the flat rise of the source function, dictated by the underlying |B|, may cause the kurtosis deficit in the C II line. However, these deviations of the spectral line from a Gaussian are similar in both CHs and QS. Thus, the similarity in spectral shapes points to similar underlying processes giving rise to CH and QS spectral profiles.

### 5.8.3   Doppler shift: Variations and correlations

Doppler measurements in all three lines demonstrate that, on average, both the chromosphere & TR are red-shifted (see. Fig. 5.16.a & d, Fig. 5.17.b, and Fig. 5.20.b). The average redshifts are found to increase with |B| and increase from Mg II k2 to Si IV for similar |B|. By studying the red-shifted and blue-shifted pixels separately, we find that in the chromospheric lines, CHs have excess upflows as well as downflows vis-á-vis QS for identical |B| and that the excess increases with increasing |B| (see Fig. 5.16, and Fig. 5.18). However, in the TR, CHs have excess (reduced) upflows (downflows) over QS for the regions with identical |B| (see Fig. 5.20. c & d). With uncertainties, the magnitudes of upflows and downflows are in approximate descending order of Si IV, Mg II cores, C II, and Mg II peak. We further note that while Mg II k3 and C II lines show similar downflows, the upflows are larger in Mg II k3.

To assess any (or otherwise) association between the flows observed in the chromosphere and TR, we perform a correlation study in the intersection of pixel sets showing flows in different lines. That is, we study the mean variation of upflows in the TR pixels, which also show upflows in the chromosphere, and so on for different combinations of flows. This analysis gives us the variation of mean TR flows with chromospheric flows and provides information on the persistence of flows in different lines. We find that the flows in the chromosphere and TR are tightly correlated, i.e., the downflows in the chromosphere with both upflow and downflows in TR and upflows in the chromosphere with those in TR. However, we did not find any correlation between chromospheric upflows with downflows in TR. Moreover, for similar downflows (upflows) in the chromosphere, the CHs show reduced TR downflows (excess upflows) over QS. Additionally, for similar downflows in the chromosphere, the CHs show excess upflows over QS in the TR.

The observations reported here lead to two questions. Firstly, what physical mechanism(s) give rise to these flows? Secondly, is it possible to explain the observed differences between the flows observed in CH and QS in the chromosphere and TR, including the difference in the intensities discussed in §5.8.1? While we deal with the former here, the latter is taken up in the §5.9.

The tight correlations between TR downflows measured using Si IV and those observed in the chromosphere measured using Mg II and C II may either be explained by field-aligned downflows due to condensations from corona to TR to chromosphere (see, e.g., Klimchuk 2006a; Tripathi et al. 2009b, 2010, 2012) or due to return flows of type-II spicules (Klimchuk 2012; Ghosh et al. 2019, 2021; Bose et al. 2021). However, we note that the observed magnitude of the TR downflows are much larger than those predicted using 1D hydrodynamic simulations of coronal impulsive heating followed by evaporation and condensation. Therefore, for the reasons elaborated in Ghosh et al. (2019, 2021), it is more likely that the observed downflows here are due to the return flows of type II spicules. Our finding that the speeds of chromospheric downflows are lower than those in TR is very likely due to the plasma flowing from lower density to higher density. However, note that the net deceleration of the plasma depends on the interplay of deceleration due to atmospheric stratification & magnetic pressure and acceleration due to gravity & plasma compression.

Our observations further show that the upflows in the chromosphere and TR are also correlated. Moreover, these upflow show an increase in magnitude with in-creasing |B| as well as atmospheric height. This may be possible if the upflows are moving through an expanding flux tube under the assumption of constant mass flux. However, the upflows may have been caused by the launch of events like Type II spicules (De Pontieu et al. 2007a; Tian et al. 2008a, 2014; Samanta et al. 2019). However, note that such upflows may also be generated due to upward propagating waves (e.g., Cranmer & Van Ballegooijen 2005). Since Alfvén waves are known to be ubiquitous in the chromosphere (De Pontieu et al. 2007b), disentangling the exact effects of Alfvén wave v/s spicule-like propagation upward is difficult (see, however, Ghosh et al. 2019, 2021).

Finally, the chromospheric downflows also bear a direct relationship with the upflows in TR. Such correlations suggest a common origin of these flows and hint towards the existence of bidirectional flows. Bidirectional flows have been observed in QS & CH (predominantly occurring in the CH) network regions by Aiouaz (2008), and in active regions by Barczynski et al. (2021) as redshift in TR and blueshifts in the low corona (see also Gupta et al. 2018a, for bidirectional flows in transient events). We propose that such bidirectional flows occurring between the chromosphere and the TR can suitably explain our observations

The scenario we propose is illustrated in Fig. 5.23. The vertical color bars changing from deep yellow in the photosphere to white in the corona indicate reducing density with increasing height in the atmosphere. The approximate formation height (or rather, a proxy for temperature) of different ions corresponding to the spectral lines studied here are labeled with horizontal lines. The 'asterisks' indicate the location of an impulsive event, while the arrows mark the direction of expected flows. The blue (red) arrows indicate upflows (downflows).

We present four different scenarios based on the same physical mechanism to explain the three sets of observations. As evidenced by similar skew & kurtosis in C II 1334 Å line (see Paper II) and non-thermal widths in Si IV (see Paper I), it is plausible to conclude that similar physical mechanisms give rise to the observed spectral profiles in CHs and QS. This mechanism, in our interpretation, is an impulsive dumping of energy in CHs and QS. For an impulsive event occurring between the formation height of Si IV and C II or Mg II, bi-directional flows will be produced in the form of upflows in Si IV and downflows in C II and Mg II. Since the chromosphere is denser than the TR, the chromospheric radiative cooling time scales are smaller. Hence, the downflows would cool down faster and be visible in cooler lines like Mg II and C II, while the upflows persist in relatively hotter lines like Si IV. Some of the upflows observed in Si IV may persist till greater heights and then fall back, similar to Type II spicule return flows (Fig. 5.23.b). The returning flows will be observed as persistent downflows in all three lines, with descending speeds. We note, however, that the persistent downflows may also be caused by the impulsive event occurring above or about the formation height of Si IV (Fig. 5.23.c) and may have similar signatures of descending speeds as in the previous case. Finally, the impulsive events may also occur either below or at the height of the formation of Mg II peaks (Fig. 5.23.d), resulting in the launch of chromospheric jets that may show persistent upflows in chromosphere and TR, followed by the downflows at later times.

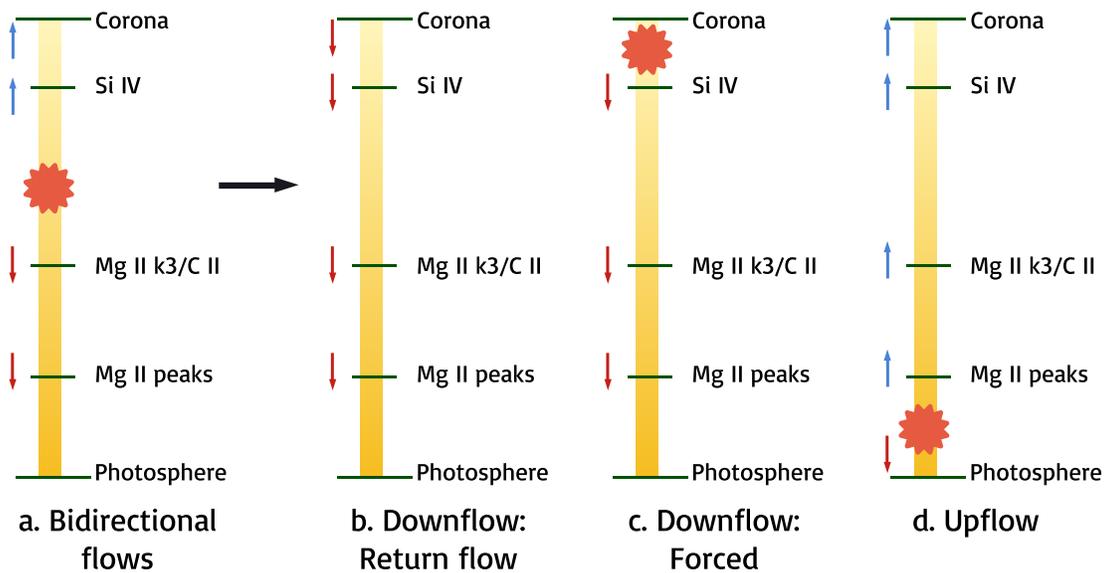

**Figure 5.23**: A schematic depicting a unified picture of flow generation, including observed correlations between flows. The vertical bar denotes density reducing with height from dark yellow to white. The red asterisk depicts an impulsive event, which gives rise to flows (arrows). Blue upward arrows depict upflows, and red downward arrows depict downflows. Panel **a** shows the basic bidirectional flow generation, which eventually gives rise to the return flow in Panel **b**. Impulsive events occurring much higher than Si IV formation height, giving rise to downflows, is shown in Panel **c**. Upflows generated due to impulsive events low in the atmosphere are depicted in Panel **d**. See in-text for details.

Bidirectional flows in an expanding, cylindrically symmetric flux tube have been observed in field-aligned 1-D simulations by He et al. (2008). In these simulations, impulsive events deposit energy at the height of $\approx 5$ Mm, which is also the location where the expansion of the flux tube starts. The results demonstrate that at the onset of the impulsive event, the plasma moves outward from the location of the impulse, showing bidirectional flows. He et al. (2008) show that the velocities in Si II, C IV and Ne VIII obtained from their setup match those observed by Tu et al.

(2005). He et al. (2008) further show that the downflows at $10^4$ K are $\approx 2$ km/s. The magnitude of these downflows increases with height, reaching up to $4$ km/s near the energy dumping heights. These velocities were obtained with a |B| of $56.5$ Gauss in the simulation. The velocities obtained in this simulation are consistent with the downflow speeds observed in Mg II k2 (see Fig. 5.16.f) at $\approx 56$ Gauss, but it is much lower than the downflow speeds observed in Mg II k3 and C II.

Hansteen et al. (2010) perform a 3D simulation of a QS region for different average |B|, spanning from the convection zone to the corona. It is found that impulsive events due to reconnection occurring at various heights give rise to bi-directional flows, seen as co-spatial blueshifts (redshifts) in the corona (TR) concentrated at loop footpoints.

Such events occurring across a range of heights can give rise to correlated flows similar to the results reported here. Hansteen et al. (2010), in their B1 model setup, were able to reproduce velocities consistent with blueshifts observed in the corona (see Fig. 11 of Hansteen et al. 2010). Note, however, that the downflows speeds near the formation temperature of Mg II and C II inferred from these simulations were between $2-4$ km/s, which are lesser than those reported in this work. Nevertheless, we note that the scenario presented by Hansteen et al. (2010) may potentially explain the correlated bidirectional flows reported in this paper.

Finally, while the above-described scenario based on impulsive events may explain the observed downflows, it is important to highlight that spicule-like flows may also be obtained due to the 'squeezing' of flux tubes near the chromosphere (see e.g. De Pontieu et al. 2007a; Martínez-Sykora et al. 2011, 2017, 2019, and also Isobe et al. (2007b) ), or through the dynamics of magnetoacoustic shocks (Kayshap et al. 2021). The rising plasma from the 'squeeze' has been observed to be heated up and detected in various IRIS lines such as Mg II and Si IV (Martínez-Sykora et al. 2017). Thus, the persistent upflows may also be explained through such spicule-like flows, while the downflows may be explained by the return of such spicule-like flows. Note that, throughout the paper, we assume that the type-II spicules are produced due to impulsive events (Moore et al. 2011, 2013; Martínez-Sykora et al. 2011; Samanta et al. 2019).

## 5.9 A unified scenario the origin of the solar wind, switch-backs and QS heating

While the occurrence of impulsive events at the interface between chromosphere and corona may explain the observed flow variations with |B| in different lines and their interrelations, the question remains as to what leads to the observed differences between the intensities as well as flows in CHs and QS. To explain the differences in the intensities, in §5.8.1, we invoked the loop statistics in CHs and QS derived by Wiegelmann & Solanki (2004). The predominant velocity differences between CH and QS are: i) reduced Si IV downflows in CHs over QS for similar downflows in the chromospheric lines, and ii) excess Si IV upflows in CHs over QS for similar upflows and downflows in the chromospheric lines. These results indicate an excess acceleration of upflows in CHs and an excess deceleration of downflows in QS. Furthermore, while the QS shows enhanced intensity over CHs for regions with similar |B|, the CHs show larger flow speeds (except Si IV downflows) over QS. Such an observation thus hints towards a unified scenario of heating the corona in QS and CH and the emergence of the solar wind. Therefore, we then ask if it is possible to combine the loop statistics and the occurrence of flows due to impulsive events illustrated in Fig. 5.23 to explain the observed differences in intensities as well as the Doppler shifts in CHs and QS, similar to that is discussed in Tripathi et al. (2021a) for Si IV.

A graphic depicting the scenario we propose is shown in Fig. 5.24. The top panel depicts the predominant topology in CHs while the bottom panel is for QS regions, based on the loop statistics of Wiegelmann & Solanki (2004), according to which both CHs and QS have an equal number of short closed loops, but QS has predominantly large closed loops, and CHs have open field lines. In CH regions, the interchange reconnection (e.g., Fisk 2005; Janardhan et al. 2008), leading to impulsive events, may occur between closed and open field lines, while in QS, the impulsive events will be due to reconnection among closed-closed loops, similar to those observed in the core of active regions during the transient formation of loops (see Tripathi 2021). The excess open and expanding flux tubes in CHs may cause preferential acceleration of upflows in CHs over QS. In principle, the scenario proposed here is similar to those employed by Tian et al. (2008b,a); He et al. (2007) to explain the Doppler shifts observed in QS-CH in coronal and TR lines, and similar to the Fig. 5 of He et al. (2010). Note that the concept of interchange reconnection has been invoked to explain active region outflows by Del Zanna et al. (2011); Barczynski et al. (2021), solar wind disappearance events by Janardhan et al. (2008), active region jets, X-ray & cool jets in polar coronal holes (Moore et al. 2011, 2013,

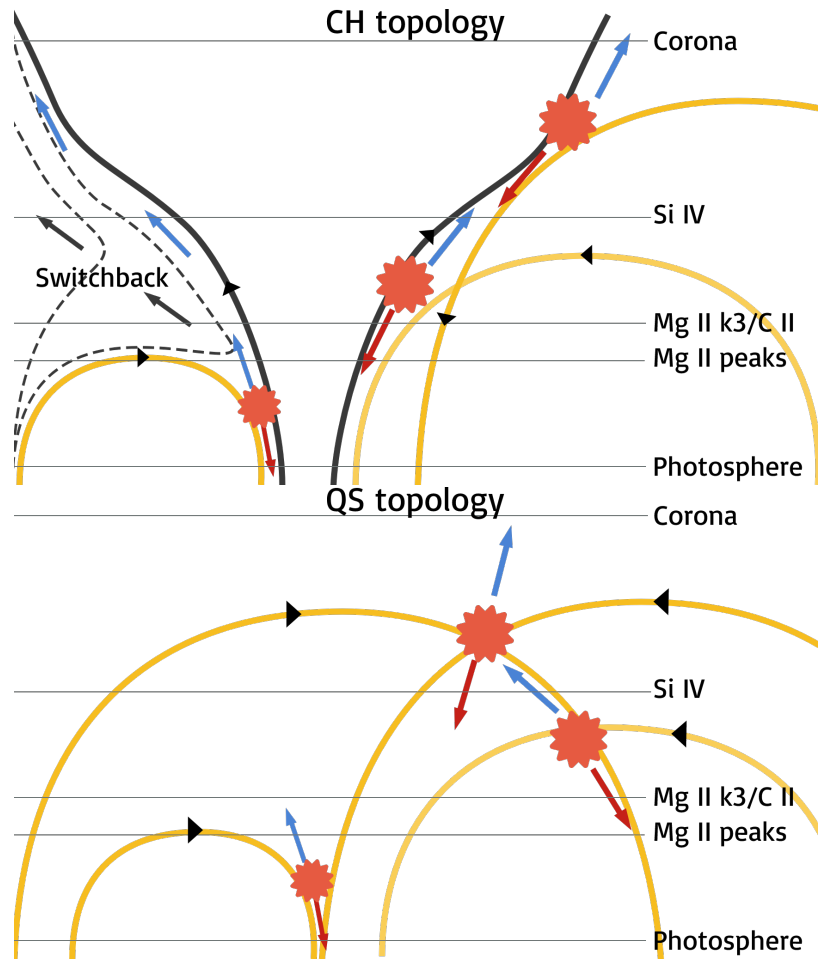

**Figure 5.24**: A schematic depicting the proposed picture of impulsive heating occurring across different magnetic field topologies. We show a CH topology in the top panel, including an open funnel-like structure (black) and closed loops of varying sizes (yellow). Impulsive events (red asterisks) due to interchange reconnection between the open and closed field lines give rise to bidirectional flows (blue and red arrows). Of these flows, the upflows are enhanced due to density stratification and the expanding flux tube in CHs. Interchange reconnection may occur over a range of heights, and the corresponding bidirectional flows may be observed across different spectral lines marked in approximate order of formation heights. An example of the reconnected field line propagating outward as a switchback is depicted as a dashed line, with the approximate propagation direction depicted by black arrows. Bottom panel: QS topology is depicted with the same terminology as CH topology. Note here that while one does expect correlated bidirectional flows in QS, the upflows are not as accelerated as CHs due to the absence of funnel-like structures. See in-text for more details.

2015) and type III radio burst (e.g., Mulay et al. 2016b).

Under the scenario presented in Fig. 5.24, the impulsive events may occur across a range of heights via magnetic reconnection among open field and closed loops of various heights in the CHs. Thus, if the energy dumping events were to occur below the formation height of Mg II, the upflowing plasma may be accelerated preferentially in CHs, and reach Si IV heights, where a strong correlation is obtained. Similarly, if the event were to occur at much greater heights, or if plasma launched from the lower heights (e.g., type II spicules-like events) are returning to the low solar atmosphere, the downflowing plasma may be falling from Si IV formation height, that will show deceleration due to increasing density towards lower atmosphere mapped by Mg II. Assuming a mass flux conserving flow, we have $\rho V = $ const, where $\rho$ is the mass density and v is the velocity. For a given downflow in Si IV, the downflows in chromospheric lines are smaller in QS over CH (Fig. 5.21.**a-c**). Hence, the density increase from Si IV formation heights to Mg II formation heights is larger in QS over CH, resulting in a larger velocity reduction in QS. However, note that since the mass flux is typically different for CHs and QS, a quantitative comparison is beyond the scope of this work.

For bidirectional flows due to reconnection event between Mg II and Si IV formation heights, the counterpart upflows will be preferentially accelerated into Si IV formation height in CHs over QS due to excess open expanding flux tubes in CHs over QS. Since the QS has predominantly closed loops, the closed loop reconnection only serves to fill the loop with plasma, raising its intensity. Thus, impulsive events occurring across a range of heights combined with loop statistics in CHs and QS elegantly tie in all our observations and provides a unified scenario for QS heating and solar wind emergence.

Finally, the scenario we present in Fig. 5.24 is also appealing to explain the switchbacks observed in the near-Sun solar wind (Balogh et al. 1999; Bale et al. 2019) using Parker Solar Probe. One of the competing scenarios for the formation of these switchbacks is through interchange reconnection events occurring in the TR and lower corona (Fisk & Kasper 2020; Mozer et al. 2021; Zank et al. 2020; Tripathi et al. 2021a; Fargette et al. 2021; Bale et al. 2021; Sterling et al. 2020; Sterling & Moore 2020, see also Liang et al. (2021) for an assessment of the viability of switchbacks from the linear theory of Zank et al. (2020)). The kinked-field lines as a result of reconnection between the closed loop and open field in the coronal holes, as shown by the black arrow in Fig. 5.24, may be transported outwards into the solar wind, which are then observed as rotations in the magnetic field. In such a scenario, the flows reported in this paper serve as constraints and modeling inputs for solar wind switchback simulations.

A straightforward association between the scenario we present and polar coronal hole jets is clearly seen. Interchange reconnection seems to play the predominant role in the generation of mass flux and plasma heating in these events. However, note that the jets observed by Moore et al. (2015) have velocity almost two orders of magnitude more than the velocities we report, and show morphological differences arising due to twist and shear in the ambient magnetic field (see also Moore et al. 2013). Since we are averaging over multiple pixels for boosting the signal, checking for such morphological signatures is beyond the scope of this work. However, newly-emerged bipoles may interact with the ambient vertical field similar to the scenario proposed by Moore et al. (2011), giving rise to Spicule-like events. The interaction height and amount of magnetic flux converted into thermal energy would then determine the lines which show correlated flow.

The observational results and the scenario presented in this paper may provide an explanation for solar wind formation including switchbacks and the dynamics observed in the QS. We, however, stress that disentangling the absolute effects of wave propagation v/s impulsive upflows is needed. Furthermore, disentangling the effect of the return of spicule-like events v/s downflows due to impulsive events occurring higher up in the TR is also difficult. Disentangling these different effects requires further high-resolution spectroscopic observations simultaneously taken at different heights, combined with numerical simulations incorporating radiative transfer & evolution of solar corona into the solar wind. Such observations may be provided with the EUV High-Throughput Spectroscopic Telescope (EUVST) on the upcoming Solar-C mission (Shimizu et al. 2020).

## 5.10 Distribution of chromospheric and transition region properties as a function of $|B|$

We present here the distribution of intensity and velocity from Mg II, C II, and Si IV lines as a function of $|B|$. The plots shown from Fig. 5.25 - 5.28 are the same as those from Fig. 5.15 - 5.20, except that the errorbars reported corresponding to $1$ and $90$ percentile of the samples in each bin.

Since the overall distribution of the various quantities look very similar in CHs and QS, the distribution within bins of $|B|$ reflects how systematic differences arise in the ensemble of pixels considered. The distribution of samples between CH and QS slowly drift apart in their mean values, depicting the transition from statistical signatures in the lower atmosphere to very clear signatures in the corona. Thus, the distribution of samples in each bin provides further constraints to the expected results from simulations.

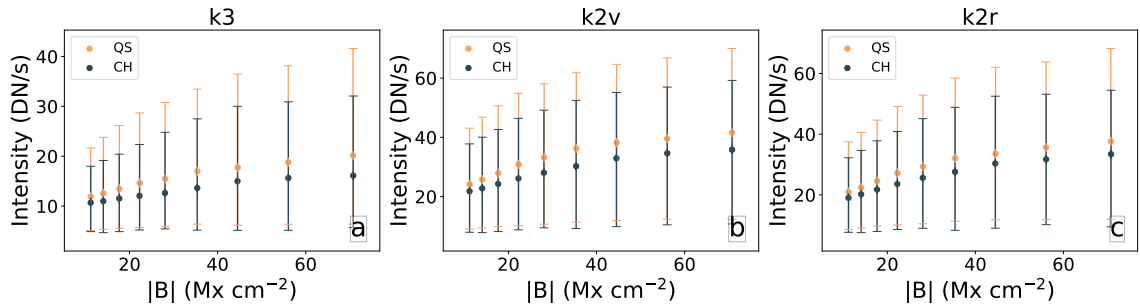

**Figure 5.25**: Same as Fig. 5.15, but the errors now represent $1$ and $90$ percentile bounds of the distribution of samples present in the bin of |B|.

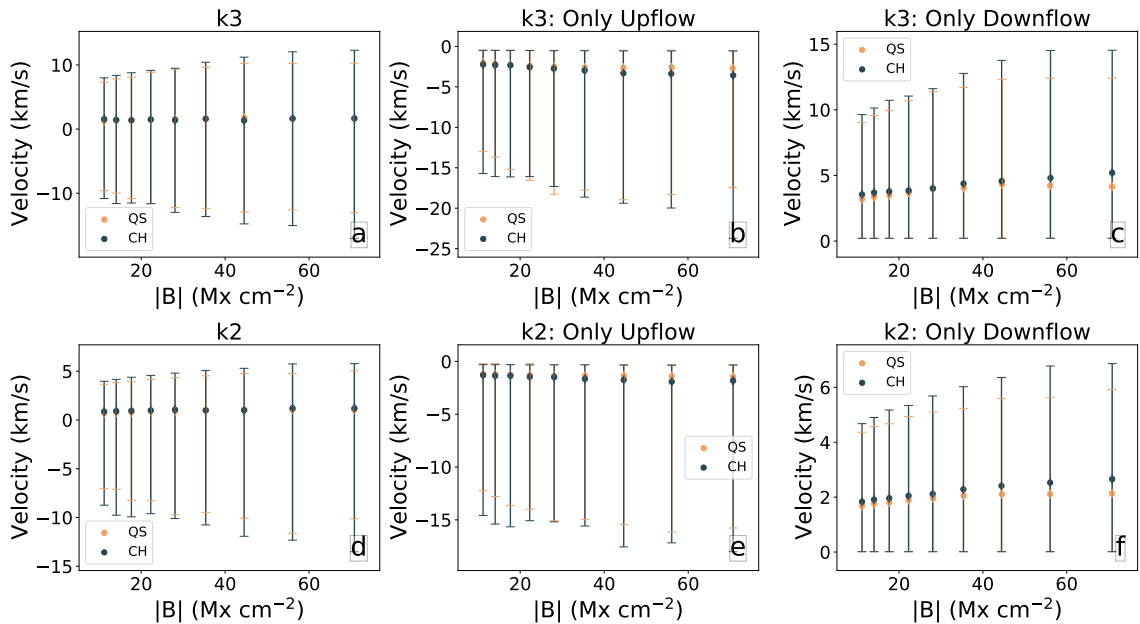

**Figure 5.26**: Same as Fig. 5.16, but the errors now represent $1$ and $90$ percentile bounds of the distribution of samples present in the bin of |B|.

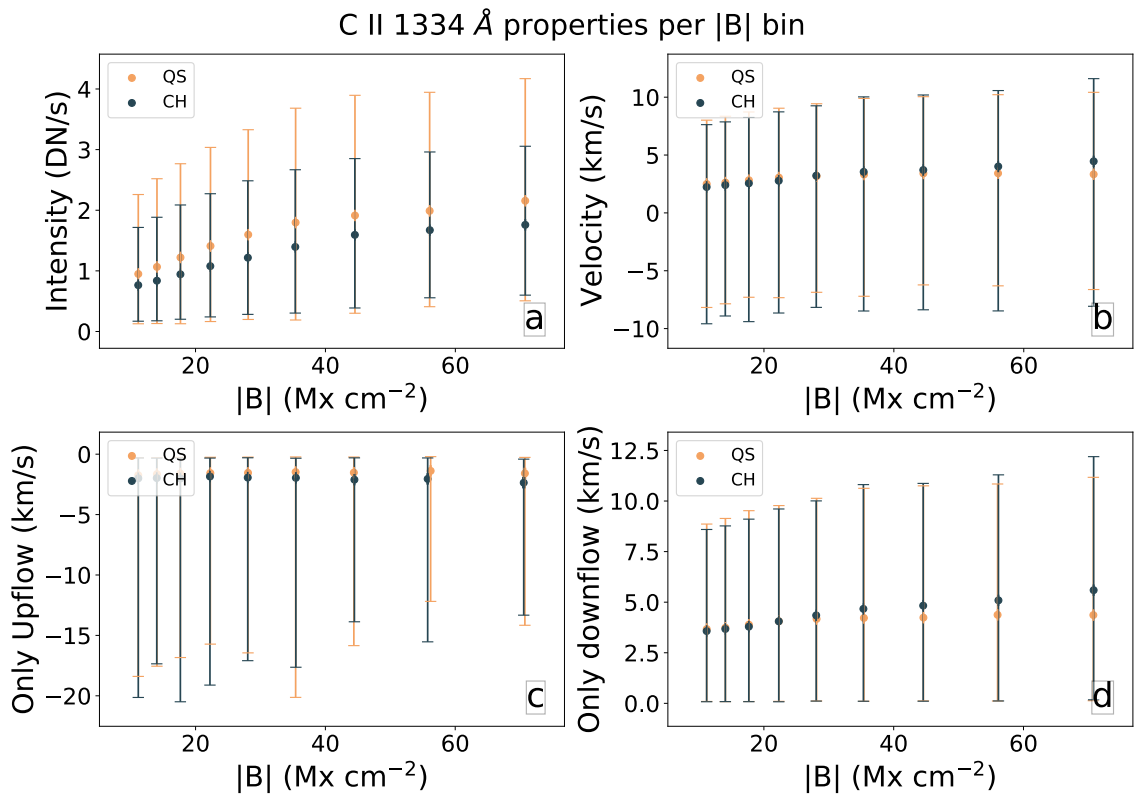

**Figure 5.27**: Same as Fig. 5.17 and 5.18, but the errors now represent 1 and 90 percentile bounds of the distribution of samples present in the bin of |B|.

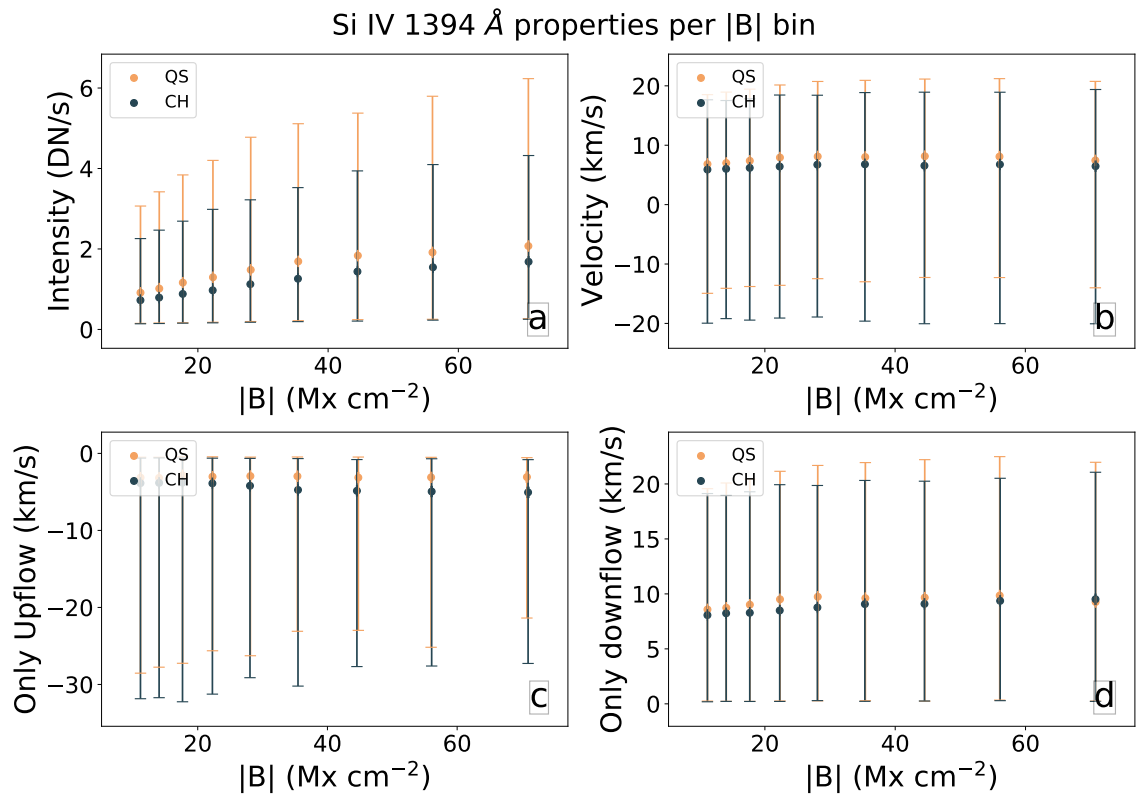

**Figure 5.28**: Same as Fig. 5.20, but the errors now represent 1 and 90 percentile bounds of the distribution of samples present in the bin of |B|.

# Chapter 6

# 2.5 D self-consistent flux emergence

Numerous differences in the dynamics and thermodynamics, along with many similarities in the underlying processes in CHs and QS, are observed. To study these dynamics in detail, we perform 2.5 D numerical experiments of self-consistent flux emergence in CH and QS topologies by incorporating localized resistivity, thermal conduction, and optically thin radiation. This work is in preparation for submission.

Our observational findings presented in Chapters. 3, 4 & 5 demonstrate that the physical processes occurring in CHs and QS are intimately tied and present us with strong evidence of a strong correspondence between the coronal heating in QS and CHs and formation of solar wind. Our results further demonstrate that a difference in magnetic field topology results in this emergence of difference resulting from similar underlying processes. To investigate this difference further, however, we would need to study the response of the plasma to similar dynamical processes in various topologies using MHD simulations.

We have found that both the CHs and QS experience local deposition of energy, which may be utilized to drive the solar wind (CH) or locally heat the plasma (QS) (Upendran & Tripathi 2022, 2021b; Tripathi et al. 2021a). Parker (1988b), for example, envisions this energy deposition mechanism through magnetic reconnection driven by photospheric motions. Indeed, footpoint driving tangles the coronal magnetic field, leading to a buildup of free energy and subsequent dissipation. Another crucial mechanism of injecting free energy into the system is the emergence of bipolar magnetic flux from the convection zone into the atmosphere.

Magnetic flux emergence has been studied in great detail (see, for example Cheung & Isobe 2014, for a comprehensive review). Nozawa et al. (1992), for in-stance, perform a theoretical study of flux emergence by perturbing an isolated flux sheet in the convection zone. Several numerical studies have been performed to study the properties of the emerging flux sheets into the atmosphere(Shibata et al. 1989a,b; Isobe et al. 2007a, to name a few). Typically, the perturbations in a flux sheet cause an imbalance of pressure in the convectively unstable convection zone, causing the perturbed region of the flux sheet to rise. As it rises, it forms a loop with the plasma draining down along the loop towards the footpoints. Finally, when the loop stops rising upwards into the corona when the net pressure balance is reached. The acceleration and evolution of these emerging loops depend mainly on the pressure scale height, speed of sound, and the plasma beta in the flux sheet (see Nozawa et al. 1992, for details).

The emergence of a flux sheet represents a transient process bringing magnetic flux and mass from within the Sun into the atmosphere. The redistribution of the mass and energy upon magnetic reconnection depends on the topology of the background magnetic field. Yokoyama & Shibata (1996) perform a suite of flux emergence simulations with different coronal magnetic field topologies to simulate emergence in different regions of the Sun. In this suite of models, the emerging flux undergoes explosive, fast magnetic reconnection due to localized, anomalous resistivity. The reconnection results in the formation of large-scale plasma motion in jets, which show differences in structures and properties depending on the magnetic field topology. However, these models do not incorporate thermal conduction or radiative losses. Miyagoshi & Yokoyama (2004) incorporate thermal conduction into the horizontal background field setup of Yokoyama & Shibata (1996) and find the formation of cool jets due to magnetic reconnection.

Simulations of flows in CHs are numerous, ranging from 1 D (e.g., He et al. 2008) to 2 D (e.g., Aiouaz et al. 2005) and 3 D (e.g., Hansteen et al. 2010) setup. Aiouaz et al. (2005) construct a 2 D setup of funnel topology mimicking a CH and study the thermodynamic response of the plasma to an assumed heating function in the funnel. Ding et al. (2010) perform a 2.5 D flux emergence experiment in a uniformly vertical background field, where the emergence was parameterized by varying the flux at the bottom boundary. They find the resultant jet temperatures and velocities to depend on the strength of the background magnetic field. Ding et al. (2011) investigate the dependence of the properties of these jets on different background atmospheres. They demonstrate that interchange reconnection in the transition region is necessary to obtain substantial emission in the Fe IX spectral line. Yang et al. (2013), on the other hand, demonstrate mass deposition in the open flux system through flux cancellation by the horizontal motion of closed-loop systems through the open flux. Yang et al. (2018) consider a funnel-like background topology, with a predefined loop undergoing emergence and interchange reconnection. They demonstrate the formation of a hot and cool jet adjacent to each other accelerated by very different mechanisms. Note that in these simulations, Ding et al. (2010, 2011); Yang et al. (2013, 2018) consider many of the physics terms, including thermal conduction and a parameterized form of optically thin radiative losses.

Hansteen et al. (2010) perform a 3 D flux emergence experiment in a CH-like setup. They find numerous reconnection events, Ohmic dissipation, and wave processes in the atmosphere. Moreno-Insertis & Galsgaard (2013) perform a 3 D flux emergence experiment with a twisted flux rope in an oblique background field representing a CH topology. Note that the simulations by Moreno-Insertis & Galsgaard (2013) do not incorporate thermal conduction or radiative cooling effects. They report the formation of hot and cool jets and a high-density 'wall' around the emerged flux and jets. A similar jet formation was seen also in simulations by Nóbrega-Siverio et al. (2016); Nóbrega-Siverio & Moreno-Insertis (2022), who also included thermal conduction and radiation effects.

A self-consistent emergence of flux into the solar atmosphere and its interaction with the background magnetic field is important to obtain an accurate sense of the dynamics and thermodynamics of reconnection and jet formation processes. Furthermore, a comparative study of differences in the processes occurring in CHs and QS is important to understand if similar processes may drive different global outcomes due to a difference in magnetic field topology. In this work, we perform self-consistent flux emergence experiments in a horizontal coronal background field depicting QS and an oblique background field depicting a CH. The atmosphere and the properties of the flux sheet are kept similar in both experiments, and we study the influence of adding different dissipation and redistribution terms in these setups. In §. 6.1, we describe the simulation setup. Then we describe the results in QS in §. 6.2.1, while we describe the CH results in §. 6.2.2. We finally conclude with a general comparison of the observations and simulations in §. 6.3.

## 6.1 Simulation setup

Our aim is to consistently model the emergence of flux sheet from the convection zone into the atmosphere and its interaction with a background field throughout its emergence. For this, we consider (i) the flux sheet in the convection zone, (ii) the ambient magnetic field, and (iii) the initial atmosphere in this work. A perturbation in the flux sheet results in its evolution and subsequent interaction with the ambient magnetic field. We solve the MHD equations (Eq. 2.1) in a 2.5 D setup, considering all three components of the variables, with the variations only along the horizontal ($x$) and vertical ($z$) directions. The derivatives and dependence along the $y$ direction (perpendicular to the paper/screen) are ignored. Our simulation grid extends from $\sim 1.55$ Mm below the photosphere to $\sim 82.15$ Mm above, in the z-direction. The grid spacing is uniform till 7.75 Mm with 200 cells, while it increases in a stretched grid with 350 cells. We use a reflective bottom boundary, while the top boundary is open (similar to the upper and lower boundaries in Shibata 1983). Horizontally along the x-direction, our domain spans 121.21 Mm, with 520 cells between $\sim 40.3$ Mm (=130 in code units) and $\sim 80.6$ Mm (=260 in code units) while having a logarithmically increasing grid towards both boundaries. We have such a horizontal grid structure since the flux emergence, and the resultant dynamics of interest occur only between $\sim 40.3$ and $\sim 80.6$ Mm. The horizontal boundary is kept periodic. The gravitational acceleration is taken to be uniform and in the negative z direction, while the gas has a specific heat ratio of $\gamma = 5/3$.

The normalization of all the physical quantities is performed using the unit density ($\rho_0 = 1.7 \times 10^{-7}$ g cm$^{-3}$), length ($L_0 = 3.1 \times 10^7$ cm) and velocity ($v_0 = 1.2 \times 10^6$ cm s$^{-1}$) as formulated by PLUTO (the normalization is also shown in §. 2.3). The derived non-dimensionalizing timescale is $\approx 26$ s (i.e., 1 code time step is $\approx 26$ s). The pressure and magnetic field are normalized as $\rho_0 v_0^2$ and $\sqrt{4\pi \rho_0 v_0^2}$, respectively. The temperature in code units is obtained as $T_C = p_C / \rho_C$, and is transformed into units of Kelvin through $T = T_C \mu m_u v_0^2 / k_B$. Here, $\mu$ is the mean molecular weight, $m_u$ is the atomic mass unit, and $k_B$ is the Boltzmann constant. This setup is similar to the simulation setup of Yokoyama & Shibata (1996).

### 6.1.1 Initial condition and background field

The atmosphere is assumed to be initially in magnetohydrostatic equilibrium. To define the atmosphere, we first specify the temperature profile of the gas. The atmosphere consists of three parts, following Yokoyama & Shibata (1996): a convection zone, a photosphere, and a corona. For $z \geq 0$, the temperature is defined as:

$$T(z) = T_{phot} + \frac{(T_{cor} - T_{phot})}{2}\left(\tanh\left(\frac{z - z_{tr}}{w_{tr}}\right) + 1\right), \qquad (6.1)$$

while for $z < 0$, it is defined as:

$$T(z) = T_{phot} - \left( a \left| \frac{dT}{dz}_{ad} \right| \right) z. \tag{6.2}$$

Here, $T_{phot}$ is the photosphere/chromospheric temperature ($\approx 17000$ Kelvin), $T_{cor}$ is the coronal temperature ($\approx 2 \times 10^6$ Kelvin), $z_{tr}$ is the height (=2480 km), $w_{tr}$ is the width (=155 km) of the transition region. $|dT/dz| = 1 - 1/\gamma$ is the adiabatic temperature gradient. Under this setup, for $a > 1$, the layer becomes convectively unstable (Nozawa et al. 1992). Following Yokoyama & Shibata (1996), we have used $a = 2$ in our simulations.

As alluded to earlier, we have two models with different background fields, one with a horizontal background field (`model_QS`) and another with a slanted background field (`model_CH`) mimicking QS and CH topologies, respectively. In `model_QS`, we consider a formulation with the flux sheet in the convection zone and a horizontal field in the corona alone, following Miyagoshi & Yokoyama (2004). However, for `model_CH`, we consider the flux sheet in the convection zone while we impose a uniform slanted field across the whole box. We define the magnetic field as:

$$B(z) = \left( \frac{2p(z)}{\beta(z)} \right), \tag{6.3}$$

where the equation is different from Yokoyama & Shibata (1996) because of different normalization in PLUTO. $\beta(z)$ is the plasma beta and is defined as:

$$\frac{1}{\beta(z)} = \frac{1}{\beta_{fs}(z)} + \frac{1}{\beta_{bg}(z)}, \tag{6.4}$$

where $\beta_{fs}(z)$ specifies the flux sheet, and $\beta_{bg}(z)$ specifies the background in the `model_QS` case. The flux sheet is defined as:

$$\frac{1}{\beta_{fs}(z)} = \frac{1}{4\beta_{fs0}} \left( 1 + \tanh\left( \frac{z - z_{fsL}}{w_{fsL}} \right) \right) \left( 1 - \tanh\left( \frac{z - z_{fsU}}{w_{fsU}} \right) \right), \tag{6.5}$$

where $\beta_{fs0}$ is the plasma beta at the center of the flux sheet (=4.0), $z_{fsL}$ is the lower end (=-1240 km), and $z_{fsU}$ (=-620 km) is the upper end of the flux sheet, while $w_{fs}$ (=155 km) determines how fast the field rises at either end. The plasma beta at the center of the flux sheet is a crucial factor in determining the emergence time scale of the flux sheet.

In `model_QS`, we define the coronal magnetic field as:

$$\frac{1}{\beta_{bg}(z)} = \frac{1}{2\beta_{bg0}} \left( 1 + \tanh\left( \frac{z - z_{cor}}{w_{cor}} \right) \right), \tag{6.6}$$

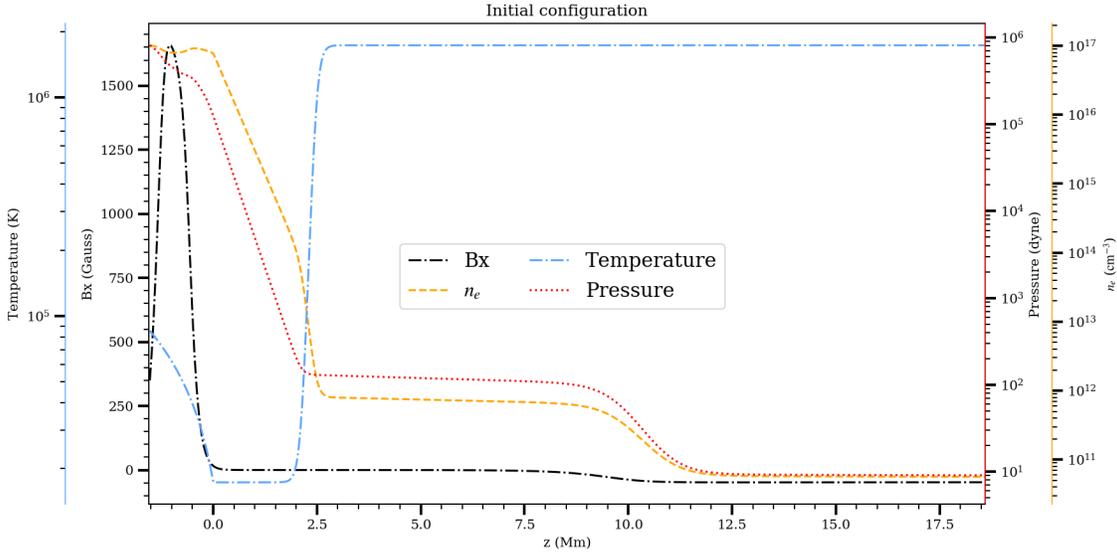

**Figure 6.1**: Initial configuration of the system.

where the coronal field starts from $z_{cor}$ (=10850 km), rises with width $w_{cor}$ (=852.5 km), and has a plasma beta of $\beta_{bg0} = 0.1$. However, note that to force reconnection between the emerging flux and the background field, the coronal field is oriented opposite to the field in the convection zone.

We then determine the initial variables by solving the magnetohydrostatic equilibrium equation:

$$\frac{d}{dz}\left(p(z) + \frac{B^2(z)}{2}\right) + \rho(z)g = 0. \tag{6.7}$$

In `model_CH`, we do not use the formulation of the coronal field from Eq. 6.6. We assume the initial atmosphere is in static equilibrium and includes only the flux sheet in the convection zone. Instead, we impose a slanted field of the form $\vec{B} = (B_{cor}\cos\theta_{cor}, 0, B_{cor}\sin\theta_{cor})$, where $B_{cor} \approx 0.0197$ and $\theta_{cor} = 135°$. Since this is a uniform field that is i) divergence-free, ii) curl-free, and iii) constant with time, we may split the magnetic field into a background field and a time-varying field (Powell 1994). This results in a modification of the MHD equations and the dependence of the energy on only the time-varying field.

The initial condition for different system parameters along a vertical column of the grid is depicted in Fig. 6.1.

## 6.1.2 Dissipation and redistribution terms

We have now described the initial conditions, the boundary conditions, and the setup of the two models *viz*. `model_CH` and `model_QS` at large. These two models also include non-ideal terms, i.e., localized resistivity, thermal conduction, and optically thin radiative loss.

We consider a localized, anomalous resistivity that depends on the drift velocity, following (Sato & Hayashi 1979; Ugai 1992; Yokoyama & Shibata 1994, 1996).The functional form is given by:

$$\eta := \left\{ \begin{array}{ll} 0 & \text{, if } v_d < v_c \\ \min\{1, \alpha(v_d/v_c - 1)^2\} & \text{, } v_d \geq v_c \end{array} \right. \tag{6.8}$$

The resistivity is defined in terms of the drift velocity as $v_d = J/\rho$, where $J$ is the current density. $v_c$ is a threshold above which the resistivity effects set in (=$10^3$), while $\alpha = 0.01$. Note that these values are not normalized by the velocity scale, and only the ratio $v_d/v_c$ has a physical meaning. The resistivity effects will set in only in regions with high current density (or alternatively low density), i.e., typically near current sheets, and will result in a fast, Petschek-like reconnection in the magnetic field setup.

We include anisotropic field-aligned thermal conduction, considering only the term along the field lines from Eq. 2.2. $\kappa_{\parallel}$ is taken to be Spitzer-type, with $\kappa_{\parallel} = \kappa_0 T^{5/2}$, and $\kappa_0 = 10^{-6}$ erg s$^{-1}$ cm$^{-1}$ K$^{-7/2}$. We ignore the conductivity across the field lines and do not impose any saturation flux in this setup (Miyagoshi & Yokoyama 2004).

We consider optically thin radiative losses in this work using the CHIANTI database (v10 Dere et al. 1997; Del Zanna et al. 2021). For computing the radiative losses, we need the characteristic density and a temperature grid. We compute the density at the base of the corona, which is $10^{11}$ cm$^{-3}$. We then compute the optically thin radiative loss function over a temperature grid of $\log T = 4$ to $\log T = 9$ over 300 points in log space, and use coronal abundances (Fludra & Schmelz 1999). In our simulation, we make the radiative loss as zero if the temperature falls below $\approx 85000$ Kelvin or if the number density grows more than $10^{13}$ cm$^{-3}$, which is typical of chromosphere or lower transition region. We perform this since the optically thin radiative cooling formalism fails in these conditions. Furthermore, we do not have any radiative loss in the convection zone (Takasao et al. 2013).

Note that in these simulations, we do not have a self-consistently generated heating in the corona like some of the 3-D models (see, for example Hansteen et al.

2010). Hence, we also add background heating terms to compensate for the radiative cooling. At t=0, we ensure there is no net dissipation or heating, and define the heating term at each grid point to be the same value as the radiative cooling at that point, following (Roussev et al. 2001).

The radiative loss function has a complicated dependence on the temperature, consisting of a general reduction and localized bumps. If due to a numerical error, the cooling term becomes slightly smaller than the heating term, the grid point will be at a higher temperature. However, at the next iteration, this results in yet lower radiative cooling, and more heating would ensue. On the other hand, a slightly higher cooling can result in a runaway cooling of the system. This instability depends on the behavior of the heating and cooling terms with temperature (see Parker 1953; Shimojo et al. 2001, for details). Thus, even after the inclusion of the heating term, numerical errors may build up over time, and result in runaway heating or cooling of the system. To mitigate this, we impose a numerical floor on the "net heating" term, i.e., if $|H + C| < 10^{-3}H$, there is no heating or cooling in that grid cell (here $H$ is the heating term and $C$ is the cooling term). This, to some extent, mitigates the effect of numerical instability in the radiation case.

With the system defined, we then perform a perturbation of the flux sheet in the vertical velocity as:

$$V_z = A \cos \left( 2\pi \frac{x - X_{max}/2}{\lambda_p} \right), \qquad (6.9)$$

where the perturbation is performed in the middle of the flux sheet, with an amplitude of $A = 0.6$ km s$^{-1}$, and a $\lambda_p = 6200$km. This wavelength is almost the most unstable wavelength for linear Parker instability. Note that the perturbation is performed only within $X_{max}/2 - \lambda_p/4 < x < X_{max}/2 + \lambda_p/4$ and $z_{fsL} < z < z_{fsU}$.

### 6.1.3 Plasmoid detection

Most of our simulations show plasmoid formation as the emerging flux interacts with the ambient magnetic field. To understand the difference in plasma temperature, density, and velocities resulting from the reconnection process, we need to understand the properties of plasmoids and jets. To this end, we develop a simple scheme to detect plasmoids from the simulations.

Visually, plasmoids show strong pressure signatures in models of both CH and QS. We consider the pressure around the reconnection region for performing the plasmoid detection. For the pressure 2-D array at any time step, we perform blob detection using the Difference of Gaussians (DoG) approach. In essence, this algorithm performs the detection of blobs based on differences between successively smoothed images (for more details, see Lowe 2004). We use a minimum sigma of 0.1, a maximum of 10, and a threshold level of $2 \times 10^{-3}$ to remove overlapping detection. A high threshold level means only very bright blobs will be detected. However, this results in numerous blobs being detected in the snapshots, not all of which may be plasmoids. Plasmoids typically have a strong maximum pressure at the center. Hence any blob with nearly uniform pressure distribution would not be a plasmoid. To quantify the presence of such outlier values of pressure, we compute the standard deviation of the pressure within the blobs and consider only those with $\sigma$ (pressure) more than $10^{-4}$ in code units. Such blobs with outlier pressure values qualify as plasmoids. An example detection using the algorithm on a snapshot of pressure from QS is shown in Fig. 6.2, where the plasmoid is shown with a blue-color dotted circle. Note that the plasmoids being depicted further on are a result of the application of this algorithm.

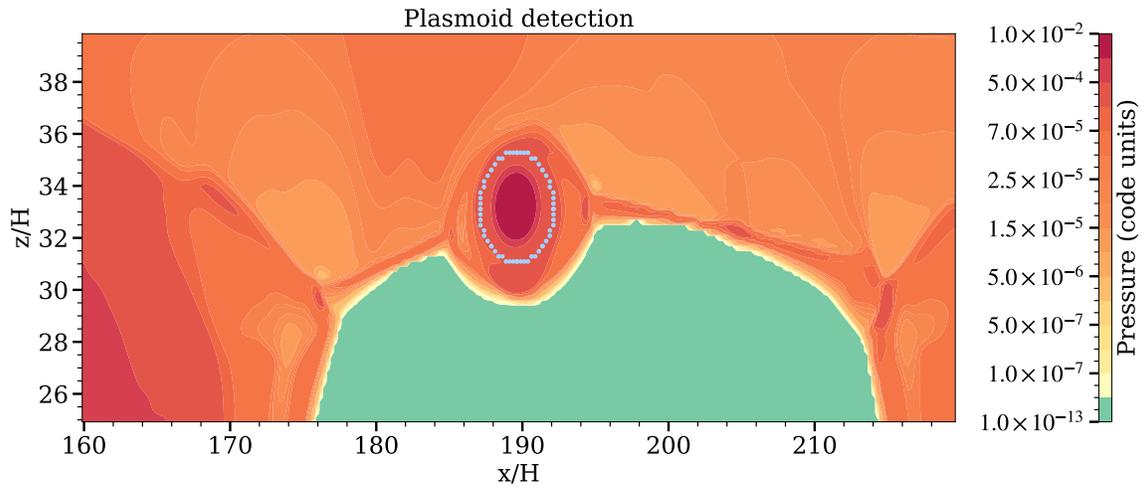

**Figure 6.2**: Graphic showcasing plasmoid detection in `model_QSR`. The color map shows pressure contours, while the blue dots encircle the detected plasmoid.

## 6.2  Results

We have performed the simulation of flux emergence for the two background magnetic topologies. We summarize the general dynamics and thermodynamics of the evolution of flux emergence in the two topologies for the case with resistivity, thermal conduction, and radiation. We emphasize that the analysis for cases with only some of the redistribution terms indicates no major difference in the dynamics while showing differences only in the thermodynamics.

## 6.2.1  QS model: Thermodynamics

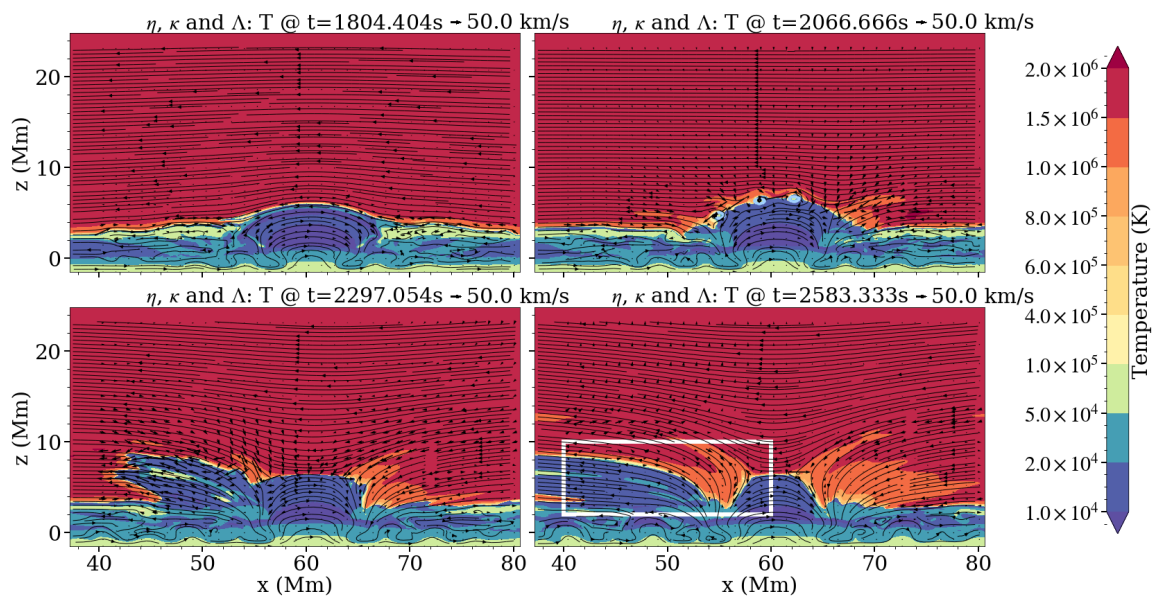

**Figure 6.3**: Temperature contours with magnetic field lines and velocity vectors for flux emergence in `model_QSRTR`. The white box in the final panel depicts the jet formed in the simulation. Note that this is not the full simulation box and only depicts the region where the dynamics of interest occur.

We consider the QS simulation with resistivity, thermal conduction, and radiation (henceforth `model_QSRTR`). We depict the evolution of the system at four time steps in Figs. 6.3 (temperature) and 6.4 (density). Over plotted are the magnetic field lines (solid black lines), and the velocity vectors at different grid points. Note that these plots only show a small part of the complete simulation box where the dynamics of interest occur.

The flux sheet, perturbed by the vertical velocity, rises up into the corona till about 7 Mm. At this height, the drift velocity $v_d$ is strong enough (due to both the increased current density and the reduced density) that localized resistivity is activated, causing magnetic reconnection. The reconnection causes plasmoids to form, which are then ejected outwards on either side. The plasmoids have a maxi-mum[1] temperature of $\approx 3.6 \times 10^5$ K, with a median[2] temperature of $\approx 3 \times 10^4$ K. Their maximum densities are typical $\approx 10^{14}$ cm$^{-3}$, while the median density within a plasmoid is $\approx 10^{13}$ cm$^{-3}$. We infer that the plasmoids are cold due to increased radiative loss from increased density. The plasmoid in `model_QSRTR` have sizes ranging from $\approx 360 - 630$ km. They travel outward with a typical maximum[3] speed of $\approx 75$ km/s, with a typical median speed of $\approx 50$ km/s. These plasmoids collide with the ambient atmosphere and result in the formation of jets. The radiative cooling results in dense, cool jets at temperatures of $\approx 2-5 \times 10^4$ K, traveling outward with a velocity of $\approx 50$ km/s. The forward edge of the jet corresponding to the region of interaction of plasmoids with the ambient atmosphere has high densities ($\approx 10^{10} - 10^{12}$ cm$^{-3}$), while the body of the jet has low densities ($\approx 10^{10}$ cm$^{-3}$), as seen in Fig. 6.5. The dense jets are reminiscent of the slingshot-like motion of plasma from Yokoyama & Shibata (1996). We also see a hot jet on top of the cool jet. This jet does not experience the same cooling as the cool jet since the density here is not enough to cool the region down. The hot jet has a characteristic density of $\approx 10^{10}$ cm$^{-3}$, and temperature of the order of $\approx 10^6$ K. Finally, the jets in `model_QSRTR` are very low-lying ($\approx 4 - 6$) Mm, and almost horizontal at late times.

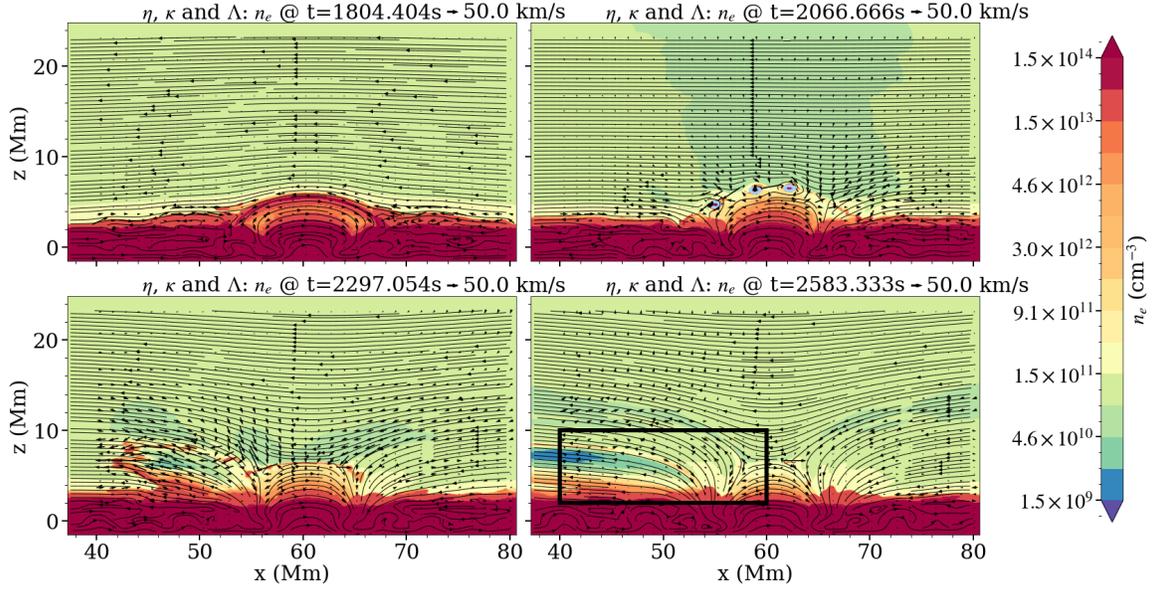

**Figure 6.4**: Similar to Fig. 6.3, but for density. Note that the jets are now enclosed by the black box.

---

[1]Calculated as mean of maximum temperature across all plasmoids

[2]Calculated as the mean of the median temperature across all plasmoids

[3]Calculated as mean of maximum velocity across all plasmoids

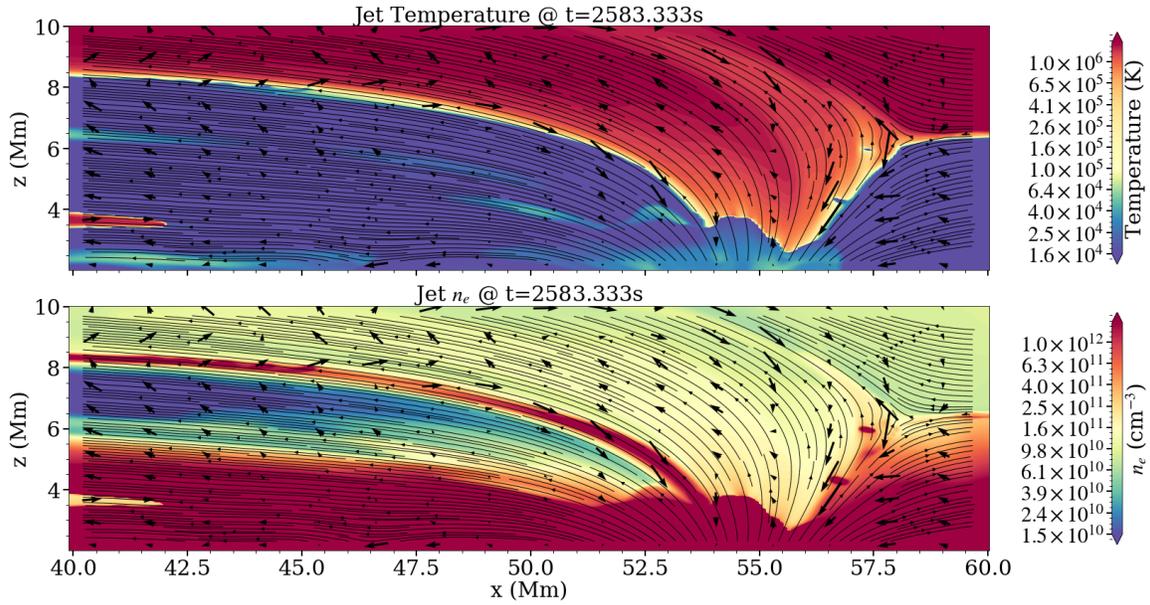

**Figure 6.5**: Temperature (top) and density (bottom) contours with magnetic field lines and velocity vectors for the cool and hot jets identified in Fig. 6.3. We have modified the colour map range to ensure the thermal and density structures in the jet are brought out well.

## 6.2.2 CH model: Thermodynamics

We now consider the CH simulation with resistivity, thermal conduction, and radiative losses (henceforth `model_CHRTR`). We depict the evolution of the system at four time steps in Fig. 6.6 and 6.7. The figures depict the evolution of temperature and density, the magnetic field lines (solid black lines), and the velocity vectors at different grid points. The blue circles contain the plasmoids in the snapshots. Note again that these plots only show a small part of the complete simulation box where the dynamics of interest occur.

The flux sheet, perturbed by the vertical velocity, rises up into the corona till about 7 Mm. We find in these simulations that the height of flux emergence is not dependent on the background topology. Similar to the QS scenario, here again, magnetic reconnection occurs and results in the formation of plasmoids. These have similar sizes as those in the QS scenario, i.e., $360 - 630$ km in radius, and have a maximum temperature of $\approx 5 \times 10^5$ K, while they have a median temperature of

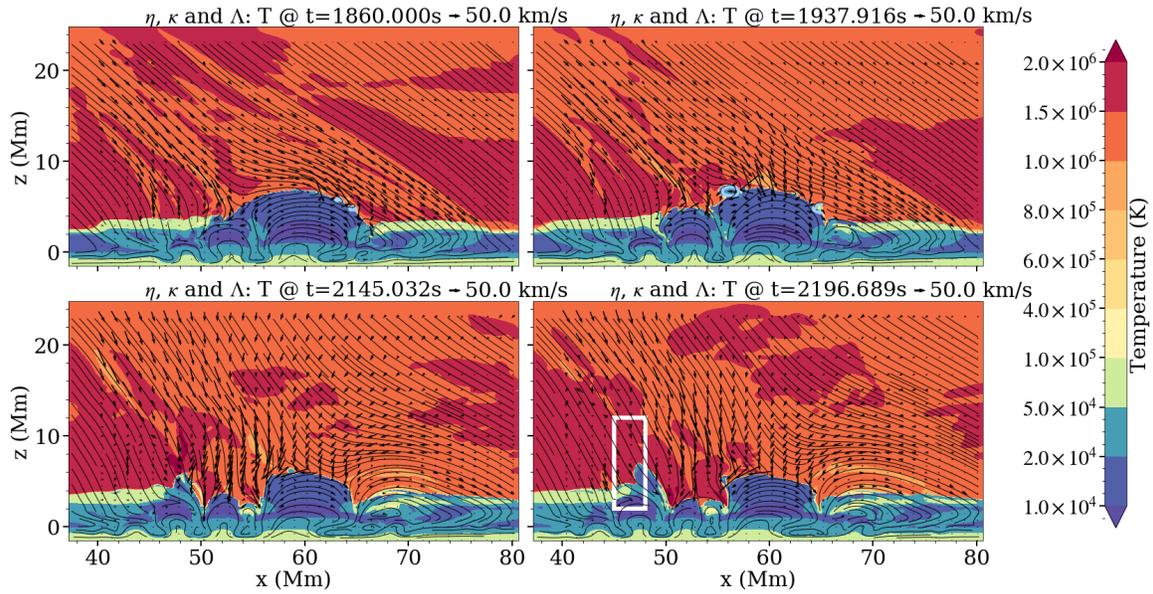

**Figure 6.6**: Similar to Fig. 6.3, but for the CH scenario.

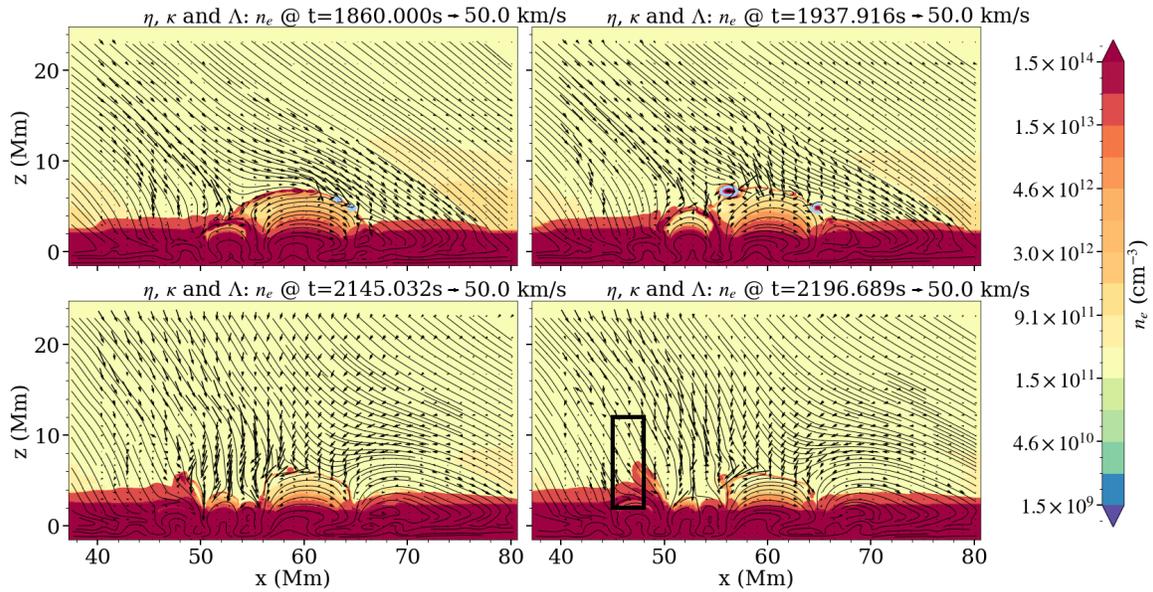

**Figure 6.7**: Similar to Fig. 6.6, but for density.

$\approx 3 \times 10^4$ K. These plasmoids have a typical maximum density of $\approx 10^{14}$ cm$^{-3}$, with a median density of $10^{13}$ cm$^{-3}$. They travel outwards at a maximum speed of $\approx 100$ km/s, while they have an average speed of $\approx 60$ km/s. The properties of plasmoids in `model_CHRTR` are thus similar to the ones we find in `model_QSRTR`. Note while the location of perturbation causes one loop to rise, the higher modes are also activated, seen here as secondary loops (near x=47 and 52 Mm). The expelled, cool plasmoids strike these higher-order modes. These are liberated along the outflow to the left of x=50 Mm. However, we do not see the formation of the cool jet in this case. Rather, the jet itself in the final snapshot of Fig. 6.6 and 6.7 suggests the formation of only the hot jet. This hot jet has characteristic temperatures of $\approx 10^6$ K and densities of $\approx 10^{11}$ cm$^{-3}$, as seen in Fig. 6.8. The jet has velocities of $\approx 50$ km/s, with minimal density stratification. We also find downflows occurring outside of the loop-background interaction region. Along with the hot jet, we also find that a hot loop forms near the cool loop due to the interchange reconnection. This loop is $\approx 7$ Mm high, at a temperature of $\approx 10^5 - 10^6$ K, and at densities of $\approx 10^{11}$ cm$^{-3}$.

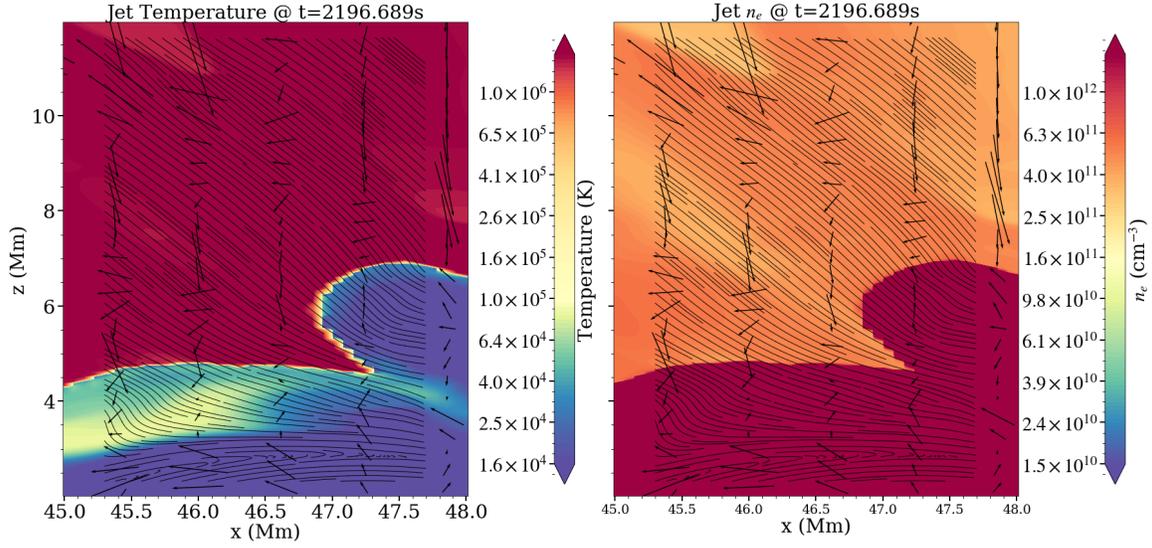

**Figure 6.8**: Temperature (left) and density (right) contours with magnetic field lines and velocity vectors for the cool and hot jets identified in Fig. 6.6. We have modified the color map range to ensure the thermal and density structures in the jet are brought out well.

## 6.3   Discussion and comparison

In this work, we developed 2.5D MHD simulations of flux emergence in the solar atmosphere with background magnetic field topology mimicking CH and QS regions. The QS box is endowed with a horizontal ambient field, while the CH box is provided with an oblique field. Thermal conduction and optically thin radiative losses are included, while localized anomalous resistivity is incorporated to cause explosive reconnection.

The rising loop undergoes reconnection with the ambient magnetic field in both setups. This gives rise to the formation of plasmoids. In both `model_CHRTR` and `model_QSRTR`, we find the plasmoids have characteristic sizes from 360-630 km. In `model_QSRTR`, we find the plasmoids move at max. speed of $\approx 70$ km/s, while in `model_CHRTR` they move with a max. speed of $\approx 100$ km/s. The plasmoids have densities of $\approx 10^{13}$ cm$^{-3}$ in both the cases. They have median temperatures of $\approx 2 - 6 \times 10^4$ K while having a maximum temperature of $\approx 10^5$ K.

As a result of this reconnection process, jets are formed in both simulations. In `model_QSRTR`, we find the existence of both a hot ($\approx 10^6$ K) and cool ($\approx 10^4$ K) jet, while `model_CHRTR` shows the formation of only the hot jets at $\approx 10^6$ K. In `model_CHRTR`, a hot loop is formed near the cool loop due to interchange reconnection. This hot loop ($\approx 10^6$ K) has low densities ($\approx 10^{11}$ cm$^{-3}$) at marginally lower temperatures than the hot jet. Such a hot loop is however not seen in `model_QSRTR`.

Plasmoids have been observed in the solar atmosphere. Patel et al. (2020) observe and track plasmoids in an off-limb flare. They find these plasmoids in the current sheet to have sizes of $\approx 5000$ km, with velocities of $\geq 190$ km/s. Mulay et al. (2023) observe and compute properties of plasmoids in AR jets. They observe plasmoids with typical size of $\approx 3000$ km (width), characteristic temperature of $\approx 10^6$ K, and densities of $\approx 10^8$ cm$^{-3}$. Chen et al. (2022), similarly find plasmoids with temperatures of $\approx 10^6$ K, but find the sizes to range from $\approx 800 - 2300$ km, with velocities from $60 - 185$ km/s. We note that these plasmoid observations are not consistent with those seen in any of the simulations. These plasmoids are formed due to reconnection either much higher in the atmosphere or due to different conditions than those in our simulations.

Singh et al. (2012), however, observe plasma ejecta in jets in Ca II H filtergrams, and demonstrate them to be $\approx 300 - 1500$ km in size while moving at a velocity of $\approx 35$ km/s. Since these plasmoids are seen in Ca II, we may assume they may be at chromospheric temperatures. Such plasmoids are similar to those observed in `model_CHRTR`, while the final jet-loop structure is also similar to that observed by Singh et al. (2012). van der Voort et al. (2017) observe signatures of plasmoids in Ca II K wing at velocities $\approx 40$ km/s. Tiwari et al. (2022) find "bright dots" of the size $\approx 675 \pm 300$ km, which are similar in size to the largest of plasmoids forming in our simulations. These dots have low velocities ($\leq 30$ km/s), with significant emissivity in the 174 Å passband, corresponding to $\approx 1 - 2 \times 10^6$ K. Tiwari et al. (2022) compare their observations with Bifrost simulations, and find that these dots form due to magnetic reconnection between emerging flux with pre-existing flux. The simulations indicate reconnection very low in the atmosphere, with a significant contribution of plasma in the transition region. Peter et al. (2019) show the genera-tion of plasmoids $\approx 300 - 600$ km in size, with maximum temperatures of $\approx 6 \times 10^4$ K and density of $\approx 10^{11}$ cm$^{-3}$. Thus, these bright dots and plasmoids appear to satisfy some aspects of the plasmoids seen in our simulations. However, the general plasmoids seen in these simulations may not be observed as individual events with current instrumentation in the first place.

The plasmoid formation gives rise to and feeds jets, which have typical velocities of $\approx 50$ km/s, densities of $\approx 10^{11}$ cm$^{-3}$, and temperature of $10^6$ K. Mulay et al. (2023) find jets with velocities of $\approx 225 - 275$ km/s at $10^6$ K, and density of $\approx 10^8$ cm$^{-3}$. Clearly, once again the jets we see are not consistent with AR jet structures. However, Singh et al. (2012) observe jets in the chromosphere, and find velocities ranging from $\approx 10 - 70$ km/s. These are similar to the jet velocities we see in our simulations. However, the observations of Singh et al. (2012) are in Ca II H, which does not form at $10^6$ K corresponding to the jet temperatures in our simulations. Whether the large-scale jets we find in our simulations provide observable signatures in Ca II H line will need to be probed with the correct treatment of radiation interaction with plasma.

We have seen in Ch. 5 how the specific intensity and velocity signatures are shown as discriminators of CHs and QS in the lower atmosphere. However, the non-thermal widths and the spectral profile shapes indicate an inherent similarity in the processes giving rise to the emission themselves. From the simulations presented in this chapter, we find similarities in the properties of the plasmoids which form in both the setups and differences in the properties of the resultant jets. The next step in this work is to compute the emission in spectral lines, which would enable a more stringent comparison in terms of the observables. We will be performing spectral response computation for lines forming in the upper transition region and the corona. Once we understand the thermodynamics of these simulations, we are next looking forward to performing flux emergence experiments in a funnel-field topology, which would be a much better representative of a CH setup. Such a comparison will enable us to understand, in some measure, the physical processes that give rise to QS heating and the emergence of solar wind.

# Chapter 7

# Parting thoughts and paths for the future

Udvayaṃ tamasaspari
jyotiṣpaśyanta uttaram
devaṃ devatrā sūryam
aganma jyotir uttamam

Beholding the up-springing light
above the darkness,
We approach the divine Sun
among the gods, the excellent light.

-Rigveda 1.50.10

---

In this thesis, we have asked four major science questions. We first ask what the specific source regions of the solar wind are and if it is possible to forecast the solar wind given changing solar coronal conditions. To address this question, we have developed a DL solar wind speed forecasting model named *WindNet* to perform the forecasts given a sequence of full-disc EUV images in the 193 Å or 211 Å passband. Following that, we ask if impulsive heating is viable to maintain the solar corona at a million degrees Kelvin. To estimate the statistical properties of these events, we employ the statistical impulsive heating model *PSM* in conjunction with a DL inversion scheme *iPSM* on QS light curves in both EUV and X-rays. We then ask how different these locally heated QS regions are with respect to CHs, and study them in the chromosphere and transition region. To discern the differences in the dynamics of these two regions, we use the Mg II h & k, C II 1334 Å and Si IV 1393 Å spectral lines, and study them as a function of the underlying magnetic flux density. Finally, we ask what sort of physical picture may be prevalent in CHs and QS, giving rise to the differences higher up in the corona. To explore this, we probe the dynamics of solar wind emergence and coronal heating in a CH and QS setup through flux emergence from the convection zone. The main results from

this thesis are:

1. Full disc EUV intensity image time series possesses information regarding the solar wind source regions and can be used to perform forecasts of solar wind speed. A non-linear DL-based model can learn associations between source regions and the corresponding wind speed to generate forecasts while outperforming baseline models, especially with large time horizons.

2. The DL model *WindNet* identifies CHs as sources of the fast solar wind in the 193 Å data, while the slow wind is primarily associated with ARs, with the timing of association in the slow wind being missed. While in a nascent stage, interpretable AI is a strong candidate for understanding physical associations between solar wind modalities and sources.

3. Impulsive events are a viable source of heating the solar corona, as seen in EUV. These EUV-impulsive events occur with a typical frequency of 2.5 events min$^{-1}$, have a slope of distribution of events $\geq 2$, and last for $\approx 15$ minutes. The events show a predominance of conduction-dominated cooling and point to an energy reservoir that may be depleted by large events occurring intermittently or small events occurring frequently.

4. In X-rays, these impulsive events are very small ($\leq 10^{23}$ ergs) and have much flatter event distribution. They have very high frequency ($\approx 25$ events min$^{-1}$), and have time scales of $\leq 11$ minutes. The high frequency of events follows from the very low amplitude of these events, again strongly pointing to an energy reservoir in the quiet corona.

5. QS and CHs, which are apparently undifferentiated in the chromosphere, show clear differences if the underlying photospheric magnetic flux density (i.e. |B|) is taken into account. The chromospheric Mg II h & k features and C II 1334 Å lines show excess intensity, reduced redshifts, and blueshifts in QS with respect to CHs, while the transition region Si IV 1393 Å line shows excess intensity, increased redshift and reduced blueshift in QS. The C II line shows excess width in CHs, while both the regions show significant skewness and flatness departures from a Gaussian. However, the spectral shapes are significantly similar in both CHs and QS. Furthermore, the chromospheric blueshifts (redshifts) are well correlated with the transition region blueshifts (redshifts), while the chromospheric redshifts and transition region blueshifts are also well correlated. Finally, the downflows were preferentially decelerated in QS, while the upflows were found to be accelerated in CHs.

6. These observations, taken together, represent a dichotomy between the chromosphere and transition region. The observations may be elegantly explained due to the prevalence of impulsive heating in different magnetic field topologies. Interchange reconnection in CHs between the majorly open field lines and closed field lines may result in bidirectional flows, which, depending on the temperature and height, would be seen as correlated upflows, downflows, or bidirectional flows as seen in this work. The closed loop reconnection in QS does not allow plasma to escape, resulting in increased deceleration of downflows due to trapped plasma within the loops. Contrarily, the outflowing plasma is accelerated in the CHs due to the open flux system. Finally, if the interchange reconnection occurring low in the atmosphere results in solar wind switchbacks, our measurements provide constraints for such models.

The numerous results that have come out of this thesis have raised many more questions regarding the origin of solar wind and the heating of the solar corona. Below, we outline a number of science questions and analyses that we would like to perform as through the understanding generated from this thesis:

1. **Localizing solar wind source regions using deep learning**: We have seen how EUV intensity data already contains information pertaining to the source regions of the solar wind and that this information may be picked up by DL methods. We now pose the following problem: given multi-wavelength images(and thus, temperatures covering from the photosphere to the corona), can a DL model classify the images as giving rise to a fast and slow wind? And can such a model be queried to localize the solar wind sources? This work is underway.

2. **Evolution of solar wind**: Major solar wind forecasting codes hit a roadblock due to the evolution of structures within the solar wind. A measure of information loss in the injected coronal structures due to the evolution towards the Earth is needed. We shall next study specific source regions on the Sun to understand their relation to solar wind measurements near Earth in terms of loss of information. Some preliminary study has been performed through the work of a master's student of Prof. Durgesh Tripathi.

3. **Updates to *iPSM***: The *PSM* model has numerous assumptions, the most critical of which are: (i). Relation between rise and fall time, (ii). Lack of constraint between event amplitude and time-to-next-event, (iii). Lack of filling factor. Similarly, the *iPSM* inversion also has some constraints: (i). Inability to perform inversions for all 5 parameters in the same pass, and (ii). Needs to be retrained for different instrument measurements to incorporate noise. These are some upgrades that need to be performed to make a truly general inversion scheme for impulsive heating models of the solar corona.  We intend to take this upgrade up in the near future.

4. ***iPSM* application to stellar light curves**: How do the typical flaring frequency, time scale, distribution of events, and amplitudes correspond for stars of different types during quiet times?  Are the coronae of these stars also impulsively heated?  How do these properties depend on the stellar type, metallicity, and surface gravity?  In essence, we seek to connect the properties of small events in the corona to the dynamics within the star and check the mechanisms that give rise to footpoint shuffling in producing these impulsive events.

5. **Mapping flows across the solar atmosphere**: We have seen the excess up- and down-flows in the chromosphere as a function of the underlying photospheric magnetic flux density.  However, the question remains as to how these flows are propagated across individual field lines across different heights. To this end, we intend to study CHs jointly using SST, IRIS, and EIS spectra using a Potential field extrapolation. Considering the flows through different open field lines would then tell us how the flows – starting from the lower chromosphere to the upper transition region – end up as solar wind flows in the corona.  Furthermore, we also intend to study the momentum transport across these temperatures and along these field lines, which would provide stronger constraints than correlations regarding mass flows in the solar atmosphere.

6. **Inference using data from Aditya-L1**: **Aditya-L1** is India's first solar mission to be put at the L1 point in space.  Many of the studies performed in this thesis tie in well with the mission objectives of Aditya-L1, especially instruments like the Solar Ultraviolet Imaging Telescope (SUIT) and Visible Emission Line Coronagraph (VELC). Typically, we intend to leverage SUIT and VELC to study possible plasma propagation in CHs from the chromosphere to the corona to estimate the solar wind mass loss and transfer.  We shall also be using SUIT to study coupling with other layers of the solar atmosphere – especially in the more quiescent regions.  Studying these different regions of the solar atmosphere as a function of the underlying magnetic field in the impulsive heating paradigm then provides us with insights into the energy budget and losses, which is critical to study the heating of the solar corona.